\documentclass[12pt,a4paper,twoside]{book}

\usepackage{amsmath}
\usepackage{epsfig}
\usepackage{graphicx}
\usepackage{color}
\usepackage{fancyheadings}
\usepackage{axodraw}

\definecolor{gray}{rgb}{0.5,0.5,0.5}
\def\newabs{\vskip2.5mm}
\def\no{\nonumber}

\def\TS{\textstyle}
\def\calO{\mathcal{O}}

\def\calP{\mathcal{P}}
\def\calS{\mathcal{S}}
\def\calD{\mathcal{D}}
\def\calC{\mathcal{C}}
\def\MSbar{\ensuremath{\overline{\text{MS}}}}
\def\MI{\text{MI}}
\def\Tr{\text{Tr}}
\def\IR{\text{\tiny IR}}
\def\UV{\text{\tiny UV}}
\def\Im{\text{Im}}
\def\Re{\text{Re}}
\def\eps{\varepsilon}
\def\om{\omega}
\def\slasha#1{#1 \hskip-0.45em /}
\def\slashb#1{#1 \hskip-0.55em /}
\def\slashc#1{#1 \hskip-0.65em /}
\def\Heff{\mathcal{H}_\text{eff}}
\def\as{\alpha_s}
\def\LQCD{\ensuremath{\Lambda_\text{QCD}}}
\def\ubar{\bar{u}}
\def\Lib{\text{Li}_2}
\def\Lic{\text{Li}_3}
\def\Lid{\text{Li}_4}
\def\Sab{\text{S}_{1,2}}
\def\Sac{\text{S}_{1,3}}
\def\Sbb{\text{S}_{2,2}}
\def\SCETI{\ensuremath{\text{SCET}_\text{I}}}
\def\SCETII{\ensuremath{\text{SCET}_\text{II}}}
\def\tot{\leftrightarrow}
\def\gev{\text{GeV}}
\def\mybinom#1#2{{\tiny\left(\!\! \begin{array}{c} #1 \\ #2
\end{array}\!\!\right)}}
\def\mybinomL#1#2{\left(\!\! \begin{array}{c} #1 \\ #2
\end{array}\!\!\right)}

\catcode`@=11

\def\@citex[#1]#2{\if@filesw\immediate\write\@auxout{\string\citation{#2}}\fi
  \def\@citea{}\@cite{\@for\@citeb:=#2\do
    {\@citea\def\@citea{,\penalty\@m}\@ifundefined
      {b@\@citeb}{{\bf ?}\@warning
       {Citation `\@citeb' on page \thepage \space undefined}}%
\hbox{\csname b@\@citeb\endcsname}}}{#1}}

\def\citer{\@ifnextchar [{\@tempswatrue\@citexr}{\@tempswafalse\@citexr[]}}

\def\@citexr[#1]#2{\if@filesw\immediate\write\@auxout{\string\citation{#2}}\fi
  \def\@citea{}\@cite{\@for\@citeb:=#2\do
    {\@citea\def\@citea{--\penalty\@m}\@ifundefined
       {b@\@citeb}{{\bf ?}\@warning
       {Citation `\@citeb' on page \thepage \space undefined}}%
\hbox{\csname b@\@citeb\endcsname}}}{#1}}

\begin{document}

\pagestyle{fancyplain} \pagestyle{empty}
\thispagestyle{empty} \color{gray} \setlength{\fboxrule}{0.4mm}

\newfont{\titlefont}{cmssq8 at 19pt}
\newfont{\namefont}{cmssq8 at 13pt}

\vspace*{-3cm}\hspace*{-1.7cm}
\fbox{\parbox{17.8cm}{\hspace{10cm}\vspace{26cm}}}

\vspace*{-25.7cm}
\begin{center}

\color{black} \sf

Fakult\"at f\"ur Physik

der

Ludwig-Maximilians-Universit\"at M\"unchen

\vspace{3mm}
\includegraphics[width=2.5cm]{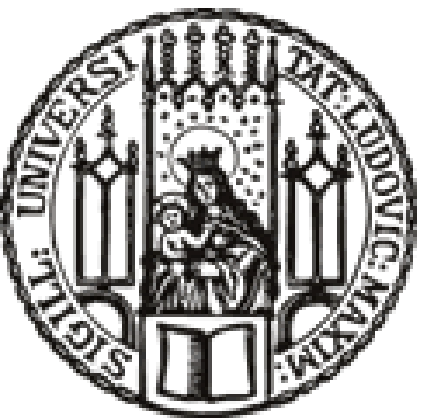}

\vspace{5.7cm}

\color{black}
{\titlefont HIGHER ORDER QCD CORRECTIONS}

\vspace{4mm} {\titlefont IN EXCLUSIVE CHARMLESS {\LARGE$B$} DECAYS}

\vspace{4cm} {\namefont GUIDO BELL}

\color{black} \vspace{8cm} {Oktober 2006}

\end{center}

\color{black}


\newpage

\vspace*{18.5cm}

\begin{center}

{\small Dissertation der Fakult\"at f\"ur Physik

der Ludwig-Maximilians-Universit\"at M\"unchen

\vspace{2mm} vorgelegt von Dipl.-Phys.~Guido Bell

aus Neuwied am Rhein

\vspace{2mm} 1.~Gutachter: Univ.-Prof.~Dr.~Gerhard Buchalla

2.~Gutachter: Priv.-Doz.~Dr.~Stefan Dittmaier

\vspace{2mm} Tag der m\"undlichen Pr\"ufung: 13.~Dezember~2006}

\end{center}

\frontmatter
\vspace*{2.2cm}
 
{\Huge\bf Abstract}

\vspace{1.4cm}

We discuss exclusive charmless $B$ decays within the Standard Model
of particle physics. These decays play a central role in the
on-going process to constrain the parameters of the CKM matrix and
to clarify the nature of CP violation. In order to exploit the rich
source of data that is currently being collected at the experiments,
a systematic theoretical treatment of the complicated hadronic
dynamics is strongly desired. QCD Factorization represents a
model-independent framework to compute hadronic matrix elements from
first principles. It is based on a power expansion in $\LQCD/m_b$
and allows for the systematic implementation of perturbative
corrections.

\newabs
In particular, we consider hadronic two-body decays as $B\to\pi\pi$
and perform a conceptual analysis of heavy-to-light form factors
which encode the strong interaction effects in semi-leptonic decays
as $B\to\pi\ell\nu$.

Concerning the hadronic decays we compute NNLO QCD corrections which
are particularly important with respect to strong interaction phases
and hence direct CP asymmetries. On the technical level, we perform
a 2-loop calculation which is based on an automatized reduction
algorithm and apply sophisticated techniques for the calculation of
loop-integrals. We indeed find that the considered quantities are
well-defined as predicted by QCD Factorization, which is the result
of a highly complicated subtraction procedure. We present results
for the imaginary part of the topological tree amplitudes and
observe that the considered corrections are substantial. The
calculation of the real part of the amplitudes is far more
complicated and we present a preliminary result which is based on
certain simplifications. Our calculation is one part of the full
NNLO analysis of nonleptonic $B$ decays within QCD Factorization
which is currently pursued by various groups.

In our conceptual analysis of the QCD dynamics in heavy-to-light
transitions we consider form factors between non-relativistic bound
states which can be addressed in perturbation theory. We perform a
NLO analysis of these form factors and discuss some open questions
of the general factorization formula which is obtained from the
heavy-quark expansion in QCD. These include the origin and
resummation of large logarithms and the non-factorization of soft
and collinear effects in the so-called soft-overlap contribution. We
show that the latter can be calculated in our set-up and address the
issue of endpoint singularities. As a byproduct of our analysis, we
calculate leading-twist light-cone distribution amplitudes for
non-relativistic bound states which can be applied for the
description of $B_c$ and $\eta_c$ mesons.

\renewcommand{\baselinestretch}{1.5}
\tableofcontents
\renewcommand{\baselinestretch}{1.05}

\mainmatter
\part{Introduction}

\pagestyle{fancyplain}
\rhead[\uppercase{Introduction}]{\fancyplain{} \thepage}
\lhead[\thepage]{}

\chapter*{Introduction}

The analyses presented in this thesis rely on the \emph{Standard
Model} of particle physics \\ \citer{Glashow:1961tr,Fritzsch:1973pi}
which reflects our current knowledge of three of the four known
fundamental forces in nature. Electromagnetic, weak and strong
interactions are described therein by a relativistic and
renormalizable quantum field theory which is based on a gauge
principle. The Standard Model represents an impressive theoretical
achievement which successfully explains phenomena from everyday
electricity to high-energetic quantum processes that are
investigated at dedicated particle accelerator facilities. It is one
of the best-tested theories of contemporary physics. This may be
illustrated by comparing the experimentally measured value of the
anomalous magnetic moment of the muon\footnote{The anomalous
magnetic moment of the electron can be measured even more precisely.
As it is less likely to be affected by physics beyond the Standard
Model, its measurement is used for the determination of one of the
Standard Model parameters, namely the fine structure
constant~\cite{Gabrielse:2006gg}.}
\begin{align}
a_\mu^\text{exp} = 11~659~208.0~(5.4)(3.3)\times 10^{-10}
\end{align}
with its theoretical prediction calculated within the Standard
Model~\cite{PDG}
\begin{align}
a_\mu^\text{SM} = 11~659~185.8~(7.2)(3.5)(0.3)\times 10^{-10}.
\end{align}
Despite its tremendous success the Standard Model has its
insufficiencies and is commonly believed to be incomplete. Severe
constraints from electroweak precision data may be interpreted as a
hint that ''something unknown'' happens at the TeV-scale, an energy
scale that has been out of the scope of todays collider experiments.
The Large Hadron Collider (LHC), which is currently being built at
CERN and is scheduled to start operation in mid-2007, has
particularly been designed to explore the physics at the TeV-scale.
Thousands of particle physicists from all around the world are
looking forward to the first data taking of the LHC, in the hope
that it will help us to reveal the limitations of the Standard Model
and give a first clue about the theory that lies beyond it.

\newabs
Apart from these direct searches, the physical effects that are
supposed to lie beyond the Standard Model (often referred to as
\emph{New Physics}) can be investigated indirectly in high-precision
measurements of low-energy observables as the anomalous magnetic
moment of the muon mentioned above. Every physical observable is in
principle sensitive to arbitrarily high energies and thus to New
Physics due to quantum effects. The quantitative investigation of
these tiny effects represents a highly challenging task, both for
experimental measurements and for theoretical calculations which
have to match the experimental accuracy.

\newabs
Whereas the gauge sector of the Standard Model has been tested to
remarkable precision in the era of the Large Electron Positron
Collider (LEP), the flavour sector is experimentally less
constrained. The flavour sector contains a large number of
parameters as the quark and the lepton masses or the four parameters
related to the Cabibbo-Kobayashi-Maskawa (CKM)
matrix~\cite{CKM:Cab,CKM:KobMas} which describes the mixing of the
quark mass eigenstates in weak interactions. The numerical values of
these parameters are not predicted by the Standard Model but rather
have to be extracted from experimental measurements before making
any theoretical prediction.

\begin{figure}[b!]
\centerline{\parbox{13cm}{
\centerline{\includegraphics[height=9.7cm]{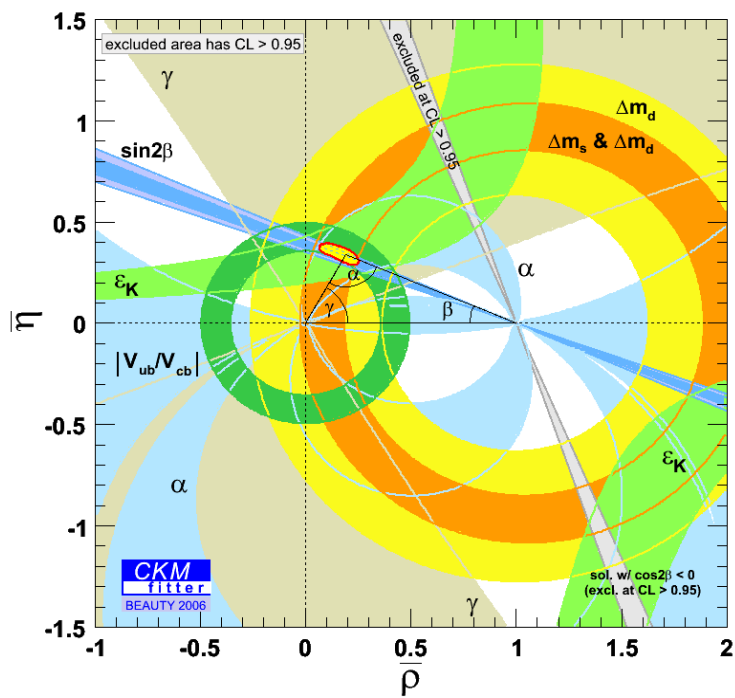}}
\vspace{-4mm} \caption{\label{fig:CKM} \small \textit{Global fit of
the unitarity triangle (status September~2006)~\cite{CKMfitter}.}}}}
\end{figure}

\newabs
The ultimate goal of $B$ physics is to precisely determine some of
these parameters and to test the CKM sector of the Standard Model.
The phenomenon of CP violation is of particular interest as it is
related to the observed matter-antimatter asymmetry in our universe.
Two CKM parameters are known to date at the percent
level~\cite{CKMfitter}
\begin{align}
\lambda =0.22717^{+0.00100}_{-0.00101}, \qquad \qquad A
=0.806^{+0.014}_{-0.014} \label{eq:CKMpara_1}
\end{align}
and the remaining two parameters $\bar \rho$ and $\bar \eta$ are
conveniently discussed in the context of a \emph{unitarity triangle}
which reflects the unitarity of the CKM matrix. The current status
of the unitarity triangle is shown in Figure~\ref{fig:CKM}. The
values of the parameters $\bar \rho$ and $\bar \eta$ correspond to
the upper tip of the triangle which is given by~\cite{CKMfitter}
\begin{align}
\bar \rho =0.195^{+0.022}_{-0.055}, \qquad \qquad \bar \eta
=0.326^{+0.027}_{-0.015}.
\end{align}
Notice that these determinations are far less accurate than the ones
of $\lambda$ and $A$ and that various independent measurements of
$\bar \rho$ and $\bar \eta$, which are indicated by the coloured
bands in Figure~\ref{fig:CKM}, are all consistent with each other.
The very fact that the area of the unitarity triangle is non-zero is
a manifestation of CP violation.

\newabs
Two dedicated $B$ factories, the BaBar experiment at SLAC and the
Belle experiment at KeK, have contributed substantially to our
current understanding of $B$ physics and CP violation. One of the
most important milestones in their physics programme was the
observation of (indirect) CP violation in the neutral $B$ system in
2001~\cite{CPV:BaBar,CPV:Belle} more than 35 years after the first
observation of CP violation by James Cronin, Val Fitch and
collaborators in the Kaon system~\cite{CPV:Kaon}.

Since the start of BaBar and Belle in 1999, the $B$ factories have
produced $\sim10^9$ pairs of $B$ mesons which correspond to
$\sim1~\text{ab}^{-1}$ of integrated luminosity. This impressive
wealth of data will continuously increase until the end of their
physics programme in 2010. It will then be complemented by the LHC-b
experiment and future plans concerning an upgrade of the $B$
factories are already envisaged. With this ongoing experimental
effort it will be possible to nicely overconstrain the unitarity
triangle and to reduce the uncertainty of the parameters $\bar \rho$
and $\bar \eta$ to a few percent.

\newabs
The reader may wonder why some observables as the anomalous magnetic
moment of the muon can be determined with an uncertainty of better
than 1 part in $10^5$ and why it is apparently so difficult to
reduce the uncertainty of the CKM parameters at the percent level.
One of the reasons is that the CKM parameters are related to quarks
which are always affected by strong interactions\footnote{Even
though the anomalous magnetic moment of the muon is a leptonic
quantity, the main limitation in its determination also stems from
small strong interaction effects.}.

\begin{figure}[t!]
\centerline{\parbox{13cm}{
\centerline{\includegraphics[height=9.7cm]{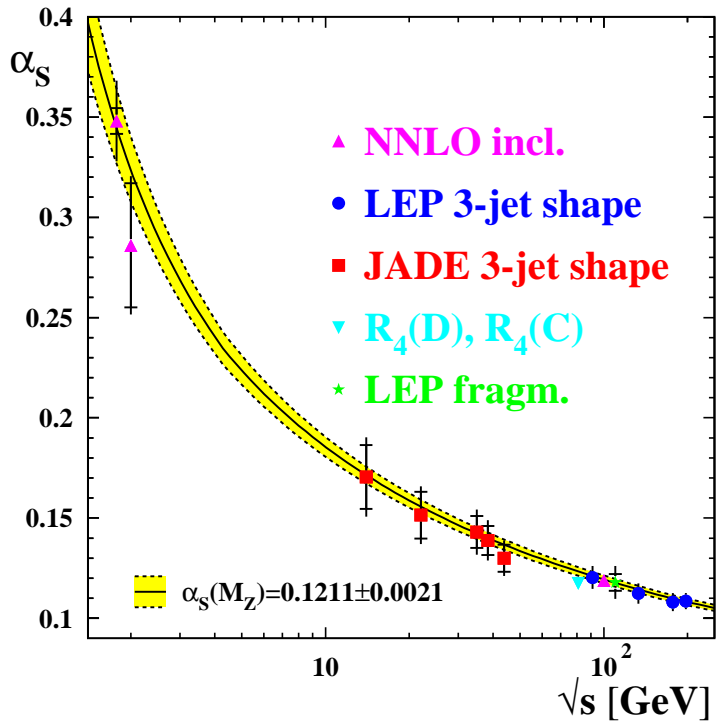}}
\vspace{-2mm} \caption{\label{fig:asymptfreedom} \small
\textit{Running of strong coupling constant~\cite{runningas}.}}}}
\end{figure}

In the Standard Model the strong interactions are described by a
non-abelian gauge theory called \emph{Quantum Chromo Dynamics}
(QCD)~\citer{Gell-Mann:1964nj,Fritzsch:1973pi}. The most important
property of QCD is \emph{asymptotic freedom}, i.e.~the fact that the
coupling of the quarks to the gluons becomes weak at large energy
scales (cf.~Figure~\ref{fig:asymptfreedom}). Strong interaction
effects from high energies as e.g.~the $Z^0$ resonance at
$M_Z=91.2$~GeV can be precisely calculated in perturbation theory as
an expansion in the coupling constant $\as(M_Z)\sim0.12$. At lower
energies of order of the QCD scale $\LQCD\sim0.5$~GeV, the
perturbative expansion in the coupling constant breaks down and the
quarks get confined into complicated colour-singlet hadrons. In our
theoretical description of any hadronic process that is observed in
experiment we thus have to deal with these non-perturbative effects.
We may sometimes be lucky and find some observables which are almost
free of these hadronic uncertainties as the ''golden'' $\sin 2\beta$
measurement from the $B\to J/\psi \, K_S$ decay which corresponds to
the thin dark blue ray in Figure~\ref{fig:CKM}. As it is not always
possible to find such clean observables, we have to look for
sophisticated methods which allow to control the hadronic dynamics.

\newabs
In this thesis we mainly deal with strong interaction effects in $B$
meson decays. The role of $B$ mesons is special as they are (apart
from the $\Upsilon$-resonances) the heaviest mesons which show an
extremely rich phenomenology including many interesting CP violating
observables. Furthermore, the hadronic dynamics in $B$ meson decays
turns out to be partly accessible within perturbation theory as the
intrinsic mass of the $b$-quark $m_b\sim5$~GeV corresponds to a
perturbative scale with $\as(m_b)\sim0.22$. More precisely, the weak
decay of the $b$-quark is accompanied by short-distance effects and
unaffected by the hadronization of the quarks which occurs at much
larger time scales. The technical procedure which disentangles
perturbative from non-perturbative effects is called
\emph{factorization}. For more details concerning factorization we
refer to the comprehensive introduction in the first chapter of this
thesis.

\newabs
We in particular consider exclusive charmless decays of $B$ mesons,
as ~{$B\to\pi\pi$} or {$B\to\pi\ell\nu$}. These decays provide
important information on the unitarity triangle, the former serve
for the determination of the CKM angle $\alpha$ and the latter for
the measurement of $|V_{ub}|$. The impact of these decays on the
unitarity triangle from Figure~\ref{fig:CKM} are reflected by the
light blue band and the dark green circle, respectively. Rather than
focussing on the phenomenological aspects of these decays, we
examine the complicated strong interaction dynamics which is encoded
in the hadronic matrix elements.

The decays considered in this thesis share the feature that the $B$
meson decays into very energetic light mesons (in the $B$ meson rest
frame). The factorization of short- and long-distance QCD effects
for these heavy-to-light transitions has first been worked out in
the framework of \emph{QCD Factorization}~\citer{QCDF:I,QCDF:III}
and has later been translated into a field theoretical formulation
which is called \emph{Soft-Collinear Effective Theory}
(SCET)~\citer{SCET:Bauer:I,SCET:Beneke}. We are mainly concerned
with the calculation of higher order perturbative QCD corrections,
but have also a general look at the factorization properties of
heavy-to-light form factors which are among the ''simplest'' objects
for studying the QCD dynamics in $B\to\pi$ transitions.

\newabs
The core of this thesis consists in a 2-loop calculation related to
the hadronic two-body decay modes. Due to the complexity of the
problem we split the calculation into two parts. We first compute
the imaginary part of the hadronic matrix elements which is
technically simpler than the real part. It is in addition of
particular interest in phenomenological applications as it is
related to a strong phase shift between the final state mesons which
''pollutes'' the interesting information about the underlying weak
CKM phases. Our calculation represents one part of the full
next-to-next-to-leading order (NNLO) analysis of the topological
tree amplitudes within QCD Factorization. Another part of this
analysis concerning (1-loop) spectator scattering has been
calculated recently by various groups~\citer{BenekeJager,Volker}. We
remark that the phenomenological impact of our corrections is beyond
the scope of this thesis as the full NNLO calculation (including
topological penguin amplitudes) is still incomplete.

Our conceptual analysis of heavy-to-light form factors is based on a
particular scenario. We consider transition form factors between
non-relativistic bound states which can be addressed in perturbation
theory. We perform a next-to-leading order (NLO) analysis of these
form factors using the same techniques that we have developed for
the aforementioned 2-loop calculation. We then compare our explicit
results with the general factorization formula for heavy-to-light
form factors and address some of its open questions concerning the
so-called soft-overlap contribution, the issue of endpoint
singularities and the resummation of large (Sudakov) logarithms. We
emphasize that this analysis can be applied for the description of
$B_c \to (\bar c c)$ transitions, although we focus on the more
conceptual aspects here.

Large parts of this thesis deal with the calculation of
loop-integrals. One important element in our calculation procedure
is an automatized reduction algorithm which allows us to reduce the
calculation of several thousands of loop-integrals to a much smaller
set of so-called Master Integrals. The calculation of these Master
Integrals represents the most difficult task of the entire
calculation. We would like to point out that the technical
difficulties that we encountered in the two considered calculations
are of rather different origins. Whereas the first calculation in
the $B\to\pi\pi$ context represents a highly challenging and complex
2-loop calculation, the other (1-loop) calculation related to the
form factors is complicated due to the presence of various physical
scales. In this case we restrict our attention to the leading power
in a mass expansion which requires different techniques than for
full loop-integrals. Consequently, our collection of loop-techniques
that we present in the second chapter of this thesis, summarizes
(almost) all of the most sophisticated techniques that have been
developed so far: the method of differential
equations~\cite{DiffEqs:Kot,DiffEqs:Rem}, the formalism of harmonic
polylogarithms~\cite{HPL}, the method of expansion by
regions~\cite{ExpByRegions}, Mellin-Barnes
techniques~\cite{MB:Smi,MB:Tau} and the method of sector
decomposition~\cite{HB}. Most of these techniques have rarely been
applied in $B$ physics so far.

\newabs
The structure of this thesis can be outlined as follows:

\newabs
In the first chapter we present the theoretical background required
for an analysis of exclusive charmless $B$ decays. We give a
comprehensive introduction to QCD Factorization and Soft-Collinear
Effective Theory and comment briefly on several alternative
approaches. The remainder of this chapter is devoted to a detailed
analysis of hadronic two-body decays and heavy-to-light form factors
which are of particular interest in this work.

In Chapter~\ref{ch:PT} we collect the techniques that we have used
in our calculations. We develop a systematic strategy which is based
on an automatized reduction algorithm. As most parts of this thesis
deal with the calculation of loop-integrals, we dedicate a sizeable
part of Chapter~2 to the presentation of several sophisticated
techniques.

\newabs
Due to this structure, the second part of this thesis is free of the
technical issues related to the calculation of loop-diagrams. In
Chapter~\ref{ch:ImPart} we consider the imaginary part of the
topological tree amplitudes in hadronic two-body decays. Apart from
the 2-loop calculation, we address the issues of Fierz symmetry,
evanescent operators, renormalization and IR subtractions. We
finally obtain our results in an analytical form and conclude this
chapter with a brief numerical analysis.

Chapter~\ref{ch:RePart} is devoted to the real part of the
topological tree amplitudes in hadronic two-body decays. The
calculation follows the same lines as in Chapter~\ref{ch:ImPart} but
is technically much more involved. So far, we have accomplished the
technical part of this calculation and present some preliminary
numerical results.

In our final analysis in Chapter~\ref{ch:NRModel} we consider
heavy-to-light form factors between non-relativistic bound states.
We first address the non-relativistic approximation in this context
and present subsequently the NLO (1-loop) calculation of the form
factors. In our conceptual analysis we investigate the origin of
endpoint singularities and comment on the resummation of logarithms.
We show that we can isolate (and calculate) the so-called
soft-overlap contribution in our set-up and calculate leading twist
light-cone distribution amplitudes of non-relativistic bound states.

\newabs
We finally conclude and give an outlook on future developments. The
results for all Master Integrals that appeared in our calculations
from Chapter~\ref{ch:ImPart}~-~\ref{ch:NRModel} as well as the
explicit expressions for the NLO form factors are summarized in the
appendix.

\part{Formalism}

\pagestyle{fancyplain}
\renewcommand{\chaptermark}[1]{\markboth{\thechapter.~\uppercase{#1}}{}}
\renewcommand{\sectionmark}[1]{\markright{\thesection.~\uppercase{#1}}}
\rhead[\leftmark]{\fancyplain{} \thepage}
\lhead[\thepage]{\rightmark}

\chapter{Exclusive charmless $B$ decays}
 
In the first chapter we present the theoretical background for the
description of exclusive charmless $B$ decays. As the perturbative
calculations in the second part of this thesis are based on QCD
Factorization and Soft-Collinear Effective Theory, we give a
profound introduction to these two developments. We paid special
attention to avoid unnecessary formulas in the introductory chapter
in order to allow for a transparent presentation of the basic ideas
behind these concepts. We comment briefly on alternative approaches
to  charmless $B$ decays and have a closer look at those decays
which are of particular interest in the work at hand. These include
hadronic two-body decays and a conceptual analysis of heavy-to-light
form factors which are important ingredients in semi-leptonic and
radiative $B$ decays.

\section{Basic concepts}

\subsection{QCD Factorization}
\label{sec:QCDF}

The phrase \emph{QCD Factorization} is closely related to the theory
of hadronic two-body decays of $B$ mesons. At the end of the 90s
Beneke, Buchalla, Neubert and Sachrajda, to which we refer as BBNS
in the following, established this novel framework which allowed for
the first time for a systematic treatment of these decays in
QCD~\citer{QCDF:I,QCDF:III}. However, QCD Factorization is a more
general framework with applications covering a wide spectrum of
semi-leptonic, radiative and hadronic $B$ decays.

\newabs
QCD Factorization basically merged two different developments that
were known at that time: the heavy quark expansion (for a review
see~\cite{NeubertReview}) and the theory of hard exclusive processes
which is also known as collinear factorization~\cite{ER,BL}. In the
course of the 90s much progress has been made in the understanding
of heavy mesons. It has been realized that the QCD dynamics of heavy
mesons simplifies substantially when it is considered in the heavy
quark expansion\footnote{The phrase \emph{heavy quark expansion} is
often used in the literature in the context of inclusive decays. We
will refer to it here more generally whenever we speak about an
expansion in $\LQCD/m_Q$.} (HQE), i.e. an expansion in the ratio
$\LQCD/m_Q$ where $m_Q$ is the mass of the heavy quark. On the other
hand the theory of hard exclusive processes can be seen as the
counterpart of deep-inelastic scattering for inclusive processes. It
was developed for the description of exclusive processes with a
large momentum transfer $Q^2\gg \LQCD^2$. Due to the large momentum
transfer, the particles in these processes are very energetic and
assumed to move collinear to light-cone directions which leads to
important simplifications.

Charmless $B$ decays naturally incorporate both of these aspects.
The $B$ meson in the initial state implies a systematic description
in terms of a HQE in $\LQCD/m_b$. The final state being charmless,
which means that it consists of light hadrons only, further implies
that the particles in the final state are very energetic in the $B$
meson rest frame and can be described to move almost on the
light-cone.

\newabs
The basic idea of factorization is the attempt to disentangle
physical effects from different length or momentum scales. This is a
very general idea that can be applied in many different fields of
physics. Concerning the dynamics of the strong interactions this
strategy is particularly suited due to the asymptotic freedom of
QCD. Any decay or scattering process involving hadrons is sensitive
to the scale $\LQCD$ which is responsible for the confinement of the
quarks into the hadrons. As the strong coupling constant at these
scales is of $\calO(1)$, the respective effects cannot be addressed
in perturbation theory and therefore we call them non-perturbative.
In $B$ physics we are confronted with an additional intrinsic scale
in form of the mass of the b-quark $m_b \gg \LQCD$ which is a
perturbative scale with $\as(m_b)\sim0.22$. We see that the idea of
disentangling  the effects from the scales $m_b$ and $\LQCD$ is
equivalent to separating perturbative from non-perturbative effects
in QCD. The predictive power of factorization lies in the fact that
we can calculate the former systematically in perturbation theory
whereas the latter typically give rise to universal hadronic
quantities which can be obtained from other methods as lattice gauge
theory or QCD sum rules or they can even be extracted from
experimental data.

The essence of the QCD Factorization prediction is summarized in a
factorization formula for a hadronic matrix element. The
factorization formula illustrates how the perturbative and
non-perturbative effects are disentangled (\emph{factorized}). In
the remainder of this section we present several examples of
factorization formulas for different classes of exclusive processes.
In the first two examples we sketch the situation that was known
before QCD Factorization was established. The first one deals with a
heavy-to-heavy transition which can be described with the help of
the HQE and the second one corresponds to collinear factorization.
In the last two examples we illustrate how QCD Factorization
combines these two pictures. We give slightly simplified
descriptions in order to concentrate on the main aspects concerning
factorization. We hope that our presentation will help to understand
the structure of the factorization formulas that we discuss in the
following sections.

\subsubsection{Example 1: $B\to D\ell\nu$}

\begin{figure}[t!]
\centerline{\parbox{13cm}{
\parbox[b]{6cm}{\includegraphics[width=5cm]{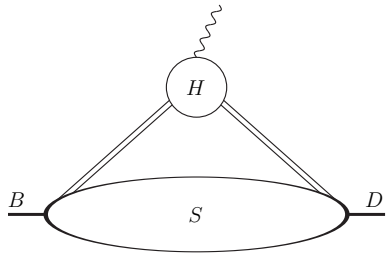}}
\parbox[b]{7cm}{
\caption{\label{fig:FFtype1} \small \textit{Factorization of short-
and long distance effects in heavy-to-heavy transitions. The former
are contained in a coefficient function $H$, the latter in a
soft-overlap contribution $S$. The double lines denote heavy quarks
with $m_Q\gg\LQCD$.}}}}} \vspace{5mm}
\end{figure}

We start with the simplest example in form of exclusive $B$ decays
into final states that do not contain light hadrons as e.g.~$B\to
D\ell\nu$. The relevant scales in these processes are the mass(es)
of the heavy quark(s) which we simply denote by $m_Q$ and the
hadronic scale $\LQCD$. The factorization formula for a generic
hadronic matrix element of a current $\mathcal{J}$ takes the
schematic form
\begin{align}
\langle D | \mathcal{J} | B \rangle = H(\mu_F) \; S(\mu_F) +
\calO(\LQCD/m_Q). \label{eq:FFtype1}
\end{align}
The factorization formula is illustrated in
Figure~\ref{fig:FFtype1}. We first notice that the factorization
formula makes a statement about the leading power in the HQE. On the
other hand it is predicted to be valid to all orders in perturbation
theory. In writing (\ref{eq:FFtype1}), \emph{hard} effects from the
scale $m_Q$ and \emph{soft} effects related to $\LQCD$ have been
disentangled, the former being contained in the coefficient function
$H(\mu_F)$ and the latter in a remnant matrix element which we
denoted by $S(\mu_F)$. Technically, factorization is achieved with
the help of a factorization scale $\mu_F$ satisfying $m_Q \gg \mu_F
\gg \LQCD$. The effects form hard gluons with virtualities $k^2>
\mu_F^2$ are contained in $H(\mu_F)$ and those from soft gluons with
$k^2< \mu_F^2$ are absorbed into $S(\mu_F)$. As the factorization
scale has been introduced artificially in the factorization formula
(\ref{eq:FFtype1}), the dependence of the functions $H(\mu_F)$ and
$S(\mu_F)$ has to cancel in their product.

\newabs
As a side remark we mention that the concept presented above applies
as well to the effective weak interactions in the description of
low-energetic hadronic processes. These processes provide the
hierarchy $M_W^2 \gg q^2$ where $M_W$ is the mass of the $W$-boson
and $q^2$ is a typical momentum scale in the process. In leading
power in an expansion in $q^2/M_W^2$ the hadronic matrix elements
factorize similar to (\ref{eq:FFtype1}) into short-distance Wilson
coefficients, which correspond to the $H(\mu_F)$ in our notation,
and remnant matrix elements $S(\mu_F)$ which can be calculated in
the Fermi theory of weak interactions. We come back to the effective
weak interactions when we consider hadronic two-body decays in
Section~\ref{sec:BtoPiPi}.

\subsubsection{Example 2: $\pi \gamma^*\to\pi$}

\begin{figure}[t!]
\centerline{\parbox{13cm}{
\parbox[b]{6cm}{\includegraphics[width=5cm]{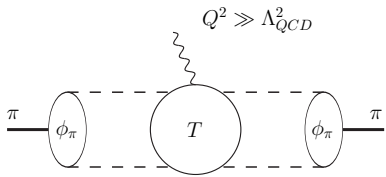}}
\parbox[b]{7cm}{
\caption{\label{fig:FFtype2} \small \textit{Collinear factorization
of short- and long distance effects. The former give rise to a
hard-scattering kernel~$T$, the latter to light-cone distribution
amplitudes $\phi_\pi$. The dashed lines denote collinear
quarks.}}}}}\vspace{5mm}
\end{figure}

In a second example we consider hard exclusive processes which are
not related to $B$ physics, as e.g.~$\pi \gamma^*\to\pi$ where the
kinematics sets a perturbative scale in form of the large momentum
transfer $Q^2\gg\LQCD^2$ . Because of the large momentum transfer,
light-cone dynamics comes into play and the energetic light mesons
can approximately be described by their two-particle quark-antiquark
Fock states. Similar to what is done in deep-inelastic scattering,
the quark and the antiquark can be assumed to move collinearly
inside the meson and share its momentum with fractions $u$ and
$\ubar\equiv1-u$, respectively. The corresponding factorization
formula is sketched in Figure~\ref{fig:FFtype2} and reads
\begin{align}
\langle \pi| \mathcal{J} | \pi \rangle =
    \int_0^1 \! du \int_0^1 \! dv \;\; \phi_\pi(v;\mu_F) \; T(v,u;\mu_F) \; \phi_\pi(u;\mu_F)  + \calO(\LQCD^2/Q^2).
\label{eq:FFtype2}
\end{align}
Again, perturbative and non-perturbative effects are systematically
disentangled in leading power. In this case the \emph{hard} effects
from the scale $Q^2$ give rise to a hard-scattering kernel
$T(v,u;\mu_F)$ which depends on the momentum fractions $v$ and $u$
of the quarks in the mesons. The \emph{collinear} effects from the
scale $\LQCD^2$ are encoded in light-cone distribution amplitudes
$\phi_\pi(v;\mu_F)$ and $\phi_\pi(u;\mu_F)$ of the initial and final
state particles.

\newabs
Let us stress two important differences between the factorization
formulas (\ref{eq:FFtype1}) and (\ref{eq:FFtype2}). First, in
(\ref{eq:FFtype2}) there is no long-distance interaction between the
initial and the final state particle at leading power. The
interaction between both mesons is entirely described by the
perturbative hard-scattering kernel. The non-perturbative input in
form of the distribution amplitudes contains information about the
structure of the participating mesons and is independent of the
considered process. This is different in (\ref{eq:FFtype1}) where
the soft matrix element depends on the overlap between the wave
functions of the initial and the final state particle and therefore
on the considered process. Second, in (\ref{eq:FFtype1}) all effects
related to the large scale $m_b$ are contained in the hard
coefficient function whereas the non-perturbative input corresponds
to a soft matrix element of a \emph{local} operator. The situation
is more complicated in (\ref{eq:FFtype2}) where all effects from
large virtualities $Q^2$ are encoded in the hard-scattering kernel,
but the large scale still enters the non-perturbative matrix
elements in form of the large energy of the mesons. The distribution
amplitudes correspond to \emph{non-local} matrix elements which are
defined on the light-cone. The non-locality in position space
translates into convolutions over the fractions $u$ and $v$ in
momentum space as illustrated in (\ref{eq:FFtype2}).

\subsubsection{Example 3: $B\to D \pi$}

\begin{figure}[t!]
\centerline{\parbox{13cm}{
\parbox[b]{6cm}{\includegraphics[width=5cm]{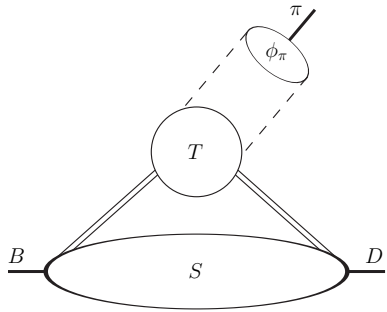}}
\parbox[b]{7cm}{
\caption{\label{fig:FFtype3} \small \textit{QCD Factorization of
short- and long distance effects in exclusive $B$ decays into
heavy-light final states. All related quantities have already been
introduced in Figure~\ref{fig:FFtype1} and
\ref{fig:FFtype2}.}}}}}\vspace{5mm}
\end{figure}

For the first example in QCD Factorization we choose exclusive $B$
decays into heavy-light final states as e.g.~$B\to D \pi$. According
to the two-particle kinematics in the final state, the pion is very
energetic in these decays with $E_\pi = \calO(m_b)$ in the $B$ meson
rest frame. Similar to what we have seen in the last example, the
pion can be described by its two particle Fock state in the
collinear approximation. The factorization formula now becomes
(cf.~Figure~\ref{fig:FFtype3})
\begin{align}
\langle D \pi | \mathcal{J} | B \rangle = S(\mu_F) \; \; \int_0^1 \!
du \; \; T(u;\mu_F) \; \phi_\pi(u;\mu_F) + \calO(\LQCD/m_Q).
\label{eq:FFtype3}
\end{align}
The soft matrix element $S(\mu_F)$ describes the long-distance
dynamics in the $B-D$ transition as in (\ref{eq:FFtype1}). The
energetic pion enters in form of its distribution amplitude
$\phi_\pi(u;\mu_F)$ and the perturbative effects are contained in a
hard-scattering kernel $T(u;\mu_F)$ similar to (\ref{eq:FFtype2}).
We see that there is no long-distance interaction between the pion
and the $B-D$ system at leading power. This corresponds to the
famous argument of colour transparency which states that the soft
gluons cannot resolve the fast moving colour-singlet
pion~\cite{ColourTransp:Bj,ColourTransp:DG}.

\newabs
Let us make one remark concerning strong phases which are important
in phenomenological applications. Strong phases arise from final
state interactions, but we have just seen that the pion decouples in
our example from the $B-D$ system in the heavy quark limit $m_b \to
\infty$. The final state interactions are entirely encoded in the
hard-scattering kernel and therefore predicted to be perturbative.
We will see in Section~\ref{sec:BtoPiPi} that a similar argument
holds for the case of $B\to\pi\pi$ where the knowledge of strong
phases is even more desirable.

\newabs
We finally point out that the factorization formula
(\ref{eq:FFtype3}) is restricted to the case where the $D$ meson
picks up the spectator antiquark from the $B$ meson. In the opposite
case when the spectator goes into the pion, factorization does not
hold.

\subsubsection{Example 4: $B\to \pi \ell \nu$}

\begin{figure}[t!]
\centerline{\parbox{13cm}{
\includegraphics[width=13cm]{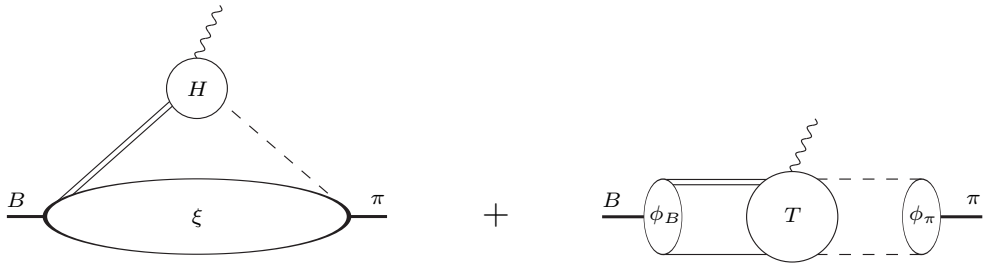}
\caption{\label{fig:FFtype4} \small \textit{QCD Factorization of
short- and long distance effects in exclusive charmless $B$ decays.
New objects enter these decays. The first term contains a
complicated overlap-contribution $\xi$ whereas the $B$ meson
distribution amplitude $\phi_B$ appears in the second one.}}}}
\vspace{2mm}
\end{figure}

In the last example we finally discuss exclusive charmless $B$
decays. Here we take the decay $B\to \pi \ell \nu$ as an example
which we reconsider in more detail in Section~\ref{sec:BtoPi}. As we
deal with a three-body decay in this case, the energy of the pion
depends on the invariant mass of the lepton pair. We restrict our
attention to the case where the pion is very energetic in the $B$
meson rest frame with $E_\pi = \calO(m_b)$. This is called the large
recoil region which is similar to the situation in the last example.
However, the factorization formula, which is illustrated in
Figure~\ref{fig:FFtype4}, turns out to be more complicated and reads
\begin{align}
\langle \pi| \mathcal{J} | B \rangle &\simeq H(\mu_F) \; \xi(\mu_F)
+ \int_0^\infty \! d\om \int_0^1 \! du \;\; \phi_B(\om;\mu_F) \;
T(\om,u;\mu_F) \; \phi_\pi(u;\mu_F). \label{eq:FFtype4}
\end{align}
It is again restricted to the leading power in $\LQCD/m_b$ which we
illustrate from now on by the symbol ''$\simeq$'' for brevity. The
second term resembles the factorized form in (\ref{eq:FFtype2}). It
consists of a perturbative hard-scattering kernel $T(\om,u;\mu_F)$
convoluted with the light-cone distribution amplitudes of the pion
$\phi_\pi(u;\mu_F)$ and the $B$ meson $\phi_B(\om;\mu_F)$. Whereas
the former already entered the factorization formulas in the last
two examples, the appearance of the latter is new. At first sight,
it might look unnatural to consider the $B$ meson on the light-cone.
Nevertheless its distribution amplitude is a well-defined object,
although more complicated and less understood than the one of the
pion.

\newabs
On the other hand the first term in (\ref{eq:FFtype4}) looks like
the factorization formula (\ref{eq:FFtype1}), but again the
situation is more complicated in this case. Whereas the hard effects
are factorized in a coefficient functions $H(\mu_F)$, the remnant
matrix element is not a soft matrix element as in
(\ref{eq:FFtype1}). We therefore wrote $\xi(\mu_F)$ for the
overlap-contribution in (\ref{eq:FFtype4}) which contains a highly
complicated interplay of soft and collinear interactions. A deeper
understanding of this overlap-contribution is the main motivation
for our analysis in Chapter~\ref{ch:NRModel}.

\subsubsection{Concluding remarks}

Factorization formulas within QCD Factorization are complicated
because of the presence of soft and collinear effects at leading
power. We have presented a simpler example in (\ref{eq:FFtype3}) and
a more complicated one in (\ref{eq:FFtype4}). In general QCD
Factorization does not \emph{prove} that a factorization formula is
correct. Typically, the factorization formula is shown to be valid
in the lowest (non-trivial) order of the perturbative expansion and
then \emph{assumed} to hold to all orders in $\as$. It was only the
development of Soft-Collinear Effective Theory which provided the
necessary tools to formulate rigorous factorization proofs.

\subsection{Soft-Collinear Effective Theory}

The origin of Soft-Collinear Effective Theory (SCET) goes back to
the beginning of this millennium. Since then SCET has sparked the
interest of many physicists which led to a fast development of this
field. If we focus on the publications which have contributed to the
formulation of the SCET-Lagrangian, we may refer to Bauer et
al.~\cite{SCET:Bauer:I,SCET:Bauer:II}, Beneke et
al.~\cite{SCET:Beneke}, Chay et al.~\cite{SCET:Chay} and Neubert et
al.~\cite{SCET:Neubert}.

For our purposes there is no need to introduce the whole concept of
SCET as only a small part of our calculations in
Chapter~\ref{ch:NRModel} is directly related to it. However, SCET
provides a deeper understanding of QCD Factorization and puts the
factorization formulas onto robust grounds. We therefore give a
brief introduction to SCET without going into the technical aspects.

\subsubsection{Preliminaries}

Effective theories in general deal with the strategy that we have
already discussed in the last section: the idea to disentangle the
physics from different length/momentum scales. They have restricted
validity up to some momentum cut-off and are designed to correctly
reproduce the IR behavior of a physical process.

There are two different ways to construct an effective theory. In
the \emph{bottom-up approach} either the underlying theory, which is
assumed to be valid at all momentum scales, is not known or it is
not understood how to derive an effective theory from the underlying
theory. It might nevertheless be possible to write down an effective
theory motivated mainly by symmetry arguments and inputs from
experimental observations. Chiral perturbation theory and the
Standard Model of particle physics fall into this class, as we all
hope that the Standard Model loses its validity above the TeV-scale.
On the other hand in the \emph{top-down approach} the effective
theory can be derived directly from the underlying theory. This is
case for SCET which stems from QCD.

\newabs
The same is true for the effective theory which describes heavy
mesons, called heavy-quark effective theory
(HQET)~\cite{HQET:EH,HQET:G}. We already introduced the phrases
\emph{hard} effects ($k^\mu\sim m_Q$) and \emph{soft} effects
($k^\mu\sim \LQCD$) in the last section, which is the appropriate
characterization in this case. E.g.~the gluon field is split into
\begin{align}
A^\mu(x) = A^\mu_h(x) + A^\mu_s(x)
\end{align}
and the hard gluons $A^\mu_h$ and the soft gluons $A^\mu_s$ are
treated as independent degrees of freedom in the effective theory.
On the other hand the heavy quark field can be decomposed into its
large ($\psi$) and small ($\chi$) components schematically  by
\begin{align}
Q(x) = \psi(x) + \chi(x) \hspace{1cm} \text{with} \hspace{1cm}
    \chi(x)=\calO(\LQCD/m_Q) \; \psi(x).
\end{align}
This is similar to what is done in non-relativistic theories where
this procedure is called the Foldy-Wouthuysen
transformation~\cite{FoldyWouthuysen}. The small components of the
heavy quark field and the hard gluons can then be eliminated as
degrees of freedom in the effective theory with their effects
encoded in coupling constants of local operators. The effective
theory is identical to QCD but benefits from the HQE at the level of
the Lagrangian. It is therefore better suited for the description of
heavy mesons than QCD and gives rise to the famous heavy quark
symmetry~\cite{HQS:I,HQS:II} in the heavy quark limit $m_Q \to
\infty$ which had a strong impact on the phenomenology of $B$ and
$D$ meson decays~\cite{NeubertReview}.

\subsubsection{Complications in SCET}

SCET is the effective theory for the description of energetic
particles and jets. It exploits the hierarchy $E\gg\LQCD$ and
follows in principle the same strategy as outlined for HQET:
introduce power counting in $\LQCD/E$, identify relevant degrees of
freedom and integrate out hard fluctuations. However, it turns out
that SCET is far more complicated than HQET.

The main reason for these complications lies in the fact that SCET
is a \emph{non-local} effective theory in contrast to HQET. We do
not go into the details related to this point here, but mention that
this was the real challenge in the formulation of SCET. Notice that
we already encountered this non-local nature of SCET in the last
section in form of the light-cone distribution amplitudes as
described at the end of \mbox{example 2}. A second complication is
related to the fact that there are many different degrees of freedom
in SCET. During the development of SCET the correct scaling of the
collinear modes was the origin of some confusion. Later it turned
out that there are two different scalings depending on the physical
process under consideration. This is the reason why there are two
different versions of SCET usually referred to as \SCETI ~and
\SCETII. Let us have a closer look at this point and at the same
time introduce our notation and terminology that we use throughout
this work.

\subsubsection{Notation and terminology}

The light-cone dynamics is conveniently formulated with the help of
two light-like vectors which we denote by $n_-$ and $n_+$ satisfying
$n_\pm^2=0$ and normalized to \mbox{$n_+ \cdot n_-=2$}. Any momentum
can be decomposed according to its projections onto these light-cone
directions and a two-dimensional transverse plane. We write
\begin{align}
k^\mu = k_- \frac{n_-^\mu}{2} + k_\perp^\mu + k_+ \frac{n_+^\mu}{2}
\hspace{1cm} \text{with} \hspace{1cm} k_\perp \cdot n_\pm=0.
\end{align}
Notice that in our notation $k_\pm \equiv k \cdot n_\mp$. For the
purpose of power counting we introduce a dimensionless parameter
$\lambda$ with $\lambda^2\equiv \LQCD/m_b$. Scaling relations are
given in the form $k\sim(k_-,k_\perp,k_+)$ where we drop factors of
$m_b$ which can be restored easily by a dimensional analysis. For
example, hard momenta with $p_h^\mu \sim m_b$ and soft momenta with
$p_s^\mu \sim \LQCD$ now become $p_h\sim(1,1,1)$ and
$p_s\sim(\lambda^2,\lambda^2,\lambda^2)$.

\newabs
In charmless $B$ decays, the energy of the fast moving light hadrons
typically scales as $E\sim m_b$ in the $B$ meson rest frame (cf.~the
examples 3-4 in the last section). There are two possible scalings
of collinear momenta given by $p_{hc}\sim(1,\lambda,\lambda^2)$ and
$p_c\sim(1,\lambda^2,\lambda^4)$, the first entry reflecting the
large energy in each case. These are the two different scalings of
collinear fields that we mentioned above. The important difference
lies in the respective virtualities $p_{hc}^2\sim m_b\LQCD\gg p_s^2$
and $p_c^2\sim \LQCD^2\sim p_s^2$. We see that the former, to which
we refer as \emph{hard-collinear} from now on, introduce a new scale
$\mu_{hc}\sim(m_b\LQCD)^{1/2}$, whereas the latter, simply called
\emph{collinear}, are related to the hadronic scale $\LQCD$ as
usual. The appearance of the hard-collinear scale is another
complication in SCET compared to HQET. In the following we treat
this intermediate scale as a perturbative scale with
$\as(\mu_{hc})\sim0.4$

\begin{table}[t!]
\centerline{
\parbox{13cm}{\setlength{\doublerulesep}{0.1mm}
\centerline{\begin{tabular}{|r|c|c|}
\hline &\hspace*{4cm}&\hspace*{4cm} \\[-0.7em]
Terminology~~~ & Momentum scaling & Virtuality \\[0.3em]
\hline\hline\hline&& \\[-0.7em]
hard~~~ & $(1,1,1)$ & $m_b^2$ \\[0.3em]
\hline&&\\[-0.7em]
~~~hard-collinear~~~ & $(1,\lambda,\lambda^2)$ & $m_b \LQCD$ \\[0.3em]
\hline&&\\[-0.7em]
soft~~~ & $(\lambda^2,\lambda^2,\lambda^2)$ & $\LQCD^2$\\[0.3em]
collinear~~~ & $(1,\lambda^2,\lambda^4)$ & $\LQCD^2$\\[0.3em]
\hline
\end{tabular}}
\vspace{4mm} \caption{\label{tab:term}\small \textit{Terminology for
the relevant momentum regions in charmless $B$ decays. The momentum
scaling corresponds to the light-cone decomposition
$k\sim(k_-,k_\perp,k_+)$ with a dimensionless parameter
$\lambda^2\equiv \LQCD/m_b$, see text. We have restored the mass
dimension in the respective virtualities for convenience.}}}}
\end{table}

We have summarized our terminology in Table~\ref{tab:term}.
Depending on the considered process some of these modes turn out to
be irrelevant. Inclusive decays as e.g.~the endpoint spectrum in
$B\to X_s \gamma$ can be described with an effective theory called
\SCETI ~which contains hard, hard-collinear and soft modes. The
invariant mass of the energetic jet is typically $M_X^2\sim
m_b\LQCD$ which explains the relevance of the hard-collinear modes
in this case. The effective theory for exclusive decays as
e.g.~$B\to D\pi$ is \SCETII ~containing hard, collinear and soft
degrees of freedom. In this case, the virtuality of the collinear
modes is related to the mass of the energetic pion with
$M_\pi^2\sim\LQCD^2$. Hard-collinear modes may appear in exclusive
decays as well induced by soft-collinear interactions, which makes a
two-step matching procedure $\text{QCD}\to\SCETI\to\SCETII$
~necessary. This is the case e.g.~in $B\to\pi\ell\nu$ but not in
$B\to D\pi$ where these effects turn out to be power suppressed.
This explains why the factorization formula for the former is much
more complex than that of the latter as we have seen in the examples
3-4 in the last section.

\subsubsection{QCD Factorization versus SCET}

What is the difference between QCD Factorization and SCET? Both
frameworks provide a systematic description of QCD in the heavy
quark limit $m_b\to\infty$, i.e.~there is none: they are equivalent!
There has been some confusion about this point in the literature. It
seems as if there were different predictions from QCD Factorization
and SCET in hadronic two-body decays of $B$ mesons. We stress that
this is \emph{not} related to the underlying frameworks but to
different treatments of power corrections and input parameters of
two groups: BBNS and Bauer, Pirjol, Rothstein, Stewart (BPRS).
Therefore, the predictions from BBNS and BPRS differ although the
ones from QCD Factorization and SCET do \emph{not}. More details
concerning this issue can be found
in~\citer{BPRSvsBBNS:I,BPRSvsBBNS:III}.

\newabs
From the conceptual point of view, QCD Factorization and SCET are
different. QCD Factorization relies on an explicit analysis of
momentum regions of Feynman diagrams. In contrast to this, SCET is a
rigorous theory in the sense that it is derived from an established
theory (QCD) in a well-defined expansion ($\LQCD/E$) on the level of
the Lagrangian. With the Lagrangian and the respective Feynman rules
at hand, it is a convenient tool to address a large variety of
inclusive and exclusive processes with applications going even
beyond the domain of $B$ physics.

\subsection{Alternative approaches}

For completeness we give an overview of alternative approaches to
exclusive charmless $B$ decays. As this section is not directly
related to the remainder of this thesis, our presentation will be
brief and concentrate on the most prominent methods.

\subsubsection{Lattice Gauge Theory}

The first method which comes into mind may be lattice field theory.
In recent years there has been considerable progress concerning the
reduction of its systematic uncertainties. Most importantly, a
steadily increasing number of unquenched lattice calculations
becomes available reducing the magnitude of its uncertainties in
some cases to the percent level (in particular in ratios of hadronic
quantities). For recent reviews on lattice results for heavy quark
systems we refer to~\cite{Lat:Wingate,Lat:Davies}.

Unfortunately, lattice gauge theory can tell us very little about
exclusive charmless $B$ decays. The main difficulty is related to
the implementation of energetic mesons in the lattice calculations
(which are usually performed in the $B$ meson rest frame). So far,
there are reliable lattice results for $B$ meson decay constants and
heavy-to-light form factors at small recoil which can be
extrapolated to the large recoil case using dispersion relations.
There are some interesting considerations to directly access the
large recoil region in an approach called \emph{moving
NRQCD}~\cite{mNRQCD}. However, if it will be possible to address
hadronic $B$ decays on the lattice remains very challenging.

\subsubsection{QCD Sum Rules}

Although intrinsically limited in their accuracy, QCD sum rules have
become a serious competitor to lattice gauge theory. In some sense
it can even be considered as complementary to QCD Factorization
since it provides important information about the non-perturbative
input parameters of the latter in form of decay constants,
heavy-to-light form factors (at large recoil) and light-cone
distribution amplitudes. For a selection of recent Light-Cone Sum
Rule (LCSR) results we refer to~\cite{BallZwicky:FF,BallZwicky:DA}.

Charmless non-leptonic $B$~decays have been studied in the LCSR
approach~\citer{LCSR:I,LCSR:IIIII}. It should be noted that this
requires a substantial extension of the standard LCSR
formalism~\citer{LCSR:I}. The outcome of the LCSR analysis is in
good overall agreement with the QCD Factorization prediction. This
technique is sometimes used to address the importance of power
corrections in the QCD Factorization framework. This issue should,
however, be treated with care as the essential QCD dynamics is
considered at \emph{finite} $m_b$ in this approach and the
heavy-quark limit $m_b \to\infty$ is only taken at the very end of
the calculation after performing the continuum subtraction. It has
been pointed out in~\cite{LCSR:SCET} that one should rather consider
LCSRs \emph{within} SCET in order to properly address distinct
contributions of the QCD Factorization framework.

\subsubsection{Perturbative QCD}

Perturbative QCD (pQCD)~\cite{pQCD:I,pQCD:II}, which is also known
as $k_\perp$-factorization, is based on a hard-scattering approach
and may at first sight look similar to QCD Factorization since decay
amplitudes are expressed as convolutions of hard-scattering kernels
with meson wave functions. However, pQCD does not share the same
systematics as QCD Factorization and SCET and relies on the
\emph{assumption} that soft contributions to heavy-to-light form
factors are suppressed by Sudakov effects (criticism concerning this
point has been raised in~\cite{SudakovCrit}). The form factor is
thus considered to be dominated by hard gluon exchange and the wave
functions have to include a dependence on transverse momenta in
order to avoid endpoint-singularities in the convolution integrals.

As a consequence pQCD is very different from QCD Factorization. This
is reflected by a different hierarchy of various contributions to
the hadronic matrix elements, different non-perturbative input
quantities and by the fact that pQCD does not recover naive
factorization (cf.~next section) in any limit. Only recently the
authors of pQCD have included next-to-leading order corrections
\texttt{}which have simply been taken over from the QCD
Factorization and SCET analyses~\cite{pQCD:NLO:I,pQCD:NLO:II}.

\subsubsection{Phenomenological approaches}

Apart from these dynamical approaches there exist many ideas how to
extract the interesting CKM information in exclusive charmless $B$
decays without explicitly calculating the hadronic dynamics. The
basic idea is to find a suitable parametrization of the decay
amplitudes and to use symmetry arguments (e.g. isospin, $SU(3)$) to
derive relations between different decay processes. A sufficiently
small set of unknown amplitudes is finally fitted to experimental
data.

These approaches should be seen as complementary to the dynamical
ones discussed above. On the one hand they require input from the
dynamical approaches in order to control their own intrinsic
uncertainties of e.g.~$SU(3)$-breaking, on the other they can serve
as a guide for the dynamical approaches and indicate where these
fail to correctly reproduce the data. Rather than presenting some
exemplary ideas here, we refer to a recent review about the
phenomenology of $B$ decays~\cite{HockerLigeti}.

\newpage
\section{Hadronic two-body decays}
\label{sec:BtoPiPi}

After this general introduction we come to the first class of
exclusive $B$ decays which we consider in detail in this work:
hadronic two-body decays. There is a very large variety of decay
channels which fall into this class as e.g.~$B\to \pi\pi$, $B\to
K\pi$, $B\to \rho\rho$ or $B_s\to \pi\pi$. From the phenomenological
viewpoint, we are mainly interested in the underlying weak
interactions which differ among these decays due to the flavour
contents of the mesons. In our analysis concerning the QCD dynamics,
the only differences lie in the pseudoscalar ($P$) or vector ($V$)
nature of the light mesons and small $SU(3)$ breaking effects. As we
do not focus on the phenomenological implications of these decays
here, it will be sufficient to concentrate on the $B\to\pi\pi$
channels.

\subsection{Preliminaries}

$B$ meson decays are mediated by weak interactions. As the typical
energy and momentum scales in these decays are much smaller than the
mass of the $W$-boson, we may work with an effective weak
Hamiltonian which consists of a sum of local operators $Q_i$
multiplied by short-distance coefficients $C_i$ and products of CKM
matrix elements $\lambda_p\equiv V_{pb}V^*_{pd}$. The effective
Hamiltonian is given by
\begin{align} \label{eq:Heff}
\Heff = \frac{G_F}{\sqrt{2}} \sum_{p=u,c} \lambda_p \left( C_1 Q_1^p
+ C_2 Q_2^p + \sum_{i=3}^6 C_i Q_i + C_8 Q_8\right) + \text{h.c.},
\end{align}
where $Q_{1,2}^p$ are the left-handed current-current operators,
$Q_{3-6}$ are the QCD penguin operators and $Q_8$ is the
chromomagnetic dipole operator. Their explicit form reads
\begin{align}
&Q_1^p = (\bar p b)_{V-A} (\bar d p)_{V-A},
    && Q_2^p  = (\bar p_i b_j)_{V-A} (\bar d_j p_i)_{V-A}, \no \\
&Q_3 = (\bar d b)_{V-A} \TS \sum_q (\bar q q)_{V-A},
    && Q_4  = (\bar d_i b_j)_{V-A} \TS \sum_q  (\bar q_j q_i)_{V-A}, \no \\
&Q_5 = (\bar d b)_{V-A} \TS \sum_q (\bar q q)_{V+A},
    && Q_6  = (\bar d_i b_j)_{V-A} \TS \sum_q  (\bar q_j q_i)_{V+A}, \no \\
&&&\hspace{-4cm} Q_8= -\frac{g_s}{8\pi^2} \, m_b \; \bar d
\sigma_{\mu\nu} (1+\gamma_5) G^{\mu\nu} b, \label{eq:Q1-8}
\end{align}
where $(\bar{q}_1 q_2)_{V\pm A}=\bar{q}_1
\gamma_\mu(1\pm\gamma_5)q_2$ and the sum runs over all active quark
flavours in the effective theory, i.e. $q=u,d,s,c,b$. If no colour
index $i,j$ is given, the two operators are assumed to be in a
colour singlet state. The definition of the dipole operator $Q_8$
corresponds to the sign convention $i D_\mu = i \partial_\mu + g_s
T^A A^A_\mu$.

In principle, more operators have to be taken into account. The
complete set of operators contains in addition electroweak penguin
operators and the electro\-magnetic dipole operator which are
important for $B\to K\pi$ decays. As we focus on the $B\to \pi\pi$
channels here, the effects from these operators can safely be
neglected.

\begin{figure}[t!]
\centerline{\parbox{13cm}{
\includegraphics[width=13cm]{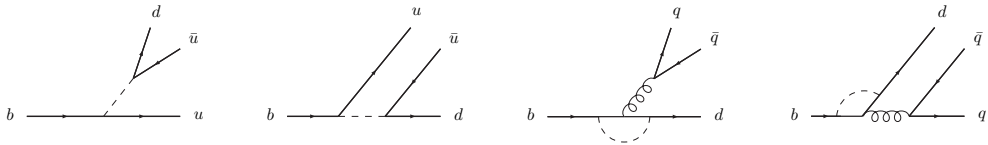}
\caption{\label{fig:alphas} \small \textit{Topological amplitudes
$\alpha_{1-4}(M_1 M_2)$ as introduced in~\cite{BenekeNeubert2003}.
The quark to the right and the spectator antiquark with flavour
$\bar{q}_s$ (not drawn) form the meson $M_1$ with flavour content
$[\bar{q}_s u]$, $[\bar{q}_s d]$, $[\bar{q}_s d]$ and $[\bar{q}_s
q]$, respectively. The two upper lines form the meson $M_2$ with
respective flavour $[\bar{u} d]$, $[\bar{u} u]$, $[\bar{q} q]$ and
$[\bar{q} d]$ where $q \in \{u,d,s\}$. The flavour-singlet penguin
amplitude $\alpha_3(M_1 M_2)$ does not contribute to the
$B\to\pi\pi$ decay amplitudes.}}}}
\end{figure}

\newpage
For the parametrization of the $B\to\pi\pi$ decay amplitudes we
follow the notation of~\cite{BenekeNeubert2003}. Neglecting some
smaller amplitudes related to weak annihilation, they can be written
as
\begin{align}
\sqrt{2} \; \langle \pi^- \pi^0 | \, \Heff \, | B^- \rangle
    &= \lambda_u   \big[\alpha_1(\pi\pi) + \alpha_2(\pi\pi) \big] \; A_{\pi\pi}, \no \\
\langle \pi^+ \pi^- | \, \Heff \, | \bar{B}^0 \rangle
    &= \Big\{ \lambda_u \big[\alpha_1(\pi\pi) + \alpha_4^u(\pi\pi)\big] + \lambda_c  \, \alpha_4^c(\pi\pi) \Big\} \; A_{\pi\pi}, \no \\
- \; \langle \pi^0 \pi^0 | \, \Heff \, | \bar{B}^0 \rangle
    &= \Big\{ \lambda_u \big[\alpha_2(\pi\pi) - \alpha_4^u(\pi\pi)\big] - \lambda_c  \, \alpha_4^c(\pi\pi) \Big\} \; A_{\pi\pi}.
\label{eq:parametr}
\end{align}
The $\alpha_i(\pi\pi)$ are called topological amplitudes and are
related to the flavour flows in the decays. More precisely,
$\alpha_1(\pi\pi)$ is the colour-allowed tree amplitude,
$\alpha_2(\pi\pi)$ is the colour-suppressed tree amplitude and
$\alpha_4^p(\pi\pi)$ is the QCD penguin amplitude as illustrated in
Figure~\ref{fig:alphas}. We see that $B^- \to \pi^- \pi^0$ is a pure
tree decay within the approximations mentioned above. It is free of
interference effects which typically introduce large uncertainties
due to our poor knowledge of the weak phases and therefore
particularly suited to test the QCD Factorization predictions. On
the other hand $\bar{B}^0 \to \pi^0 \pi^0$ is a colour-suppressed
decay as can be seen in (\ref{eq:parametr}) by the absence of
$\alpha_1(\pi\pi)$. Any prediction concerning this decay is expected
to be accompanied by large uncertainties.

\newabs
For later convenience it is useful to write the normalization of the
amplitudes as
\begin{align}
A_{\pi\pi} = i \, \frac{G_F}{\sqrt{2}} \; m_B^2 F_+^{B\to\pi}(0)
f_\pi,
\end{align}
where $F_+^{B\to\pi}(0)$ is a transition form factor at maximum
recoil and $f_\pi$ is the pion decay constant. Their precise
definitions read
\begin{align}
\langle \pi^+(p') | \, \bar{u} (\slasha{p} - \slasha{p}') b \, |
\bar{B}^0(p) \rangle \big{|}_{(p-p')^2=0} \;
    &\simeq m_B^2 F_+^{B\to\pi}(0), \no \\
\langle \pi^-(q) | \, \bar d \gamma_\mu \gamma_5 u\, | 0 \rangle &=
    -i f_\pi q_\mu.
\end{align}
As in the examples that we discussed at the beginning of this
chapter, the light-cone distribution amplitudes of the pion and the
$B$ meson enter the factorization formula for hadronic two-body
decays. Their explicit definitions will be given in
Chapter~\ref{ch:NRModel} (for light mesons the definition can be
found in (\ref{eq:NRdefLCDAEta}) and for the $B$ meson in
(\ref{eq:NRdefLCDABc})).

\subsection{Factorization Formula}

\begin{figure}[b!]
\centerline{\parbox{13cm}{
\includegraphics[width=12.5cm]{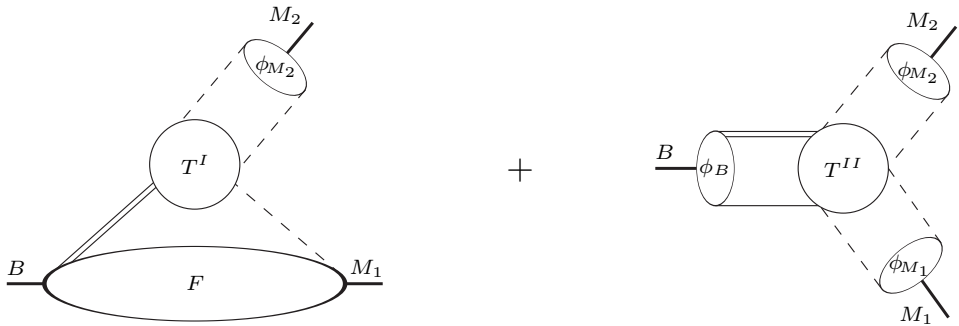}
\caption{\label{fig:QCDF} \small \textit{QCD Factorization of short-
and long distance effects in hadronic two-body decays. The former
are contained in perturbative hard-scattering kernels~ $T^{I}$,
$T^{II}$, the latter in light-cone distribution amplitudes $\phi_B$,
$\phi_{M_{1}}$, $\phi_{M_{2}}$. The heavy-to-light form factor $F$
is not factorized in the BBNS approach.}}}}
\end{figure}

According to the QCD Factorization
framework~\citer{QCDF:I,QCDF:III}, the hadronic matrix elements of
the operators in the effective weak Hamiltonian take the form
\begin{align} \label{eq:QCDF}
\langle M_1 M_2 | Q_i | \bar{B} \rangle &\simeq m_B^2 \; F_+^{B \to
M_1}(0) \; f_{M_2}
\int du \; \; T_{i}^I(u) \; \phi_{M_2}(u) \\
&  \quad + f_{B} \, f_{M_1} \, f_{M_2} \int d\om dv du \; \;
T_{i}^{II}(\om,v,u) \; \phi_B(\om)  \; \phi_{M_1}(v) \;
\phi_{M_2}(u), \no
\end{align}
where $M_1$ denotes the light meson which picks up the spectator
quark of the $\bar B$ meson and $M_2$ is a second light meson. The
factorization formula is illustrated in Figure~\ref{fig:QCDF}. Let
us compare this factorization formula with the examples that we
discussed in the beginning of this chapter. First, we notice that
the factorization formula is again restricted to the leading power
in the HQE in $\LQCD/m_b$. Further, we have suppressed the
dependence on the factorization scale $\mu_F$ for simplicity here.
The perturbative information in (\ref{eq:QCDF}) is encoded in the
hard-scattering kernels $T^{I}$ and $T^{II}$ which also appeared in
the examples 2-4. The same is true for the light-cone distribution
amplitudes of the light mesons $\phi_{M_1},\phi_{M_2}$ whereas the
one of the $B$ meson $\phi_{B}$ entered the factorization formula in
example 4. The distribution amplitudes always come in combination
with decay constants $f_B$, $f_{M_1}$, $f_{M_2}$ which have been
suppressed in Section~\ref{sec:QCDF} for simplicity.

The last piece in (\ref{eq:QCDF}) is the heavy-to-light form factor
$F_+^{B \to M_1}$ at maximum recoil. The form factor still contains
perturbative \emph{and} non-perturbative effects and therefore does
not fit into the pattern that we developed in
Section~\ref{sec:QCDF}. The form factor itself can be factorized as
we discuss in Section~\ref{sec:BtoPi}, but this is not needed here.
We follow the BBNS analysis which treats the form factor as an
external input with its numerical value taken from a Light-Cone QCD
Sum Rule calculation.

\newabs
In leading order (LO) of the perturbative expansion, the
factorization formula simplifies tremendously since $T^I =
\text{const} + \calO(\as)$ and $T^{II}=\calO(\as)$. The second term
in (\ref{eq:QCDF}) is thus absent in this case and the convolution
in the first term simply gives the normalization of the distribution
amplitude.  This yields
\begin{align}
\langle M_1 M_2 | Q_i | \bar{B} \rangle \sim F_+^{B \to M_1}(0) \;
f_{M_2}. \label{eq:naive}
\end{align}
This approximation corresponds to \emph{naive factorization}. With
naive factorization we mean that a hadronic matrix element of a
local four-quark operator is split into two matrix elements of
bilinear quark currents, as e.g.~in
\begin{align}
\langle M_1 M_2 | Q_1^u | \bar{B} \rangle \sim \langle M_1 | (\bar u
b)_{V-A} | \bar{B} \rangle \; \langle M_2 | (\bar d u)_{V-A} |0
\rangle
\end{align}
with the first matrix element on the right-hand side giving the form
factor and the second one the decay constant as in (\ref{eq:naive}).
Naive factorization was used before QCD Factorization was
established and gave fairly good predictions which was surprising at
that time~\cite{NaiveFact:I,NaiveFact:II}. We now understand why
this is the case: Naive factorization corresponds to the leading
term in the combined expansion in $\as$ and $\LQCD/m_b$. The fact
that QCD Factorization reproduces naive factorization in this
expansion is non-trivial.

One obvious problem with naive factorization is that it does not
give rise to strong rescattering phases because of the lack of final
state interactions. In (\ref{eq:QCDF}) we see how this problem is
cured in QCD Factorization. At leading power all final state
interactions are encoded in the hard-scattering kernels $T^I$ and
$T^{II}$. The strong phases are thus predicted to be perturbative in
the QCD Factorization framework. Moreover, we just have seen that
strong phases are absent in LO of the perturbative expansion. To
summarize, strong phases are of $\calO(\as)$ and $\calO(\LQCD/m_b)$
i.e. generically small  in QCD Factorization.

\newabs
We conclude this section with a comment on the factorization proof
of (\ref{eq:QCDF}). The explicit calculation
in~\citer{QCDF:I,QCDF:III} showed that the factorization formula
holds to $\calO(\as)$. A recent analysis~\citer{BenekeJager,Volker}
in combination with our calculations in Chapter~\ref{ch:ImPart} and
\ref{ch:RePart} extends this explicit proof to $\calO(\as^2)$. So
far a rigorous proof to all orders in $\as$ is still missing. For
the similar (but simpler) case of $B\to D\pi$, which we introduced
in example 3 of Section~\ref{sec:QCDF}, a factorization proof has
been formulated in SCET and can be found in~\cite{Proof B->D Pi}. A
first step towards an all-order proof of (\ref{eq:QCDF}) has been
undertaken in~\cite{Proof B->Pi Pi}.

\subsection{Perturbative corrections}
\label{sec:BtoPiPi:PC}

The power of the factorization formula (\ref{eq:QCDF}) lies in the
fact that it allows for a systematic calculation of perturbative
corrections. They are contained in the hard-scattering kernels $T^I$
and $T^{II}$ which describe two different mechanisms to which we
refer as \emph{vertex corrections} and \emph{spectator
interactions}, respectively. Whenever the spectator antiquark in the
$\bar B$ meson enters the perturbative subgraph, this contribution
is assigned to $T^{II}$ otherwise to $T^{I}$. We now present the
status of the perturbative calculation and identify the
contributions that we address in Chapter~\ref{ch:ImPart} and
\ref{ch:RePart}.

\subsubsection{Tree Level}

\begin{figure}[t!]
\centerline{\parbox{13cm}{
\parbox[b]{4cm}{\includegraphics[width=3cm]{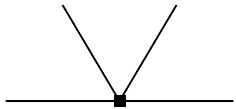}}
\parbox[b]{9cm}{
\caption{\label{fig:BtoPiPiTree} \small \textit{Tree level diagram.
The line to the left (right) of the vertex denotes the $b$ quark
(light quark which goes into $M_1$), the upper lines the light
quark/antiquark which form $M_2$. When the spectator does not
participate in the scattering, it is not drawn.}}}}}\vspace{5mm}
\end{figure}

We have already anticipated in the last section that QCD
Factorization reproduces naive factorization in LO of the
perturbative expansion. There is only one diagram that contributes
at this order which is depicted in Figure~\ref{fig:BtoPiPiTree}. As
the spectator antiquark does not participate in this case, this
corresponds to a contribution to $T^I$.

We quote the result in form of the topological amplitudes introduced
in (\ref{eq:parametr})
\begin{align}
\alpha_i^p(M_1 M_2) &= C_i + \frac{C_{i\pm1}}{N_c} + \calO(\as),
    \qquad \qquad i=1,2,4
\end{align}
where the upper (lower) signs apply when $i$ is odd (even) and the
superscript $p$ is to be omitted for $i=1,2$. To illustrate the
phrases \emph{colour-allowed} ($\alpha_1$) and
\emph{colour-suppressed} ($\alpha_2$), we have a look at the
numerical values of the Wilson coefficients $C_1(m_b)\sim1.1$ and
$C_2(m_b)\sim-0.2$ giving $\alpha_1\sim1.0$ and $\alpha_2\sim0.2$.
Notice that $\alpha_2$ is particularly small due to a cancellation
in the two terms.

\subsubsection{Next-to-leading order}

\begin{figure}[b!]
\centerline{\parbox{13cm}{\includegraphics[width=13cm]{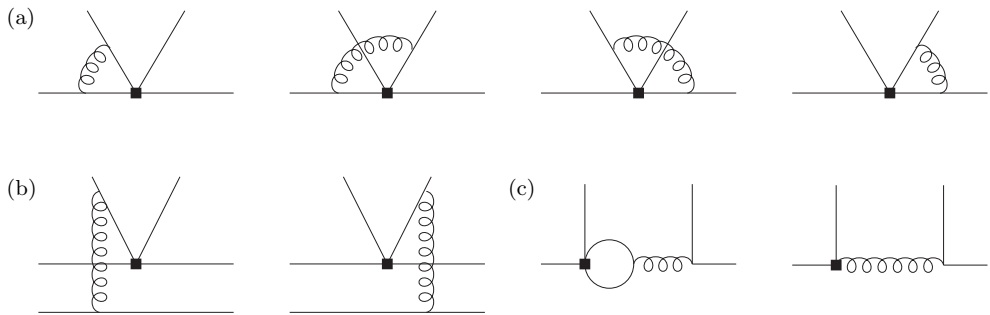}
\vspace{-2mm} \caption{\label{fig:BtoPiPiNLOnf} \small
\textit{Non-factorizable diagrams in NLO: Vertex corrections (a),
hard spectator interactions (b) and penguin contractions (c). The
last diagram involves an insertion of the dipole operator $Q_8$.}}}}
\end{figure}

\begin{figure}[t!]
\centerline{\parbox{13cm}{\includegraphics[width=13cm]{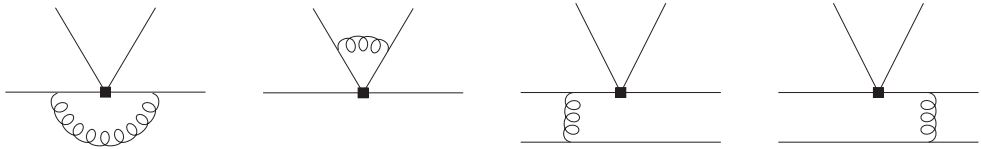}
\vspace{0mm} \caption{\label{fig:BtoPiPiNLOf} \small
\textit{Factorizable diagrams in NLO that need not to be
calculated.}}}}
\end{figure}

The next-to-leading order (NLO) corrections have been calculated by
BBNS~\citer{QCDF:I,QCDF:III}. They consist in the calculation of the
(naively) non-factorizable diagrams in Figure~\ref{fig:BtoPiPiNLOnf}
whereas the (naively) factorizable diagrams in
Figure~\ref{fig:BtoPiPiNLOf} turn out to be irrelevant. The
terminology is such that the former contain interactions between the
$B - M_1$ system and $M_2$ whereas the latter do not. An explanation
for the fact that only non-factorizable diagrams have to be
considered in this context is relegated to
Section~\ref{sec:BtoPiPi:FFinNNLO}. In addition to the vertex
corrections in Figure~\ref{fig:BtoPiPiNLOnf}a and the spectator
interactions in Figure~\ref{fig:BtoPiPiNLOnf}b, the penguin
contractions in Figure~\ref{fig:BtoPiPiNLOnf}c have to be taken into
account for the calculation of the penguin amplitude $\alpha_4^p$.
As the spectator is not involved in these diagrams, the respective
contribution is again assigned to $T^I$.

\newabs
At NLO the general form of the topological amplitudes $\alpha_1$,
$\alpha_2$ and $\alpha_4^p$ becomes
\begin{align}
\alpha_i^p(M_1 M_2) &= C_i + \frac{C_{i\pm1}}{N_c} + \frac{\as
C_F}{4\pi N_c} \Big[ C_{i\pm1} \, V^{(1)} + N_H \,C_{i\pm1} \,
H^{(1)} + \delta_{i4} \, P_p^{(1)} \Big] +\calO(\as^2),
\label{eq:alphasNLO}
\end{align}
where the functions $V^{(1)}/H^{(1)}/P_p^{(1)}$ stem from the
diagrams in Figure~\ref{fig:BtoPiPiNLOnf}a/b/c, respectively. They
contain convolutions of the hard-scattering kernels with light-cone
distribution amplitudes according to the factorization formula
(\ref{eq:QCDF}). The respective arguments $V^{(1)}(M_2)$,
$H^{(1)}(M_1 M_2)$ and $P_p^{(1)}(M_2)$ have been suppressed in
(\ref{eq:alphasNLO}) for simplicity. The normalization of the hard
spectator interactions reads
\begin{align}
N_H = 4\pi^2 \; \frac{f_B f_{M_1}}{N_c \, m_B \lambda_B F_+^{B\to
M_1}(0)}. \label{eq:NH}
\end{align}
In contrast to the remnant NLO diagrams, the spectator interactions
correspond to tree level diagrams at this order which is the origin
of the factor $4\pi^2$. The ratio of hadronic quantities involves
the quantity $\lambda_B=\calO(\LQCD)$ which is related to the $B$
meson distribution amplitude (the definition can be found in
(\ref{eq:NRLCDABmom})). Notice that the normalization scales as $N_H
\sim 1$ in the HQE and in the large $N_c$-expansion. One might
therefore expect the spectator interactions to dominate over the
vertex corrections and the penguin contributions because of the
factor $4\pi^2\sim40$. However, for realistic values of the hadronic
quantities this is not the case as $N_H\sim0.5-1.5$.

\newabs
In the following we quote the explicit NLO results for the vertex
corrections $V^{(1)}$ and the spectator interactions $H^{(1)}$. As
our calculations in Chapter~\ref{ch:ImPart} and \ref{ch:RePart} are
related to the topological tree amplitudes, the penguin
contributions $P_p^{(1)}$ are not needed for our purposes (they can
be found in~\cite{BenekeNeubert2003}). The NLO vertex corrections
can be written in the form
\begin{align}
V^{(1)}(M_2) &= - 6 \ln \frac{\mu^2}{m_b^2} - 18 + \int_0^1 du \;
g(u) \; \phi_{M_2}(u). \label{eq:V1trad}
\end{align}
The scale dependence exactly matches the one of the Wilson
coefficients in the LO terms in (\ref{eq:alphasNLO}) as desired. The
function $g(u)$ is found to be
\begin{align}
g(u) &=\frac{3(1-2u)}{\ubar} \ln u - 3 i \pi + \left[ 2 \Lib(u) -
\ln^2u -\frac{1-3u}{\ubar}\ln u -2i\pi\ln u - (u\tot \ubar) \right],
\end{align}
where we wrote $\ubar\equiv1-u$. The appearance of an imaginary part
in $g(u)$ is particularly interesting since it is the origin of a
strong phase shift between the final state mesons. The penguin
contributions $P_p^{(1)}$ are another source of an imaginary part
whereas the tree-level spectator scattering gives a real
contribution which reads
\begin{align}
H^{(1)}(M_1 M_2) &= \lambda_B \int_0^\infty \! d\om \int_0^1 du
\int_0^1 dv \;\;
    \frac{\phi_B(\om)}{\om}  \; \frac{\phi_{M_1}(v)}{\bar v} \; \frac{\phi_{M_2}(u)}{\ubar}.
\end{align}
Let us make one remark concerning the typical scales of the
perturbative corrections. Following our terminology in
Table~\ref{tab:term}, the perturbative effects in $T^I$ stem from
hard modes ($k^2\sim m_b^2$) whereas $T^{II}$ contains hard as well
as hard-collinear effects \mbox{($k^2\sim\mu_{hc}^2\sim m_b\LQCD$)}.
The tree level spectator scattering in
Figure~\ref{fig:BtoPiPiNLOnf}b is in particular most naturally
associated to the hard-collinear scale which enhances this
contribution by a factor of $\as(\mu_{hc})/\as(m_b) \sim 2$ with
respect to the other contributions. This issue is hidden in our
notation in (\ref{eq:alphasNLO}). A correct treatment of the
perturbative scales favours a SCET analysis of the spectator
interactions in order to resum large logarithms $\ln\mu/m_b$ into
short-distance coefficient functions.

\subsubsection{Next-to-next-to-leading order}

The next-to-next-to-leading order (NNLO) calculation is to date
incomplete. It has first been addressed in the so-called \emph{large
$\beta_0$-limit} in~\cite{beta0:NeuPec,beta0:BurWil}. In this
approximation, only a very small subset of NNLO diagrams has to be
calculated, namely the ones with massless fermion-loops which give
rise to a factor $n_f$. At the end of the calculation, the fairly
motivated substitution $n_f \to -3\beta_0/2$ is made in the hope to
catch the main contribution of the full calculation in this way. It
must be admitted that the large $\beta_0$-limit sometimes gives a
good approximation. However, whether this is the case or not can
only be judged a posteriori when the full result is known.
Concerning the parts of our NNLO calculation, we will answer this
question in Chapter~\ref{ch:ImPart} and \ref{ch:RePart}.

\begin{figure}[b!]
\centerline{\parbox{13cm}{\includegraphics[width=13cm]{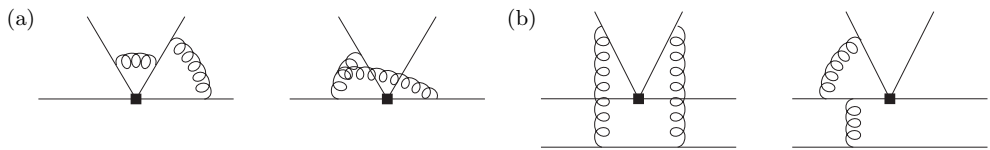}
\caption{\label{fig:BtoPiPiNNLO} \small \textit{Sample of NNLO
diagrams: Vertex corrections (a) and spectator interactions (b).}}}}
\end{figure}

\newabs
Up to the work in~\cite{LiYang}, which also considered a small and
incomplete subset of NNLO diagrams, the penguin amplitude
$\alpha_4^p$ has not yet been addressed. We therefore focus on the
tree amplitudes in the following and disregard all diagrams with
penguin contractions as well as insertions of the operators
$Q_{3-6}$ and  $Q_8$. We extend the general form of the tree
amplitudes $\alpha_1$ and $\alpha_2$ in NNLO to
\begin{align} \label{eq:alpha12NNLO}
\alpha_i(M_1 M_2) &= C_i + \frac{C_{i\pm1}}{N_c} + \frac{\as
C_F}{4\pi N_c} \Big[ C_{i\pm1} \, V^{(1)} + N_H \,C_{i\pm1} \,
H^{(1)}  \Big]  \\
&\hspace{5mm} + \frac{\as^2 C_F}{(4\pi)^2 N_c} \Big[ C_i V_1^{(2)} +
C_{i\pm1} V_2^{(2)} + N_H \Big( C_i H_1^{(2)} + C_{i\pm1}
H_2^{(2)}\Big) \Big] +\calO(\as^3). \no
\end{align} A first systematic study
of the NNLO corrections addressed the hard spectator interactions
$H_{1,2}^{(2)}$ which were calculated within SCET
in~\cite{BenekeJager}. A subsequent SCET calculation can be found
in~\cite{Kivel}. There is a tiny discrepancy\footnote{The corrected
version of~\cite{Kivel} is in agreement with~\cite{BenekeJager}.}
between both calculations which is claimed to be of minor numerical
importance~\cite{Kivel}. A third independent calculation of the
spectator interactions in full QCD may help to resolve the origin of
this discrepancy~\cite{Volker}. The QCD calculation consists in the
computation of $\sim50$ 1-loop diagrams as those depicted in
Figure~\ref{fig:BtoPiPiNNLO}b with its main complication due to the
presence of two perturbative scales $\mu_{hc}$ and $m_b$. This
problem is circumvented in SCET where the two scales are
systematically disentangled in the two-step matching procedure
$\text{QCD}\to\SCETI\to\SCETII$. Another advantage of the SCET
calculation lies in the fact that the hard-collinear effects in
$T^{II}$ are exactly the same as those entering the $B\to M_1$ form
factor~\cite{BPRSvsBBNS:I}. As they were already known at the 1-loop
level from a different analysis~\citer{BPiJet:BH,BPiJet:BY},
Beneke/J\"{a}ger and Kivel calculated the missing hard
contributions~\cite{BenekeJager,Kivel}.

\newabs
No such complication arises in the calculation of the vertex
corrections $V_{1,2}^{(2)}$ which we address in detail in this work.
However, the vertex corrections represent a challenging 2-loop
calculation including $\sim80$ diagrams as the ones in
Figure~\ref{fig:BtoPiPiNNLO}a. We divide the calculation into two
parts: We first focus on the imaginary part of $V_{1,2}^{(2)}$ in
Chapter~\ref{ch:ImPart} and calculate the real part subsequently in
Chapter~\ref{ch:RePart}. This is done for several reasons. First of
all, each part is independently well-defined and represents a
lengthy calculation on its own. It is natural to focus on a simpler
part of this highly demanding calculation in a first step. Second,
the imaginary part enters the topological amplitudes at $\calO(\as)$
which means that the imaginary part of $V_{1,2}^{(2)}$ has the
complexity of an effective NLO calculation. Conceptual complications
that arise in NNLO can therefore  be relegated to our analysis in
Chapter~\ref{ch:RePart}. Finally, the knowledge of the imaginary
part is particularly important in phenomenological applications as
it is the origin of a strong phase shift between the final state
mesons.

\newabs
One important remark is in order concerning the general form
(\ref{eq:alpha12NNLO}) of the tree amplitudes in NNLO. The reason
for this compact form which implies $\alpha_{1}\tot\alpha_2$ by the
exchange $C_1\tot C_2$ is related to Fierz symmetry arguments. A
Fierz reordering of a four-quark operator reshuffles the contracted
fields and thus the related flavours. Notice that $\alpha_1$ and
$\alpha_2$ interchange their roles under the exchange of the quark
flavours $u\tot d$ as shown in Figure~\ref{fig:alphas}. However, it
is no longer obvious if Fierz symmetry is preserved in the effective
theory at the level of the Wilson coefficients. The NLO calculation
showed the Fierz symmetric form (\ref{eq:alphasNLO}) to hold and we
therefore expect the same for the imaginary part of $V_{1,2}^{(2)}$
which has NLO complexity. In their SCET analysis of the spectator
interactions, Beneke and J\"{a}ger paid special attention to this
point and found that their result can indeed be written in the Fierz
symmetric form (\ref{eq:alpha12NNLO}). We have not yet solved this
question for the real part of $V_{1,2}^{(2)}$ which requires a Fierz
symmetric definition of evanescent operators at the 2-loop level. As
a consequence, we can only give a result for the real part of the
tree amplitude $\alpha_1$ in Chapter~\ref{ch:RePart} but so far we
cannot relate it to that of $\alpha_2$. More details concerning
Fierz symmetry and evanescent operators can be found in
Section~\ref{eq:BtoPiPi:OpBasis}.

\subsection{Power corrections}
\label{sec:power}

The factorization formula (\ref{eq:QCDF}) is exact in the formal
heavy quark limit $m_b\to\infty$. Unfortunately, power corrections
do not factorize in general and cannot be calculated systematically.
They represent the main limitation of QCD Factorization with a
generic size of $\calO(\LQCD/m_b)\sim5-15\%$. The reader may wonder
why we perform highly non-trivial NNLO calculations that are
accompanied by substantial systematic uncertainties. Our ultimate
goal is to look for deviations between experimental data and the QCD
Factorization prediction and to address them to effects from New
Physics rather than to presumably sizeable power corrections.

\newabs
From our point of view QCD Factorization is nevertheless the most
convincing theory to study non-leptonic $B$ decays and we should
therefore reduce all uncertainties as much as possible. Calculable
higher order perturbative corrections can also turn out to be
sizeable and it would be erroneous to associate them with unknown
power corrections. The rich phenomenology and the interplay with
alternative approaches to non-leptonic $B$ decays may help us to
test QCD Factorization and to get an estimate of the size of power
corrections on the one hand and to focus on the interesting CP
violating observables on the other. Rather than aiming at a perfect
global description of all decay channels, we are convinced that we
can achieve the necessary precision in selected observables. We also
refer to our discussion after (\ref{eq:parametr}) in this context,
concerning the question why some decay channels are better suited
for a QCD Factorization prediction than others.

\newabs
Obviously, it would be a significant improvement if we could get
some handle about power corrections. In the remainder of this
section we briefly recapitulate how BBNS implement certain classes
of power corrections. The price to pay for this procedure is that
their predictions become model-dependent in this way. For a
model-independent analysis we refer to~\cite{FeldmannHurth} which
contains a formal classification of power corrections according to
factorizable and non-factorizable operators in SCET.

\subsubsection{Chirally enhanced contributions}

One class of power corrections, which is related to the projection
on higher twist distribution amplitudes of the light mesons, appears
to be enhanced by large numerical coefficients. These corrections
are typically proportional to
\begin{align}
r_\chi^\pi(\mu) = \frac{2m_\pi^2}{m_b(\mu)(m_u+m_d)(\mu)},
\end{align}
which is formally of $\calO(\LQCD/m_b)$ but numerically close to
unity. Fortunately, an important part of this contribution, the
so-called \emph{scalar penguins}, turn out to be factorizable (at
least to NLO) and can be calculated systematically within QCD
Factorization. This results in an additional contribution to the
penguin amplitude $\alpha_4^p$ which is currently known to NLO and
can be found in~\cite{BenekeNeubert2003}.

\newabs
In contrast to this the twist-3 projections related to the spectator
interactions in Figure~\ref{fig:BtoPiPiNLOnf}b do not factorize
which can be seen as follows. In this case, the contributions to
$\alpha_1$, $\alpha_2$ and $\alpha_4^p$ can be written as
\begin{align}
\delta \alpha_i^p(M_1 M_2) &= \frac{\as C_F}{4\pi N_c} \; C_{i\pm1}
\, N_H \; r_\chi^{M_1} \, \lambda_B\,  \int_0^\infty \! d\om
\int_0^1 du \int_0^1 dv \;\;  \frac{\phi_B(\om)}{\om}  \;
\frac{\phi_{m_1}(v)}{\bar v} \; \frac{\phi_{M_2}(u)}{u},
\label{eq:alphaChi}
\end{align}
which involves the twist-3 distribution amplitude of the meson $M_1$
with its asymptotic form $\phi_{m_1}(v)=1$. As it does not vanish at
$v=1$, the contribution in (\ref{eq:alphaChi}) is divergent which
states that factorization does not hold for this power-suppressed
contribution. BBNS propose a parametrization of the logarithmically
divergent integral in the form
\begin{align}
\int_0^1 dv \;\; \frac{\phi_{m1}(v)}{\bar v} \equiv X_H \qquad \to
\qquad X_H = \left(1 + \rho_H \, e^{i\phi_H}\right) \; \ln
\frac{m_B}{\Lambda_h} \label{eq:XH}
\end{align}
as the integral is expected to be regulated by a hadronic scale
$\Lambda_h =\calO(\LQCD)$. In addition, they allow for a complex
coefficient with $\rho_H=\calO(1)$ and an arbitrary phase $\phi_H$
which might be generated due to soft rescattering effects. Notice
that $X_H$ is treated to be \emph{universal}, i.e it does not depend
on the meson $M_1$ and is assumed to be the same for all topological
amplitudes in (\ref{eq:alphaChi}). In our analysis of the tree
amplitudes in Chapter~\ref{ch:ImPart} and \ref{ch:RePart}, we adopt
the BBNS treatment of non-factorizable chirally enhanced
contributions.

\subsubsection{Weak annihilation}

\begin{figure}[h!]
\vspace{10mm}
\centerline{\parbox{13cm}{\includegraphics[width=13cm]{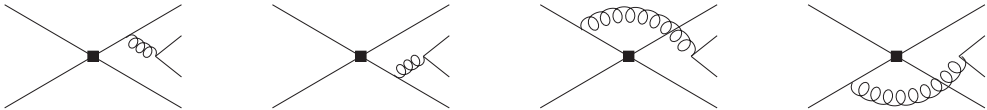}
\caption{\label{fig:BtoPiPiWA} \small \textit{Annihilation
diagrams.}}}}
\end{figure}

We briefly comment on a second class of power-suppressed effects:
weak annihilation. The annihilation diagrams from
Figure~\ref{fig:BtoPiPiWA} turn out to be suppressed by one power in
$\LQCD/m_b$. They exhibit similar endpoint divergences as the
chirally enhanced contributions and are therefore non-factorizable.
In the BBNS approach, they are parameterized in analogy to
(\ref{eq:XH}) by
\begin{align}
\qquad X_A = \left(1 + \rho_A \, e^{i\phi_A}\right) \; \ln
\frac{m_B}{\Lambda_h}.
\end{align}
Beneke and Neubert make a more relaxed assumption on the
universality of annihilation contributions
in~\cite{BenekeNeubert2003} and examine a specific scenario with
three different phases $\phi_A$ for the final states $PP$, $PV$ and
$VV$.

\newabs
Our final remark is related to recent investigations of the
annihilation amplitudes. They have been considered
in~\cite{RapFact:Ann} using a new type of factorization formula
which includes zero-bin subtractions in \SCETII~\cite{RapFact}.
These subtractions render the convolution integrals finite which
implies that the annihilation amplitudes become calculable. We will
come back to the formal aspects of this approach in
Section~\ref{sec:xi} and focus on the outcome for the annihilation
amplitudes here. Strong phases from annihilation are found
in~\cite{RapFact:Ann} to be of $\calO(\as^2(\mu_{hc}) \LQCD/m_b)$
and are therefore expected to be small in qualitative agreement with
predictions from a Light-Cone QCD Sum Rule
analysis~\cite{LCSR:IIIII}. However, whether or not this new type of
factorization formula represents a model-independent prediction for
the annihilation amplitudes is still a matter of debate.

\newpage
\section{Heavy-to-light form factors}
\label{sec:BtoPi}

Our calculation in Chapter \ref{ch:NRModel} is related to an
analysis of heavy-to-light form factors. These form factors encode
the strong interaction effects in exclusive semi-leptonic decays
(e.g.~$B\to\pi\ell\nu$), hadronic two-body decays
(e.g.~$B\to\pi\pi$) and exclusive radiative decays (e.g.~$B\to
K^*\gamma$). Rather than aiming at a quantitative description, our
analysis focuses on the conceptual aspects concerning factorization.
Heavy-to-light form factors are among the ''simplest'' objects for
studying the complicated QCD dynamics in exclusive charmless $B$
decays.

\subsection{Factorization Formula}

We concentrate on the three independent form factors in $B\to\pi$
transitions which can be defined as
\begin{align}
\langle \pi (p')\,|\,\bar q \gamma^\mu b \,|\, \bar B (p) \rangle &=
    F_+(q^2) \; (p^\mu + p'~\!^\mu) + F_-(q^2) \; q^\mu, \no \\
\langle \pi (p')\,|\,\bar q \sigma^{\mu\nu} q_\nu b \,|\, \bar B (p) \rangle &=
    \frac{i \, F_T(q^2)}{m_B + m_\pi} \; \left[ q^2(p^\mu + p'~\!^\mu)-(m_B^2-m_\pi^2)\, q^\mu\right]
\label{eq:defFF}
\end{align}
with $q=p-p'$. We are interested in the large recoil region $q^2\ll
m_b^2$ where the pion is very energetic with $E_\pi=\calO(m_b)$ in
the $B$ meson rest frame. The factorization formula has already been
introduced in example 4 at the beginning of this chapter. In a
slightly different notation, it reads
\begin{align}
F_i(q^2) &\simeq H_i(q^2) \; \xi(q^2) + \int_0^\infty \! d\om
\int_0^1 \! du \;\; \phi_B(\om) \; T_i(\om,u; q^2) \; \phi_\pi(u),
\label{eq:BtoPiFF}
\end{align}
where we made the kinematical dependence on the momentum transfer
explicit and suppressed the one on the factorization scale. We
recapitulate its interpretation in terms of our SCET terminology
from Table~\ref{tab:term}. The perturbative information is contained
in the hard coefficient function $H_i(q^2)$ and the hard-scattering
kernel $T_i(u,\om; q^2)$; the former including hard effects ($\sim
m_b^2$), the latter hard and hard-collinear effects ($\sim
m_b\LQCD$). The light-cone distribution amplitudes $\phi_B(\om)$ and
$\phi_\pi(u)$ encode soft and collinear effects, respectively, from
the scale $\LQCD$. Finally, the overlap-contribution $\xi(q^2)$ is
poorly understood so far and the main subject of our analysis. The
important point to notice is that the \emph{same} function
$\xi(q^2)$ enters the three $B\to\pi$ form factors (there are two
other overlap functions $\xi_\|(q^2)$ and $\xi_\perp(q^2)$ for the
seven $B\to V$ form factors). This is the basis of approximate
symmetry relations between different form factors~\cite{CharlesEtAl}
which are broken by perturbative corrections and power corrections.
Even if we consider the overlap-function $\xi(q^2)$ as unknown, the
factorization formula (\ref{eq:BtoPiFF}) represents a useful
simplification as the number of unknown hadronic quantities has been
reduced from three to one (counting the distribution amplitudes as
known hadronic quantities in this context as they are universal
quantities that can be analyzed in different processes).

\newabs
The structure of the factorization formula has first been seen to
emerge in~\cite{MBTF2000}. A rigorous factorization proof in SCET
has been formulated later in~\citer{Proof B->Pi:BPS,Proof B->Pi:LN}.
The proof is usually performed in two steps, matching first
$\text{QCD}\to\SCETI$~at the hard scale and subsequently
$\SCETI\to\SCETII$~at the hard-collinear scale. The two terms in
(\ref{eq:BtoPiFF}) are related to two different \SCETI~currents
which contribute in leading power to the form factors. What remains
to be shown is that one of them exhibits the symmetry properties
mentioned above and that the other gives rise to a finite
convolution integral of leading twist distribution amplitudes as in
the second term of the factorization formula (\ref{eq:BtoPiFF})
(more details will be given in Section~\ref{sec:NRFF}).

\newabs
We finally remark that the overlap-contribution $\xi(q^2)$ still
contains hard-collinear effects as only hard effects have been
factorized in the first matching step into $H_i(q^2)$. As long as we
treat the hard-collinear scale $(m_b \LQCD)^{1/2}$ as a perturbative
scale, the factorization of short- and long-distance effects is thus
incomplete.

\subsection{Closer look at $\xi(q^2)$}
\label{sec:xi}

We follow the notation of~\cite{Proof B->Pi:MBTF} by defining the
overlap-contribution as a matrix element of a \SCETI~current in the
form
\begin{align}
\langle \pi (p')\,|\,(\bar{\xi}_{hc} W_{hc}) h_v \,|\, \bar B (p)
\rangle &\equiv 2E_\pi \; \xi(q^2), \label{eq:xidef}
\end{align}
where $h_v$ denote the large components of the heavy quark field
satisfying $\slasha{v} h_v=h_v$ and $\xi_{hc}$ the ones of the
hard-collinear field with $\slashb{n}_- \xi_{hc}=0$. The definition
of the hard-collinear Wilson line $W_{hc}$ can be found
in~\cite{Proof B->Pi:MBTF}.

\newabs
The attempt to factorize $\xi(q^2)$ in \SCETII~into light-cone
distribution amplitudes leads to several complications: Sub-leading
twist projections and three-particle Fock states of the $B$ meson
and the pion are found to contribute at leading power. Even worse,
some convolutions turn out to be divergent at the endpoints which
renders the overlap-contribution incalculable similar to what we
have seen in our discussion about power corrections in the context
of non-leptonic decays in Section~\ref{sec:power}.

\newabs
The non-factorization of soft and collinear effects in \SCETII~and
the related question about the correct treatment and interpretation
of endpoint singularities is currently not completely understood.
The problem does not only appear in energetic $B\to\pi$ transitions
but also in the perturbative description of the pion form factor. In
this case the soft-overlap contribution (\emph{Feynman mechanism})
appears at sub-leading power. A solution within the effective theory
approach would be a significant breakthrough on the conceptual
level. Moreover, \emph{if} the soft-overlap contribution could be
expressed in terms of a few fundamental hadronic parameters, this
would immediately lead to many interesting applications. This is to
some extent realized in a recent approach from Manohar and
Stewart~\cite{RapFact} which we will discuss below. In any case, the
problem of endpoint-divergences in SCET deserves further study which
is the main motivation for our analysis in Chapter~\ref{ch:NRModel}.

\newabs
Let us briefly comment on two analyses of the overlap-contribution
$\xi(q^2)$ in SCET. The first one has been performed by Neubert et
al.~\cite{Proof B->Pi:LN} and is based on an extended formulation of
\SCETII~\cite{soft-coll messengers} which includes additional
degrees of freedom, so-called soft-collinear messenger modes, with
momentum scaling $k_{sc}\sim(\lambda^2,\lambda^3,\lambda^4)$. These
modes allow for a cross talk between the soft and the collinear
sector of the theory as they can be exchanged between soft and
collinear fields without bringing them far off-shell. Technically,
these modes serve as a regulator of the divergent convolution
integrals which leads to the following conclusion: If matrix
elements involving messenger fields contribute at leading power to a
hadronic matrix element, factorization into soft and collinear
effects is spoilt.

While messenger modes provide a technical solution to calculate
otherwise ill-defined quantities in perturbation theory, their
physical interpretation in non-perturbative matrix elements remains
obscure as virtualities with $k_{sc}^2\sim \LQCD^3/m_b$ are below
the confinement scale.

\newabs
A second analysis of the overlap-contribution has been performed by
Manohar and Stewart~\cite{RapFact}. They propose a new type of
factorization in \SCETII~which separates modes in their virtuality
$k^2$ and in their rapidity $k_-$. Their method is based on a
subtraction of zero-bin modes which is claimed to avoid double
counting of soft and collinear modes in the effective theory. As a
result a divergent convolution integral
\begin{align}
\int_0^1 du \;\; \frac{\phi_\pi(u)}{u^2}
\end{align}
is replaced by
\begin{align}
\int_0^1 du \;\; \frac{\phi_\pi(u)}{(u^2)_\emptyset} \;\equiv\;
\int_0^1 du \;\; \frac{\phi_\pi(u)-u \, \phi_\pi'(0)}{u^2} \;+\;
\phi_\pi'(0) \; \ln \frac{m_b}{\mu_-}, \label{eq:MS-distr}
\end{align}
where $\mu_-$ appears as the factorization scale in rapidity space.
In~\cite{RapFact} Manohar and Stewart present a factorization
formula for the overlap-contribution $\xi(q^2)$ which involves
leading and sub-leading twist as well as three-particle distribution
amplitudes together with their derivatives at the endpoints (without
the need to introduce soft-collinear messenger modes).

A still unresolved issue in this approach concerns the cancellation
of the $\mu_-$ dependence which requires a more careful analysis of
higher-order perturbative corrections beyond fixed-order
perturbation theory. It is also to be emphasized that the quantities
$\phi_\pi'(0)$, which naturally appear in this framework due to the
$\emptyset$-distributions in (\ref{eq:MS-distr}), have to be
interpreted as independent hadronic parameters since they cannot be
derived from a finite number of moments of the distribution
amplitude $\phi_\pi(u)$, see the argument in~\cite{LCSR:SCET}.
Unfortunately, the factorization formula loses some of its power in
this way since it involves a larger number of new hadronic
parameters. However, the approach from Manohar and Stewart has
appeared very recently and deserves further investigations.

\newabs
In our analysis in Chapter~\ref{ch:NRModel} we start from a
different viewpoint. We consider heavy-to-light form factors between
non-relativistic bound states which can be addressed in fixed-order
perturbation theory. As a consequence, all quantities in the
factorization formula (\ref{eq:BtoPiFF}), in particular the
overlap-contribution $\xi(q^2)$, can be calculated explicitly in our
set-up. Our results shed further light on the physics of the
soft-overlap contribution and also allow to address the proposals
mentioned above.

\chapter{Perturbative corrections}
\label{ch:PT}
 
In this chapter we address the technical aspects of the calculations
that we discuss in the second part of this thesis. Most of the
techniques that we present in the following have been developed
within the last ten years, reflecting the steadily increasing effort
in performing high precision calculations. We do not intent to give
a review on multi-loop techniques but rather give a presentation
from the perspective of our calculations in
Chapter~\ref{ch:ImPart},~\ref{ch:RePart} and~\ref{ch:NRModel} in the
hope that the reader can get an idea what these calculations looked
like in detail.

\section{Strategy}
\label{sec:strategy}

We start with an outline of the general strategy that we have used
to tackle the 1- and 2-loop calculations in
Chapter~\ref{ch:ImPart},~\ref{ch:RePart} and~\ref{ch:NRModel}. The
calculations have been performed on the basis of the computer
algebra system {\sc Mathematica}. Unless otherwise stated, all
routines have been written by ourselves. We emphasize that our
algorithm is not restricted to calculations in the Standard Model.
However, for calculations that are even more complex than our 2-loop
calculation in Chapter~\ref{ch:ImPart} and~\ref{ch:RePart}, the
efficiency of our algorithm should be improved. Our strategy
consists of the following four steps:

\subsubsection{Step 1: Set-up for loop calculation}

We deal with up to $80$ Feynman diagrams in our calculations. We
first examine the kinematics of the process and calculate the colour
factor of each diagram. In a 1-loop (2-loop) calculation we denote
the loop momentum by $k$ ($k,l$). The diagrams are expressed with
the help of a minimal set of denominators $\calP_i$ of propagators
which we specify in each of our calculations in the second part of
this thesis.

\newabs
We use a general tensor decomposition to express all tensor
integrals as linear combinations of scalar integrals $S_i$
multiplied by linearly independent tensors that can be formed out of
the external momenta $p_i$ and the metric tensor $g$. For example in
the case of two external momenta $p_1$ and $p_2$, a 1-loop integral
of rank $r=2$ becomes
\begin{align}
\int d^d k  \quad
        \frac{k^\mu k^\nu}
        {\calP_1^{n_1}\ldots \calP_p^{n_p}}\;\;\equiv\; m^2 g^{\mu \nu} \; S_1+
        p_1^\mu p_1^\nu \; S_2 + p_2^\mu p_2^\nu \; S_3 + (p_1^\mu p_2^\nu + p_1^\nu
        p_2^\mu) \; S_4,
\label{eq:examtens1}
\end{align}
reflecting the symmetry in $\mu \tot \nu$ of the tensor integral
which reduces the number of independent tensor structures from five
to four in this example. In contrast to this a 2-loop tensor
integral of rank $(r_k,r_l)=(1,1)$ is decomposed into
\begin{align}
\int d^d k  \, d^d l \quad
        \frac{k^\mu l^\nu}
        {\calP_1^{n_1}\ldots \calP_p^{n_p}}\;\;\equiv\; m^2 g^{\mu \nu} \; S'_1+
        p_1^\mu p_1^\nu \; S'_2 + p_2^\mu p_2^\nu \; S'_3 + p_1^\mu p_2^\nu \; S'_4 + p_1^\nu
        p_2^\mu \; S'_5,
\label{eq:examtens2}
\end{align}
since the tensor integral has no manifest symmetry in $\mu \tot \nu$
in this case. As the tensor decomposition becomes less trivial for
higher ranks of the loop momenta, we develop a systematic
decomposition in Section~\ref{sec:TensorDecomp}.

\newabs
On the other hand we will see below that the reduction algorithm
requires scalar projections $P_i$ of the tensor integrals as an
input. In our first example from (\ref{eq:examtens1}), we can form
four different projections given by
\begin{align}
\{P_1,P_2,P_3,P_4\}\;\equiv\;\;\int d^d k \quad
        \frac{\{(kp_1)^2,(kp_1 kp_2),(kp_2)^2,(k^2)\}}
        {\calP_1^{n_1}\ldots \calP_p^{n_p}},
\label{eq:examproj1}
\end{align}
whereas in the second example (\ref{eq:examtens2}), there are five
projections $P'_i$ which read
\begin{align}
\{P'_1,P'_2,P'_3,P'_4,P'_5\}\;\equiv\;\int d^d k \, d^d l \;\;
        \frac{\{(kp_1)(lp_1),(kp_1)(lp_2),(kp_2)(lp_1),(kp_2)(lp_2),(kl)\}}
        {\calP_1^{n_1}\ldots \calP_p^{n_p}}.
\label{eq:examproj2}
\end{align}
The scalar integrals $S_i$ and the projections $P_i$ are related by
a linear system of equations which can be solved easily. This allows
us to express each tensor integral in terms of projections $P_i$
which will be the subject of the second step.

\subsubsection{Step 2: Reduction to Master Integrals}

Our calculations are performed with the help of an automatized
reduction algorithm which allows us to express the projections $P_i$
of the general form
\begin{align}
\int d^d k \, d^d l \quad
        \frac{\calS_1^{m_1} \ldots \calS_s^{m_s}}
        {\calP_1^{n_1}\ldots \calP_p^{n_p}}
\label{eq:intRedtoMI}
\end{align}
in terms of a small set of so-called Master Integrals (MIs). The
$\calS_i$ in (\ref{eq:intRedtoMI}) denote scalar products of a loop
momentum with an external momentum or of two loop momenta as in the
examples in (\ref{eq:examproj1}) and (\ref{eq:examproj2}). The
reduction algorithm will be described in detail in
Section~\ref{sec:RedToMI}. We emphasize that it makes the use of
Dimensional Regularization (DR) mandatory for the calculation as
anticipated by our notation writing $d$-dimensional integration
measures.

Our interest to implement a reduction algorithm is threefold: First,
multi-loop calculations as the one that we present in
Chapter~\ref{ch:ImPart} and~\ref{ch:RePart} are extremely complex
including typically several thousands of integrals making an
automatization indispensable. Second, even in less complicated
1-loop calculations as the one in Chapter~\ref{ch:NRModel} the
algorithm is very helpful to avoid errors in the calculation as it
is basically reduced to the computation of some MIs. Finally, the
algorithm provides a powerful tool that can be used in the future
for any kind of loop calculation as it is not based on certain
Feynman rules but on particle propagators.

\subsubsection{Step 3: Manipulation of Dirac structures}

With the loop integrations reduced to a minimal form, we consider
the Dirac structures of the diagrams in the third step. We use the
programme {\sc Tracer}~\cite{Tracer} for all manipulations
concerning the Dirac algebra. We express all diagrams in terms of a
minimal set of irreducible Dirac structures $D_i$. A generic diagram
$\calD_i$ takes the form
\begin{align}
\calD_i = \sum_{j,k} \; A_{ijk}(d) \; \MI_j \; D_k, \label{eq:calD}
\end{align}
where we indicated that the coefficients $A_{ijk}$ depend on the
dimension $d$. We do not go into the details concerning the
reduction of the Dirac structures here as it depends on the
considered calculation and represents in general only a minor
problem.

\newabs
The treatment of $\gamma_5$ in $d\neq4$ dimensions requires special
care. The matrix $\gamma_5$ enters our QCD calculations because of
the weak vertices and/or the projection on pseudoscalar meson
states. We apply the \emph{Naive Dimensional Regularization} (NDR)
scheme which treats $\gamma_5$ as a completely anticommutating
object. Despite the fact that this scheme is algebraically
inconsistent~\citer{NDR:Gamma5:I,NDR:Gamma5:III}, it leads to
correct results provided that we can avoid traces as $\Tr(\gamma^\mu
\gamma^\nu \gamma^\rho \gamma^\sigma \gamma_5)$~\cite{BurasWeisz}.

\subsubsection{Step 4: Calculation of Master Integrals}

The most difficult part finally consists in the calculation of the
MIs. We have devoted Section~\ref{sec:CalcMI} to this subject where
we present several advanced techniques that we have found useful in
our calculation. We are looking for a solution of the MIs in form of
an expansion in $\eps\equiv(4-d)/2$
\begin{align}
\text{MI}_{\,i} = \sum_{j} \; \frac{c_{ij}}{\eps^j}.
\end{align}
Expanding (\ref{eq:calD}) then determines the maximal order in the
expansion of the MI that is required for the calculation. In our
calculations, the expansion starts at most with double (quartic)
poles due to soft and collinear IR singularities at the 1-loop
(2-loop) level.

\section{Decomposition of Tensor Integrals}
\label{sec:TensorDecomp}

The tensor decomposition has already been illustrated by means of a
simple example in the last section. The procedure for more
complicated tensor integrals is straight-forward but starts to
become involved for higher ranks $(r_k,r_l)$ in the 2-loop case. We
therefore find it useful to present a systematic algorithm for the
decomposition of 1- and 2-loop tensor integrals which can easily be
extended to the multi-loop case.

\subsubsection{1-loop integrals}

We consider the decomposition of a 1-loop tensor integral of rank
$r$
\begin{align}
\int d^d k  \quad
        \frac{k^{\mu_1} \ldots k^{\mu_r}}
        {\calP_1^{n_1}\ldots \calP_p^{n_p}}
\label{eq:examtensgen}
\end{align}
according to $n$ external momenta $p_1, \ldots, p_n$. Notice that
the integral is totally symmetric under the exchange of any pair of
indices $\mu_i\tot\mu_j$.

\newabs
We start with all tensors that can be formed out of $r$ external
momenta $p_i$. In total there are $n^r$ tensors of this type with
$\mybinom{n+r-1}{r}$ totally symmetric combinations. E.g. for $n=2$
and $r=3$ this results in $\mybinom{4}{3}=4$ totally symmetric
tensors which read
\begin{align}
k^{\mu_1} k^{\mu_2} k^{\mu_3} \quad\to\qquad & p_1^{\mu_1}
p_1^{\mu_2} p_1^{\mu_3}, \qquad p_1^{\mu_1} p_1^{\mu_2}
p_2^{\mu_3}+p_1^{\mu_1} p_2^{\mu_2} p_1^{\mu_3}+p_2^{\mu_1}
p_1^{\mu_2} p_1^{\mu_3},
\no \\
& p_1^{\mu_1} p_2^{\mu_2} p_2^{\mu_3}+p_2^{\mu_1} p_1^{\mu_2}
p_2^{\mu_3}+p_2^{\mu_1} p_2^{\mu_2} p_1^{\mu_3}, \qquad p_2^{\mu_1}
p_2^{\mu_2} p_2^{\mu_3}.
\end{align}
If $r\geq2$, we continue with all tensors that can be formed out of
$r-2$ external momenta $p_i$ and one insertion of the metric tensor
$g$. From the external momenta, we obtain $n^{r-2}$ combinations.
Further, there are $\mybinom{r}{r-2}$ possible insertions of the
metric tensor giving $\mybinom{r}{r-2}n^{r-2}$ tensors of this type.
From these, $\mybinom{n+r-3}{r-2}$ are totally symmetric. In our
example with $n=2$ and $r=3$ we thus find $\mybinom{2}{1}=2$ totally
symmetric tensors given by
\begin{align}
k^{\mu_1} k^{\mu_2} k^{\mu_3} \quad\to\qquad & g^{\mu_1 \mu_2}
p_1^{\mu_3} +  g^{\mu_3 \mu_1} p_1^{\mu_2} + g^{\mu_2 \mu_3}
p_1^{\mu_1}, \no \\
& g^{\mu_1 \mu_2} p_2^{\mu_3} +  g^{\mu_3 \mu_1} p_2^{\mu_2} +
g^{\mu_2 \mu_3} p_2^{\mu_1}.
\end{align}
The procedure continues if $r\geq4$ with $r-4$ external momenta
$p_i$ and two insertions of the metric tensor $g$. We then find
$\mybinom{n+r-5}{r-4}$ totally symmetric combinations etc.

\newabs
We conclude that the tensor integral can be decomposed in this way
according to
\begin{align}
T(n;r) = \sum_{j=0}^{[r/2]} \;\; \mybinomL{n+r-1-2j}{r-2j}
\end{align}
totally symmetric tensors where $[x]$ denotes the greatest integer
less than or equal to $x$. The proof follows by complete induction,
but we refrain from presenting it here. We refer to
Table~\ref{tab:tensdecomp1} for examples that are relevant in
typical 1-loop calculations.

\begin{table}[t!]
\centerline{
\parbox{13cm}{\setlength{\doublerulesep}{0.1mm}
\centerline{\begin{tabular}{|c||c|c|c|c|c|} \hline
\hspace*{1.1cm}&\hspace*{1.1cm}&\hspace*{1.1cm}&\hspace*{1.1cm}&\hspace*{1.1cm}&\hspace*{1.1cm} \\[-0.7em]
$n\,\backslash\,r$ & 0 & 1 & 2 & 3 & 4 \\[0.3em]
\hline\hline&&&&& \\[-0.7em]
1 & 1 & 1 & 2 & 2 & 3\\[0.3em]
\hline&&&&&\\[-0.7em]
2 & 1 & 2 & 4 & 6 & 9 \\[0.3em]
\hline&&&&&\\[-0.7em]
3 & 1 & 3 & 7 & 13 & 22 \\[0.3em]
\hline
\end{tabular}} \vspace{4mm} \caption{\label{tab:tensdecomp1}\small
\textit{Number $T(n;r)$ of totally symmetric tensors of rank $r$
that can be formed out of $n$ external momenta and the metric tensor
(relevant for the decomposition of 1-loop tensor integrals). The
cases $T(2;2)$ and $T(2;3)$ have been discussed explicitly in
Section~\ref{sec:strategy} and~\ref{sec:TensorDecomp},
respectively.}}}}
\end{table}

\subsubsection{2-loop integrals}

We come to the decomposition of a 2-loop tensor integral of rank
$(r_k,r_l)$
\begin{align}
\int d^d k  \, d^d l \quad
        \frac{k^{\mu_1} \ldots k^{\mu_{r_k}} \; l^{\nu_1} \ldots
        l^{\nu_{r_l}}}{\calP_1^{n_1}\ldots \calP_p^{n_p}}
\label{eq:examtensgen2}
\end{align}
according to $n$ external momenta $p_1, \ldots, p_n$. In this case,
the integral is totally symmetric under the exchange of any pair of
indices $\mu_i\tot\mu_j$ and $\nu_i\tot\nu_j$.

\newabs
We start with all products that can be formed out of a
$r_k$-dimensional tensor with indices $\mu_1,\ldots,\mu_{r_k}$ and
$r_l$-dimensional tensor with $\nu_1,\ldots,\nu_{r_l}$. From our
analysis of 1-loop tensor integrals we conclude that we can form
$T(n;r_k)\,T(n;r_l)$ tensors in this way which have the desired
symmetry properties. To illustrate what we have obtained so far, we
reconsider the case $n=2$ and $(r_k,r_l)=(1,1)$ which we introduced
in Section~\ref{sec:strategy}. In this case, there are five
independent tensors given by
\begin{align}
k^\mu l^\nu \quad\to\qquad p_1^\mu p_1^\nu, \quad p_2^\mu p_2^\nu,
\quad p_1^\mu p_2^\nu, \quad p_1^\nu p_2^\mu, \quad g^{\mu \nu},
\end{align}
whereas $T(2;1)\,T(2;1)=4$ which corresponds to the first four
tensors, only. We see that we have to add all contributions with
metric tensors connecting the $\mu_i$ with the $\nu_j$. There are
$T(n;r_k-1)\,T(n;r_l-1)$ combinations with one metric tensor
respecting the symmetry constraint, $T(n;r_k-2)\,T(n;r_l-2)$ with
two metric tensors etc.

\newabs
We conclude that the tensor integral in (\ref{eq:examtensgen2}) can
be decomposed according to
\begin{align}
T(n;r_k,r_l) = \sum_{j=0}^{\text{min}(r_k,r_l)} \;\; T(n;r_k-j) \,
T(n;r_l-j),
\end{align}
tensors respecting the symmetry in $\mu_i\tot\mu_j$ and
$\nu_i\tot\nu_j$. We again refrain from presenting a proof and give
an explicit example for illustration. Let us consider the case $n=2$
and $(r_k,r_l)=(2,1)$. We find 10 independent tensors given by
\begin{align}
k^{\mu_1} k^{\mu_2} l^\nu \quad\to\qquad & p_1^{\mu_1} p_1^{\mu_2}
p_1^\nu, \qquad p_1^{\mu_1} p_1^{\mu_2} p_2^\nu, \qquad p_1^{\mu_1}
p_2^{\mu_2} p_1^\nu + p_2^{\mu_1} p_1^{\mu_2} p_1^\nu, \no \\
& p_1^{\mu_1} p_2^{\mu_2} p_2^\nu + p_2^{\mu_1} p_1^{\mu_2} p_2^\nu,
\qquad p_2^{\mu_1} p_2^{\mu_2} p_1^\nu, \qquad p_2^{\mu_1}
p_2^{\mu_2}
p_2^\nu, \no \\
& g^{\mu_1 \nu} p_1^{\mu_2}+ g^{\mu_2 \nu} p_1^{\mu_1}, \qquad
g^{\mu_1 \mu_2} p_1^{\nu},\no \\
& g^{\mu_1 \nu} p_2^{\mu_2}+ g^{\mu_2 \nu} p_2^{\mu_1}, \qquad
g^{\mu_1 \mu_2} p_2^{\nu}
\end{align}
according to $T(2;2,1)=4\cdot2+2\cdot1=10$. We refer to
Table~\ref{tab:tensdecomp2} for further examples that are relevant
in typical 2-loop calculations.

\begin{table}[h!]
\vspace{12mm} \centerline{
\parbox{13cm}{\setlength{\doublerulesep}{0.1mm}
\centerline{\begin{tabular}{|c||c|c|c|c|c|} \hline
\hspace*{1.1cm}&\hspace*{1.1cm}&\hspace*{1.1cm}&\hspace*{1.1cm}&\hspace*{1.1cm}&\hspace*{1.1cm} \\[-0.7em]
$r_k\,\backslash\,r_l$ & 0 & 1 & 2 & 3 & 4 \\[0.3em]
\hline\hline&&&&& \\[-0.7em]
0 & 1 & 1 & 2 & 2 & 3\\[0.3em]
\hline&&&&&\\[-0.7em]
1 & 1 & 2 & 3 & 4 & 5 \\[0.3em]
\hline&&&&&\\[-0.7em]
2 & 2 & 3 & 6 & 7 & 10 \\[0.3em]
\hline
\end{tabular}}\vspace{8mm}
\centerline{\begin{tabular}{|c||c|c|c|c|c|} \hline
\hspace*{1.1cm}&\hspace*{1.1cm}&\hspace*{1.1cm}&\hspace*{1.1cm}&\hspace*{1.1cm}&\hspace*{1.1cm} \\[-0.7em]
$r_k\,\backslash\,r_l$ & 0 & 1 & 2 & 3 & 4 \\[0.3em]
\hline\hline&&&&& \\[-0.7em]
0 & 1 & 2 & 4 & 6 & 9\\[0.3em]
\hline&&&&&\\[-0.7em]
1 & 2 &5  & 10 & 16 & 24 \\[0.3em]
\hline&&&&&\\[-0.7em]
2 & 4 & 10 & 21 & 34 & 52 \\[0.3em]
\hline
\end{tabular}} \vspace{4mm} \caption{\label{tab:tensdecomp2}\small
\textit{Number $T(n;r_k,r_l)$ of tensors that can be formed out of
$n$ external momenta and the metric tensor which are totally
symmetric in the first $r_k$ and in the last $r_l$ indices. The
tables correspond to $n=1$ and $n=2$ and are relevant for the
decomposition of 2-loop tensor integrals. The cases $T(2;1,1)$ and
$T(2;2,1)$ have been discussed explicitly in
Section~\ref{sec:strategy} and~\ref{sec:TensorDecomp},
respectively.}}}}
\end{table}

\newpage
\section{Reduction to Master Integrals}
\label{sec:RedToMI}

The reduction algorithm is a very important element in our strategy
for the calculation of higher order perturbative corrections. In
principle it can be used for the reduction of any (scalar) loop
integral but it reveals its full power only in the multi-loop case.
To give an illustration, it served in our 2-loop calculation from
Chapter~\ref{ch:ImPart} and~\ref{ch:RePart} to express about 6.000
integrals in terms of 36 MIs which are in addition much simpler than
the original integrals. Needless to say that this 2-loop calculation
would have been (almost) impossible without the implementation of
the reduction algorithm.

\newabs
The starting point is the representation of a scalar loop integral
in the form
\begin{align}
\int d^d k \, d^d l \quad
        \frac{\calS_1^{m_1} \ldots \calS_s^{m_s}}
        {\calP_1^{n_1}\ldots \calP_p^{n_p}}
\label{eq:intRedtoMICopy}
\end{align}
where the $\calS_i$ are scalar products of a loop momentum with an
external momentum or of two loop momenta. The $\calP_i$ denote the
denominators of propagators and the exponents fulfil $n_i, m_i \geq
0$. We focus on 2-loop integrals in the following with obvious
simplifications (extensions) in the 1-loop (multi-loop) case. Notice
that an integral can have different representations in terms of
$\{\calS,\calP,n,m\}$ because of the freedom to shift loop momenta
in DR. In fact the underlying topology, i.e.~the interconnection of
propagators and external momenta, uniquely defines the integral. In
the following we (loosely) use the word topology in order to
classify the integrals. An integral with $t$ different propagators
$\calP_i$ with $n_i>0$ is called a $t$-topology.

\subsubsection{Preliminaries}

We first classify the scalar products $\calS_i$ into reducible and
irreducible scalar products. The reducible scalar products can be
written as linear combinations of denominators $\calP_i$ which leads
to the first simplification. This is sometimes called
\emph{Passarino-Veltman Reduction}~\cite{PassVelt} which represents
a standard tool for the computation of loop integrals. Since it
appears as a preliminary step in our reduction algorithm, we briefly
illustrate the idea by means of a simple example.

\newabs
We consider the following $5$-topology
\begin{align}
\int d^d k \, d^d l \quad
        \frac{kp}
        {k^2 (k+p)^2 (k-l)^2 (l+p)^2 l^2} \quad \equiv \quad kp \;\;
        \parbox[c]{2cm}{\psfig{file=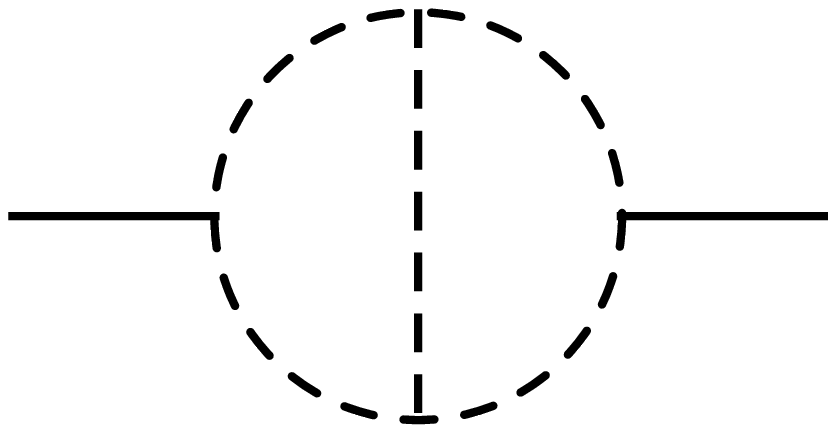,width = 2cm}}
\end{align}
where the right-hand side shows the underlying topology of the
integral. We use dashed internal lines for massless propagators. The
solid external line corresponds to the virtuality $p^2$ of the
incoming momentum $p$ in this case. As the figure on the right-hand
side reflects the denominator of the integral only, the scalar
products in the numerator appear explicitly in our notation.

\pagebreak The scalar product turns out to be reducible in this
case, as
\begin{align}
kp \;=\; \frac12 \Big[ (k+p)^2 - k^2 - p^2 \Big]
\end{align}
which gives rise to
\begin{align}
kp \;\; \parbox[c]{2cm}{\psfig{file=pic01.eps,width = 2cm}} \quad =
\quad \frac12 \;\;\parbox[c]{2cm}{\psfig{file=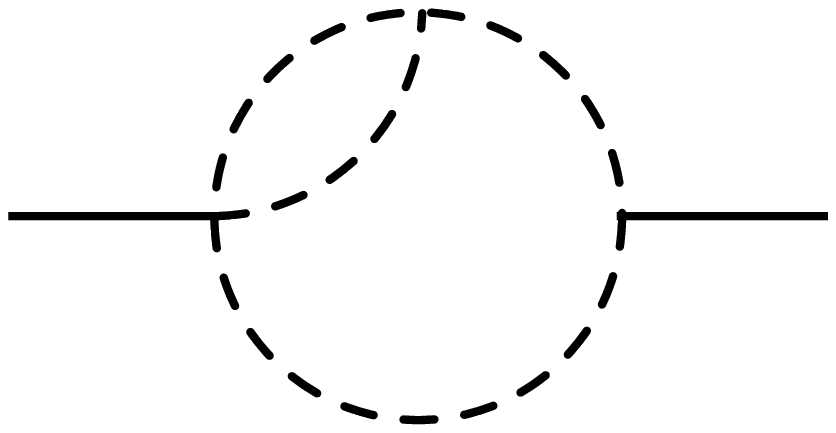,width =
2cm}}\quad - \frac12\;\;
\parbox[c]{2cm}{\psfig{file=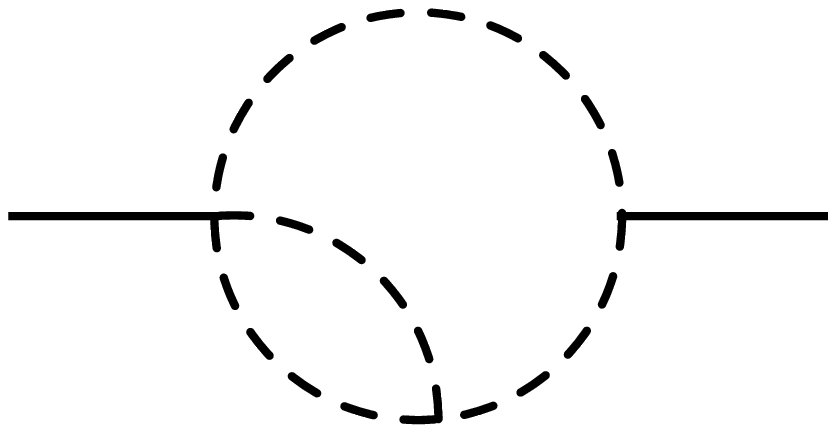,width = 2cm}}\quad - \frac{p^2}{2}\;\;
\parbox[c]{2cm}{\psfig{file=pic01.eps,width = 2cm}}
\label{eq:PassVeltaux}
\end{align}
Notice that the first two integrals on the right-hand side
correspond to simpler $4$-topologies and that the last one has a
trivial numerator. Moreover, the advantage of our notation becomes
obvious in (\ref{eq:PassVeltaux}) as it allows us to recognize
easily that the first two integrals are topologically equivalent.
This brings us to
\begin{align}
kp \;\; \parbox[c]{2cm}{\psfig{file=pic01.eps,width = 2cm}} \quad =
\quad - \frac{p^2}{2}\;\;
\parbox[c]{2cm}{\psfig{file=pic01.eps,width = 2cm}}
\end{align}
All reducible scalar products can be rewritten along these lines. We
may therefore concentrate on integrals of the form
(\ref{eq:intRedtoMICopy}) with irreducible scalar products in the
numerator. In the 1-loop case there are $n+1$ scalar products $S_i$
which can be formed out of $n$ external momenta and the loop
momentum: $\{kp_1, \ldots, kp_n,k^2\}$. For a 1-loop $t$-topology we
thus find $n+1-t$ irreducible scalar products. In the 2-loop case
the irreducible scalar products amount to $2n+3-t$. This implies
that in our example discussed above ($n=1$, $t=5$) all scalar
products $S_i$ are in fact reducible.

\subsubsection{Integration-by-parts identities}

The core of the reduction algorithm consists in the use of
integration-by-parts (IBP) identities~\cite{IBP:I,IBP:II} which
follow from the fact that surface terms vanish in DR
\begin{align}
\int d^d k \, d^d l \quad
        \frac{\partial}{\partial v^\mu} \; \;
        \frac{\calS_1^{m_1} \ldots \calS_s^{m_s}}
        {\calP_1^{n_1}\ldots \calP_p^{n_p}}  \;= \;0,
        \qquad \qquad v\in\{k,l\}.
\label{eq:IBP}
\end{align}
In order to obtain scalar identities one may contract (\ref{eq:IBP})
with any loop or external momentum under the integral before
performing the derivative. In the 1-loop (2-loop) case this
generates $n+1$ ($2n+4$) identities from each integral.

\newabs
We illustrate the use of IBP identities by means of the famous
example from Chetyrkin and Tkachov~\cite{IBP:I,IBP:II}. Let us
consider the integral
\begin{align}
\int d^d k \, d^d l \quad
        \frac{1}
        {k^2 (k+p)^2 (k-l)^2 (l+p)^2 l^2} \;\; \equiv \quad
        \parbox[c]{2cm}{\psfig{file=pic01.eps,width = 2cm}}
\end{align}
which can be reduced with the help of the following IBP identity
\begin{align}
\int d^d k \, d^d l \quad
        \frac{\partial}{\partial k^\mu} \; \; (k-l)^\mu \; \;
        \frac{1}{k^2 (k+p)^2 (k-l)^2 (l+p)^2 l^2}  \;=\; 0.
\end{align}
We perform the derivatives explicitly on the integrand which leads
to
\begin{align}
&d \;\; \parbox[c]{2cm}{\psfig{file=pic01.eps,width = 2cm}} \quad
-2k\cdot(k-l) \;\; \parbox[c]{2cm}{\psfig{file=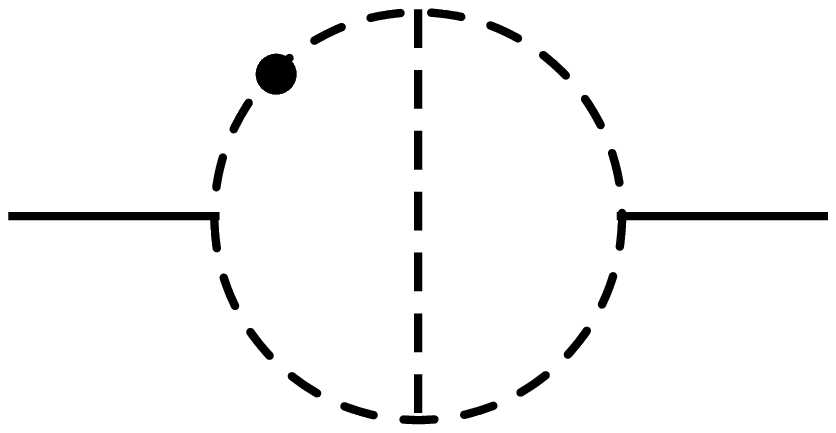,width =
2cm}} \no  \\[0.8em]
& \hspace{8mm} -2(k+p)\cdot(k-l)\;\;
\parbox[c]{2cm}{\psfig{file=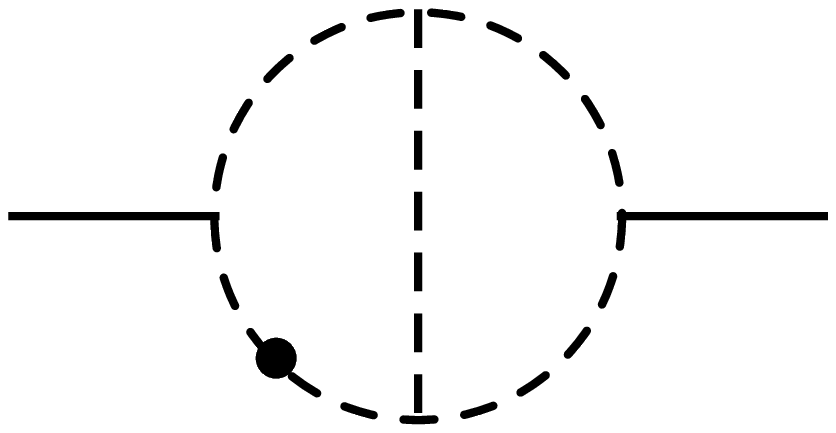,width = 2cm}} \quad
-2(k-l)^2\;\; \parbox[c]{2cm}{\psfig{file=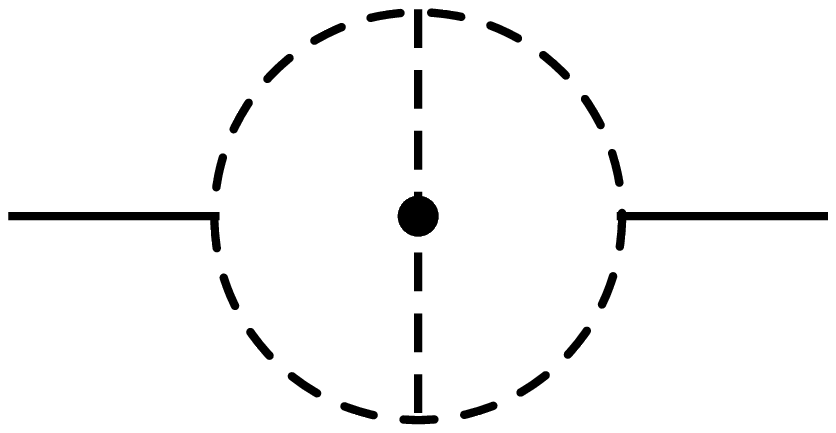,width = 2cm}}
\quad = \;\; 0 \label{eq:IBPaux1}
\end{align}
where the dotted propagators are taken to be squared. The scalar
products can be reduced with the help of the Passarino-Veltman
Reduction to give
\begin{align}
-2k\cdot(k-l) \;\; \parbox[c]{2cm}{\psfig{file=pic03a.eps,width =
2cm}} \;\; &= \;\; -2(k+p)\cdot(k-l)\;\;
\parbox[c]{2cm}{\psfig{file=pic03b.eps,width = 2cm}}\no \\[0.8em]
&= \;\; \parbox[c]{2cm}{\psfig{file=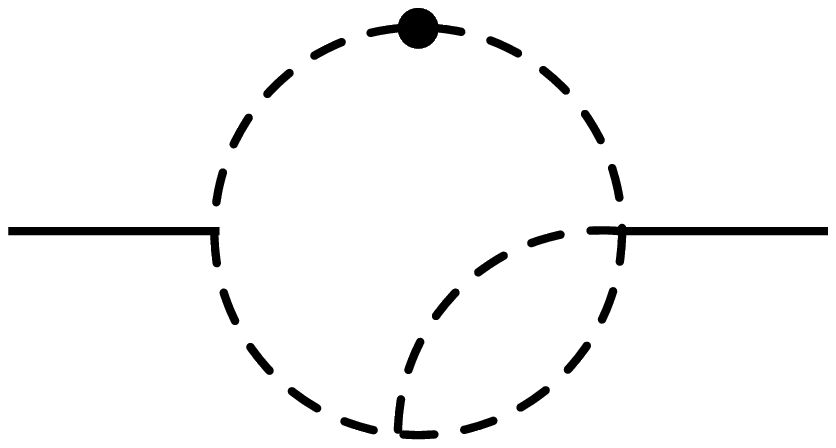,width = 2cm}} \quad -
\;\; \parbox[c]{2cm}{\psfig{file=pic01.eps,width = 2cm}} \quad -
\;\; \parbox[c]{2cm}{\psfig{file=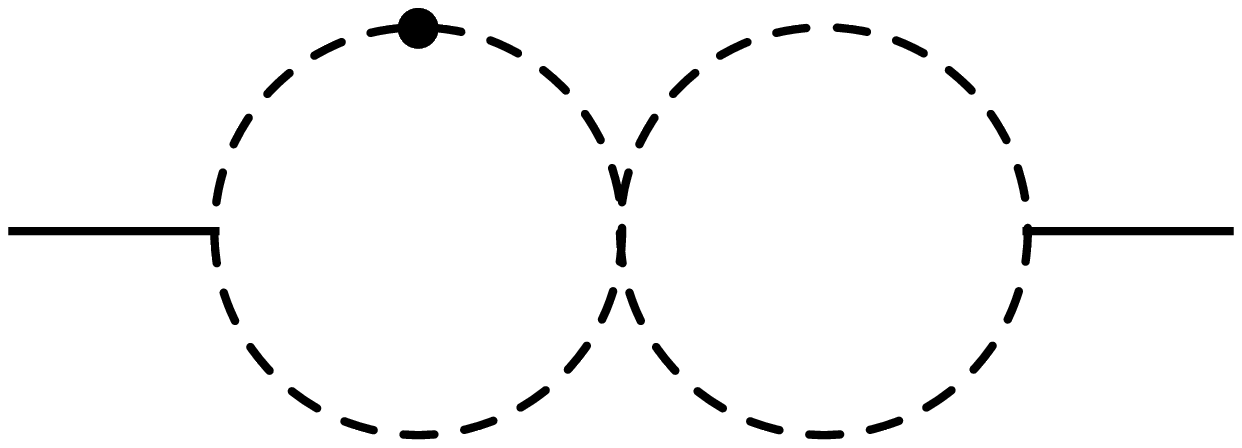,width = 2cm}} \no \\[0.8em]
-2(k-l)^2\;\; \parbox[c]{2cm}{\psfig{file=pic04.eps,width = 2cm}}
\;\;&= -2 \;\; \parbox[c]{2cm}{\psfig{file=pic01.eps,width = 2cm}}
\label{eq:IBPaux2}
\end{align}
Combining (\ref{eq:IBPaux1}) and (\ref{eq:IBPaux2}) yields the
famous result
\begin{align}
\parbox[c]{2cm}{\psfig{file=pic01.eps,width = 2cm}} \;\;&= \;\; \frac{2}{d-4} \bigg[ \quad
\parbox[c]{2cm}{\psfig{file=pic06.eps,width = 2cm}} \quad - \quad \parbox[c]{2cm}{\psfig{file=pic05.eps,width =
2cm}} \quad \bigg].
\end{align}
The original $5$-topology has thus been reduced to two
$4$-topologies, the first of them being particulary simple as it
represents a product of two 1-loop integrals.

\subsubsection{Lorentz-invariance identities}

It has been pointed out in~\cite{LI} that similar Lorentz-invariance
(LI) identities can be formulated which become important in problems
with many external momenta (multi-leg integrals). The LI identities
exploit the fact that the integrals in (\ref{eq:intRedtoMICopy})
transform as scalars under a Lorentz-transformation of the external
momenta. In this way one may generate up to six identities from each
integral depending on the number of external momenta.

As the LI identities are of minor importance in our calculations, we
refer to~\cite{LI} for a detailed description. In our applications
in Chapter~\ref{ch:ImPart},~\ref{ch:RePart} and~\ref{ch:NRModel}, we
deal with two external momenta $p, q$ which leads to one LI identity
per integral given by
\begin{align}
\int d^d k \, d^d l \;
        \left[ p \cdot q \left(p^\mu \frac{\partial}{\partial p^\mu} - q^\mu \frac{\partial}{\partial q^\mu} \right)
        + q^2 \, p^\mu \frac{\partial}{\partial q^\mu}
        - p^2 \, q^\mu \frac{\partial}{\partial p^\mu}
        \right] \;
        \frac{\calS_1^{m_1} \ldots \calS_s^{m_s}}
        {\calP_1^{n_1}\ldots \calP_p^{n_p}} \;=\; 0.
\label{eq:LI}
\end{align}

\subsubsection{Reduction algorithm}

As we have seen in our explicit example, the identities relate
various integrals of the form (\ref{eq:intRedtoMICopy}) with
different exponents $\{n,m\}$. In general each of the identities
contains the so-called \emph{seed integral} which has been used to
generate the identity, simpler integrals with smaller $\{n,m\}$ and
more complicated integrals with larger $\{n,m\}$. The identity is
then used to reduce the most complicated integral rather than the
seed integral itself.

\newabs
We do not want to inspect single identities but are looking for an
automatized reduction algorithm which exploits the full power of IBP
and LI identities. This can be achieved in a bottom-up approach
generating systematically all identities from all sub-topologies of
a given integral. In the 2-loop case, we thus start with the
identities from all $2$-topologies, $3$-topologies etc.~which leads
to a large set of identities which contains the integral that we are
interested in beneath simpler and more complicated ones. If we were
to use all unknown integrals that appear in these identities again
as seed integrals, the procedure would be without end as we generate
identities with more and more complicated integrals in each step. It
was an important observation from Laporta that the number of
identities grows faster than the number of unknown integrals in this
procedure. At some point we may therefore stop the generation of new
identities and solve the (apparently) over-constrained system of
equations by expressing more complicated integrals in terms of
simpler ones. Not all of the identities being linearly independent,
some integrals finally turn out to be irreducible to which we refer
as MIs.

The choice of the point where to stop the outlined procedure
requires thorough experimentation. On the one hand we want the
system of equations to be large enough to be sure that the remnant
integrals in the reduction procedure are indeed irreducible. On the
other hand the system should be as small as possible to assure an
optimized realization of the reduction algorithm. Based on his own
experiences Laporta proposed a ''golden rule'' for the choice of
such a cutoff in~\cite{LaportaAlg} which we have found very useful.

\newabs
In our calculation we typically deal with systems of equations made
of several thousands equations. The solution being straight-forward,
the runtime of the reduction algorithm depends strongly on the order
in which the equations are solved. As a guideline for an efficient
implementation we have followed the algorithm described
in~\cite{LaportaAlg}.

\newpage
\section{Calculation of Master Integrals}
\label{sec:CalcMI}

We now present a collection of techniques that we have used for the
calculation of the MIs. This last step in our strategy from
Section~\ref{sec:strategy} represents the most difficult part of the
perturbative calculation. We always intended to compute all MIs with
two independent methods in order to become sure of the correctness
of our results.

\newabs
We illustrate the calculation techniques with two explicit examples:
The first one corresponds to a 1-loop integral that appears in our
calculation from Chapter~\ref{ch:NRModel}, the second one to a
2-loop integral from Chapter~\ref{ch:ImPart} and~\ref{ch:RePart}.
Albeit this section has become somewhat lengthy in this way, we
think that it is instructive to present the calculation of these
integrals with different methods. In both cases we have chosen
simple (but non-trivial) examples which allow for a transparent
presentation of the central aspects of the techniques. Despite being
simple, the examples cover most of the conceptually interesting
features.

\subsubsection{1-loop example: Definition}

In the first example we consider a $3$-topology
\begin{align}
\int [d k] \;\;
        \frac{1}
        {[(m w'-m w + k)^2-m^2][k^2-m^2](k-m w')^2} \;\; \equiv \;\;
        \parbox[c]{2.5cm}{\vspace{-6mm}\psfig{file=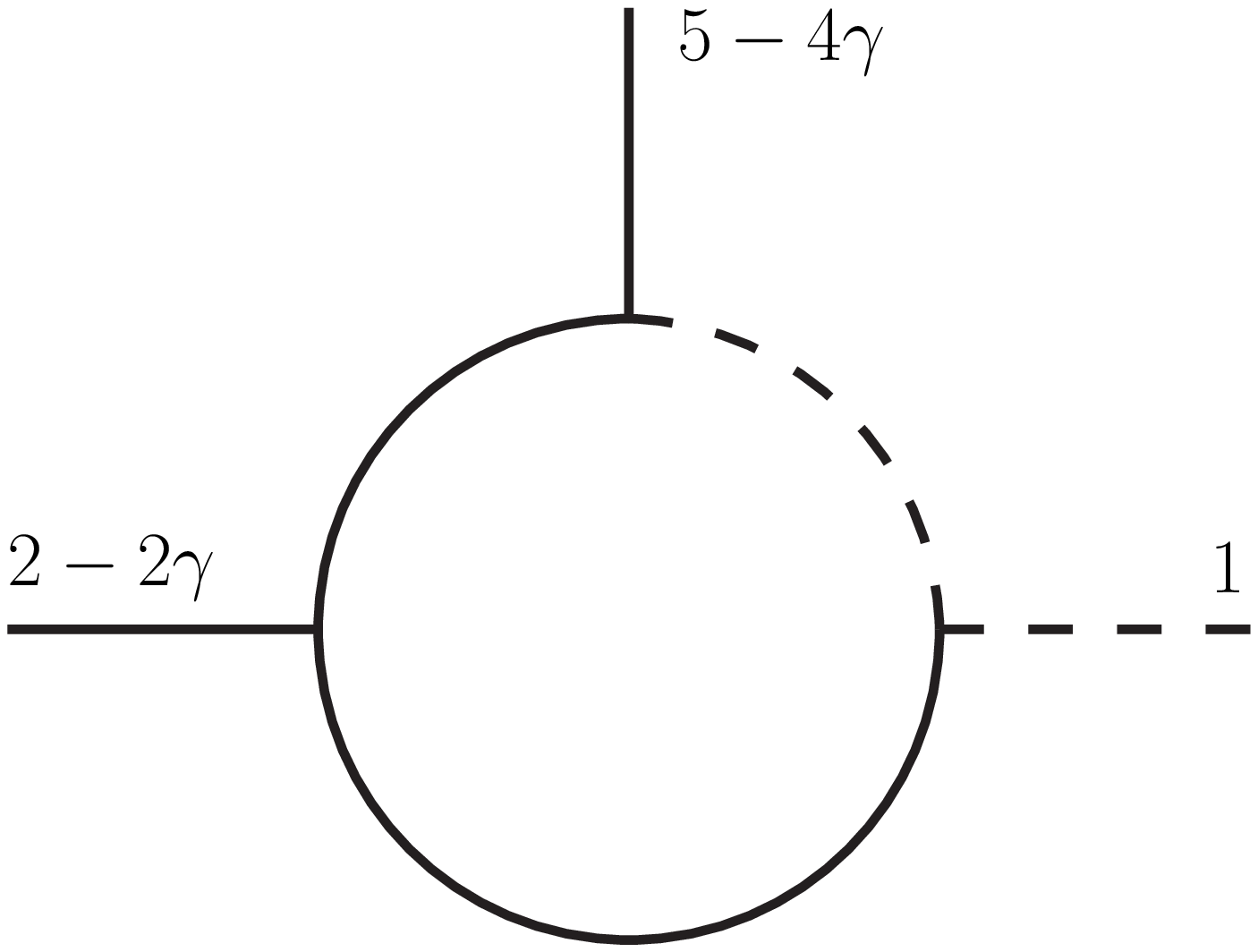,width = 2.5cm}}
\label{eq:1loopex}
\end{align}
with $w^2={w'}^2=1$ and $w\cdot w'=\gamma>1$. In our analysis in
Chapter~\ref{ch:NRModel}, we focus on the leading power in an
expansion in $m/M$ where $M$ is the mass of a heavy quark and $m$
the mass of a light (non-relativistic) quark. The parameter $\gamma$
corresponds to the large boost between the rest frames of the bound
states and scales as $\gamma=\calO(M/m)$. In (\ref{eq:1loopex}) we
used dashed/solid internal lines for massless/massive propagators.
Dashed/solid external lines denote virtualities of
$\calO(m^2)$/$\calO(m M)$ which are written explicitly here as
multiples of $m^2$. The normalization of the integral reads
\begin{align}
[dk] \equiv  \frac{\Gamma(1-\eps)}{i\pi^{d/2}}\; d^d k.
\label{eq:intmeasure}
\end{align}

\subsubsection{2-loop example: Definition}

In the 2-loop case we consider a massless $4$-topology
\begin{align}
\int [d k] \, [d l] \;\;
        \frac{1}
        {(p-q-l)^2(u q + l)^2(p-q-k-l)^2(\ubar q + k)^2} \;\; \equiv \;\;
        \parbox[c]{2cm}{\vspace{-5mm}\psfig{file=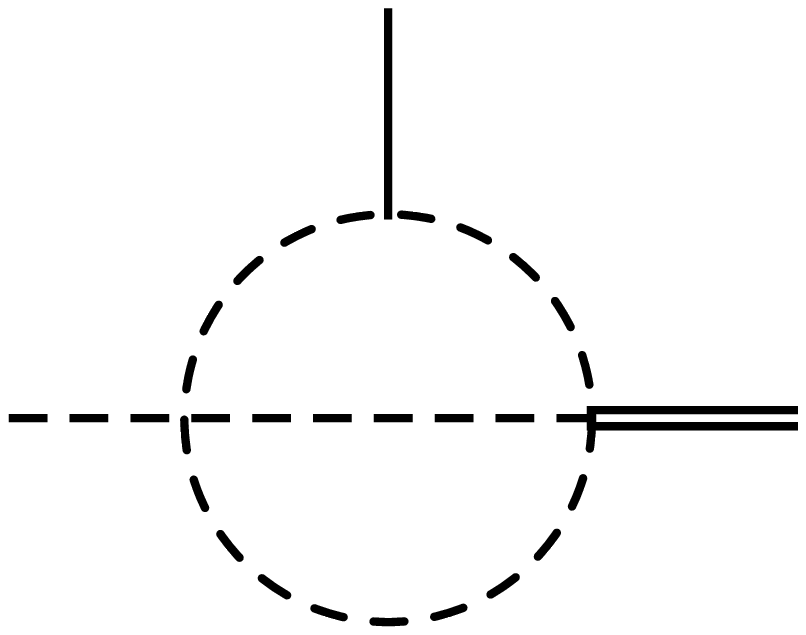,width = 2cm}}
\label{eq:2loopex}
\end{align}
with $0\leq u\leq1$, $\ubar=1-u$, $p^2=2p\cdot q=m^2$ and $q^2=0$.
The usual $i\eps$-prescription of propagators is understood.
Internal dashed lines correspond to massless propagators and
dashed/solid/double external lines to virtualities $0/u m^2/m^2$,
respectively\footnote{Notice that $m$ denotes the mass of the
$b$-quark in this case, whereas the light quarks are considered
massless.}.

\subsection{Feynman Parameters}
\label{sec:Feynman}

The standard method to compute loop integrals introduces Feynman
parameters. As this approach is not practicable for most of our MIs,
we do not describe this method here in detail. However, it turns out
that our simple examples can be solved with this technique. We
comment briefly on the calculations and quote their results which
will serve as a reference for the other calculations.

\subsubsection{1-loop example: Feynman Parameters}

As we deal with a $3$-topology, we introduce two Feynman parameters
and obtain
\begin{align} \no\\[-0.5em]
\parbox[c]{2cm}{\vspace{-4.5mm}\psfig{file=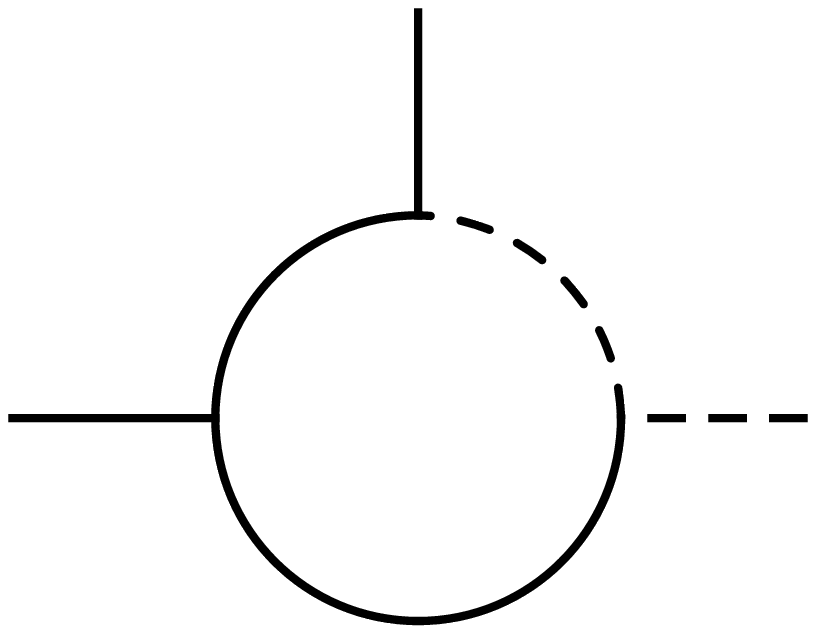,width = 2cm}}
\;\;&=\;\; \int_0^1 dx\int_0^{1-x}\!\!\!dy \;\; \frac{-
(m^2)^{-1-\eps} \; \Gamma(1-\eps)\, \Gamma(1+\eps)
}{\left[(5-4\gamma)x^2+y^2+2(2-\gamma)x y
+4(\gamma-1)x\right]^{1+\eps}}. \label{eq:1loopexSA}
\end{align}
The integral is ultraviolet (UV) and infrared (IR) finite. We focus
on the leading term in the $\eps$-expansion which allows us to set
$\eps=0$. The integral can then be solved in a closed form in $m/M$,
but we refrain from presenting the exact result since it looks
rather complicated. We are only interested in the leading power of
the $m/M$ expansion which takes a simple form
\begin{align} \no\\[-0.5em]
\parbox[c]{2cm}{\vspace{-4.5mm}\psfig{file=pic08.eps,width = 2cm}}
\;\; &\simeq \;\;-\frac{1}{4\gamma m^2} \left[2\ln 2 \ln \gamma +
\ln^2 2 +\frac{\pi^2}{3} + \calO(\eps) \right].
\label{eq:1loopexSARes}
\end{align}
Notice that we cannot expand the integrand in (\ref{eq:1loopexSA})
in $m/M$ before performing the parameter integrals as this would
invoke IR singularities for $x\to0$. In other words, the two
expansions in $\eps$ and $m/M$ do not commute with each other which
complicates the extraction of the leading power.

\subsubsection{2-loop example: Feynman Parameters}

In this example, we introduce three Feynman parameters but one of
these integrations can be done trivially. We obtain
\begin{align} \no\\[-1em] \label{eq:2loopexSA}
\parbox[c]{2cm}{\vspace{-5mm}\psfig{file=pic07.eps,width = 2cm}}
\;\;&=\;\; (m^2)^{-2\eps} \; e^{2i\pi \eps} \; \frac{\Gamma(2\eps)\,
\Gamma(1-\eps)^4}{\Gamma(2-2\eps)}  \\
& \hspace{2.2cm} \int_0^1 dx\int_0^{1-x}\!\!\!dy \quad x^{\eps-1}
(1-x-y)^{-2\eps} (x+u y)^{-2\eps}. \no
\end{align}
The integral exhibits a double pole in $\eps$ reflected by the
factor $\Gamma(2\eps)$ and a second singularity from $x\to0$. The
solution can be found in a closed form in $\eps$ in terms of an
hypergeometric function and reads
\begin{align} \no \\[-1em]
\parbox[c]{2cm}{\vspace{-5mm}\psfig{file=pic07.eps,width = 2cm}}
\;\;&=\;\; (m^2)^{-2\eps} \; e^{2i\pi \eps} \;
\frac{2\Gamma(-2\eps)\, \Gamma(1-\eps)^5}{\Gamma(2-2\eps)\,
\Gamma(2-3\eps)}\\
& \hspace{2.2cm}  \Big[ \Gamma(2\eps) \; _2F_1(1,2\eps;1+\eps;u) -
\eps \Gamma(\eps)^2 \, u^{-\eps} \ubar^{-\eps} \Big]. \no
\end{align}
The integral enters our calculations from Chapter~\ref{ch:ImPart}
and~\ref{ch:RePart} up to $\calO(\eps^2)$, but we restrict our
attention to the first three coefficients here. Notice that the
calculation of the imaginary part essentially requires one
coefficient less in the $\eps$-expansion. The hyper\-geometric
function can be expanded with the help of the {\sc Mathematica}
package {\sc HypExp}~\cite{HypExp} which yields
\begin{align}\no \\[-1em] \label{eq:2loopexSARes}
\parbox[c]{2cm}{\vspace{-5mm}\psfig{file=pic07.eps,width = 2cm}}
\;\;&=\;\; (m^2)^{-2\eps} \; e^{2i\pi \eps} \;   \bigg[
\frac{1}{2\eps^2} + \left( \frac52 - \ln u \right) \frac{1}{\eps} +
\frac{19}{2} - \frac{\pi^2}{6} \\
& \hspace{3.4cm}  + \Lib(u) +\frac12 \ln^2 u + \ln u \ln \ubar -5
\ln u + \calO(\eps) \bigg]. \no
\end{align}

\subsection{Method of Differential Equations}

The method of differential equations~\cite{DiffEqs:Kot,DiffEqs:Rem}
in combination with the formalism of harmonic polylogarithms
(HPLs)~\cite{HPL} turned out to be extremely useful for our 2-loop
calculation from Chapter~\ref{ch:ImPart} and~\ref{ch:RePart}. The
method is based on the reduction algorithm that we presented in
Section~\ref{sec:RedToMI} and can therefore only be applied with
such an algorithm at hand.

In the following we start with a general description of the method,
introduce the HPLs and comment briefly on the calculation of the
boundary conditions to the differential equations. Finally, we apply
this method to our explicit 2-loop example. We do not use this
technique in our 1-loop calculation from Chapter~\ref{ch:NRModel}
where we focus on the extraction of the leading power in $m/M$. We
therefore refrain from calculating our 1-loop example with this
method.

\subsubsection{Solving loop-integrals with differential equations}

The MIs are functions of the physical scales of the process which
are given by scalar products of the external momenta and masses of
the particles. In our calculations most of our MIs depend on two
scales which give rise to one dimensionless ratio: $\gamma$ ($u$) in
our 1-loop (2-loop) example. We therefore restrict our attention to
the case that there are only two distinct scales in the process with
obvious modifications for the general case. We denote the ratio of
the two scales by $u$ and recall the general form
(\ref{eq:intRedtoMICopy}) of a MI
\begin{align}
\MI_i(u)=\int d^d k \, d^d l \quad
        \frac{\calS_1^{m_1} \ldots \calS_s^{m_s}}
        {\calP_1^{n_1}\ldots \calP_p^{n_p}}
\label{eq:intRedtoMICopy2}
\end{align}
We perform the derivative with respect to $u$ and interchange the
order of integration and derivation
\begin{align}
\frac{\partial}{\partial u} \; \MI_i(u)
        =
        \int d^d k \, d^d l \quad
        \frac{\partial}{\partial u} \; \;
        \frac{\calS_1^{m_1} \ldots \calS_s^{m_s}}
        {\calP_1^{n_1}\ldots \calP_p^{n_p}}.
\end{align}
The right-hand side being of the same type as the IBP and LI
identities in (\ref{eq:IBP}) and (\ref{eq:LI}), this procedure again
leads to a sum of various integrals with different exponents
$\{n,m\}$. With the help of the reduction algorithm, these integrals
can be expressed in terms of MIs which yields a differential
equation of the form
\begin{align}
\frac{\partial}{\partial u} \; \text{MI}_{\,i}(u)
        =
        a(u;d) \; \text{MI}_{\,i}(u)
        + \sum_{j \neq i} b_j(u;d) \; \text{MI}_{\,j}(u),
\label{eq:DiffEq}
\end{align}
where we indicated that the coefficients $a$ and $b_j$ depend on the
dimension $d$. The inhomogeneity of the differential equation
typically contains MIs of sub-topologies which are supposed to be
known in a bottom-up approach. In some cases few MIs in the
inhomogeneous part are of the same topology as the MI on the left
hand side of (\ref{eq:DiffEq}) and thus unknown. Writing down the
differential equations for these MIs, we find that we are left with
a coupled system of linear, first order differential equations.

\newabs
We are looking for a solution of the differential equation in form
of an expansion
\begin{align}
\text{MI}_{\,i}(u) = \sum_j \; \frac{c_{ij}(u)}{\eps^j}.
\label{eq:Laurent}
\end{align}
Expanding (\ref{eq:DiffEq}) then gives much simpler differential
equations for the coefficients $c_{ij}$ which can be solved order by
order in $\eps$.  The solution of the homogeneous equations is in
general straight-forward. The inhomogeneous equations can then be
addressed with the method of the variation of the constant. This in
turn leads to indefinite integrals over the inhomogeneities which
typically contain products of rational functions with logarithms or
related functions as dilogarithms. With the help of the formalism of
HPLs these integrations simplify substantially.

\subsubsection{Harmonic Polylogarithms}

The HPLs have been introduced in~\cite{HPL} and several extensions
of the formalism have been considered
in~\citer{HPLext:I,HPLext:III}. We briefly summarize their basic
features here, focussing on the properties that are relevant for our
2-loop calculation in Chapter~\ref{ch:ImPart} and~\ref{ch:RePart}.

\newabs
The HPLs, denoted by $H(\vec{m}_w;x)$, are described by a
$w$-dimensional vector $\vec{m}_w$ of parameters and by its argument
$x$. In their simplest form, the parameters can take the values $0$
and $\pm1$. The basic definitions of the HPLs are for weight $w=1$
\begin{align}
H(0;x) &\equiv \ln x, \no \\
H(1;x) &\equiv - \ln (1-x), \no \\
H(-1;x) &\equiv \ln (1+x)
\end{align}
and for weight $w>1$
\begin{align}
H(a, \vec{m}_{w\text{--}1};x) &\equiv
        \int_0^x dx' \; f(a;x') \; H(\vec{m}_{w\text{--}1};x'),
\label{eq:HPL}
\end{align}
where the basic functions $f(a;x)$ are given by
\begin{align}
f(0;x) &\equiv \frac{d}{d x}\, H(0;x) = \frac{1}{x}, \no \\
f(1;x) &\equiv \frac{d}{d x}\, H(1;x) = \frac{1}{1-x}, \no \\
f(-1;x) &\equiv \frac{d}{d x}\, H(-1;x) = \frac{1}{1+x}.
\end{align}
In the case $\vec{m}_w=\vec{0}_w$, the definition in (\ref{eq:HPL})
does not apply and the HPLs read
\begin{align}
H(0, \ldots,0;x) &\equiv \frac{1}{w!} \ln^w x.
\end{align}
The HPLs form a closed, linearly independent set under integrations
over the basic functions $f(a;x)$ and fulfil an algebra such that a
product of two HPLs of weight $w_1$ and $w_2$ gives a linear
combination of HPLs of weight $w=w_1+w_2$.

\newabs
As anticipated above, the solution of the differential equation
typically leads to integrals over products of rational functions
with transcendental functions as logarithms or dilogarithms. More
precisely, we often encounter integrals of the type
\begin{align}
\int^x dx' \; \left\{ \frac{1}{x'}, \frac{1}{1-x'},
\frac{1}{1+x'}\right\} \; H(\vec{m}_{w};x') \label{eq:HPLints1}
\end{align}
which become trivial within the formalism of HPLs as they simply
correspond to an HPL of weight $w+1$ according to (\ref{eq:HPL}).
Further integrals take e.g. the form
\begin{align}
\int^x dx' \; \left\{ 1, \frac{1}{x'^2}, \frac{1}{(1-x')^2},
\frac{1}{(1+x')^2}\right\}  \; H(\vec{m}_{w};x').
\label{eq:HPLints2}
\end{align}
The solution of these integrals is also straight-forward within the
formalism of HPLs as an integration-by-parts relates them either to
an HPL of weight $w+1$ or gives rise to a simple recurrence
relation. Not surprisingly, the pattern in (\ref{eq:HPLints1}) and
(\ref{eq:HPLints2}) does not apply to all integrals that we
encounter in our 2-loop calculation from Chapter~\ref{ch:ImPart}
and~\ref{ch:RePart}. However, a large part of this lengthy
calculation can be performed along these lines.

\newabs
Furthermore, the formalism of HPLs helps us to avoid the problem of
''hidden zeros''. Let us illustrate this point with a simple
example. In our calculation we may obtain the result
\begin{align}
\left[ \Lib(u) + \Lib(\ubar) + \ln u \ln \ubar - \frac{\pi^2}{6}
\right] \frac{1}{\eps} \label{eq:hiddenzero}
\end{align}
and be surprised that it contains a singularity. A closer look
reveals that it is indeed free of any singularities since
(\ref{eq:hiddenzero}) vanishes identically. This can be seen easily
when we rewrite the second dilogarithm in the form
\begin{align}
\Lib(\ubar) = \frac{\pi^2}{6} - \ln u \ln \ubar - \Lib(u).
\label{eq:hiddenzeroaux}
\end{align}
We therefore call (\ref{eq:hiddenzero}) a ''hidden zero''. In
complicated multi-loop calculations with HPLs of higher weight and
various arguments, it might become difficult to identify ''hidden
zeros''. In order to avoid this problem, it is important to use a
formulation in terms of a minimal set of transcendental functions.
The HPLs provide such a minimal set. To give an example, in our
2-loop calculation from Chapter~\ref{ch:ImPart} we were able to
express our results in terms of the following HPLs
\begin{align}
H(0;u) &= \ln u, \no \\
H(1;u) &= - \ln (1-u), \no \\
H(0,1;u) &= \Lib(u), \no \\
H(0,0,1;u) &= \Lic(u), \no \\
H(0,1,1;u) &= \Sab(u). \label{eq:HPLminiIm}
\end{align}

\subsubsection{Comment on boundary conditions}

A unique solution of a differential equation requires the knowledge
of its boundary conditions. In the considered case the boundary
conditions typically represent single-scale integrals corresponding
to $u=0,1$. It is of crucial importance that the integral has a
smooth limit at the chosen point such that setting $u=0$ or $u=1$
does not modify the divergence structure introduced in
(\ref{eq:Laurent}).

The boundary conditions can be addressed with any method presented
in this section. In some cases the calculation becomes trivial as
they correspond to simpler topologies which may turn out to be
reducible. If so, the integral can be expressed in terms of known
MIs with the help of the reduction algorithm. For the more
complicated integrals this is not the case and the calculation of
the boundary conditions represents sometimes the most difficult part
in the computation of the MIs.

\subsubsection{2-loop example: Method of Differential Equations}

With the help of the reduction algorithm we derive the following
differential equation
\begin{align}\no \\[-1em]
\frac{\partial}{\partial u} \;\;
\parbox[c]{2cm}{\vspace{-5mm}\psfig{file=pic07.eps,width = 2cm}}
\;\;&=\;\;
    \frac{1-2u}{2u\ubar} \; (d-4) \;\;
    \parbox[c]{2cm}{\vspace{-5mm}\psfig{file=pic07.eps,width = 2cm}}
    \;\;
 + \;\;\frac{3d-8}{2u\ubar m^2}\;\;
    \parbox[c]{2cm}{\psfig{file=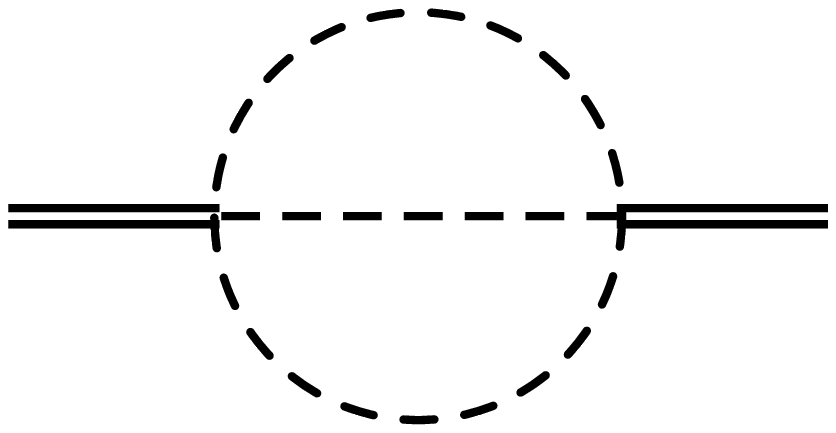,width = 2cm}}
    \label{eq:2loopexDiffEq}
\end{align}
where the inhomogeneous part contains a $3$-topology which is
supposed to be known
\begin{align}
\parbox[c]{2cm}{\psfig{file=pic09.eps,width = 2cm}} \;\;&=\;\;
    (m^2)^{1-2\eps} \; e^{2i\pi \eps} \;
    \frac{\Gamma(-1+2\eps)\Gamma(1-\eps)^5}{\Gamma(3-3\eps)}.
\end{align}
Concerning the boundary conditions, we examine if the integral has a
smooth limit for $u\to0,1$. From (\ref{eq:2loopexSA}) we see that
setting $u=0$ modifies the divergence structure of the integral in
the form $x^{\eps-1} \to x^{-\eps-1}$, whereas $u=1$ does not. We
therefore use the latter boundary condition which corresponds to a
reducible topology
\begin{align} \no \\[-1em]
\parbox[c]{2cm}{\vspace{-5mm}\psfig{file=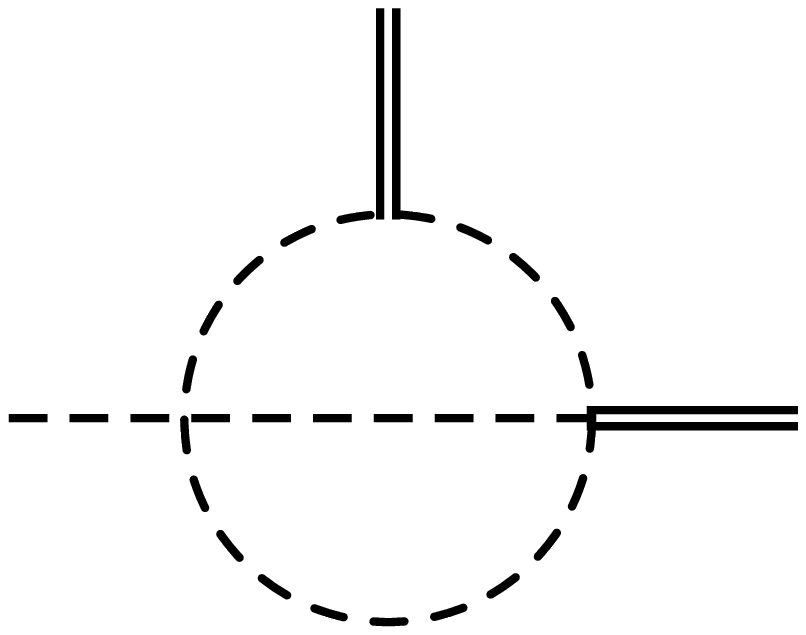,width = 2cm}} \;\;&=\;\;
    \frac{3d-8}{(d-4) m^2} \;\;
    \parbox[c]{2cm}{\psfig{file=pic09.eps,width = 2cm}} \;\;=\;\;
    (m^2)^{-2\eps} \; e^{2i\pi \eps}
    \bigg[ \frac{1}{2\eps^2} + \frac{5}{2\eps} +\frac{19}{2} +
    \calO(\eps) \bigg]
\label{eq:2loopexDiffEqBC}
\end{align}
We solve the differential equation (\ref{eq:2loopexDiffEq}) order by
order in $\eps$ with the ansatz
\begin{align} \no \\[-1em]
\parbox[c]{2cm}{\vspace{-5mm}\psfig{file=pic07.eps,width = 2cm}}
\;\;&=\;\;  (m^2)^{-2\eps} \; e^{2i \pi \eps} \; \left\{
\sum_{i=-4}^{0} c_i(u) \, \eps^i + \calO(\eps) \right\}.
\label{eq:2loopexDiffEqansatz}
\end{align}
Notice that the differential equation has a particular simple
structure in our example because of the explicit factor
$(d-4)=\calO(\eps)$. However, the non-trivial element in the
solution of the differential equation consists in the integration
over the inhomogeneity which can be illustrated here. In the lowest
order, we simply find
\begin{align}
\frac{\partial}{\partial u} \; c_{-4}(u) = 0
    \qquad \to \qquad c_{-4}(u) = \text{const}
\end{align}
and the comparison with (\ref{eq:2loopexDiffEqBC}) leads to
$c_{-4}(u) =0$. The computation of $c_{-3}$ follows exactly the same
lines. The first non-vanishing coefficient is $c_{-2}$ which again
obeys
\begin{align}
\frac{\partial}{\partial u} \; c_{-2}(u) = 0
    \qquad \to \qquad c_{-2}(u) = \text{const},
\end{align}
but the boundary condition (\ref{eq:2loopexDiffEqBC}) yields
$c_{-2}(u) =\frac12$. From now on, the differential equations take a
non-trivial form which perfectly fit into the formalism of HPLs
\begin{align}
\frac{\partial}{\partial u} \; c_{-1}(u) = -\frac1u
    \qquad \to \qquad c_{-1}(u) = - H(0;u) + \text{const}.
\end{align}
The boundary condition results in $c_{-1}(u) = \frac52- H(0;u)$. In
the next step, we have
\begin{align}
\frac{\partial}{\partial u} \; c_{0}(u) &= \left( \frac1u
-\frac{1}{\ubar} \right)H(0;u) -\frac5u \no \\
\to \qquad c_{0}(u) &= H(0,0;u) - H(1,0;u) - 5 H(0;u) +
\text{const}.
\end{align}
The comparison with (\ref{eq:2loopexDiffEqBC}) gives $c_{0}(u) =
\frac{19}{2}-\frac{\pi^2}{6}+ H(0,0;u) - H(1,0;u) - 5 H(0;u)$ which
can be expressed in terms of the minimal set (\ref{eq:HPLminiIm})
with the help of
\begin{align}
H(0,0;u) &= \frac12 H(0;u)^2,\no\\
H(1,0;u) &= H(0;u)H(1;u) - H(0,1;u).
\end{align}
Collecting all coefficients $c_i$ and using
(\ref{eq:2loopexDiffEqansatz}) with (\ref{eq:HPLminiIm}), we finally
reproduce our result from (\ref{eq:2loopexSARes}). It is evident
that this MI has been particularly simply but it nevertheless
illustrated the usefulness of the HPLs in this context.

\subsection{Expansion by Momentum Regions}
\label{sec:ExpByRegions}

The method of expansion by regions~\cite{ExpByRegions} is one of the
most powerful techniques to determine loop integrals to leading
power in some small expansion parameter. We have already pointed out
that the extraction of the leading power would simplify
substantially if we could interchange the order of loop integration
and power expansion. However, we have seen in the analysis of our
1-loop example after (\ref{eq:1loopexSARes}) that this procedure may
generate artificial IR singularities which invalidates this naive
approach.

We will see in this section that we \emph{can} do so if we split the
loop integration into individual momentum regions and expand the
integrands in each region separately. This in turn generates
artificial singularities in each region which cancel when we
reconstruct the full integral as the sum over all regions. When the
calculation is performed within DR, no further cutoffs are required
to separate the momentum regions due to the fact that scaleless
integrals vanish in DR. We now switch to our explicit 1-loop example
to demonstrate how this technique works in detail.

\subsubsection{1-loop example: Expansion by Momentum Regions}

We first address a power-counting to the external momenta and
particle masses in the problem. We adopt the SCET terminology from
Table~\ref{tab:term} in our example by identifying $\lambda^2\equiv
m/M$ in the non-relativistic set-up. The momentum of the spectator
antiquark in the initial bound state corresponds to a soft momentum
with $m w = (m,0,m) \sim (\lambda^2,0,\lambda^2)$ whereas the quark
and antiquark in the final bound state are in a collinear
configuration with $m w'=(2\gamma m,0,m/2\gamma)\sim
(1,0,\lambda^4)$. Notice that this parametrization fulfils the
conditions $w^2={w'}^2=1$, $w\cdot w'\simeq\gamma\sim1/\lambda^2$
(transverse components are suppressed in the non-relativistic
approach).

\newabs
We want to compute the leading power in $\lambda$ of the integral
\begin{align}\no\\[-1em]
\parbox[c]{2cm}{\vspace{-5mm}\psfig{file=pic08.eps,width = 2cm}}
    \;\; = \;\; \int [d k] \;\;
        \frac{1}
        {[(m w'-m w + k)^2-m^2][k^2-m^2](k-m w')^2}
\end{align}
Potential leading contributions come from hard, hard-collinear, soft
and collinear momentum configurations of the particles in the loop.
We may also look for contributions from soft-collinear messenger
modes as discussed in Section~\ref{sec:xi}. We perform a power
counting of the integral in the individual momentum regions of the
loop momentum $k$. This gives rise to the following pattern
\begin{align}
&&\text{hard}           &\qquad(1,1,1)
    &&\,~0 - 0 - 0 - 0 =0,\no \\
&&\text{hard-collinear} &\qquad(1,\lambda,\lambda^2)
    &&\,~4 - 2 - 2 - 2 =-2,\no \\
&&\text{soft}           &\qquad(\lambda^2,\lambda^2,\lambda^2)
    &&\,~8 - 2 - 4 - 2 =0,\no \\
&&\text{collinear}      &\qquad(1,\lambda^2,\lambda^4)
    &&\,~8 - 2 - 4 - 4 =-2,\no \\
&&\text{soft-collinear} &\qquad(\lambda^2,\lambda^3,\lambda^4)
    &&12 - 2 - 4 - 4 =2. \hspace{1cm}
\end{align}
The first term in each region comes from the scaling of the
integration measure and the other terms from the three propagators.
We find a leading contribution of $\calO(1/\lambda^2)$ which stems
from the hard-collinear and the collinear momentum region.

\newabs
We first address the hard-collinear region. As anticipated above, we
expand the integrand in this region counting
$k\sim(1,\lambda,\lambda^2)$ and do not need to introduce additional
cutoffs in DR. The hard-collinear contribution becomes
\begin{align}\no\\[-1.8em]
\parbox[c]{2cm}{\vspace{-3mm}\psfig{file=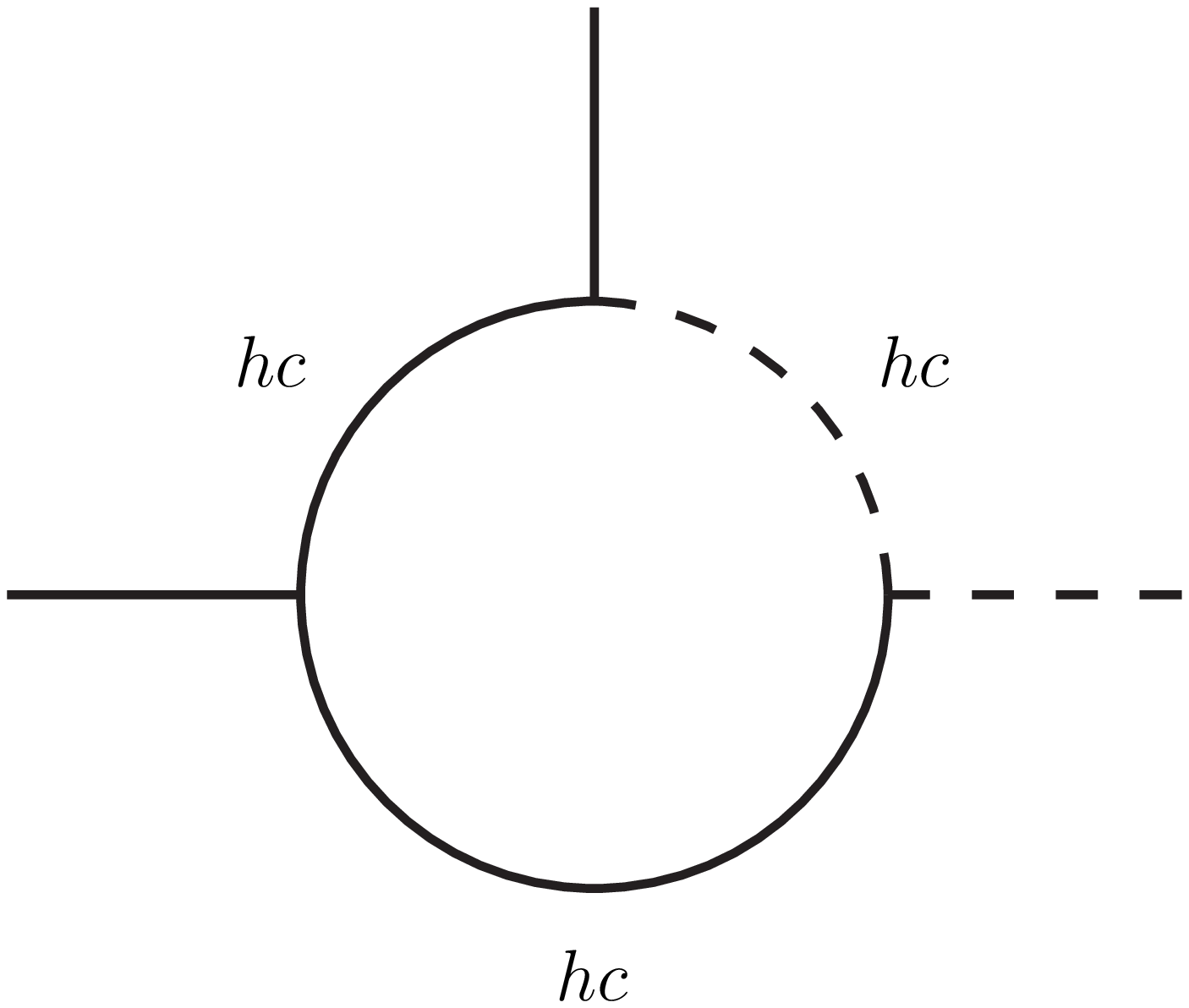,width = 2cm}}
    \;\; &\simeq \;\; \frac{\Gamma(1-\eps)}{2i\pi^{d/2}}\; \int dk_- \,
dk_+ \, d^{d-2} k_\perp \quad \frac{1}{[k_+ k_-
+k_\perp^2-2 \gamma m k_++i\eps]}\no \\
&  \hspace{5mm} \frac{1}{[k_+ k_- + k_\perp^2 +2\gamma m(k_+-m)-m
k_-+i\eps][k_+ k_- + k_\perp^2+i\eps]}
\end{align}
where we have made the $i\eps$-prescription of the propagators
explicit and translated the $d$-dimensional integration measure from
(\ref{eq:intmeasure}) into light-cone coordinates. We first perform
the $k_-$ integration by contour methods. As all poles lie in the
upper (lower) half plane for $k_+<0$ ($k_+>m$), the only
non-vanishing contribution comes from $0<k_+<m$. In this case one of
the poles lies in the upper half plane and the two others in the
lower one which yields
\begin{align}\no\\[-1.8em]
\parbox[c]{2cm}{\vspace{-3mm}\psfig{file=pic08HC.ps,width = 2cm}}
    \;\; &\simeq \;\;
    \frac{\Gamma(1-\eps)}{\pi^{d/2-1}}\; \int_0^m dk_+ \quad
    \frac{k_+-m}{m^2} \no \\
& \hspace{5mm} \int d^{d-2} k_\perp \quad \frac{1}
        {[k_\perp^2 -4\gamma k_+(m-k_+)][k_\perp^2 -2\gamma
        k_+(m-k_+)]}.
\end{align}
The remnant integrations are straight-forward and give
\begin{align}\no\\[-1.8em]
\parbox[c]{2cm}{\vspace{-3mm}\psfig{file=pic08HC.ps,width = 2cm}}
    \;\; &\simeq \;\;
    - \frac{\Gamma(1-\eps)\, \Gamma(1+\eps)}{m^2} \;
\frac{1-2^{-\eps}}{\eps} \; \int_0^m dk_+ \;
    (2\gamma k_+)^{-1-\eps}\;(m-k_+)^{-\eps} \no \\
&= - (2\gamma m^2)^{-1-\eps} \; \frac{4^\eps-2^{\eps}}{\eps} \;
    \frac{\Gamma(1-\eps)\, \Gamma(1+\eps) \, \Gamma(1/2)\,
    \Gamma(-\eps)}{\Gamma(1/2-\eps)} \no \\
&= -\frac{1}{4\gamma m^2} \left[ -\frac{2\ln2}{\eps} +2\ln2
\ln(\gamma m^2) +3 \ln^2 2 + \calO(\eps) \right].
\label{eq:1loopexHC}
\end{align}

Now we consider the collinear region counting
$k\sim(1,\lambda^2,\lambda^4)$
\begin{align}\no\\[-1.8em]
\parbox[c]{2cm}{\vspace{-3mm}\psfig{file=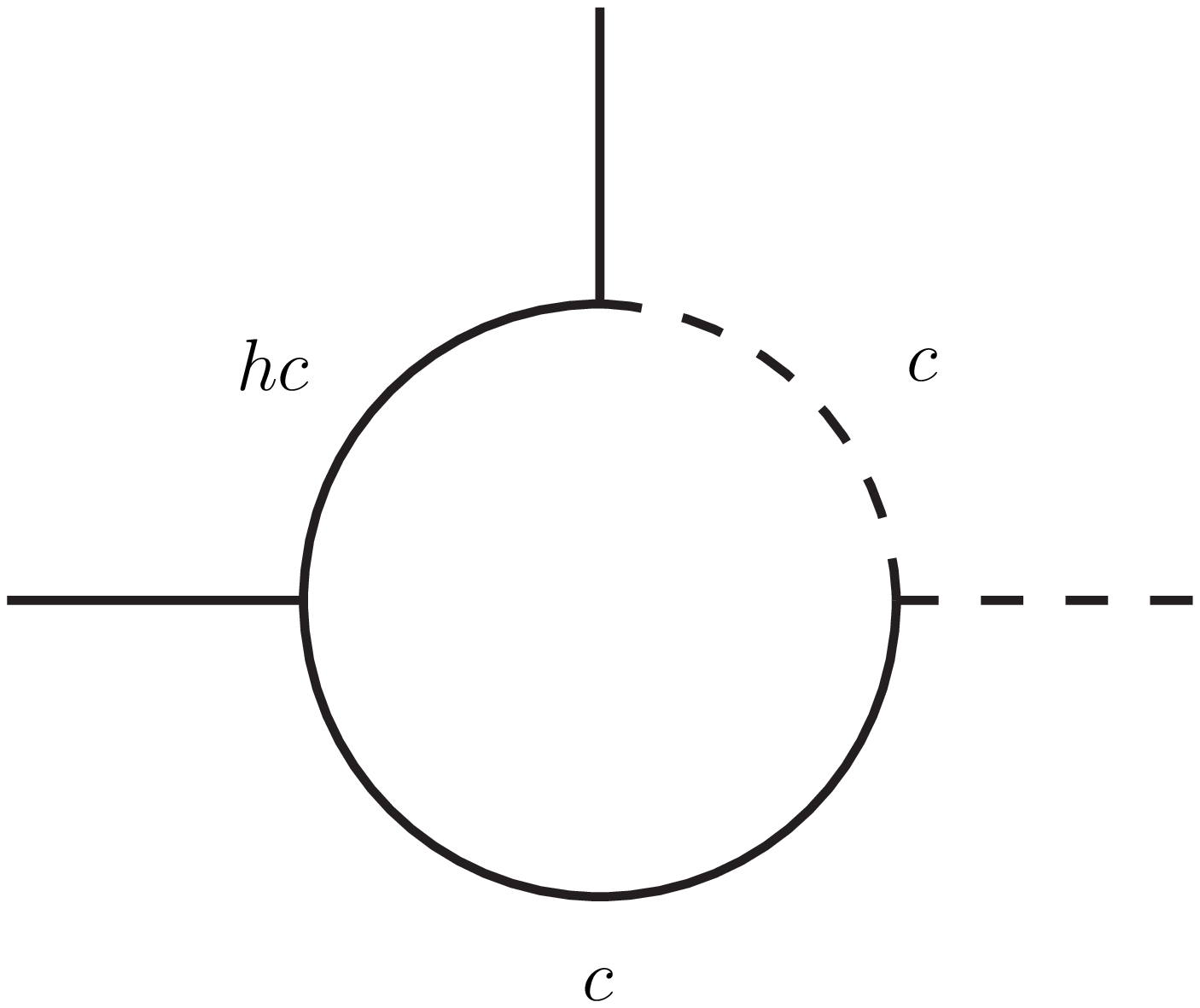,width = 2cm}}
    \;\; &\simeq \;\;
     \frac{\Gamma(1-\eps)}{2i\pi^{d/2}}\; \int dk_- \,
dk_+ \, d^{d-2} k_\perp \quad \frac{1}{[k_+ k_- +
k_\perp^2-m^2+i\eps]}\\
&  \frac{1}{[-m k_- -2\gamma m^2+i\eps][k_+ k_- +k_\perp^2-2 \gamma
m k_+-mk_-/(2\gamma) +m^2+i\eps]} \no
\end{align}
We perform the $k_+$-integration with contour methods and find a
contribution for $0<k_-<2\gamma m$. The other integrations are again
straight-forward.
\begin{align}
\no\\[-1.8em]
\parbox[c]{2cm}{\vspace{-3mm}\psfig{file=pic08C.ps,width = 2cm}}
    \;\; &\simeq \;\;
    \frac{\Gamma(1-\eps)}{\pi^{d/2-1}}\; \int_0^{2\gamma
m} \!\!\! dk_- \;
    \frac{1}{2\gamma m^2(k_-+2\gamma m)}\;\; \frac{d^{d-2} k_\perp}
        {[k_\perp^2 -(m-k_-/(2\gamma))^2]} \no \\
&= - \frac{\Gamma(1-\eps)\, \Gamma(\eps)}{2\gamma m^2}  \;
\int_0^{2\gamma m} \!\!\! dk_- \;
    (k_-+2\gamma m)^{-1}\;(m-k_-/(2\gamma))^{-2\eps} \no \\
&= - \frac{\Gamma(1-\eps)\, \Gamma(\eps)}{2\gamma m^2}  \;
(m^2)^{-\eps} \; \frac{_2F_1(1,1;2-2\eps;-1)}{1-2\eps}  \no \\
&= -\frac{1}{4\gamma m^2} \left[ \frac{2\ln2}{\eps} -2 \ln 2 \ln m^2
-2\ln^2 2 + \frac{\pi^2}{3} + \calO(\eps) \right].
\label{eq:1loopexC}
\end{align}
Combining the two contributions from (\ref{eq:1loopexHC}) and
(\ref{eq:1loopexC}), we see that the artificial divergences that
appeared in each region cancel each other. We finally reproduce our
result from (\ref{eq:1loopexSARes})
\begin{align}\no\\[-1.8em]
\parbox[c]{2cm}{\vspace{-3mm}\psfig{file=pic08HC.ps,width = 2cm}}
    \;\; + \;\; \parbox[c]{2cm}{\vspace{-3mm}\psfig{file=pic08C.ps,width = 2cm}}
    \;\; &\simeq \;\;-\frac{1}{4\gamma m^2} \left[ 2\ln2 \ln \gamma+
\ln^2 2 + \frac{\pi^2}{3}+ \calO(\eps) \right].
\end{align}
Notice that (\ref{eq:1loopexHC}) and (\ref{eq:1loopexC}) determine
the leading power of the integral in a closed form in $\eps$ whereas
we were only able to determine the first coefficient in the
$\eps$-expansion with the help of Feynman parameters in
Section~\ref{sec:Feynman}.

\subsection{Mellin-Barnes Techniques}
\label{sec:MB}

In the last two sections we have introduced two sophisticated
techniques for the calculation of loop integrals. The first one was
particularly suited for our 2-loop calculation from
Chapter~\ref{ch:ImPart} and~\ref{ch:RePart}, the second one for the
1-loop calculation in Chapter~\ref{ch:NRModel}. We now present a
third method which we used in both calculations. In our 2-loop
calculation we typically applied Mellin-Barnes
techniques~\cite{MB:Smi,MB:Tau} for the calculation of the boundary
conditions, i.e.~for single-scale integrals. As the boundary
condition in our 2-loop example is rather trivial, we present the
computation of the full 2-loop integral in the following.
Mellin-Barnes techniques are also a comfortable tool for the
calculation of loop integrals in a power expansion as we will see in
our 1-loop example below.

\newabs
The method is based on the following representation
\begin{align}
\frac{1}{(X+Y)^\lambda} =
        \frac{1}{\Gamma(\lambda)} \;\;
        \frac{1}{2\pi i} \;\int_{-i \infty}^{+i \infty}
        \!\! dz \; \; \Gamma(\lambda+z) \; \Gamma(-z) \;\;
            Y^z \;  X^{-z-\lambda},
\label{eq:MB}
\end{align}
where the contour separates \emph{left poles} from \emph{right
poles} (we specify the meaning of this phrases in our examples) and
$z$ is called a Mellin-Barnes parameter. We see that (\ref{eq:MB})
basically converts a sum into a product at the cost of a contour
integration.

We typically introduce Feynman parameter first and use the
Mellin-Barnes representation (\ref{eq:MB}) to simplify the
corresponding integrations. With all integrations over Feynman
parameters done, we examine the analytical properties of the
integrand and perform the Mellin-Barnes integrations with contour
methods. This leads to infinite sums over residues which often
represent the main obstacle in this approach.

\subsubsection{2-loop example: Mellin-Barnes Techniques}

\begin{figure}[t!]
\centerline{\parbox{13cm}{
\includegraphics[width=13cm]{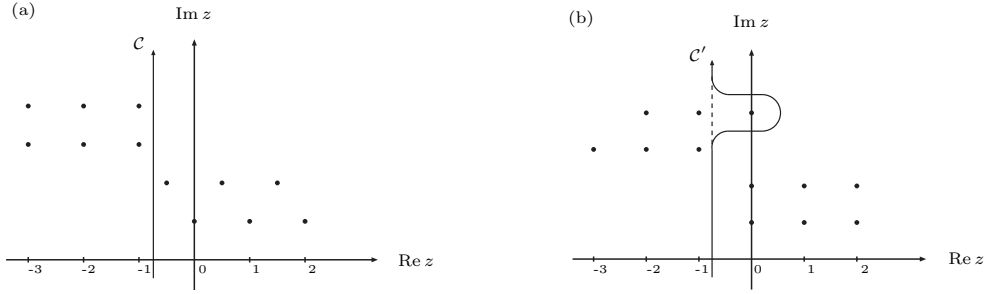}
\caption{\label{fig:Contour} \small \textit{Possible integration
contours in (\ref{eq:2loopexMBaux}) that separate left poles from
right poles. The figures show the position of the poles for
$\eps=1/2$ (a) and $\eps=0$ (b). All poles lie on the real axis, but
we have assigned a positive imaginary part for illustrative
reasons.}}}}
\end{figure}

We first address our 2-loop integral from (\ref{eq:2loopex}). The
starting point is equation (\ref{eq:2loopexSA}) which we rewrite
with the help of the substitution $x=st$, $y=\bar{s}t$ with the
usual bar-notation understood. This yields
\begin{align} \no\\[-1em]
\parbox[c]{2cm}{\vspace{-5mm}\psfig{file=pic07.eps,width = 2cm}}
\;\;&=\;\; (m^2)^{-2\eps} \; e^{2i\pi \eps} \; \frac{\Gamma(2\eps)\,
\Gamma(1-\eps)^4}{\Gamma(2-2\eps)} \;\int_0^1 ds\int_0^1 dt \;\;
s^{\eps-1} \,t^{-\eps} \,\bar{t}^{-2\eps} \,(s+u\bar{s})^{-2\eps}
\label{eq:2loopexMB}
\end{align}
We introduce one Mellin-Barnes parameter to split the last factor
into a product. The integrations over the Feynman parameters give
rise to $\Gamma$-functions and we obtain
\begin{align} \no\\[-1em] \label{eq:2loopexMBaux}
\parbox[c]{2cm}{\vspace{-5mm}\psfig{file=pic07.eps,width = 2cm}}
\;\;&=\;\; (m^2)^{-2\eps} \; e^{2i\pi \eps} \; \frac{
\Gamma(1-\eps)^4}{(1-2\eps)\,\Gamma(2-3\eps)} \\
& \hspace{10mm} \frac{1}{2\pi i} \;\int_{\calC} \,
        dz \quad \Gamma(2\eps+z) \, \Gamma(1+z)
         \, \Gamma(-\eps-z) \, \Gamma(-z)\; u^z. \no
\end{align}
The integrand has simple poles which we classify into \emph{left
poles} which stem from $\Gamma(\ldots+z)$ and \emph{right poles}
from $\Gamma(\ldots-z)$. We find
\begin{align}
&\text{left poles:}&&z=-2\eps,\,-1-2\eps,\,-2-2\eps,\,\ldots &&z=-1,\,-2,\,-3,\,\ldots\no \\
&\text{right poles:}&&z=-\eps,\,1-\eps,\,2-\eps,\,\ldots
&&z=0,\,1,\,\,2,\,\ldots
\end{align}
The contour $\calC$ separates these two types of poles. Notice that
we cannot simply set $\eps=0$ as this would glue left and right
poles together at $z=0$. There is no such problem for $0<\eps<1$;
let us take $\eps=1/2$ for the moment. We see that we can choose the
contour $\calC$ as a straight line in this case (cf.
Figure~\ref{fig:Contour}a). For $\eps \to 0$ the contour is deformed
into $\calC'$ as shown in Figure~\ref{fig:Contour}b. Alternatively,
we may stick to the contour $\calC$ and add the contributions from
the poles that crossed the straight line for $\eps\to0$ explicitly.
The crucial point is that all singularities of the integral stem
from such contour crossings. In our example we pick up the pole for
$z=-2\eps$ and may safely set $\eps=0$ in the remnant contour
integral which results in
\begin{align} \no\\[-1em]
\parbox[c]{2cm}{\vspace{-5mm}\psfig{file=pic07.eps,width = 2cm}}
\;\;&=\;\; (m^2)^{-2\eps} \; e^{2i\pi \eps} \; \frac{
\Gamma(1-\eps)^4}{(1-2\eps)\,\Gamma(2-3\eps)}
\bigg[ \Gamma(1-2\eps) \Gamma(\eps) \Gamma(2\eps) u^{-2\eps} \no \\
& \hspace{10mm} + \frac{1}{2\pi i} \;\int_{-3/4-i \infty}^{-3/4+i
\infty}
        \!\!\!\!\! dz \quad \bigg( \Gamma(z) \, \Gamma(1+z)
         \, \Gamma(-z)^2\; u^z + \calO(\eps) \bigg) \bigg]
         \label{eq:2loopexMBaux2}
\end{align}

We close the contour to the left and sum the infinite residues from
$z=-1,-2,-3,\ldots$
\begin{align}         \label{eq:2loopexMBaux3}
\frac{1}{2\pi i} \;\int_{-3/4-i \infty}^{-3/4+i \infty}
        \!\!\!\!\! dz \quad \Gamma(z) \, \Gamma(1+z)
         \, \Gamma(-z)^2\; u^z
         &\;=\; \sum_{n=1}^{\infty} \; \frac{-(1+n \ln u)}{n^2 u^n} \\
         &\;=\; \Lib(u) + \ln u \ln \ubar -\frac12 \ln^2 u
         -\frac{\pi^2}{3}. \no
\end{align}
The expansion of (\ref{eq:2loopexMBaux2}) in $\eps$ yields together
with (\ref{eq:2loopexMBaux3}) our result from
(\ref{eq:2loopexSARes}).

\subsubsection{1-loop example: Mellin-Barnes Techniques}

In our 1-loop example we start from (\ref{eq:1loopexSA}) and perform
the substitution $x=st$, $y=\bar{s}t$ as in the last example
\begin{align} \no\\[-0.5em]
\parbox[c]{2cm}{\vspace{-4.5mm}\psfig{file=pic08.eps,width = 2cm}}
\;\;&=\;\; \int_0^1 ds\int_0^{1} dt \;\; \frac{- (m^2)^{-1-\eps} \;
\Gamma(1-\eps)\, \Gamma(1+\eps) \; t^{-\eps} }{\left[t+2s(2\bar t +t
\bar s)(\gamma-1)\right]^{1+\eps}}.
\end{align}
We introduce two Mellin-Barnes parameters in this case to obtain
standard integrations over Feynman parameters in terms of
$\Gamma$-functions. We find
\begin{align} \no\\[-0.5em]
\parbox[c]{2cm}{\vspace{-4.5mm}\psfig{file=pic08.eps,width = 2cm}}
\;\;&=\;\; -(m^2)^{-1-\eps} \; \frac{
\Gamma(1-\eps)}{\Gamma(1-2\eps)} \;
    \frac{1}{2\pi i} \;\int_{\calC_z}
        dz \;\; \Gamma(1+\eps+z) \, \Gamma(1+z)
         \;2^{2z} (\gamma-1)^z \no \\
& \hspace{0mm} \frac{1}{2\pi i} \;\int_{\calC_w}
        dw \;\; 2^{-w} \;\frac{\Gamma(w-z)  \, \Gamma(-w)
         \, \Gamma(w-2\eps-z) \, \Gamma(1+z-w) \, \Gamma(1+w)}{\Gamma(2+z+w)}.
\label{eq:1loopexMBaux}
\end{align}
The contours separate left poles from right poles if the arguments
of the $\Gamma$-functions are positive. This can be achieved by
integrating along straight lines with $w_0=-1/2$ and $z_0=-3/4$ for
$\eps=0$ (we already know that the integral is finite and set
$\eps=0$ in the following). The poles in $w$ can be classified
according to
\begin{align}
&\text{left poles:}&&w=z,\,-1+z,\,-2+z,\,\ldots &&w=-1,\,-2,\,-3,\,\ldots\no \\
&\text{right poles:}&&w=1+z,\,2+z,\,3+z\,\ldots
&&w=0,\,1\,\,2,\,\ldots
\end{align}
We close the contour to the right and find
\begin{align}
& \frac{1}{2\pi i} \;\int_{\calC_w}
        dw \;\; 2^{-w} \;\frac{\Gamma(w-z)^2  \, \Gamma(-w)
         \, \Gamma(1+z-w) \,
         \Gamma(1+w)}{\Gamma(2+z+w)} \no \\
& \qquad = \sum_{n=0}^{\infty} \; (-1)^{n}
    \bigg[ 2^{-n} \; \frac{\Gamma(n-z)^2\,
    \Gamma(1-n+z)}{\Gamma(2+n+z)} \no \\
& \hspace{4cm} +
        2^{-n-1-z} \; \frac{\Gamma(-1-n-z)\,
    \Gamma(2+n+z)\, \Gamma(n+1)}{\Gamma(3+n+2z)} \bigg] \no \\
& \qquad = 2(1+z)\; \Gamma(-1-z)^2 ~_2F_1(1,2+2z;2+z;-1) \no \\
& \hspace{2cm} - 2^{-1-z} \; \frac{\Gamma(1-z)\,
\Gamma(z)}{\Gamma(2+2z)} \; \big[\psi(1+z)-\psi(3/2+z)\big]
\end{align}
where $\psi(z)$ is the digamma function given by
$\psi(z)\equiv\Gamma'(z)/\Gamma(z)$. So far our computation has been
exact in $m/M$. The important point to notice is that the remnant
contour integral has poles for $z\in \mathbf{Z}$. If we close the
contour to the left, \emph{the leading contribution in $m/M$ stems
from a single pole at $z=-1$} because of the explicit factor
$(\gamma-1)^z\sim\gamma^z$ in (\ref{eq:1loopexMBaux}). This yields
\begin{align} \no\\[-0.5em]
\parbox[c]{2cm}{\vspace{-4.5mm}\psfig{file=pic08.eps,width = 2cm}}
\;\; &\simeq \;\;-\frac{1}{4\gamma m^2} \left[2\ln 2 \ln \gamma +
\ln^2 2 +\frac{\pi^2}{3} + \calO(\eps) \right]
\end{align}
in agreement with (\ref{eq:1loopexSARes}). We see that the
Mellin-Barnes technique is particularly suited to compute the
subleading contributions in $m/M$ which stem from subsequent poles
in $z$ in our example.

\subsection{Sector Decomposition}

Let us summarize what we have obtained so far. Concerning our 1-loop
calculation from Chapter~\ref{ch:NRModel}, we presented two
analytical methods for the calculation of the MIs: the method of
expansion by regions and Mellin-Barnes techniques. The comparison
between these two independent calculations is extremely helpful to
avoid errors in this part of the calculation. So far, we have found
only one method for our 2-loop calculation from
Chapter~\ref{ch:ImPart} and~\ref{ch:RePart}: the method of
differential equations in combination with Mellin-Barnes techniques
for the calculation of the boundary conditions. A second independent
method would also be very helpful in this case.

\newabs
We use a numerical method, the method of sector
decomposition~\cite{HB}, in order to check our analytical results of
the 2-loop integrals from Chapter~\ref{ch:ImPart}
and~\ref{ch:RePart}. We refrain from presenting this technique here
in detail as we will only use it as an (important) check. The basic
idea corresponds to a decomposition procedure of Feynman parameter
integrals which systematically disentangles overlapping IR
divergences. These divergences can then be isolated with the help of
an adequate subtraction procedure. At this step the MI is given in
form of a Laurent expansion in $\eps$ with coefficients that
correspond to \emph{regular} parameter integrals that can be
evaluated numerically.

The power of the considered algorithm lies in the fact that it can
easily be automatized and be applied, at least in principle, for any
multi-loop integral. Its numerical precision is mainly limited by
the potential presence of thresholds in the parameter integrals. The
obtained precision therefore varies strongly in our applications
between 1 part in $10^{2}$-$10^{8}$ depending on the considered
integral. This situation could, however, be improved by performing
the numerical integrations with more efficient routines.

\part{Applications}

\chapter{Hadronic two-body decays I: Imaginary part}
\label{ch:ImPart}

In the second part of this thesis we consider several perturbative
calculations in the framework of QCD Factorization. We first compute
the imaginary part of the NNLO vertex corrections in charmless
hadronic two-body decays as e.g.~$B\to\pi\pi$ (the real part will be
considered in Chapter~\ref{ch:RePart}). The results presented in
this chapter will be published in \cite{GBImPart}.

\newabs
The outline of this chapter is as follows: We first introduce a new
operator basis of the effective Hamiltonian which is particularly
suited for multi-loop calculations. We then present some details of
the 2-loop calculation following our strategy from
Section~\ref{sec:strategy}. As the factorization formula reveals a
rather complicated divergence structure at NNLO, we elaborate the
subsequent UV and IR subtractions in some detail. The imaginary part
of the topological tree amplitudes is finally obtained in an
analytic form. We conclude with a brief analysis of the numerical
impact of the NNLO vertex corrections.

\section{Change of operator basis}
\label{eq:BtoPiPi:OpBasis}

In view of the calculation of topological tree amplitudes, we
restrict our attention to the current-current operators $Q_{1,2}$ of
the effective weak Hamiltonian (\ref{eq:Heff}). Due to the fact that
we work within Dimensional Regularization (DR), we also have to
consider \emph{evanescent
operators}~\cite{BurasWeisz,Evan:DuganGrin,Evan:HerrNier}. These
non-physical operators vanish in $d=4$ dimensions but contribute at
intermediate steps of the calculation in $d\neq4$ dimensions. We
have emphasized in Section~\ref{sec:BtoPiPi:PC} that the imaginary
part considered here has effectively NLO complexity. We will indeed
see that the calculation of the imaginary part only requires 1-loop
evanescent operators (2-loop evanescent operators will contribute to
the real part which we consider in Chapter~\ref{ch:RePart}). For our
purposes the complete operator basis is thus given by
\begin{align}
\tilde Q_1 &=  \left[\bar u \gamma^\mu L\, b\right] \;
        \left[\bar d \gamma_\mu \,L\, u\right],\no\\
\tilde Q_2 &=  \left[\bar u_i \gamma^\mu L\, b_j\right] \;
        \left[\bar d_j \gamma_\mu \,L\, u_i\right],\no\\
\tilde E_1 &=  \left[\bar u \gamma^\mu\gamma^\nu\gamma^\rho L\,
b\right] \;
        \left[\bar d \gamma_\mu\gamma_\nu\gamma_\rho \,L\, u\right]
        -(16-4\eps) \,\tilde Q_1,\no\\
\tilde E_2 &=  \left[\bar u_i \gamma^\mu\gamma^\nu\gamma^\rho L\,
b_j\right] \;
        \left[\bar d_j \gamma_\mu\gamma_\nu\gamma_\rho \,L\, u_i\right]
        -(16-4\eps) \,\tilde Q_2,
\label{eq:Basis:QCDF}
\end{align}
with $L\equiv1-\gamma_5$. We refer to this basis as the (standard)
QCDF basis for convenience and denote the corresponding Wilson
coefficients and operators with a tilde.

\newabs
It has been argued by Chetyrkin, Misiak and M\"unz (CMM) that one
should use a different operator basis in order to perform multi-loop
calculations~\cite{CMM}. Although the deeper reason is related to
the penguin operators which we do not consider here, we prefer to
introduce the CMM basis in view of future extensions of our work.
The CMM basis allows to consistently use DR with a naive
anticommuting $\gamma_5$ to all orders in perturbation theory. In
the CMM basis the current-current operators and corresponding 1-loop
evanescent operators read  (denoted with a hat)
\begin{align}
\hat Q_1 &=  \left[\bar u \gamma^\mu L\, T^A b\right] \;
        \left[\bar d \gamma_\mu \,L\, T^A u\right],\no\\
\hat Q_2 &=  \left[\bar u \gamma^\mu L\, b\right] \;
        \left[\bar d \gamma_\mu \,L\, u\right],\no\\
\hat E_1 &=  \left[\bar u \gamma^\mu\gamma^\nu\gamma^\rho L\, T^A
b\right] \;
        \left[\bar d \gamma_\mu\gamma_\nu\gamma_\rho \,L\, T^A u\right]
        -16 \,\hat Q_1,\no\\
\hat E_2 &=  \left[\bar u \gamma^\mu\gamma^\nu\gamma^\rho L\,
b\right] \;
        \left[\bar d \gamma_\mu\gamma_\nu\gamma_\rho \,L\, u\right]
        -16 \,\hat Q_2.
\label{eq:Basis:CMM}
\end{align}
Comparing (\ref{eq:Basis:QCDF}) and (\ref{eq:Basis:CMM}) we observe
two differences: First, the two bases use different colour
structures which is a  rather trivial point. More importantly, they
contain slightly different definitions of evanescent operators which
we will examine now in detail.

\begin{figure}[b!]
\centerline{\parbox{13cm}{\centerline{
\includegraphics[height=2.5cm]{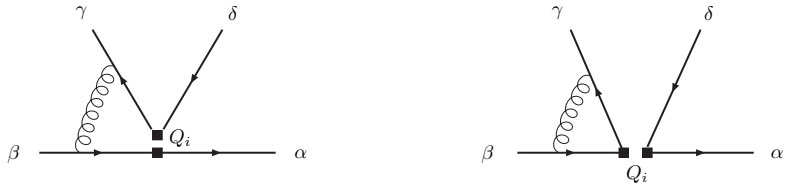}}\vspace{-2mm}
\caption{\label{fig:Fierz} \small \textit{Generic 1-loop diagram
with different contractions of fields in a four-quark operator
$Q_i$. The two insertions are related by a Fierz reordering, see
text ($\alpha, \beta, \gamma, \delta$ are spinor indices).}}}}
\end{figure}

\newabs
The issue is related to Fierz symmetry. We will see below that the
definition of evanescent operators in the CMM basis explicitly
breaks Fierz symmetry which relates the two diagrams in
Figure~\ref{fig:Fierz}. This can be seen by considering the UV part
of the left diagram which involves the following combination of
Dirac matrices
\begin{align}
\left(\gamma^\mu L \gamma^\nu \gamma^\rho \right)_{\alpha \beta}\!
    \; \left(\gamma_\rho \gamma_\nu \gamma_\mu L\right)_{\gamma \delta}
&= -\left(\gamma^\mu \gamma^\nu \gamma^\rho L \right)_{\alpha
\beta}\!
    \; \left(\gamma_\mu \gamma_\nu \gamma_\rho L\right)_{\gamma
    \delta}
    + (6d-4) \left(\gamma^\mu L \right)_{\alpha \beta}\!
    \; \left(\gamma_\mu L\right)_{\gamma \delta}
\label{eq:FierzL}
\end{align}
which we reshuffled into a more convenient form using an
anticommuting $\gamma_5$. On the other hand the right diagram gives
\begin{align}
\left(\gamma^\rho\gamma^\nu\gamma^\mu L \gamma_\nu \gamma_\rho
\right)_{\gamma \beta}\!
    \; \left(\gamma_\mu L\right)_{\alpha \delta}
&=
     (d-2)^2 \left(\gamma^\mu L \right)_{\gamma \beta}\!
    \; \left(\gamma_\mu L\right)_{\alpha \delta}\overset{\text{\tiny Fierz}}{=}
     (d-2)^2 \left(\gamma^\mu L \right)_{\alpha \beta}\!
    \; \left(\gamma_\mu L\right)_{\gamma \delta},
\label{eq:FierzR}
\end{align}
where we have performed a Fierz reordering in the second step. If we
now \emph{impose} Fierz symmetry, the expressions (\ref{eq:FierzL})
and (\ref{eq:FierzR}) have to be equal and we arrive at
\begin{align}
\left(\gamma^\mu \gamma^\nu \gamma^\rho L \right)_{\alpha \beta}\!
    \; \left(\gamma_\mu \gamma_\nu \gamma_\rho L\right)_{\gamma
    \delta} &=
    (16-4\eps-4\eps^2) \left(\gamma^\mu L \right)_{\alpha \beta}\!
    \; \left(\gamma_\mu L\right)_{\gamma \delta}
\label{eq:Fierz}
\end{align}
If we look at the remnant 1-loop diagrams we find that
(\ref{eq:Fierz}) is in fact the only constraint from Fierz symmetry.
In NLO the terms of $\calO(\eps^2)$ do not contribute as the
loop-integrals have at most $1/\eps$ (UV) divergences. They can
therefore simply be neglected here (we stress that this will be
different in our analysis in Chapter~\ref{ch:RePart}). We conclude
that the QCDF basis from (\ref{eq:Basis:QCDF}) is Fierz symmetric
whereas the CMM basis from (\ref{eq:Basis:CMM}) is \emph{not}. In
other words, the freedom in the definition of evanescent operators
has been used in the QCDF basis to properly adjust the $\eps$-terms
into a Fierz symmetric form.

\newabs
Why do we care about Fierz symmetry? In the considered calculation
the contraction depicted in the left diagram from
Figure~\ref{fig:Fierz} is related to the colour-allowed tree
amplitude ($\alpha_1$), whereas the one from the right diagram leads
to the colour-suppressed tree amplitude ($\alpha_2$). On the
technical level these two possible insertions of a four-quark
operator correspond to two completely different calculations. It
would be very tedious if we had to perform both calculations
explicitly, in particular in the considered 2-loop case. We have
pointed out in Section~\ref{sec:BtoPiPi:PC} that $\alpha_1$ and
$\alpha_2$ can naturally be related by Fierz symmetry. For this, it
is of crucial importance that we \emph{preserve} Fierz symmetry when
we work in the effective theory which factorizes the amplitudes into
Wilson coefficients and matrix elements within DR. As we have argued
above, this is indeed the case in the QCDF basis which allows us to
\emph{derive} $\alpha_2$ from $\alpha_1$ by simply interchanging
$\tilde C_1\leftrightarrow \tilde C_2$.

\newabs
We conclude that the CMM basis is the appropriate choice for a
2-loop calculation whereas the QCDF basis provides a short-cut for
the derivation of the colour-suppressed amplitude. We therefore
propose the following strategy for the calculation of the NNLO
vertex corrections: We perform the explicit 2-loop calculation in
the CMM basis using the first type of insertion in the left diagram
from Figure~\ref{fig:Fierz}. In this way we obtain $\alpha_1(\hat
C_i)$. We then transform this expression into the QCDF basis which
yields $\alpha_1(\tilde C_i)$ and finally apply Fierz symmetry
arguments to derive $\alpha_2(\tilde C_i)$ from $\alpha_1(\tilde
C_i)$ under the exchange $\tilde C_1\leftrightarrow \tilde C_2$.

\section{2-loop calculation}
\label{sec:BtoPiPi:2loop}

In this section we present a brief overview of the technical aspects
of the considered 2-loop calculation. Herein, we follow the
systematics of our strategy from Section~\ref{sec:strategy}.

\subsubsection{Step 1: Set-up for loop calculation}

\begin{figure}[]
\centerline{\parbox{15cm}{\includegraphics[width=15cm]{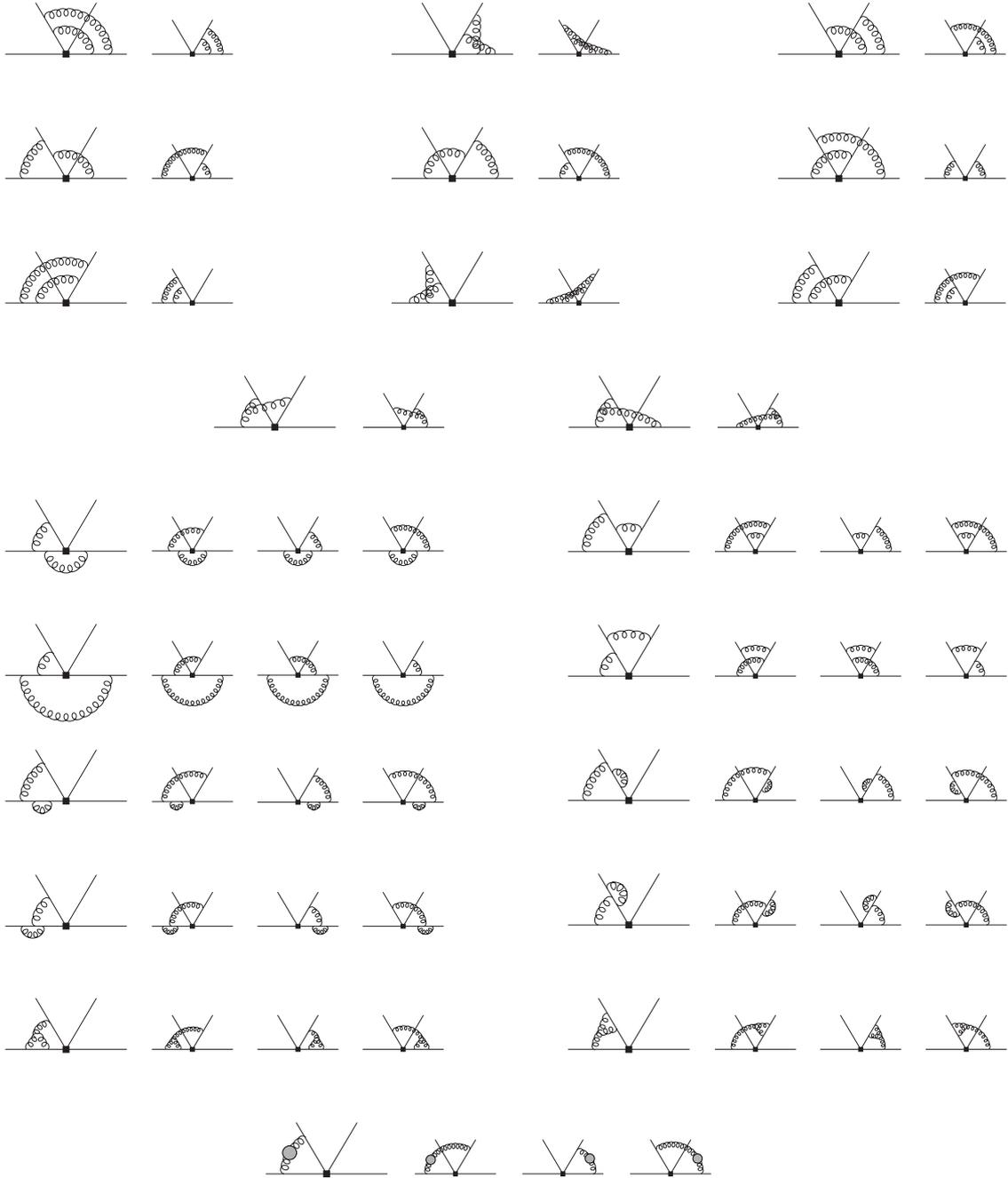}
\vspace{4mm}
\caption{\label{fig:BtoPiPi:NNLODiags} \small
\textit{Full set of non-factorizable 2-loop diagrams. In each
diagram the fermion line to the left of the four-quark vertex
denotes the massive $b$ quark, all other quarks are massless. The
bubble in the last four diagrams represents the 1-loop gluon
self-energy. Only diagrams with final state interactions, i.e. with
at least one gluon connecting the line to the right of the vertex
with one of the upper lines, give rise to an imaginary part.}}}}
\end{figure}

We will see explicitly in Section~\ref{sec:BtoPiPi:FFinNNLO} that we
may restrict our attention to (naively) non-factorizable diagrams,
similar to what we have seen in the NLO analysis from
Section~\ref{sec:BtoPiPi:PC}. These diagrams contain at least one
gluon which connects the two currents in the left diagram of
Figure~\ref{fig:Fierz}. The full set of 2-loop diagrams to be
considered here is depicted in Figure~\ref{fig:BtoPiPi:NNLODiags},
but only about half of these diagrams give rise to an imaginary
part. It is an easy task to identify this subset of diagrams as the
generation of an imaginary part is always related to final state
interactions.

\newabs
The colour factors of the 2-loop diagrams from
Figure~\ref{fig:BtoPiPi:NNLODiags} can be found in
Table~\ref{tab:BtoPiPi:Colour}. The diagrams can be written in terms
of the following denominators of propagators
\begin{align}
\calP_1&=\left(p-q-k\right)^2,      &&\calP_{13}=\left(p+k\right)^2-m^2,\no\\
\calP_2&=\left(p-q-l\right)^2,      &&\calP_{14}=\left(p+l\right)^2-m^2,\no\\
\calP_3&=\left(u q +k\right)^2,     &&\calP_{15}=\left(p+k+l\right)^2-m^2,\no\\
\calP_4&=\left(u q +l\right)^2,     &&\calP_{16}=\left(u q +k+l\right)^2,\no\\
\calP_5&=\left(\ubar q +k\right)^2, &&\calP_{17}=\left(\ubar q +k+l\right)^2,\no\\
\calP_6&=\left(\ubar q +l\right)^2, &&\calP_{18}=\left(p+k-l\right)^2-m^2,\no\\
\calP_7&=k^2,                       &&\calP_{19}=\left(p-q+k-l\right)^2,\no\\
\calP_8&=l^2,                       &&\calP_{20}=\left(p-q+k\right)^2,\no\\
\calP_9&=\left(k-l\right)^2,        &&\calP_{21}=\left(u q -l\right)^2,\no\\
\calP_{10}&=\left(u q +k-l\right)^2,&&\calP_{22}=\left(k-l\right)^2-z_f^2 m^2,\no\\
\calP_{11}&=\left(\bar u q +k-l\right)^2,     &&\calP_{23}=l^2-z_f^2 m^2,\no\\
\calP_{12}&=\left(p-q-k-l\right)^2, \label{eq:BtoPiPi:Props}
\end{align}
where $p$ denotes the momentum of the $b$-quark (with mass $m\equiv
m_b$), $p-q$ the one of the quark to the right of the weak vertex
and the quark/antiquark of the emitted meson have $u q/\ubar q$,
respectively. The on-shell kinematics is reflected by $q^2=0$ and
$p^2 = 2 p \cdot q = m^2$. The variable $z_f=m_f/m_b$ is related to
the diagrams with a closed fermion loop. For massless quarks in the
loop we simply have $z_q=0$, for an internal b-quark $z_b=1$ and for
the case of a charm quark we write $z\equiv z_c=m_c/m_b$.

\begin{table}[t!]\vspace*{0mm}
\centerline{
\parbox{15cm}{\setlength{\doublerulesep}{0.1mm}
\centerline{\begin{tabular}{|c||c|c|} \hline
\hspace*{2.5cm}&\hspace*{2.5cm}&\hspace*{2.5cm}  \\[-0.7em]
Operator & $\hat Q_1$ & $\hat Q_2$ \\[0.3em]
\hline\hline&& \\[-0.7em]
line 1-3 (L) & $C_F^2-\frac{C_FN_c}{2}$ & $\frac{C_F}{2}$ \\[0.3em]
\hline&& \\[-0.7em]
line 1-3 (S) & $C_F^2-\frac{C_FN_c}{4}$ & $\frac{C_F}{2}$ \\[0.3em]
\hline&& \\[-0.7em]
line 6-7 & $\frac{C_F^2}{2}$ & $0$ \\[0.3em]
\hline&& \\[-0.7em]
line 5,8 & $\frac{C_F^2}{2}-\frac{C_FN_c}{4}$ & $0$ \\[0.3em]
\hline&& \\[-0.7em]
line 4,9 & $\frac{C_FN_c}{4}$ & $0$ \\[0.3em]
\hline&& \\[-0.7em]
line 10 (A) & $\frac{C_F}{4}$ & $0$ \\[0.3em]
\hline&& \\[-0.7em]
line 10 (NA) & $-\frac{C_FN_c}{2}$ & $0$ \\[0.3em]
\hline
\end{tabular}} \vspace{6mm} \caption{\label{tab:BtoPiPi:Colour}\small
\textit{Colour factors of the diagrams in
Figure~\ref{fig:BtoPiPi:NNLODiags}. With ''L'' and ''S'' we refer to
the large and small figures, with ''A'' and ''NA'' to abelian and
non-abelian diagrams. The normalization is chosen such that the tree
diagram from Figure~\ref{fig:BtoPiPiTree} gives $N_c$ for $\hat
Q_2$.}}}}\vspace{1mm}
\end{table}

\begin{figure}[h!]\vspace{8mm}
\centerline{\parbox{15cm}{\centerline{
\includegraphics[width=15cm]{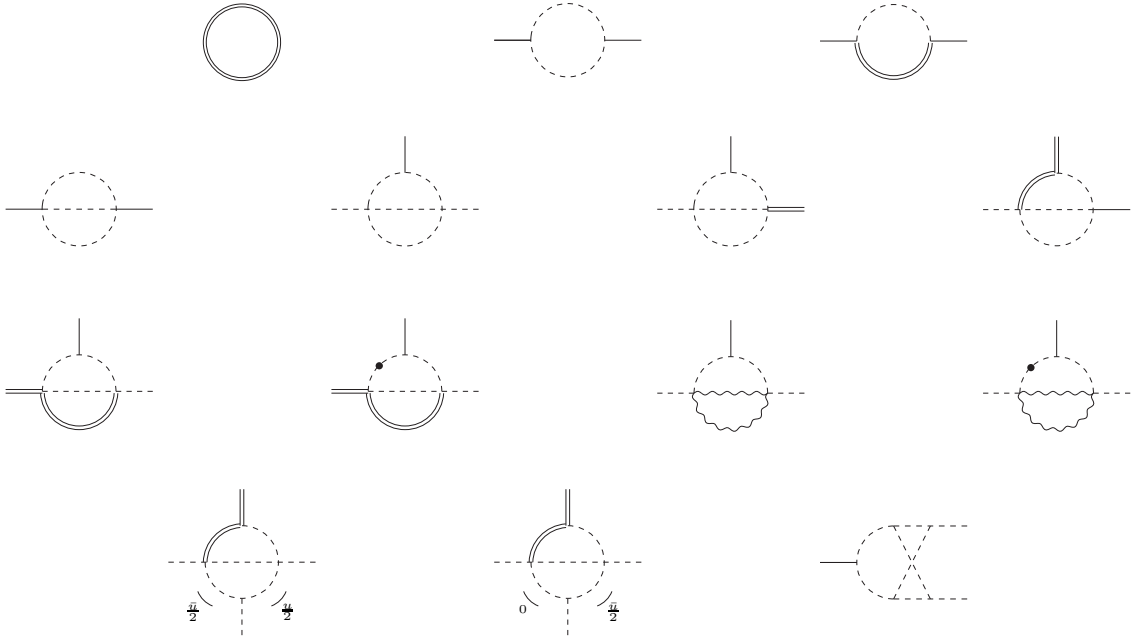}}\vspace{3mm}
\caption{\label{fig:BtoPiPi:MIs} \small \textit{Scalar Master
Integrals that appear in our calculation. We use dashed lines for
massless propagators and double (wavy) lines for the ones with mass
$m$ ($z_f m$). Dashed/solid/double external lines correspond to
virtualities $0/u m^2/m^2$, respectively. Dotted propagators are
taken to be squared.}}}}
\end{figure}

\subsubsection{Step 2: Reduction to Master Integrals}

The reduction algorithm represents an indispensable tool for the
considered calculation. It enables us to express all diagrams from
Figure~\ref{fig:BtoPiPi:NNLODiags} as linear combinations of MIs
which are multiplied by some Dirac structures. As the coefficients
in these linear combinations are real, we may extract the imaginary
part of a diagram at the level of the MIs which is a much simpler
task than for the full diagrams. The calculation of the imaginary
part involves 14 MIs which are depicted in
Figure~\ref{fig:BtoPiPi:MIs}.

\subsubsection{Step 3: Manipulation of Dirac structures}

We do not perform the bound state projections at this level of the
calculation as this would yield unwanted traces with $\gamma_5$. We
instead treat the two currents independently and make use of the
equations of motion for the on-shell quarks in order to simplify the
Dirac structures. In this way we end up with three irreducible
structures which are given by
\begin{align}
\left[\bar u \gamma^\mu L\, b\right] &\;
        \left[\bar d \gamma_\mu \,L\, u\right],\no\\
\left[\bar u \gamma^\mu\gamma^\nu\gamma^\rho L\, b\right] &\;
        \left[\bar d \gamma_\mu\gamma_\nu\gamma_\rho \,L\, u\right],\no\\
\left[\bar u \gamma^\mu\gamma^\nu\gamma^\rho\gamma^\sigma\gamma^\tau
L\, b\right] &\;
        \left[\bar d \gamma_\mu\gamma_\nu\gamma_\rho\gamma_\sigma\gamma_\tau \,L\, u\right].
\label{eq:BtoPiPi:Dirac}
\end{align}
The second structure gives rise to 1-loop evanescent operators
according to (\ref{eq:Basis:CMM}). As the last structure only enters
in the finite piece of the considered calculation, it can simply be
evaluated in $d=4$ dimensions without the need to introduce 2-loop
evanescent operators.

\subsubsection{Step 4: Calculation of Master Integrals}

Some MIs in Figure~\ref{fig:BtoPiPi:MIs} can be solved easily with
the help of Feynman parameters. This direct approach could be
improved with the help of the {\sc Mathematica} package {\sc
HypExp}~\cite{HypExp} which allows to expand a special class of
hyper-geometric functions to arbitrary order in $\eps$. However, as
discussed in detail in Section~\ref{sec:CalcMI}, wide parts of the
2-loop calculation were performed with the help of the method of
differential equations in combination with the formalism of Harmonic
Polylogarithms (HPLs). We further applied Mellin-Barnes techniques
for the calculation of the boundary conditions to the differential
equations. Apart from the two MIs with an internal charm quark (wavy
line), we were able to express all MIs with the help of a minimal
set of five HPLs given in (\ref{eq:HPLminiIm}).

The situation is more complicated for the MIs with an internal charm
quark which introduces a new scale to the problem. However, a closer
look reveals that these MIs depend on two physical scales only,
namely $u m_b^2$ and $m_c^2 \equiv z^2 m_b^2$. The MIs can then be
solved within the formalism of HPLs in terms of the ratio $\xi
\equiv z^2/u$ if we allow for more complicated arguments of the HPLs
as $\eta \equiv\frac12 \left( 1- \sqrt{1+4\xi} \right)$. As an
independent check of our results we evaluated the MIs numerically
using the method of sector decomposition. The results of the MIs can
be found in Appendix~\ref{app:MIsIm}.

\newpage
\section{Renormalization and IR subtractions}
\label{sec:BtoPiPi:UVIR}

So far we have computed the (unrenormalized) matrix elements
\begin{align}
\langle \hat{Q}_{1,2} \rangle \equiv \langle M_1 M_2 | \hat{Q}_{1,2}
| \bar{B} \rangle
\end{align}
to NNLO in perturbation theory (without spectator scattering). These
matrix elements are UV and IR divergent. In this section we discuss
adequate subtractions which will lead to a finite result for the
NNLO vertex corrections.

\subsection{Renormalization}

\begin{figure}[b!]
\centerline{\parbox{13cm}{
\centerline{\includegraphics[width=2.5cm]{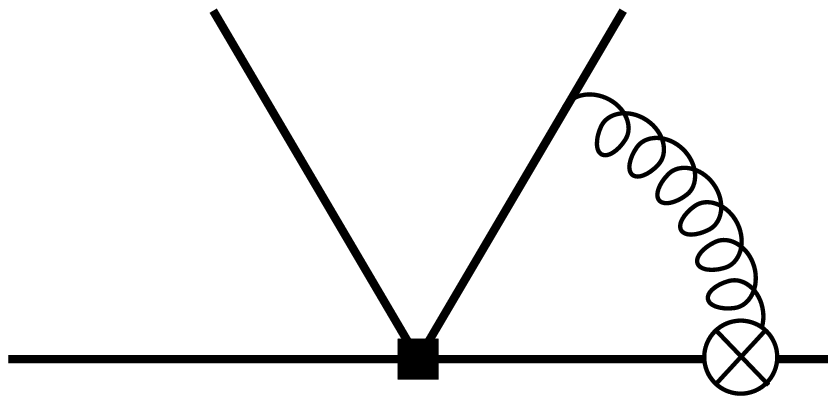}\hspace{15mm}\includegraphics[width=2.5cm]{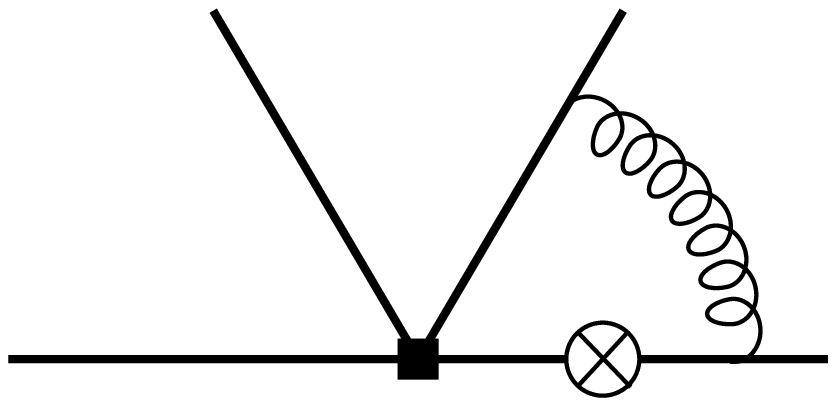}\hspace{15mm}\includegraphics[width=2.5cm]{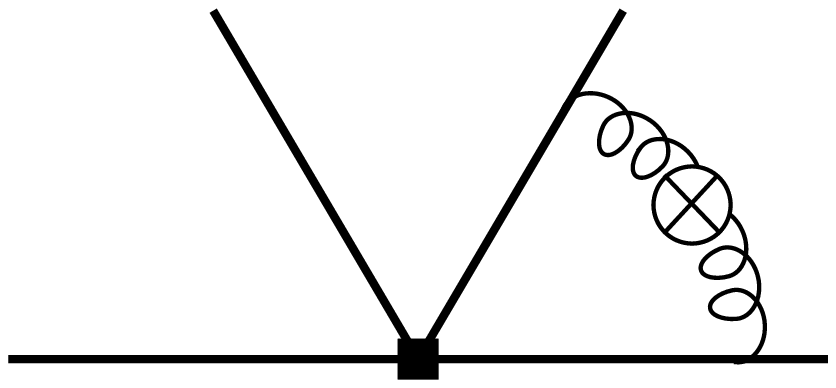}}
\caption{\label{fig:BtoPiPi:CTs} \small \textit{Sample of
counterterm diagrams.}}}}
\end{figure}

The renormalization procedure involves standard QCD counterterms,
which amount to the calculation of various 1-loop diagrams as the
ones depicted in Figure~\ref{fig:BtoPiPi:CTs}, as well as
counterterms from the effective Hamiltonian. We write the
renormalized matrix elements as
\begin{align}
\langle \hat{Q}_{i} \rangle_\text{ren} &= Z_\psi \, \hat{Z}_{i j} \,
\langle \hat{Q}_{j} \rangle_\text{bare}, \label{eq:defQren}
\end{align}
where $Z_\psi = Z_b^{1/2} Z_q^{3/2}$ contains the wave-function
renormalization factors of the massive b-quark $Z_b$ and the
massless quarks $Z_q$, whereas $\hat{Z}$ is the operator
renormalization matrix in the effective theory. We introduce the
following notation for the perturbative expansions of these
quantities
\begin{align}
\langle \hat{Q}_{i} \rangle_\text{ren/bare} =
    \sum_{k=0}^\infty \left( \frac{\as}{4\pi} \right)^k \langle \hat{Q}_{i} \rangle_\text{ren/bare}^{(k)}, \hspace{2.2cm} \no \\
Z_\psi = 1 + \sum_{k=1}^\infty  \left( \frac{\as}{4\pi} \right)^k
Z_\psi^{(k)}, \hspace{1.5cm} \hat{Z}_{i j} = \delta_{ij} +
\sum_{k=1}^\infty  \left( \frac{\as}{4\pi} \right)^k
\hat{Z}_{ij}^{(k)}
\end{align}
and rewrite (\ref{eq:defQren}) in perturbation theory up to NNLO
which yields
\begin{align}
\label{eq:expQren} \langle \hat{Q}_{i} \rangle_\text{ren}^{(0)} &=
    \langle \hat{Q}_{i} \rangle_\text{bare}^{(0)}, \no \\
\langle \hat{Q}_{i} \rangle_\text{ren}^{(1)} &=
    \langle \hat{Q}_{i} \rangle_\text{bare}^{(1)}
    + \left[ \hat{Z}_{ij}^{(1)} + Z_\psi^{(1)} \delta_{ij} \right]  \langle \hat{Q}_{j} \rangle_\text{bare}^{(0)}, \\
\langle \hat{Q}_{i} \rangle_\text{ren}^{(2)} &=
    \langle \hat{Q}_{i} \rangle_\text{bare}^{(2)}
    + \left[ \hat{Z}_{ij}^{(1)} + Z_\psi^{(1)} \delta_{ij} \right] \langle \hat{Q}_{j} \rangle_\text{bare}^{(1)}
    + \left[ \hat{Z}_{ij}^{(2)} + Z_\psi^{(1)} \hat{Z}_{ij}^{(1)} + Z_\psi^{(2)} \delta_{ij} \right] \langle \hat{Q}_{j} \rangle_\text{bare}^{(0)}. \no
\end{align}
The full calculation thus requires the operator renormalization
matrices $\hat{Z}^{(1,2)}$. For the calculation of the imaginary
part, the terms proportional to the tree level matrix elements do
not contribute and $\hat{Z}^{(2)}$ drops out in (\ref{eq:expQren})
as expected for an effective NLO calculation.

\newabs
Mass and wave function renormalization are found to be higher order
effects. For the renormalization of the coupling constant we use
\begin{align}
Z_g = 1 - \frac{\as}{4\pi \eps} \left( \frac{11}{6} N_c - \frac13
n_f \right) + \calO(\as^2).
\end{align}
The 1-loop renormalization matrix $\hat{Z}^{(1)}$ can be found
e.g.~in \cite{GGH} and reads
\renewcommand{\arraycolsep}{2mm}
\begin{align}
\hat{Z}^{(1)} &= \left(
\begin{array}{c c c c}
\rule[-2mm]{0mm}{7mm} -2 & \frac43 & \frac{5}{12} & \frac{2}{9} \\
\rule[-2mm]{0mm}{7mm} 6 & 0 & 1 & 0
\end{array}
\right) \, \frac{1}{\eps}, \label{eq:Z1}
\end{align}
where the two lines correspond to the basis of physical operators
$\{ \hat{Q}_1, \hat{Q}_2\}$ and the four columns to the extended
basis $\{ \hat{Q}_1, \hat{Q}_2, \hat{E}_1, \hat{E}_2 \}$ including
the mixing of the non-physical evanescent operators into the
physical ones.

\subsection{Factorization in NNLO}
\label{sec:BtoPiPi:FFinNNLO}

In this section it will be convenient to introduce the following
short-hand notation for the factorization formula (\ref{eq:QCDF})
\begin{align}
\langle \hat{Q}_{i} \rangle_\text{ren} &= F \cdot T_i \otimes \Phi +
\ldots \label{eq:FF}
\end{align}
where $F$ denotes the $B\to M_1$ form factor, $T_i$ the
hard-scattering kernels and $\Phi$ the product of the decay constant
$f_{M_2}$ and the distribution amplitude $\phi_{M_2}$. The
convolution in (\ref{eq:QCDF}) has been represented by the symbol
$\otimes$ and the ellipses contain the terms from spectator
scattering which we disregard in the following.

\newabs
Formally, we may introduce the perturbative expansions
\begin{align}
F = \sum_{k=0}^\infty \left( \frac{\as}{4\pi} \right)^k F^{(k)},
\hspace{1cm} T_i = \sum_{k=0}^\infty \left( \frac{\as}{4\pi}
\right)^k T_i^{(k)}, \hspace{1cm} \Phi = \sum_{k=0}^\infty \left(
\frac{\as}{4\pi} \right)^k \Phi^{(k)}.
\end{align}
Up to NNLO the expansion of (\ref{eq:FF}) yields
\begin{align}
\label{eq:expFF} \langle \hat{Q}_{i} \rangle_\text{ren}^{(0)} &=
    F^{(0)} \cdot T_i^{(0)} \otimes \Phi^{(0)}, \no \\
\langle \hat{Q}_{i} \rangle_\text{ren}^{(1)} &=
    F^{(0)} \cdot T_i^{(1)} \otimes \Phi^{(0)} + F^{(1)} \cdot T_i^{(0)} \otimes \Phi^{(0)} + F^{(0)} \cdot T_i^{(0)} \otimes \Phi^{(1)}, \no \\
\langle \hat{Q}_{i} \rangle_\text{ren}^{(2)} &=
    F^{(0)} \cdot T_i^{(2)} \otimes \Phi^{(0)} + F^{(1)} \cdot T_i^{(1)} \otimes \Phi^{(0)} + F^{(0)} \cdot T_i^{(1)} \otimes \Phi^{(1)} \no \\
    & \quad + F^{(2)} \cdot T_i^{(0)} \otimes \Phi^{(0)} + F^{(1)} \cdot T_i^{(0)} \otimes \Phi^{(1)} + F^{(0)} \cdot T_i^{(0)} \otimes \Phi^{(2)}.
\end{align}
In LO the comparison of (\ref{eq:expQren}) and (\ref{eq:expFF})
gives the trivial relation
\begin{align}
    \langle \hat{Q}_{i} \rangle^{(0)} \equiv
    \langle \hat{Q}_{i} \rangle_\text{bare}^{(0)} =
    F^{(0)} \cdot T_i^{(0)} \otimes \Phi^{(0)}
\end{align}
which states that $T_i^{(0)}$ can be computed from the tree level
diagram in Figure~\ref{fig:BtoPiPiTree}. In order to address higher
order terms we split the bare matrix elements into contributions
from (naively) factorizable (f) and non-factorizable (nf) diagrams
\begin{align}
\langle \hat{Q}_{i} \rangle_\text{bare}^{(k)} &\equiv
    \langle \hat{Q}_{i} \rangle_\text{f}^{(k)} +
    \langle \hat{Q}_{i} \rangle_\text{nf}^{(k)}.
\end{align}
In NLO the corresponding diagrams have been shown in
Figure~\ref{fig:BtoPiPiNLOf} and Figure~\ref{fig:BtoPiPiNLOnf},
respectively. To this order (\ref{eq:expQren}) and (\ref{eq:expFF})
lead to
\begin{align}
    &\langle \hat{Q}_{i} \rangle_\text{f}^{(1)} + \langle \hat{Q}_{i} \rangle_\text{nf}^{(1)}
    + \left[ \hat{Z}_{ij}^{(1)} + Z_\psi^{(1)} \delta_{ij} \right]  \langle \hat{Q}_{j} \rangle^{(0)} \no \\
    & \qquad = F^{(0)} \cdot T_i^{(1)} \otimes \Phi^{(0)} + F^{(1)} \cdot T_i^{(0)} \otimes \Phi^{(0)} + F^{(0)} \cdot T_i^{(0)} \otimes \Phi^{(1)},
\end{align}
which splits into
\begin{align}
    \langle \hat{Q}_{i} \rangle_\text{nf}^{(1)}
    + \hat{Z}_{ij}^{(1)} \langle \hat{Q}_{j} \rangle^{(0)}
    &= F^{(0)} \cdot T_i^{(1)} \otimes \Phi^{(0)}
\label{eq:NLOFact}
\end{align}
for the calculation of the NLO kernels $T_i^{(1)}$ and
\begin{align}
    \langle \hat{Q}_{i} \rangle_\text{f}^{(1)} + Z_\psi^{(1)} \langle \hat{Q}_{i} \rangle^{(0)}
    &= F^{(1)} \cdot T_i^{(0)} \otimes \Phi^{(0)} + F^{(0)} \cdot T_i^{(0)} \otimes \Phi^{(1)},
\end{align}
which shows that the factorizable diagrams and the wave-function
renormalization are absorbed by the form factor and wave function
corrections $F^{(1)}$ and $\Phi^{(1)}$.

\newabs
This suggests in NNLO the following structure
\begin{align}
    & \langle \hat{Q}_{i} \rangle_\text{f}^{(2)} + Z_\psi^{(1)} \langle \hat{Q}_{i} \rangle_\text{f}^{(1)} + Z_\psi^{(2)} \langle \hat{Q}_{i} \rangle^{(0)} \no \\
    & \qquad = F^{(2)} \cdot T_i^{(0)} \otimes \Phi^{(0)} + F^{(1)} \cdot T_i^{(0)} \otimes \Phi^{(1)} + F^{(0)} \cdot T_i^{(0)} \otimes \Phi^{(2)}.
\end{align}
These terms are thus irrelevant for the calculation of the NNLO
kernels $T_i^{(2)}$ which justifies that we could restrict our
attention to the non-factorizable 2-loop diagrams from
Figure~\ref{fig:BtoPiPi:NNLODiags}. In NNLO the remaining terms from
(\ref{eq:expQren}) and (\ref{eq:expFF}) contain non-trivial IR
subtractions
\begin{align}
    & \langle \hat{Q}_{i} \rangle_\text{nf}^{(2)} + Z_\psi^{(1)} \langle \hat{Q}_{i} \rangle_\text{nf}^{(1)}
    + \hat{Z}_{ij}^{(1)} \left[ \langle \hat{Q}_{j} \rangle_\text{nf}^{(1)} + \langle \hat{Q}_{j} \rangle_\text{f}^{(1)} \right]
    + \left[ \hat{Z}_{ij}^{(2)} + Z_\psi^{(1)} \hat{Z}_{ij}^{(1)} \right] \langle \hat{Q}_{j} \rangle^{(0)}  \no \\
    & \qquad = F^{(0)} \cdot T_i^{(2)} \otimes \Phi^{(0)} + F^{(1)} \cdot T_i^{(1)} \otimes \Phi^{(0)} + F^{(0)} \cdot T_i^{(1)} \otimes \Phi^{(1)}.
\label{eq:preMaster}
\end{align}
This equation can be simplified further when we make the wave
function renormalization factors in the form factor and the
distribution amplitude explicit
\begin{align}
    F = Z_b^{1/2} Z_q^{1/2} F_\text{amp}, \hspace{2cm}
    \Phi = Z_q \, \Phi_\text{amp}.
\label{eq:amputated}
\end{align}
Notice that the resulting amputated form factor $F_\text{amp}$ and
wave function $\Phi_\text{amp}$ contain UV divergences by
construction. We recall that $Z_\psi = Z_b^{1/2} Z_q^{3/2}$ and find
\begin{align}
& F^{(1)} \cdot T_i^{(1)} \otimes \Phi^{(0)} + F^{(0)} \cdot T_i^{(1)} \otimes \Phi^{(1)} \no \\
& \qquad =F_\text{amp}^{(1)} \cdot T_i^{(1)} \otimes \Phi^{(0)} +
F^{(0)} \cdot T_i^{(1)} \otimes \Phi_\text{amp}^{(1)}
          + Z_\psi^{(1)} \; F^{(0)} \cdot T_i^{(1)} \otimes \Phi^{(0)}.
\label{eq:aux1}
\end{align}
Combining (\ref{eq:preMaster}) with (\ref{eq:aux1}) and
(\ref{eq:NLOFact}), we arrive at the Master Formula for the
calculation of the hard-scattering kernels $T_i^{(2)}$ in NNLO
\begin{align}
    & \langle \hat{Q}_{i} \rangle_\text{nf}^{(2)}
    + \hat{Z}_{ij}^{(1)} \left[ \langle \hat{Q}_{j} \rangle_\text{nf}^{(1)} + \langle \hat{Q}_{j} \rangle_\text{f}^{(1)} \right]
    + \hat{Z}_{ij}^{(2)}  \langle \hat{Q}_{j} \rangle^{(0)}  \no \\
    & \qquad = F^{(0)} \cdot T_i^{(2)} \otimes \Phi^{(0)}
        + F_\text{amp}^{(1)} \cdot T_i^{(1)} \otimes \Phi^{(0)}
        + F^{(0)} \cdot T_i^{(1)} \otimes \Phi_\text{amp}^{(1)}.
\label{eq:Master}
\end{align}
The 2-loop matrix elements on the left-hand side of
(\ref{eq:Master}) have been considered in
Section~\ref{sec:BtoPiPi:2loop} and the 1-loop renormalization
matrix has been given in (\ref{eq:Z1}). The 1-loop matrix elements
of the non-factorizable diagrams involve the calculation of the
diagrams in Figure~\ref{fig:BtoPiPiNLOnf}a. The tree level matrix
elements as well as the 1-loop matrix elements of the factorizable
diagrams from Figure~\ref{fig:BtoPiPiNLOf} can be disregarded here
as they do not give rise to an imaginary part. Hence, the only
missing pieces for the calculation of the imaginary part of the NNLO
kernels $T_i^{(2)}$ are the IR subtractions on the right-hand side
of (\ref{eq:Master}) which we consider in the following section.

\subsection{IR subtractions}

Let us first address the NLO kernels $T_i^{(1)}$ which can be
determined from equation (\ref{eq:NLOFact}). The renormalization in
the evanescent sector implies that the left hand side of
(\ref{eq:NLOFact}) is free of contributions from evanescent
operators up to the \emph{finite order} $\eps^0$. However, as the
NLO kernels enter (\ref{eq:Master}) in combination with the form
factor correction $F^{(1)}$ which contains double (soft and
collinear) IR divergences, the NLO kernels are required here up to
$\calO(\eps^2)$. Concerning the subleading terms of $\calO(\eps)$,
the evanescent operators do not drop out on the left hand side of
(\ref{eq:NLOFact}) and we therefore have to extend the factorization
formula on the right hand side to include these evanescent
structures as well. Schematically,
\begin{align}
    \langle \hat{Q}_{i} \rangle_\text{nf}^{(1)}
    + \hat{Z}_{ij}^{(1)} \langle \hat{Q}_{j} \rangle^{(0)}
    &= F^{(0)} \cdot T_i^{(1)} \otimes \Phi^{(0)} +
    F_E^{(0)} \cdot T_{i,E}^{(1)} \otimes \Phi_E^{(0)}
\end{align}
with a kernel $T_{i,E}^{(1)}=\calO(1)$ and an evanescent matrix
element $F_E^{(0)} \Phi_E^{(0)}=\calO(\eps)$. Similarly, the right
hand side of (\ref{eq:Master}) has to be modified to include these
evanescent structures.

From the calculation of the 1-loop diagrams in
Figure~\ref{fig:BtoPiPiNLOnf}a, we find that the NLO kernels vanish
in the colour-singlet case, $T_2^{(1)}=T_{2,E}^{(1)}=0$, whereas the
imaginary part of the colour-octet kernels is given by
\begin{align}
\frac{1}{\pi} \; \Im \;T_1^{(1)}(u)  &= \frac{C_F}{2N_c}
\bigg\{ (-3-2\ln u+2\ln \ubar) \Big( 1+\eps L +\frac12 \eps^2 L^2\Big) \no \\
& \hspace{1.6cm} - ( 11 - 3 \ln \ubar  - \ln^2 u+ \ln^2 \ubar ) \Big( \eps +\eps^2 L \Big) \no \\
& \hspace{1.6cm} + \bigg[ \frac{3\pi^2}{4} - 26 + \Big( 2 +
\frac{\pi^2}{2} \Big) \ln u
+ \Big( 9 - \frac{\pi^2}{2} \Big) \ln \ubar \no \\
& \hspace{2.2cm} - \frac32 \ln^2 \ubar - \frac13 \left( \ln^3 u -
\ln^3 \ubar \right) \bigg] \eps^2 + \calO(\eps^3) \bigg\}, \nonumber
\end{align}
\begin{align}
\frac{1}{\pi} \; \Im \;T_{1,E}^{(1)}(u)  &= - \frac{C_F}{4N_c}
\bigg\{ 1+\eps L + \Big( \frac83 - \frac12 \ln u -\frac 12 \ln \ubar
\Big) \eps +  \calO(\eps^2) \bigg\}, \label{eq:T1}
\end{align}
where $L \equiv \ln \mu^2/m_b^2$ and we recall that
$\ubar\equiv1-u$.

\subsubsection*{Form factor subtractions}

\begin{figure}[b!]
\centerline{\parbox{13cm}{
\centerline{\includegraphics[width=4cm]{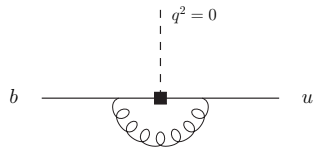}}
\caption{\label{fig:BtoPiPi:FFcorr}  \small \textit{1-loop
contribution to the form factor correction $F^{(1)}_\text{amp}$.}}}}
\end{figure}

We now address the form factor corrections which require the
calculation of the diagram in Figure~\ref{fig:BtoPiPi:FFcorr} (for
on-shell quarks) and its counterterm. According to the definition of
$F_\text{amp}$ in (\ref{eq:amputated}), we do not have to consider
the wave function renormalization of the quark fields here.

We again have to compute the corrections for physical and evanescent
operators. Concerning the physical operators with Dirac structure
$[\gamma^\mu L]\;[\gamma_\mu L]$ the counterterm is found to vanish
and we get
\begin{align}
F_\text{amp}^{(1)} \,\Phi^{(0)}= - C_F \left(
  \frac{e^{\gamma_E}\mu^2}{m_b^2} \right)^\eps \Gamma(\eps) \;
\frac{1-\eps+2\eps^2}{\eps(1-2\eps)}  \,\; F^{(0)}  \,\Phi^{(0)}
\label{eq:BtoPiPi:FFphys}
\end{align}
reflecting the $1/\eps^2$--singularities mentioned at the beginning
of this section. On the other hand, the evanescent operators with
$[\gamma^\mu \gamma^\nu \gamma^\rho L]\;[\gamma_\mu \gamma_\nu
\gamma_\rho L] - 16 [\gamma^\mu L]\;[\gamma_\mu L]$ yield a
contribution proportional to the evanescent \emph{and} the physical
operators. We find a non-vanishing contribution from the counterterm
diagram in this case and obtain
\begin{align}
F_\text{amp,E}^{(1)} \,\Phi_E^{(0)}&= C_F \left[ \left(
    \frac{e^{\gamma_E}\mu^2}{m_b^2} \right)^\eps  \Gamma(\eps) \;
  \frac{24 \eps (1+\eps)}{(1-\eps)^2} - 24 \right] \;  F^{(0)}
\,\Phi^{(0)}  \nonumber\\
& \hspace{4mm} -  C_F \left( \frac{e^{\gamma_E}\mu^2}{m_b^2}
\right)^\eps  \Gamma(\eps) \;
\frac{1-3\eps+\eps^2+3\eps^3+2\eps^4}{\eps(1-2\eps)(1-\eps)^2} \,\;
F_E^{(0)} \,\Phi_E^{(0)}. \label{eq:BtoPiPi:FFevan}
\end{align}
The first subtraction term in (\ref{eq:Master}) then follows from
combining the form factor corrections in (\ref{eq:BtoPiPi:FFphys})
and (\ref{eq:BtoPiPi:FFevan}) with the NLO kernels in (\ref{eq:T1}).
We emphasize that the corrections related to the evanescent
operators do \emph{not} induce a contribution to the physical NNLO
kernel $T_1^{(2)}$ in this case since
\begin{align}
\frac{1}{\pi} \; F_\text{amp,E}^{(1)} \; \Im \;T_{1,E}^{(1)} \;
\Phi_E^{(0)} \quad \to \quad \bigg[\calO(\eps) \bigg] \;  F^{(0)}
\,\Phi^{(0)}.
\end{align}

\subsubsection*{Wave function subtractions}

\begin{figure}[t!]
\centerline{\parbox{13cm}{
\centerline{\includegraphics[width=11cm]{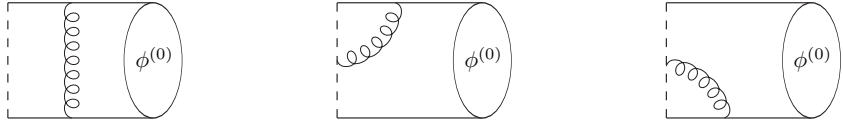}}
\caption{\label{fig:WFcorr}  \small \textit{1-loop contributions to
the wave function correction $\Phi^{(1)}_\text{amp}$. The dashed
line indicates the Wilson-line connecting the quark and antiquark
    fields.}}}}
\end{figure}

Concerning the wave function corrections we are left with the
calculation of the diagrams in Figure~\ref{fig:WFcorr} for collinear
and on-shell partons with momenta $u q$ and $\ubar q$. However, as
in our set-up $q^2=0$ all these diagrams vanish due to scaleless
integrals in DR. We conclude that the wave function corrections are
determined entirely by the counter\-terms. We compute these
counterterms by calculating the diagrams from
Figure~\ref{fig:WFcorr} with an off-shell regularization
prescription in order to isolate the UV-divergences. The
counter\-term for the physical operators is found to be
\begin{align}
F^{(0)} \, \Phi_\text{amp}^{(1)} (u) = - \frac{2 C_F}{\eps} \int_0^1
dw \; V(u,w) \; F^{(0)} \, \Phi^{(0)} (w) \label{eq:Phiphys}
\end{align}
with the familiar Efremov-Radyushkin-Brodsky-Lepage (ERBL)
kernel~\cite{ER,BL}
\begin{align}
V(u,w) = \left[ \theta(w-u) \frac{u}{w} \left(1 + \frac{1}{w-u}
\right) + \theta(u-w) \frac{\bar u}{\bar w} \left(1 + \frac{1}{\bar
w-\bar u} \right) \right]_+
\end{align}
where the plus-distribution is defined as $\left[ f(u,w) \right]_+ =
f(u,w) - \delta(u-w) \int_0^1 dv \; f(v,w)$. For the evanescent
operators we obtain
\begin{align}
F_E^{(0)} \, \Phi_\text{amp,E}^{(1)} (u) = - \frac{2 C_F}{\eps}
\int_0^1 dw \, \bigg[ 24 \eps \,V_E(u,w) \; F^{(0)} \, \Phi^{(0)}
(w) + V(u,w) \; F_E^{(0)} \, \Phi_E^{(0)} (w) \bigg]
\label{eq:Phievan}
\end{align}
where $V_E(u,w)$ denotes the spin-dependent part of the ERBL kernel
given by
\begin{align}
V_E(u,w) = \theta(w-u) \frac{u}{w}  + \theta(u-w) \frac{\bar u}{\bar
w}
\end{align}
Notice that the evanescent operators \emph{do} induce a finite
contribution to the physical kernel $T_1^{(2)}$ in this case as the
convolution with the corresponding NLO kernel implies
\begin{align}
\frac{1}{\pi} \; F_E^{(0)} \; \Im \;T_{1,E}^{(1)} \;
\Phi_\text{amp,E}^{(1)} \quad \to \quad \bigg[ \frac{6C_F^2}{N_c} +
\calO(\eps) \bigg] \;  F^{(0)} \,\Phi^{(0)}.
\end{align}
We finally quote the result for the convolution with the physical
NLO kernel
\begin{align}
&\frac{1}{\pi} \, F^{(0)} \, \Im \,T_1^{(1)} \,
\Phi_\text{amp}^{(1)} = \frac{C_F^2}{N_c} \bigg\{ \bigg[
\frac{\pi^2}{3} + \frac{\ln
  u}{\ubar} - \frac{\ln \ubar}{u}  + \ln^2 u - 2 \ln u \ln \ubar - \ln^2
\ubar -4 \Lib (u) \bigg] \no \\
&\hspace{3.8cm} \bigg( \frac{1}{\eps} + L \bigg) + \frac{\pi^2}{2} -
\frac{15}{4} - 2 \zeta_3 + \frac{5u-4}{2} \bigg( \frac{\ln u}{\ubar}
+ \frac{\ln
  \ubar}{u} \bigg)- \frac{\pi^2}{3} \ln \ubar    \no \\
& \hspace{3.8cm} - 3 \Lib(u) - \frac{1}{2 \ubar} \ln^2 u +
\frac{1-3u}{2u}\ln^2 \ubar -
\frac23 \ln^3 u  + \ln^2 u \ln \ubar  \no \\
& \hspace{3.8cm} + \frac23 \ln^3 \ubar+ 2 \ln \ubar \, \Lib (u) + 2
\Lic(u) + 2 \Sab (u) +\calO(\eps) \bigg\} \; F^{(0)} \, \Phi^{(0)}
\end{align}

\section{Tree amplitudes in NNLO}

The NNLO kernels $T_i^{(2)}$ can now be determined from the Master
Formula (\ref{eq:Master}). We indeed observe that all UV and IR
singularities cancel in the NNLO kernels which provides a very
important and highly non-trivial cross-check of the
calculation\footnote{As we do not distinguish between UV and IR
singularities in our calculation, we cannot verify their
cancellation independently. However, the aforementioned
renormalization and IR subtraction procedure can be organized in a
way that allows to control the cancellation of the leading poles in
several intermediate steps of the calculation.}. In analogy to
Section~\ref{sec:BtoPiPi:PC}, we quote our results in terms of the
tree amplitudes $\alpha_{1,2}$.

\subsection{$\alpha_1$ in CMM basis}
\label{sec:BtoPiPi:a1CMM}

The procedure outlined so far leads to the colour-allowed tree
amplitude in the CMM operator basis from (\ref{eq:Basis:CMM}). We
write
\begin{align}
\alpha_1(M_1 M_2) &= \hat{C}_2 +\frac{\as}{4\pi} \, \frac{C_F}{2
N_c} \bigg[ \hat{C}_1 \hat{V}^{(1)} + \frac{\as}{4\pi} \left(
\hat{C}_{1} \, \hat{V}_1^{(2)} + \hat{C}_{2} \,
\hat{V}_2^{(2)}\right) + \calO(\as^2)  \bigg] + \ldots
\label{eq:alpha1Mod}
\end{align}
where the ellipses denote the terms from spectator scattering which
are irrelevant for our purposes. In the CMM basis, the imaginary
part of the vertex corrections $\hat{V}^{(1,2)}$ can be written in
the form
\begin{align}
\frac{1}{\pi} \; \Im \;\hat{V}^{(1)}   &\equiv
    \int_0^1 du \; g_1(u)  \; \phi_{M_2}(u), \no\\
\frac{1}{\pi} \; \Im \;\hat{V}_1^{(2)} &\equiv
    \int_0^1 du \; \bigg\{ \Big[ \Big( \frac{29}{3} N_c - \frac23 n_f \Big) g_1(u) + C_F \,h_1(u) \Big] \ln \frac{\mu^2}{m_b^2}  \no \\
    & \hspace{3mm} + C_F \, h_2(u) + N_c \, h_3(u) + (n_f-2) \, h_4(u;0) + h_4(u;z) + h_4(u;1) \bigg\}   \phi_{M_2}(u), \no \\
\frac{1}{\pi} \; \Im \;\hat{V}_2^{(2)} &\equiv
    \int_0^1 du \; \bigg\{ -6\, g_1(u)\, \ln \frac{\mu^2}{m_b^2} + h_0(u) \bigg\}
    \phi_{M_2}(u).
\label{eq:V12mod}
\end{align}
We stress that these quantities do \emph{not} correspond to the
$V^{(1,2)}$ from (\ref{eq:alpha12NNLO}) which will be given in the
following section after the transformation to the QCDF basis. In
writing (\ref{eq:V12mod}), we have made the dependence on the
renormalization scale explicit and we have disentangled
contributions that belong to different colour structures. The
function $h_4(u;z_f)$ stems from diagrams with a closed fermion loop
and depends on the mass of the internal quark through $z_f=m_f/m_b$.
We write $z\equiv z_c = m_c/m_b$ for simplicity.

\newabs
In NLO we find
\begin{align}
g_1(u) &=
    -3 -2 \ln u + 2 \ln \ubar.
\label{eq:BtoPiPi:kernels:g}
\end{align}
The NNLO kernels were so far unknown. They are found in this work to
be
\begin{align}
h_0(u)  &=  \bigg[ \frac{155}{4} + 4 \zeta_3 + 4 \Lic(u) - 4\Sab(u) -12 \ln u\, \Lib(u) +\frac43 \ln^3 u -6\ln^2 u\ln \ubar \no \\
        &\hspace{8mm} + \frac{2-u^2}{\ubar}\Lib(u) -\frac{5-3u+3u^2-2u^3}{2\ubar}\ln^2 u +\frac{3-2u^4}{2u\ubar}\ln u \ln\ubar \no \\
        &\hspace{8mm} - \Big( \frac{4-11u+2u^2}{\ubar}+\frac{4\pi^2}{3}\Big) \ln u - \frac{(5+6u^2-12u^4)\pi^2}{24u \ubar} + (u\leftrightarrow\ubar)\bigg] \no \\
        &\hspace{5mm} + \bigg[ \frac{3-u+7u^2}{2\ubar}\ln^2 u - \frac{11-10u^2}{4u\ubar} \Lib(u)+ \frac{1-14u^2}{4\ubar}\ln u \ln \ubar \no \\
        &\hspace{12mm} + \frac{46-51u}{\ubar}\ln u - \frac{(41-42u^2)\pi^2}{24\ubar} - (u\leftrightarrow\ubar) \bigg],\no
        \\
h_1(u) &=
    36 + \bigg[ 2 \ln^2 u  - 4 \Lib(u) + \frac{2(13-12u)}{1-u} \ln u -(u\leftrightarrow\ubar) \bigg],\no\\
h_2(u) &= \bigg[ 79 + 32 \zeta_3 -16 \Lic(u) - 32 \Sab(u) + \frac83 \ln^3 u +\frac{2(4-u^2)}{\ubar} \Lib(u) \no \\
        &\hspace{8mm} - \frac{13-9u+6u^2-4u^3}{2\ubar}\ln^2 u + \frac{17-6u^2-8u^4}{4u\ubar}\ln u \ln \ubar \no \\
        &\hspace{8mm} - 2\Big(\frac{5-11u+2u^2}{\ubar}+\frac{4\pi^2}{3} \Big) \ln u- \frac{(1+14u^2-8u^4)\pi^2}{8u\ubar} + (u\leftrightarrow\ubar)\bigg] \no \\
        &\hspace{5mm} + \bigg[ 4 \Lic(u) + 4\Sab(u) -\frac23 \ln^3 u +2 \ln^2 u \ln \ubar -\frac{9-14u^2}{u\ubar} \Lib(u) \no \\
        &\hspace{12mm}+ \frac{13-11u+14u^2}{2\ubar} \ln^2 u+\frac{5-7u^2}{\ubar}\ln u \ln \ubar  \no \\
        &\hspace{12mm}+ 4 \Big( \frac{24-23u}{\ubar} + \frac{\pi^2}{3} \Big) \ln u - \frac{(26-21u^2)\pi^2}{6\ubar} - (u\leftrightarrow\ubar)
        \bigg],\no \\
h_3(u) &= \bigg[ - \frac{1379}{24} - 12 \zeta_3 + 6 \Lic(u) +12 \Sab(u) -\ln^3 u - \frac{4-u^2}{\ubar} \Lib(u) \no \\
        &\hspace{8mm} + \frac{9-2u+6u^2-4u^3}{4\ubar}\ln^2 u- \frac{7+4u^2-4u^4}{4u\ubar}\ln u \ln \ubar \no \\
        &\hspace{8mm} + \Big( \frac{41-66u + 8u^2}{4\ubar}+ \pi^2 \Big) \ln u + \frac{(1+6u^2-4u^4)\pi^2}{8u\ubar} + (u\leftrightarrow\ubar)\bigg]\no\\
        &\hspace{5mm} + \bigg[ -2 \Lic(u) + 4 \Sab(u) + 4 \ln u \, \Lib(u) + \frac13 \ln^3 u + \frac{15-26u^2}{4u\ubar} \Lib(u) \no \\
        &\hspace{12mm}+ \frac{11-14u-42u^2}{12\ubar} \ln^2 u - \frac{11-14u^2}{4\ubar} \ln u \ln \ubar \no
\label{eq:BtoPiPi:kernels:h}
\end{align}

\newpage
\vspace*{-1.2cm}
\begin{align}
        &\hspace{12mm}- \Big( \frac{2165-2156u}{36\ubar} - \frac{\pi^2}{3} \Big) \ln u + \frac{(53-42u^2)\pi^2}{24\ubar} - (u\leftrightarrow\ubar) \bigg],\no\\
h_4(u;z) &= \bigg[ \frac{17}{6} + \frac{7\xi}{\ubar} -2 \xi^2 \ln^2 \frac{x_1}{x_2} + 2 \Big( (1+4\xi) x_1 + (1+6\xi) x_2 \Big) \ln x_1 -4 \xi \, x_1 \ln x_2 \no \\
        &\hspace{6mm} + \Big(\frac{\xi}{\ubar} - 2(1+4\xi)x_2 \Big) \ln z^2 + \Big((1+2\xi)x_1+(1+6\xi)x_2 \Big) \ln u  + (u\leftrightarrow\ubar)\bigg] \no \\
        &\hspace{0mm} + \bigg[ \frac{94z^2}{9\ubar} - \frac{2(1-3\xi^2)}{3} \ln^2 \frac{x_1}{x_2} - \frac{2[ (6+29\xi+20\xi^2)x_1+(29+38\xi)\xi\, x_2]}{9\xi} \ln x_1 \no \\
        &\hspace{6mm} - \frac{4(1-3\xi^2)}{3\xi} x_1 \ln x_2+\frac{2u \ubar (6+29\xi+20\xi^2)x_2+(1-2u) (6 \ubar- u \xi^2)}{9u \ubar\xi}  \ln z^2\no \\
        &\hspace{6mm} -\frac43 \ln u \ln z^2 - \frac{ (12+29\xi+2\xi^2)x_1+(29+38\xi)\xi\, x_2}{9\xi} \ln u - (u\leftrightarrow\ubar) \bigg].
\end{align}
The last kernel $h_4(u;z)$ has been given in terms of
\begin{align}
x_{1} \equiv \frac12 \left( \sqrt{1+4\xi} - 1 \right), \qquad x_{2}
\equiv \frac12 \left( \sqrt{1+4\xi} + 1 \right), \qquad \xi \equiv
\frac{z^2}{u}.
\end{align}
In the massless limit $h_4(u;z)$ simply becomes
\begin{align}
h_4(u;0) &=
    \frac{17}{3} - \frac23 \ln^2 u + \frac23 \ln^2 \ubar +\frac{20}{9} \ln u - \frac{38}{9} \ln \ubar.
\end{align}

\subsection{$\alpha_1$ and $\alpha_2$ in QCDF basis}

We now perform the transformation of the colour-allowed tree
amplitude $\alpha_1$ into the QCDF operator basis from
(\ref{eq:Basis:QCDF}). As discussed in
Section~\ref{eq:BtoPiPi:OpBasis}, manifest Fierz symmetry in the
QCDF basis allows us to derive the colour-suppressed amplitude
$\alpha_2$ directly from $\alpha_1$ under the exchange $\tilde
C_1\leftrightarrow \tilde C_2$.

\newabs
The colour-allowed tree amplitude has been given in the CMM basis in
(\ref{eq:alpha1Mod}) and in the QCDF basis in
(\ref{eq:alpha12NNLO}). If we focus on the imaginary part and
disregard contributions from spectator scattering, these relations
become
\begin{align}
\frac{1}{\pi} \; \Im \; \alpha_1 \big {|}_V &=
    \frac{\as C_F}{4\pi \, N_c}\;\; \frac{1}{\pi} \;\Im\bigg[ \frac12
    \hat{C}_{1} \, \hat{V}^{(1)} + \frac{\as}{4\pi} \bigg( \frac12
    \hat{C}_{1} \, \hat{V}_1^{(2)} + \frac12 \hat{C}_{2} \,
    \hat{V}_2^{(2)} \bigg) + \calO(\as^2) \bigg] \no \\
&=
    \frac{\as C_F}{4\pi
    \, N_c}\;\; \frac{1}{\pi} \;\Im\bigg[
     \tilde{C}_{2} \, V^{(1)} +
    \frac{\as}{4\pi} \bigg( \tilde{C}_{1} \, V_1^{(2)} + \tilde{C}_{2}
    \, V_2^{(2)} \bigg) + \calO(\as^2) \bigg]. \label{eq:alpha1Im}
\end{align}
In order to compute $V^{(1,2)}$, we need the relation between the
Wilson coefficients in the CMM basis $\hat{C}_{1,2}$ and the ones in
the QCDF basis $\tilde{C}_{1,2}$ to NLL approximation.

\newabs
The Wilson coefficients can be found e.g.~in~\cite{CMM}, where
the transformation between both bases has been studied in detail.
From this, we derive
\begin{align}
\hat{C}_1 &= 2 \tilde{C}_2 + \frac{\as}{4\pi} \left( 4 \tilde{C}_1 + \frac{14}{3} \tilde{C}_2 \right) + \calO(\as^2),\no \\
\hat{C}_2&= \tilde{C}_1 + \frac13 \tilde{C}_2 + \calO(\as).
\label{eq:WilsonTransform}
\end{align}
Combining (\ref{eq:alpha1Im}), (\ref{eq:WilsonTransform}) and
(\ref{eq:V12mod}) we obtain
\begin{align}
\frac{1}{\pi} \; \Im \;V^{(1)}   &=
    \frac{1}{\pi} \; \Im \;\hat{V}^{(1)} \no\\
    &= \int_0^1 du \; g_1(u)  \; \phi_{M_2}(u), \no\\
\frac{1}{\pi} \; \Im \;V_1^{(2)} &=
    \frac{1}{\pi} \; \Im \bigg[ \frac12 \,\hat{V}^{(2)}_2 +2 \,\hat{V}^{(1)} \bigg] \no\\
    &= \int_0^1 du \; \bigg\{ -3\, g_1(u) \ln \frac{\mu^2}{m_b^2} +2 \,g_1(u) +\frac12 \, h_0(u)  \bigg\}   \phi_{M_2}(u), \no \\
\frac{1}{\pi} \; \Im \;V_2^{(2)} &=
    \frac{1}{\pi} \; \Im \bigg[ \hat{V}^{(2)}_1 + \frac16 \,\hat{V}^{(2)}_2 + \frac73 \,\hat{V}^{(1)} \bigg] \no\\
    &= \int_0^1 du \; \bigg\{ \Big[ \Big( 28- \frac23 n_f \Big) g_1(u) + \frac43 h_1(u) \Big] \ln \frac{\mu^2}{m_b^2} + \frac73 \,g_1(u) + \frac16 \, h_0(u)\no \\
    & \hspace{3mm} + \frac43 \, h_2(u) + 3 \, h_3(u) + (n_f-2) \, h_4(u;0) + h_4(u;z) + h_4(u;1) \bigg\}   \phi_{M_2}(u).
\label{eq:V12trad}
\end{align}
Notice that these expressions determine the vertex corrections for
the colour-allowed amplitude $\alpha_1$ \emph{and} the
colour-suppressed amplitude $\alpha_2$ according to
(\ref{eq:alpha12NNLO}). The equations in (\ref{eq:V12trad})
represent the central result of our analysis. The expression for
$V^{(1)}$ is in agreement with (\ref{eq:V1trad}), whereas the
expressions for $V^{(2)}_{1,2}$ are new. The kernels $g_1$ and
$h_{0-4}$ can be found in (\ref{eq:BtoPiPi:kernels:g}) and
(\ref{eq:BtoPiPi:kernels:h}). The terms proportional to $n_f$ have
already been considered in the analysis of the large $\beta_0$-limit
in \cite{beta0:NeuPec,beta0:BurWil}. Our results are in agreement
with these findings.

\subsection{Convolution with distribution amplitude}

The NNLO vertex corrections have been given in (\ref{eq:V12trad}) as
convolutions of hard-scattering kernels with the light-cone
distribution amplitude of the meson $M_2$. We may explicitly perform
the convolution integrals by expanding the distribution amplitude
into the eigenfunctions of the 1-loop evolution kernel
\begin{align}
\phi_{M_2}(u) &=
    6 u \ubar \left[ 1 + \sum_{n=1}^\infty \, a_n^{M_2} \; C_n^{(3/2)}(2u-1) \right],
    \label{eq:Gegenbauer}
\end{align}
where $a_n^{M_2}$ and $C_n^{(3/2)}$ are the Gegenbauer moments and
polynomials, respectively. It is convenient to truncate this
expansion at $n=2$. The convolution integrals with the kernels $g_1$
and $h_{0-3}$ from (\ref{eq:BtoPiPi:kernels:g}) and
(\ref{eq:BtoPiPi:kernels:h}) then give
\begin{align}
\int_0^1 du \; g_1(u)  \; \phi_{M_2}(u) &=
    -3 -3 \, a_1^{M_2}, \hspace{7cm}\no \\
\int_0^1 du \; h_0(u)  \; \phi_{M_2}(u) &=
    \frac{1333}{12} + \frac{47\pi^2}{45} -16 \zeta_3 +\left( \frac{15}{4} + \frac{23\pi^2}{5}  \right) a_1^{M_2}  \no\\
    & \hspace{6mm} - \left(  \frac{173}{30} + \frac{18\pi^2}{35}\right) a_2^{M_2} ,\no \\
\int_0^1 du \; h_1(u)  \; \phi_{M_2}(u) &=
    36 + 28 \, a_1^{M_2},\no \\
\int_0^1 du \; h_2(u)  \; \phi_{M_2}(u) &=
    \frac{1369}{6} + \frac{139\pi^2}{45}  -32 \zeta_3 - \left( \frac{17}{6} - \frac{51\pi^2}{5}  \right) a_1^{M_2} \no \\
    & \hspace{6mm} - \left(\frac{103}{15} + \frac{71\pi^2}{35} \right) a_2^{M_2},\no \\
\int_0^1 du \; h_3(u)  \; \phi_{M_2}(u) &=
    - \frac{481}{3} + \frac{7\pi^2}{30} +12 \zeta_3 - \left( \frac{643}{12} +\frac{11\pi^2}{10} \right) a_1^{M_2} \no \\
    & \hspace{6mm} - \left( \frac{1531}{80} - \frac{169\pi^2}{70} \right) a_2^{M_2}.
\label{eq:convol1}
\end{align}
The convolution with $h_4(u;z)$ from (\ref{eq:BtoPiPi:kernels:h})
can also be performed analytically
\begin{align}
H_4(z) &\equiv \int_0^1 du \; h_4(u;z)  \; \phi_{M_2}(u) \no\\
&= \frac{22}{3} +148 z^2-96 z^4 F(z) -36 z^4 \ln^2 \frac{y_1}{y_2} \no \\
& \hspace{1.2cm} -2\Big[1-(2y_1+1)(1+22z^2)\Big] \ln \frac{y_1}{y_2} -4\ln y_2 \no \\
& \quad +\bigg( 7 + 164 z^2 + 180z^4 + 144 z^6 - 288z^4 F(z) +12z^4 (3 + 16 z^2 + 12 z^4) \ln^2 \frac{y_1}{y_2} \no \\
& \hspace{1.2cm}    -2\Big[1 - (2y_1 + 1)(1 + 22z^2 + 84z^4 + 72z^6)\Big] \ln \frac{y_1}{y_2} -4 \ln y_2\bigg) a_1^{M_2}\no \\
& \quad +\bigg( \frac35 + 244 z^2 + \frac{148}{3} z^4 - 640 z^6 - 960 z^8  +24 z^4(1 - 30 z^4 - 40z^6) \ln^2  \frac{y_1}{y_2} \no \\
& \hspace{1.2cm}    - 576 z^4 F(z) +8z^2(2y_1 + 1)(6 + 11z^2 - 70z^4
- 120 z^6) \ln \frac{y_1}{y_2} \bigg) a_2^{M_2},
\label{eq:convol2}
\end{align}
where we defined
\begin{align}
y_{1} \equiv \frac12 \left( \sqrt{1+4z^2} - 1 \right), \qquad
y_{2} \equiv \frac12 \left( \sqrt{1+4z^2} + 1 \right), \hspace{2cm}\no \\
F(z) \equiv \Lic(-y_1)-\Sab(-y_1)-\ln y_1 \Lib(-y_1)+\frac12 \ln y_1
\ln^2 y_2 -\frac{1}{12} \ln^3 z^2 + \zeta_3.
\end{align}
In the massless limit the function $H_4(z)$ simply becomes
\begin{align}
H_4(0) &=\frac{22}{3}+7 a_1^{M_2} +\frac35 a_2^{M_2}.
\end{align}
The finiteness of all convolution integrals in (\ref{eq:convol1})
and (\ref{eq:convol2}) completes the explicit factorization proof of
the imaginary part of the NNLO vertex corrections.

\newabs
We summarize our results for the vertex corrections in the
considered representation of the light-cone distribution amplitude
of the emitted meson $M_2$
\begin{align}
\frac{1}{\pi} \; \Im \;V^{(1)}
    &=-3-3 a_1^{M_2},\no \\
\frac{1}{\pi} \; \Im \;V_1^{(2)}
    &=9\left(1+ a_1^{M_2}\right) \, \ln \frac{\mu^2}{m_b^2} + \frac{1189}{24} + \frac{47\pi^2}{90} -8 \zeta_3 -\left( \frac{33}{8} - \frac{23\pi^2}{10}  \right) a_1^{M_2} \no \\
    &\quad - \left( \frac{173}{60} + \frac{9\pi^2}{35}\right) a_2^{M_2} ,\no \\
\frac{1}{\pi} \; \Im \;V_2^{(2)}
    &=-\left(26+ \frac{110}{3} a_1^{M_2}\right) \,  \ln \frac{\mu^2}{m_b^2} -\frac{10315}{72} +\frac{674\pi^2}{135} -\frac{28}{3} \zeta_3 \no \\
    &\quad  -\left( \frac{10793}{72} - \frac{166\pi^2}{15} \right) a_1^{M_2}  -\left( \frac{3155}{48} -\frac{187\pi^2}{42} \right) a_2^{M_2} + H_4(z) +
    H_4(1),
\label{eq:V12:Anas}
\end{align}
with $H_4(z)$ given in (\ref{eq:convol2}). In order to illustrate
the relative importance of the individual contributions, we set
$\mu=m_b$ and $z=m_c/m_b=0.3$ which yields
\begin{align}
\Im \;V^{(1)}
    &= -9.425 - 9.43 a_1^{M_2},\no \\
\Im \;V_1^{(2)}
    &= 141.621 + 58.36  \; a_1^{M_2} - 17.03 \; a_2^{M_2} ,\no \\
\Im \;V_2^{(2)}
    &= -317.940 - 115.62 \; a_1^{M_2} - 68.31 \; a_2^{M_2}.
\label{eq:V12:Nums}
\end{align}
We find large coefficients for the NNLO vertex corrections and
expect only a minor impact of the higher Gegenbauer moments (in
particular in the symmetric case with $a_1^{M_2}=0$). Notice that
all contributions add constructively in $\alpha_{1,2}$ due to the
relative signs of the accompanying Wilson coefficients. In the case
of $\alpha_1$ the contribution from $V_1^{(2)}$ is found to exceed
the formally leading contribution $V^{(1)}$  due to the fact that
the latter is multiplied by the small Wilson coefficient
$\tilde{C}_2$. For $\alpha_2$ the impact of the NNLO vertex
corrections is also substantial, roughly saying they amount to a
$50\%$ correction. A more detailed numerical analysis including the
contributions from spectator scattering will be given in the
following section.

\section{Numerical analysis}

We conclude this chapter with a brief numerical analysis (an
extended version will be given in~\cite{GBImPart}). We first
consider the vertex corrections solely without the spectator
scattering contributions which have been computed recently
in~\cite{BenekeJager,Kivel}. These will be added in the second part
of our analysis which will lead us to the full NNLO result for the
imaginary part of the topological tree amplitudes in QCD
Factorization.

\subsection{Vertex corrections}

\begin{figure}[b!]
\centerline{\parbox{13cm}{
\centerline{\includegraphics[width=6cm]{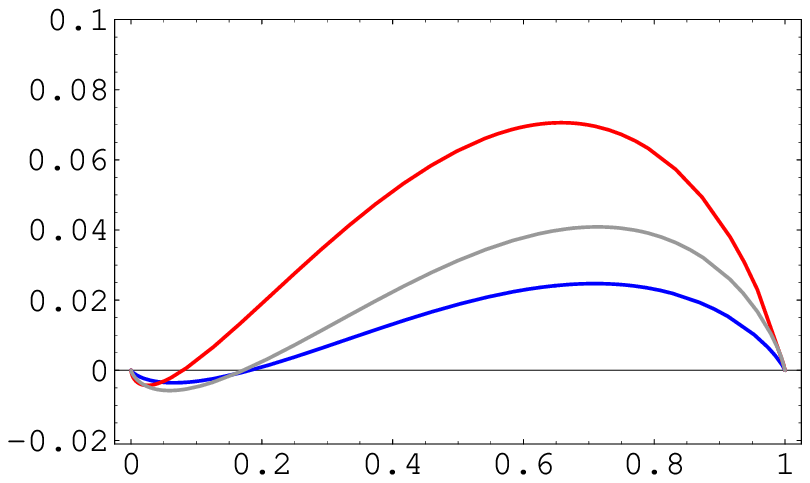}\hspace{10mm}
\includegraphics[width=6cm]{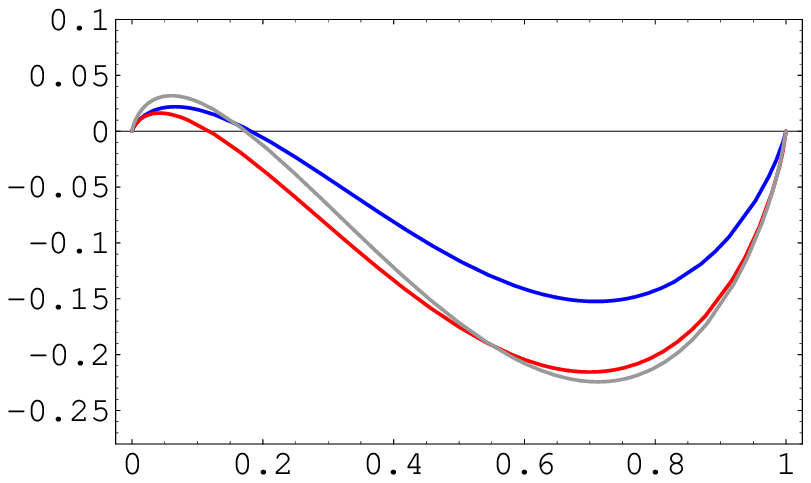}}
\caption{\label{fig:BtoPiPi:beta0} \small \textit{Imaginary part of
$\zeta_1(u)$ and $\zeta_2(u)$ as introduced in (\ref{eq:defzeta}).
The graphs show the 1-loop vertex corrections (blue), the large
$\beta_0$-approximation (gray) and our new results including the
2-loop vertex corrections (red) for $\mu=m_b$ (with asymptotic
distribution amplitude).}}}}
\end{figure}

We come back to the question if the large $\beta_0$-limit considered
in~\cite{beta0:NeuPec,beta0:BurWil} represents a good approximation
for the imaginary part of the NNLO vertex corrections. For
illustration, we introduce two functions $\zeta_{i}(u)$ defined by
\begin{align}
\alpha_1 \big {|}_V &\equiv \int_0^1 du \;\; \zeta_i(u).
\label{eq:defzeta}
\end{align}
The $\zeta_{i}(u)$ correspond to a combination of Wilson
coefficients and hard-scattering kernels multiplied by the
distribution amplitude of the emitted meson $M_2$. The imaginary
part of the functions $\zeta_{i}(u)$ are shown in
Figure~\ref{fig:BtoPiPi:beta0} with the asymptotic form of the
distribution amplitude, for simplicity.

As we have stated at the end of the last section, we find that the
NNLO corrections add constructively to the NLO results. They provide
the dominant contribution to the imaginary part of $\alpha_1$ and a
substantial contribution to the one of $\alpha_2$. We further see
that the large $\beta_0$-limit is a good approximation in the case
of $\alpha_2$ but not for $\alpha_1$. This can be traced back to the
fact that the imaginary part of $V_2^{(2)}$ is reproduced well in
this approximation whereas the one of $V_1^{(2)}$, which provides
the most important contribution to $\alpha_1$, is missed completely.

\begin{figure}[t!]
\centerline{\parbox{13cm}{
\centerline{\includegraphics[width=6cm]{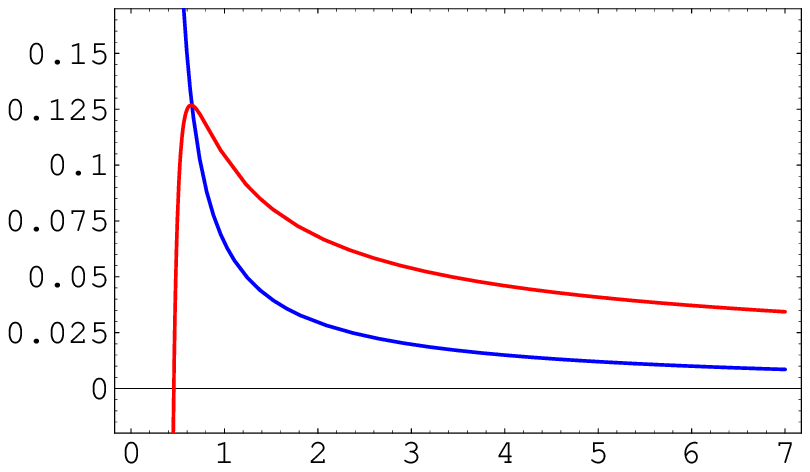}\hspace{10mm}
\includegraphics[width=6cm]{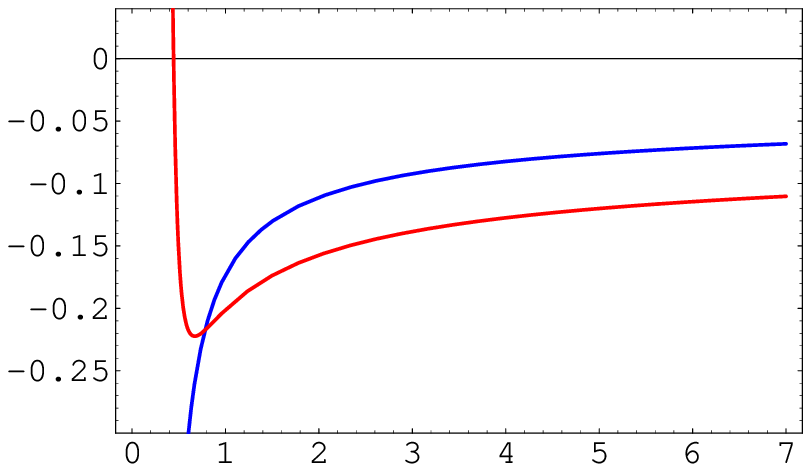}}
\caption{\label{fig:BtoPiPi:scale} \small \textit{Renormalization
scale dependence of the imaginary part of $\alpha_{1}$ and
$\alpha_{2}$ (without spectator scattering). Blue: NLO result. Red:
NNLO result.}}}}
\end{figure}

\newabs
Concerning the scale dependence we recall that the imaginary part
has only NLO complexity at the considered order in perturbation
theory. We therefore use the 2-loop expression for the running
coupling constant ($n_f=5$,
$\Lambda_\text{\tiny\MSbar}^{(5)}=0.225$~\gev) and consider the
Wilson coefficients in NLL approximation which we take
from~\cite{WeakDecays}. Contrary to the second Gegenbauer moment,
the first moment would also be required in NLL approximation as it
enters the expression for $V^{(1)}$ in (\ref{eq:V12:Anas}). However,
in the following analysis we focus on the $B\to\pi\pi$ decays where
the first moment is absent and we only implement the LL evolution of
the second moment which is given by
\begin{align}
a_2^{M_2}(\mu) = \left( \frac{\as(\mu)}{\as(\mu_0)}
\right)^{-\gamma_2/2\beta_0}a_2^{M_2}(\mu_0), \quad
\quad\gamma_2=-\frac{100}{9}.
\end{align}
The scale dependence of the imaginary part of $\alpha_{1,2}$ is
shown in Figure~\ref{fig:BtoPiPi:scale}. We observe only a minor
reduction of the scale dependence if we vary the scale between
$m_b/2\sim2.4$~\gev~and $2m_b\sim9.6$~\gev, in particular for
$\alpha_1$ where the NNLO correction dominates over the NLO result.

\subsection{Full NNLO result}

We finally combine our results with the spectator scattering
contributions from~\cite{BenekeJager}. One remark is in order
concerning the scale dependence of the spectator term. The
respective kernel $T^{II}$ receives hard and hard-collinear
contributions which are encoded in a hard coefficient function and a
jet function, respectively (notice that the hard coefficient
function represents the only source for an imaginary part). In the
following discussion we simply evaluate all quantities related to
the hard coefficient function at the hard scale $\mu_h$ and all
other quantities at the hard-collinear scale $\mu_{hc}$ (this
corresponds to equation (58) of~\cite{BenekeJager} with
$U_\|(\mu_h,\mu_{hc})=1$). A more sophisticated treatment of the
scale issues in the spectator term is relegated to~\cite{GBImPart}.

\begin{table}[t!]
\centerline{
\parbox{13cm}{\setlength{\doublerulesep}{0.1mm}
\centerline{\begin{tabular}{|c|c||c|c|} \hline
\hspace*{2.5cm}&\hspace*{3cm}&\hspace*{2.5cm}&\hspace*{3cm} \\[-0.7em]
Parameter & Value & Parameter & Value\\[0.3em]
\hline\hline\hline&&& \\[-0.7em]
$\Lambda_\text{\tiny\MSbar}^{(5)}$ & $0.225$ & $f_B$& $0.2\pm 0.03$\\[0.3em]
$m_b$ & $4.8$ & $F_+^{B\to\pi}(0)$&$0.26 \pm 0.04$\\[0.3em]
$m_c$ & $1.3\pm 0.2$ & $\lambda_B$&$0.35 \pm 0.15$\\[0.3em]
$f_\pi$ & $0.131$ & $a_2^\pi(1\gev)$& $0.2\pm 0.2$\\[0.3em]
\hline
\end{tabular}}
\vspace{4mm} \caption{\label{tab:nums}\small \textit{Theoretical
input parameters (in units of \gev~or dimensionless).}}}}
\end{table}

\newabs
Our input parameters for the $B\to\pi\pi$ amplitudes are summarized
in Table~\ref{tab:nums}. They correspond to the values from previous
analysis in QCD Factorization \cite{BenekeJager,BenekeNeubert2003}
with updated values for the form factor and the Gegenbauer moment
based on recent LCSR analyses~\cite{BallZwicky:FF,BallZwicky:DA}. In
order to estimate unknown perturbative corrections we vary the hard
scale in the range $\mu_h=4.8^{+4.8}_{-2.4}~\gev$ and the
hard-collinear scale independently between
$\mu_{hc}=1.5^{+0.7}_{-0.5}~\gev$.

\newpage
The complete NNLO result for the imaginary part of the topological
tree amplitude is found to be
\begin{align}
\Im\; \alpha_1(\pi\pi)& =\,~~ 0.012 \big{|}_{V^{(1)}}
                            + 0.031 \big{|}_{V^{(2)}}
                            - 0.019 \big{|}_{H^{(2)}} \no \\
                      & =\,~~ 0.025 \pm 0.021,\no\\
\Im\; \alpha_2(\pi\pi)& =   - 0.077 \big{|}_{V^{(1)}}
                            - 0.052 \big{|}_{V^{(2)}}
                            + 0.031 \big{|}_{H^{(2)}} \no \\
                      & =   - 0.098 \pm 0.035,
\label{eq:alpha12:result}
\end{align}
where we disentangled the contributions from $V^{(1)}$, $V^{(2)}$
and $H^{(2)}$ according to (\ref{eq:alpha12NNLO}). In the case of
$\alpha_1$ the NNLO corrections exceed the NLO result which can be
explained by the fact that the latter is multiplied by the small
Wilson coefficient $\tilde{C}_2$. In both cases the individual NNLO
corrections are found to be sizeable, but we observe a large
cancellation in their sum. The NNLO vertex corrections considered in
this work turn out to dominate over the spectator terms resulting in
a moderate additive contribution to the NLO (BBNS) result.

\newabs
The uncertainties quoted in (\ref{eq:alpha12:result}) stem from the
variation of the parameters shown in Table~\ref{tab:uns}. As the
dominant sources we identify the hadronic parameters $\lambda_B$ and
$a_2^\pi$. Moreover, the sensitivity to the renormalization scale
remains sizeable at NNLO as we have mentioned at the end of the last
section. We finally emphasize that we have not yet assigned an error
estimate to unknown power corrections which will be included
in~\cite{GBImPart}.

\begin{table}[b!]
\centerline{
\parbox{13cm}{\setlength{\doublerulesep}{0.1mm}
\centerline{\begin{tabular}{|c||c|c|c|c|c|c|c|}\hline
\hspace*{1.2cm}&\hspace*{1.2cm}&\hspace*{1.2cm}&\hspace*{1.2cm}&\hspace*{1.2cm}&\hspace*{1.2cm}&\hspace*{1.2cm}&\hspace*{1.2cm} \\[-0.7em]
&$\mu_h$ & $\mu_{hc}$ & $m_c$ & $f_B$ & $F_+^{B\pi}$& $\lambda_B$ & $a_2^\pi$\\[0.3em]
\hline\hline\hline&&&&&&& \\[-0.7em]
$\alpha_1$ & $0.011$ & $0.006$& $0.000$& $0.003$& $0.003$& $0.014$& $0.008$\\[0.3em]
$\alpha_2$ & $0.019$ & $0.009$& $0.000$& $0.005$& $0.006$& $0.023$& $0.013$\\[0.3em]
\hline
\end{tabular}}
\vspace{4mm} \caption{\label{tab:uns}\small \textit{Uncertainties in
our predictions of the imaginary part of $\alpha_1(\pi\pi)$ and
$\alpha_2(\pi\pi)$ from the scale variation and the input parameters
in Table~\ref{tab:nums}.}}}}
\end{table}

\chapter{Hadronic two-body decays II: {Real~part}}
\label{ch:RePart}
 
The calculation of the real part of the NNLO vertex corrections in
hadronic two-body decays proceeds along the same lines as the one of
the imaginary part which we presented in the previous chapter.
However, we will see below that the calculation of the real part is
far more complicated involving many additional MIs and the full NNLO
complexity concerning e.g.~the issue of renormalization and the
treatment of evanescent operators.

In this chapter we present a preliminary result for the real part of
the colour-allowed tree amplitude $\alpha_1$ (in the CMM operator
basis). Similar to what we have seen in the previous chapter, the
colour-suppressed tree amplitude $\alpha_2$ can then be derived from
$\alpha_1$ after the transformation into a Fierz-symmetric operator
basis (which we called QCDF basis in~Chapter~\ref{ch:ImPart}). In
NNLO we thus have to extend the basis from (\ref{eq:Basis:QCDF}) to
include 2-loop evanescent operators with an appropriate definition
of $\eps$- and $\eps^2$-terms which guarantees manifest Fierz
symmetry in this operator basis. As we have not yet worked out the
details of this last step, we refer to~\cite{GBRePart} for the
analysis of the colour-suppressed tree amplitude.

\section{2-loop calculation}

The 2-loop calculation will be performed in the CMM operator basis
(cf.~Section~\ref{eq:BtoPiPi:OpBasis}). Apart from the operators in
(\ref{eq:Basis:CMM}), we have to take into account 2-loop evanescent
operators which are defined by
\begin{align}
\hat E_1' &=  \left[\bar u
\gamma^\mu\gamma^\nu\gamma^\rho\gamma^\sigma\gamma^\tau L\, T^A
b\right] \;
        \left[\bar d \gamma_\mu\gamma_\nu\gamma_\rho\gamma_\sigma\gamma_\tau \,L\, T^A u\right]
        -20 \hat E_1-256 \,\hat Q_1,\no\\
\hat E_2' &=  \left[\bar u
\gamma^\mu\gamma^\nu\gamma^\rho\gamma^\sigma\gamma^\tau L\, b\right]
\;
        \left[\bar d \gamma_\mu\gamma_\nu\gamma_\rho\gamma_\sigma\gamma_\tau \,L\, u\right]
        -20 \hat E_2-256 \,\hat Q_2.
\label{eq:2loopEvan:CMM}
\end{align}
We turn to a brief characterization of the considered 2-loop
calculation following our recipe from Section~\ref{sec:strategy}.

\subsubsection{Step 1: Set-up for loop calculation}

In contrast to the calculation of the imaginary part
from~Chapter~\ref{ch:ImPart}, we now have to consider the whole set
of non-factorizable 2-loop diagrams from
Figure~\ref{fig:BtoPiPi:NNLODiags}. Notice that the most complicated
diagrams only enter the calculation of the real part. Whereas the
diagrams that we considered for the calculation of the imaginary
part contained at most one massive ($b$-quark) propagator, we now
have to deal with up to three massive propagators.

\subsubsection{Step 2: Reduction to Master Integrals}

\begin{figure}[b!]\vspace{3mm}
\centerline{\parbox{15cm}{\centerline{
\includegraphics[width=15cm]{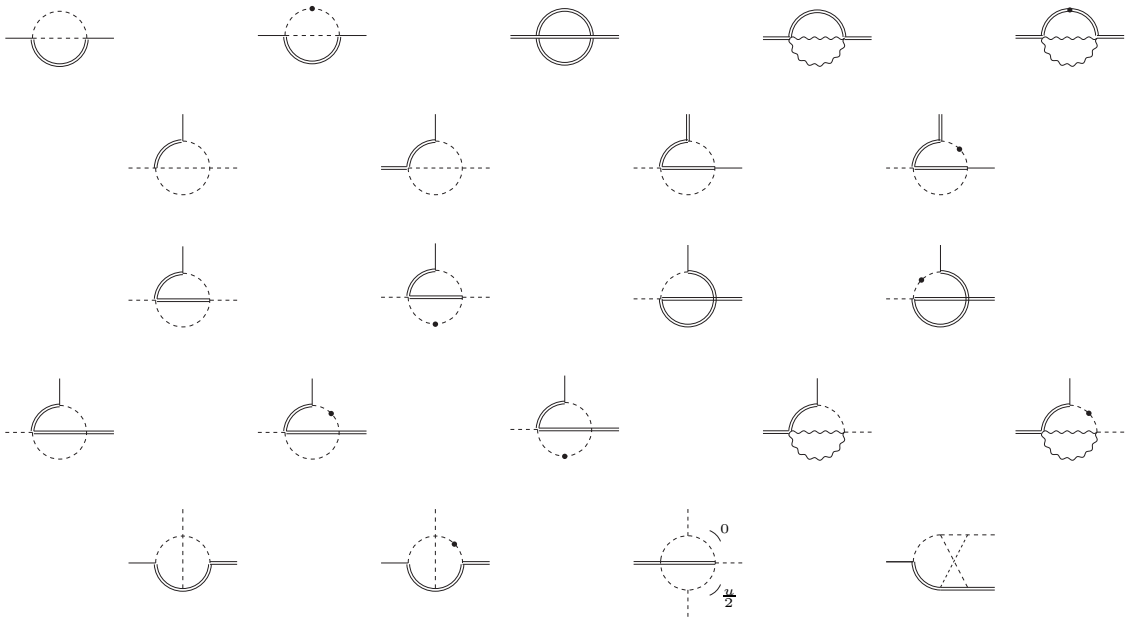}}\vspace{5mm}
\caption{\label{fig:BtoPiPi:MIsRe} \small \textit{Additional Master
Integrals that have to be considered for the calculation of the real
part of the NNLO vertex corrections. For details
cf.~Figure~\ref{fig:BtoPiPi:MIs}.}}}}
\end{figure}

The fact that the diagrams involve between $0$-$3$ massive
propagators immediately leads to many distinct topologies and a
large number of MIs. In addition to the 14 MIs from
Figure~\ref{fig:BtoPiPi:MIs}, we find 22 (real) MIs which are shown
in Figure~\ref{fig:BtoPiPi:MIsRe}. We further remark that our {\sc
Mathematica} implementation of the reduction algorithm hardly
succeeds to reduce the most complicated diagrams of the considered
calculation within a reasonable amount of CPU time.

\subsubsection{Step 3: Manipulation of Dirac structures}

Concerning the irreducible Dirac structures we find the same set
(\ref{eq:BtoPiPi:Dirac}) as in the calculation of the imaginary
part. In contrast to our analysis from Chapter~\ref{ch:ImPart}, the
last structure now enters the divergent piece of the calculation
giving rise to 2-loop evanescent operators according to
(\ref{eq:2loopEvan:CMM}).

\subsubsection{Step 4: Calculation of Master Integrals}

In the last step of the calculation it becomes obvious that the
analysis of the real part of the NNLO vertex corrections is much
more complex than the one of the imaginary part. We now have to
calculate the real parts of the MIs from
Figure~\ref{fig:BtoPiPi:MIs} as well as the MIs from
Figure~\ref{fig:BtoPiPi:MIsRe}, in general up to five orders in the
$\eps$-expansion. In order to tackle this highly challenging task we
applied the same techniques as in our calculation from
Chapter~\ref{ch:ImPart}. It turns out that almost all MIs can be
expressed in terms of the set (\ref{eq:HPLminiIm}) supplemented by
the following HPLs of weight $w=4$
\begin{align}
H(0,0,0,1;u) &= \Lid(u), \no \\
H(0,0,1,1;u) &= \Sbb(u), \no \\
H(0,1,1,1;u) &= \Sac(u).
\end{align}
Unfortunately, some MIs with two and three massive propagators do
not fit into this pattern. We further have to include HPLs related
to the parameter $-1$ and to the argument $u-1$,
cf.~(\ref{eq:MI:HPLext1})~--~(\ref{eq:MI:HPLext2}). Still, we find
two functions which we could not express in terms of Nielsen
polylogarithms: $H(0,-1,0,1;u)$ and ${\cal{A}}(u)$ from
(\ref{eq:auxA})\footnote{The appearance of the function
${\cal{A}}(u)$ seems to be an artefact of our calculation, i.e.~of
our special choice of the MIs. Though it enters two of our MIs from
Figure~\ref{fig:BtoPiPi:MIsRe}, it drops out in their sum in the
respective Feynman diagrams and is therefore irrelevant for our
purposes.}.

\newabs
Similar to what we have seen in Chapter~\ref{ch:ImPart}, the
situation is more complicated for the MIs which stem from the
diagrams with a closed fermion loop. So far we have not yet
calculated these MIs with a massive internal quark (wavy lines),
i.e.~we work in an approximation which treats all quarks in these
loops as massless. We stress that this is certainly inconsistent for
an internal $b$-quark with respect to the remainder of the
calculation. However, we expect that this will only have a minor
impact on our (preliminary) result. A consistent treatment of these
diagrams is relegated to~\cite{GBRePart}. The analytical results for
all other MIs can be found in Appendix~\ref{app:MIsRe}.

\section{Renormalization and IR subtractions}

We have shown in detail in Section~\ref{sec:BtoPiPi:UVIR} how to
perform appropriate UV and IR subtractions in order to extract the
NNLO kernels $T_i^{(2)}$ from the hadronic matrix elements. The
essence is summarized in the Master Formula (\ref{eq:Master}). We
briefly address the issues that go beyond our analysis of the
imaginary part from Chapter~\ref{ch:ImPart}.

\subsection{Renormalization}

The renormalization of the $b$-quark mass becomes relevant in the
considered calculation. We treat the $b$-quark in the on-shell
scheme according to
\begin{align}
Z_m(m) = 1 - \frac{\as C_F}{4\pi} \left( \frac{\mu^2
e^{\gamma_E}}{m^2} \right)^\eps \Gamma(\eps) \;
\frac{3-2\eps}{1-2\eps} + \calO(\as^2).
\end{align}
Due to the full NNLO complexity, we require the 2-loop
renormalization matrix $\hat{Z}^{(2)}$ which we take from~\cite{GGH}
\renewcommand{\arraycolsep}{2mm}
\begin{align}
\hat{Z}^{(2)} &= \left(
\begin{array}{c c c c c c}
\rule[-2mm]{0mm}{7mm} \frac{41}{3} & -\frac{58}{9} & -\frac{125}{36} & - \frac{73}{54} & \frac{19}{96} & \frac{5}{108} \\
\rule[-2mm]{0mm}{7mm} -29 & 4 & -\frac{73}{12} & 0 & \frac{5}{24} &
\frac19
\end{array}
\right) \, \frac{1}{\eps^2} \no \\
& \quad + \left(
\begin{array}{c c c c c c}
\rule[-2mm]{0mm}{7mm} \frac{317}{36} & -\frac{515}{54} & \frac{4493}{864} & -\frac{49}{648} & \frac{1}{384} & -\frac{35}{864}  \\
\rule[-2mm]{0mm}{7mm} \frac{349}{12} & 3 &\frac{1031}{144} & \frac89
& -\frac{35}{192} & -\frac{7}{72}
\end{array}
\right) \, \frac{1}{\eps}, \label{eq:Z1}
\end{align}
where the two lines correspond to the physical operators $\{
\hat{Q}_1, \hat{Q}_2\}$ and the columns to the extended operator
basis $\{ \hat{Q}_1, \hat{Q}_2, \hat{E}_1, \hat{E}_2, \hat{E}_1',
\hat{E}_2' \}$ including evanescent operators.

\subsection{IR subtractions}

In order to perform the IR subtractions from the right hand side of
(\ref{eq:Master}), we must compute the real parts of the NLO kernels
$T_i^{(1)}$ to $\calO(\eps^2)$. The kernels vanish in the
colour-singlet case $T_2^{(1)}=T_{2,E}^{(1)}=0$, whereas the
colour-octet kernels are found to be
\begin{align}
\Re \;T_1^{(1)}(u)  &= \frac{C_F}{2N_c} \bigg\{ \Big(-6L+t_0(u)\Big)
\Big( 1+\eps L +\frac12 \eps^2 L^2\Big) \no \\
& \hspace{2cm} + \Big(3L^2+t_1(u)\Big) \Big( \eps +\eps^2 L \Big)  \no \\
& \hspace{2cm} + \Big( -L^3 + t_2(u)\Big) \,\eps^2 + \calO(\eps^3)
\bigg\}, \no \\
\Re \;T_{1,E}^{(1)}(u)  &= -\frac{C_F}{4N_c} \bigg\{
\Big(2L+t_{E,0}(u)\Big)\Big( 1+\eps L \Big) \no \\
& \hspace{2.2cm} + \Big(-L^2+t_{E,1}(u)\Big)\eps + \calO(\eps^2)
\bigg\},
\end{align}
with $L\equiv\ln \mu^2/m_b^2$ and
\begin{align}
t_0(u) &=  4 \Lib(u) - \ln^2 u +2 \ln u \ln \ubar + \ln^2 \ubar  +(2-3u) \Big( \frac{\ln u}{\ubar} - \frac{\ln \ubar}{u} \Big) -\frac{\pi^2}{3} -22, \no \\
t_1(u) &= -2 \Lic(u) -2 \Sab(u) -2 \ln \ubar \, \Lib(u) + \ln^3 u -2 \ln^2 u \ln \ubar + \ln u \ln^2 \ubar - \ln^3 \ubar \no \\
& \quad \,+ \frac{2-3u^2}{u\ubar} \Lib(u) - \frac{2-3u}{\ubar} \Big(\ln^2 u - \ln u \ln \ubar \Big) + \frac{6-11u +2\ubar \pi^2}{\ubar} \ln u \no \\
& \quad \,+\frac{4-3u}{2u} \ln^2 \ubar - \frac{18-33u+5u
\pi^2}{3u}\ln \ubar  +\frac{(7-6u)\pi^2}{6\ubar} +2 \zeta_3 -52, \no
\\
t_2(u) &= 10 \Lid(u) -8 \Sbb(u) +10 \Sac(u) -8 \ln \ubar \, \Lic(u)
+10 \ln \ubar \, \Sab(u) - \frac{7}{12} \ln^4 u  \no
\end{align}
\begin{align}
& \quad \, +5 \ln^2 \ubar \, \Lib(u) + \frac43 \ln^3 u \ln \ubar -\ln^2 u \ln^2 \ubar +\frac13 \ln u \ln^3 \ubar +\frac{7}{12} \ln^4\ubar \no \\
& \quad \, + \frac{2-6u+6u^2}{u \ubar} \Lic(u) -\frac{4-6u+3u^2}{u\ubar} \Big( \Sab(u) +\ln \ubar\,\Lib(u) \Big) -\frac{8-3u}{6u} \ln^3 \ubar \no \\
& \quad \, + \frac{2-3u}{6\ubar} \Big( 4 \ln^3 u - 6\ln^2 u \ln\ubar
+3 \ln u \ln^2 \ubar \Big) -
\frac{60(1-2u)+17\ubar\pi^2}{12\ubar}\ln^2 u \no \\
 & \quad \, +
\frac{3(6-4u-7u^2)+u\ubar \pi^2}{3u\ubar} \Lib(u)
+\frac{24-54u+5\ubar \pi^2}{6\ubar} \ln u \ln \ubar
+\frac{(29-24u)\pi^2}{6\ubar} \no \\
& \quad \, + \frac{6(12-13u)+7u\pi^2}{12u} \ln^2 \ubar +\frac{24(7-13u)+(10-15u)\pi^2}{12\ubar} \ln u - \frac{23\pi^4}{180}\no \\
& \quad \, - \frac{24\ubar(7-13u)+(2+23u-27u^2)\pi^2+24u\ubar
\zeta_3}{12u\ubar} \ln \ubar + \frac{10-11u}{\ubar} \zeta_3-112, \no
\\
t_{E,0}(u) &=-\frac{1-2u}{2} \Big( \frac{\ln u}{\ubar} - \frac{\ln
\ubar}{u} \Big) + \frac{16}{3}, \no \\
t_{E,1}(u) &= -\frac{1-2u}{2u\ubar} \Lib(u) + \frac{1-3u}{4\ubar}
\ln^2 u + \frac{u}{2\ubar} \ln u \ln \ubar -\frac{2-3u}{4u} \ln^2
\ubar \no \\
& \quad \, -\frac{4(1-2u)}{3} \Big( \frac{\ln u}{\ubar} - \frac{\ln
\ubar}{u} \Big) - \frac{(6-5u)\pi^2}{12\ubar}+12. \label{eq:T1:Re}
\end{align}
In combination with the form factor corrections from
(\ref{eq:BtoPiPi:FFphys}) and (\ref{eq:BtoPiPi:FFevan}), this
determines the first subtraction term in (\ref{eq:Master}). In the
second subtraction we require the convolution of the NLO kernels
with the wave function corrections from (\ref{eq:Phiphys}) and
(\ref{eq:Phievan}). We find
\begin{align}
F^{(0)} \; \Re \;T_1^{(1)}\;  \Phi_\text{amp}^{(1)} & \;=\;
\frac{C_F^2}{N_c} \bigg\{ t_3(u) \, \bigg( \frac{1}{\eps} + L \bigg)
+  t_4(u) +\calO(\eps) \bigg\} \;
F^{(0)} \;\Phi^{(0)}, \no \\
F_E^{(0)} \; \Re \;T_{1,E}^{(1)}\;  \Phi_\text{amp,E}^{(1)} &
\;\to\; \frac{C_F^2}{N_c} \bigg\{ 12L+ t_{E,2}(u) +\calO(\eps)
\bigg\} \; F^{(0)} \;\Phi^{(0)},
\end{align}
where
\begin{align}
t_3(u) &= 4 \Lic(u) + 4 \Sab(u) -4 \ln u \, \Lib(u) +\frac23 \ln^3 u
-2 \ln^2 u \ln \ubar -\frac23 \ln^3 \ubar - \frac{\Lib(u)}{u\ubar} \no \\
& \quad - \frac{1-3u}{2u\ubar} \Big( u \ln^2u +2 \ubar \ln u \ln
\ubar - \ubar \ln^2 \ubar \Big)\! - \frac{3}{2u} \ln \ubar +
\frac{(4-3u)\pi^2}{6\ubar}-\frac{15}{2}-4 \zeta_3 ,\no \\
t_4(u) &= 12 \Lid(u) -20 \Sbb(u) +12 \Sac(u) -8 \Big(\ln u + \ln
\ubar \Big) \Lic(u)  +12 \ln u \Sab(u) \no \\
& \quad +4 \ln \ubar \, \Sab(u) + \! \Big(4\ln^2 u +4 \ln u \ln
\ubar +2 \ln^2 \ubar \Big) \Lib(u)  - \frac34 \ln^4 u +\frac73 \ln^3
u \ln \ubar \no \\
& \quad -\frac12 \ln^2 u \ln^2 \ubar -\frac13 \ln u \ln^3 \ubar
+\frac34 \ln^4 \ubar - \frac{4-11u+3u^2}{u\ubar} \Lic(u) +
\frac{5-12u}{6\ubar} \ln^3 u\no
\end{align}
\begin{align}
& \quad + \frac{1+u-3u^2}{u\ubar} \Sab(u) +
\frac{2-10u+6u^2}{u\ubar} \ln u \, \Lib(u) -
\frac{1-5u+3u^2}{u\ubar} \ln \ubar \, \Lib(u) \no \\
 & \quad +
\frac{2-10u+9u^2}{2u\ubar} \ln^2 u \ln \ubar -
\frac{1-2u}{2u\ubar} \ln u \ln^2 \ubar - \frac{5-6u}{6u} \ln^3 \ubar  \no \\
& \quad - \frac{18-24u+15u^2-10u\ubar \pi^2}{3u\ubar} \Lib(u) -
\frac{16-27u+4\ubar\pi^2}{4\ubar} \ln^2 u \no \\
& \quad - \frac{6-36u+27u^2-4u\ubar\pi^2}{2u\ubar} \ln u \ln \ubar +
\frac{3(14-17u) +8u\pi^2}{12u} \ln^2 \ubar  \no \\
& \quad + \frac{8-15u-4\pi^2-48\ubar \zeta_3}{4\ubar} \ln u  + \frac{3(2-3u)}{\ubar} \zeta_3 - \frac{23\pi^4}{60} + \frac{(23-17u)\pi^2}{12\ubar}  \no \\
& \quad -
\frac{81-126u+45u^2-(14-22u+6u^2)\pi^2-192u\ubar\zeta_3}{12u\ubar}
\ln \ubar - \frac{137}{4}, \no \\
t_{E,2}(u) &= -\frac{6(1-2u)}{u\ubar}\Lib(u) -\frac{6}{u} \ln u \ln
\ubar -6\ln u-6\ln \ubar-\frac{\pi^2}{\ubar}+50.
\end{align}

\newpage
\section{Tree amplitudes in NNLO}

The NNLO kernels $T_i^{(2)}$ follow from (\ref{eq:Master}) and are
indeed found to be free of UV and IR singularities. We emphasize
that this provides a very powerful check of our calculation which
involves the cancellation of poles up to $1/\eps^4$ ($1/\eps^3$) for
the calculation of the real (imaginary) part.

\subsection{$\alpha_1$ in CMM basis}

We now present preliminary results for the real parts of the NNLO
vertex corrections. Our results are still preliminary in the sense
that the calculation is not yet complete (massive fermion loops are
still missing) and we have not yet performed numerical checks of all
MIs. In analogy to (\ref{eq:V12mod}) we write
\begin{align}
\Re \;\hat{V}^{(1)}   &\equiv
    \int_0^1 du \; \bigg\{ -6 \ln \frac{\mu^2}{m_b^2} +g_2(u)  \bigg\} \phi_{M_2}(u), \no\\
\Re \;\hat{V}_1^{(2)}  &\equiv
    \int_0^1 du \; \bigg\{ \Big( 36 C_F -29 N_c +2 n_f \Big) \ln^2 \frac{\mu^2}{m_b^2} \no \\
    & \hspace{3mm} +  \Big[ \Big( \frac{29}{3} N_c - \frac23 n_f \Big) g_2(u) - \frac{91}{6} N_c - \frac{10}{3} n_f + C_F \,h_6(u) \Big] \ln \frac{\mu^2}{m_b^2}  \no \\
    & \hspace{3mm} + C_F \, h_7(u) + N_c \, h_8(u) + (n_f-2) \, h_9(u;0) + h_9(u;z) + h_9(u;1) \bigg\}   \phi_{M_2}(u), \no \\
\Re \;\hat{V}_2^{(2)} &\equiv
    \int_0^1 du \; \bigg\{ 18 \ln^2 \frac{\mu^2}{m_b^2} +\Big(21-6\, g_2(u)\Big) \ln \frac{\mu^2}{m_b^2} + h_5(u) \bigg\}
    \phi_{M_2}(u).
\end{align}
The NLO kernel is found to be
\begin{align}
g_2(u) &= -22 + \frac{3(1-2u)}{\ubar} \ln u + \bigg[2\Lib(u) - \ln^2
u - \frac{1-3u}{\ubar} \ln u -(u\to\ubar) \bigg].
\end{align}
Concerning the NNLO kernels $h_{5-9}$ we do not quote the
expressions for $h_5$, $h_7$ and $h_8$ here, as they are extremely
complicated and we have not yet expressed them in terms of a minimal
set of HPLs. On the other hand, the expressions for $h_6$ and for
$h_9$ (in the massless case) are much simpler and given by
\begin{align}
h_6(u)  &=  \bigg[  \frac{327}{2} - \frac{3(1-2u)}{2\ubar} \ln^2 u + \frac{3(1-2u^2)}{2u\ubar} \ln u \ln \ubar - \frac{3(13-24u)}{2\ubar} \ln u  \no \\
        &\hspace{8mm}  + \frac{(1-2u^2)\pi^2}{4u\ubar} + (u\leftrightarrow\ubar)\bigg] \no \\
        &\hspace{5mm} + \bigg[  8 \Lic(u) -8 \ln u \, \Lib(u) + \frac43 \ln^3 u -4 \ln^2 u \ln \ubar - \frac{13-24u^2}{u\ubar} \Lib(u) \no \\
        &\hspace{12mm}  + \frac{25-24u}{2\ubar} \ln^2 u + \frac{13}{\ubar} \ln u \ln \ubar - \frac{9}{2\ubar} \ln u - \frac{11\pi^2}{6\ubar} - (u\leftrightarrow\ubar) \bigg],\no
\end{align}
\begin{align}
h_9(u;0)  &=  \bigg[ \frac{125}{12} + \frac{\Lib(u) }{\ubar} + \frac{1-3u}{2\ubar} \ln^2 u + \frac{1+u}{2\ubar} \ln u \ln \ubar -\frac{17(1-2u)}{6\ubar} \ln u  \no \\
        &\hspace{8mm} - \frac{(1+u)\pi^2}{12\ubar} + (u\leftrightarrow\ubar)\bigg] \no \\
        &\hspace{5mm} + \bigg[ \frac43 \Lic(u) -\frac23 \ln^3 u + \frac43
        \ln^2 u \ln \ubar -
        \frac{32-29u}{9\ubar} \Lib(u)  + \frac{35-29u}{18\ubar} \ln^2 u \no \\
        &\hspace{12mm} -
        \frac{1}{3\ubar} \ln u \ln \ubar - \frac{13+24\ubar
        \pi^2}{18\ubar} \ln u + \frac{\pi^2}{18\ubar} - (u\leftrightarrow\ubar)
        \bigg].
\end{align}

\subsection{Convolution with distribution amplitude}

We now perform the convolution integrals by expressing the
distribution amplitude of the emitted meson $M_2$ in terms of its
Gegenbauer expansion (\ref{eq:Gegenbauer}). We obtain analytical
results for the convolutions with the kernels $g_2$, $h_6$ and $h_9$
\begin{align}
\int_0^1 du \; g_2(u)  \; \phi_{M_2}(u) &=
    -\frac{45}{2} + \frac{11}{2} a_1^{M_2} -\frac{21}{20} a_2^{M_2}\no \\
\int_0^1 du \; h_6(u)  \; \phi_{M_2}(u) &=
    348-\frac{154}{3} a_1^{M_2}  + \frac{329}{40} a_2^{M_2}\no \\
\int_0^1 du \; h_9(u;0)  \; \phi_{M_2}(u) &=
    \frac{493}{18} - \frac{2\pi^2}{3} - \left( \frac{40}{3} + 2\pi^2
    \right) a_1^{M_2}+ \left( \frac{8059}{600} - \pi^2 \right) a_2^{M_2},
\end{align}
and computed the remaining convolution integrals numerically
\begin{align}
\int_0^1 du \; h_5(u)  \; \phi_{M_2}(u) &=
    322 - 213 \,a_1^{M_2} + 3.8\,a_2^{M_2} \no \\
\int_0^1 du \; h_7(u)  \; \phi_{M_2}(u) &=
    731  - 348 \,a_1^{M_2} -a_2^{M_2} \no \\
\int_0^1 du \; h_8(u)  \; \phi_{M_2}(u) &=
    -409 + 412 \,a_1^{M_2} -32\,a_2^{M_2}.
\end{align}
The cancellation of all singularities and the finiteness of all
convolution integrals completes the explicit factorization proof of
the NNLO vertex corrections.

\newabs
We finally collect all contributions and illustrate the relative
importance of the individual vertex corrections setting $\mu=m_b$
\begin{align}
\Re \;\hat{V}^{(1)}
    &= -22.5 +5.5 \,a_1^{M_2} - 1.1 \,a_2^{M_2},\no \\
\Re \;\hat{V}_1^{(2)}
    &= -148 +606\,a_1^{M_2}-80\,a_2^{M_2},\no \\
\Re \;\hat{V}_2^{(2)}
    &= 322-213\,a_1^{M_2}+3.8\,a_2^{M_2}.
\label{eq:V12:Nums:Re}
\end{align}

\subsection{Preliminary numerical result}

We conclude this chapter with the presentation of a numerical result
for the real part of the NNLO vertex corrections. We stress again
that this corresponds to a preliminary result which treats the
$c$-quark and the $b$-quark in the closed fermion loops as massless
quarks. If we reconsider our results for the imaginary part in this
approximation, we find deviations of $\sim5\%$ of the individual
NNLO contributions. As the NNLO terms are subleading for the real
part of the colour-allowed tree amplitude, we expect that this
approximation will have only a minor impact here.

\newabs
Our preliminary result for the real part of the NNLO vertex
corrections reads
\begin{align}
\Re\; \alpha_1(\pi\pi)& =\,~~ 1.01 \big{|}_{V^{(0)}}
                            + 0.03 \big{|}_{V^{(1)}}
                            + 0.03 \big{|}_{V^{(2)}} \no \\
                      & =\,~~ 1.06,
\end{align}
where we have used Wilson coefficients in NLL approximation for
simplicity (they are indeed known to the required NNLL accuracy and
can be found in \cite{GorHai}). As expected, the contribution is of
minor importance in absolute terms. However, the NNLO corrections
are found to be as important as the NLO terms which are numerically
suppressed by the small Wilson coefficient. Interestingly, the
vertex corrections add again constructively and, as can be seen in
comparison with the results from~\cite{BenekeJager}, come again with
the opposite of the spectator interactions.

\newabs
The colour-suppressed tree amplitude $\alpha_2(\pi\pi)$ is
phenomenologically more interesting as the respective QCD
Factorization prediction is rather low for a satisfactory
description of the experimental data. In order to derive the NNLO
result for $\alpha_2$ we still have to solve some conceptual aspects
concerning a Fierz-symmetric definition of (2-loop) evanescent
operators. We therefore relegate the discussion of the
colour-suppressed tree amplitude to~\cite{GBRePart}.

\chapter[Heavy-to-light form factors for NR bound states]
    {Heavy-to-light form factors for non-relativistic bound states}
\label{ch:NRModel}
 
In this chapter we investigate transition form factors between
non-relativistic QCD bound states at large recoil energy. Assuming
the decaying quark to be much heavier than its decay product, the
relativistic dynamics can be treated according to the factorization
formula (\ref{eq:BtoPiFF}) for heavy-to-light form factors obtained
from the HQE in QCD. In contrast to the $B\to\pi$ transition, the
form factors can be calculated entirely in perturbation theory in
the non-relativistic approximation which allows for an explicit
analysis of the factorization formula. We perform a NLO calculation
based on the methods that we developed in Chapter~\ref{ch:PT} and
look for an interpretation of the results from the viewpoint of QCD
Factorization.

The basic idea of this work has already been presented
in~\cite{GBTF05}. We emphasize that the formalism which we develop
in this chapter can be applied for $B_c\to (\bar c c)$ transitions
as all these particles can be considered approximatively as
non-relativistic bound states. Notice that we treat the charm quark
as a light quark in this case, $m_c\ll m_b$. We will adopt this
terminology throughout this chapter although we consider this
analysis rather as a toy model for the $B\to\pi$ transition. A
phenomenological study of various $B_c$ decays in QCD Factorization
will be given in~\cite{ABBinprep}.

\section{Non-relativistic approximation}
\label{sec:NRapprox}

\begin{figure}[b!]
\centerline{\parbox{13cm}{\centerline{\includegraphics[width=10cm]{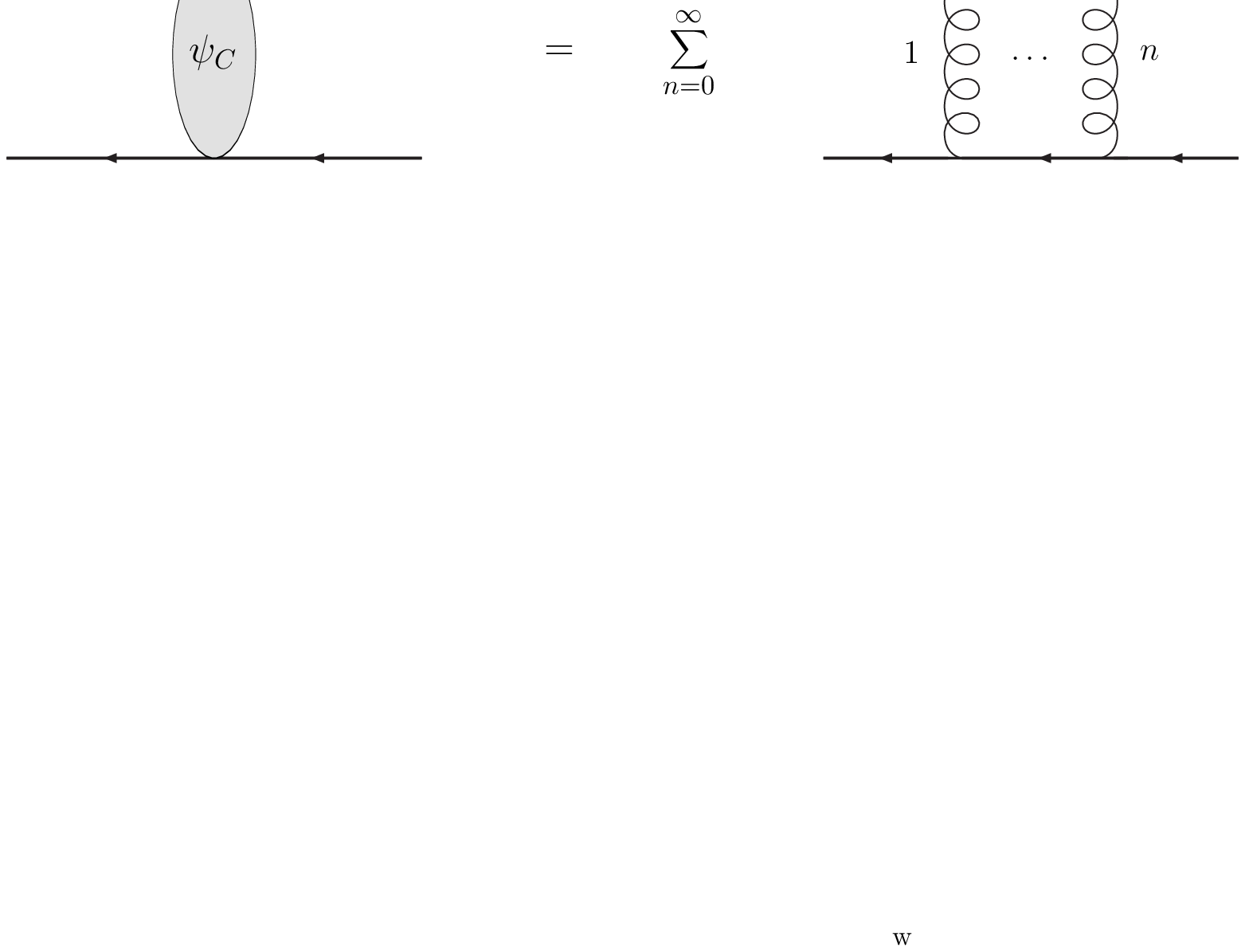}}
\caption{\label{fig:CoulombResum} \small \textit{Resummation of
potential gluons into a non-relativistic Coulomb wave-function
(details of the resummation can be found e.g. in \cite{GB03}).}}}}
\end{figure}

The wave function for a non-relativistic (NR) bound state of a quark
and an anti-quark with respective masses $m_1$ and $m_2$ can be
obtained from the resummation of NR (\emph{potential}) gluon
exchange as sketched in Figure~\ref{fig:CoulombResum}. The solution
of the corresponding Schr\"odinger equation with Coulomb potential
yields
\begin{align}
\psi_\text{C}(\vec{p}) \propto
\frac{\kappa^{5/2}}{(\kappa^2+\vec{p}^2)^2},
\label{eq:NRwavefunction}
\end{align}
where $\kappa=m_r\alpha_s C_F$ and $m_r=m_1 m_2/(m_1+m_2)$ is the
reduced mass. The normalization of the wave function gives the
(non-relativistic) meson decay constant
\begin{align}
f_\text{NR} =  \frac{2 \sqrt{N_c}}{\pi} \;
\frac{\kappa^{3/2}}{(m_1+m_2)^{1/2}}. \label{eq:NRdecayconst}
\end{align}

\newabs
In this approximation, the $B_c$ meson is entirely dominated by the
two-particle Fock state built from a bottom quark with mass $M\equiv
m_b$ and a charm antiquark with mass $m\equiv m_c$. Consequently to
first approximation in the NR expansion, the $B_c$ meson consists of
a quark with momentum $M w_\mu$ and an antiquark with momentum $m
w_\mu$, where $w_\mu$ is the four-velocity of the $B_c$ meson
($w^2=1$). The spinor degrees of freedom for the $B_c$ meson in the
initial state are represented by the Dirac projector
$\calP_H=\frac12 (1+\slashb{w})\gamma_5$.

Similarly, a pseudoscalar $\eta_c$ meson is interpreted as a $c\bar
c$ bound state where both constituents have approximately equal
momenta $m w'_\mu$, where $w'_\mu$ is the four-velocity of the
$\eta_c$ meson (${w'}^2=1$). The Dirac projector of the $\eta_c$
meson in the final state is given by $\calP_L=\frac12
(1-\slashb{w}')\gamma_5$.

\newabs
In the following we consider heavy-to-light transitions at large
recoil energy assuming $M\gg m$ and working in leading power of the
HQE in $m/M$. The QCD dynamics is then described by the SCET degrees
of freedom from Table~\ref{tab:term}, where we identify the typical
hadronic scale with the mass of the $\eta_c$ meson
$M_\eta\simeq2m=\calO(m)$ in the NR approach. The relevant scales in
the process are thus given by
\begin{align}
\mu_h\sim M \quad \gg \quad \mu_{hc}\sim \sqrt{M m} \quad \gg \quad
\mu_{s,c} \sim m \quad \gg \quad \mu_\text{NR} \sim m v,
\end{align}
where the NR scale refers to the virtuality of the potential gluons
from Figure~\ref{fig:CoulombResum} and $v\ll1$ is a NR velocity.
Notice that all relativistic scales are perturbative in our set-up
as $M,m\gg \LQCD$. The relativistic dynamics in the heavy-to-light
transition can therefore be analyzed in perturbation theory.

\newabs
The momentum transfer can be approximated as
\begin{align}
q^2=(M_B w-M_\eta w')^2\simeq M^2-4M m \; w\cdot w'
\end{align}
which implies a large relativistic boost
\begin{align}
\gamma \equiv w\cdot w' = \frac{M^2-q^2}{4M m} = \calO(M/m)
\label{eq:boost}
\end{align}
for large recoil energies with $M^2-q^2= \calO(M^2)$.

\section{Perturbative calculation}
\label{sec:NRPT}

For simplicity, we focus on the $B_c\to\eta_c$ transition and
consider the form factors $F_+(q^2)$, $F_-(q^2)$ and $F_T(q^2)$
which can be defined in analogy to (\ref{eq:defFF}). According to
the general discussion for heavy-to-light decays at large recoil, we
have to consider hard, hard-collinear, collinear and soft gluon
exchange in order to describe the relativistic dynamics of the
transition form factors, while the non-relativistic modes are
contained in the bound state wave functions of the initial and final
state mesons.

\subsection{Tree level}
\label{sec:NRTree}

\begin{figure}[b!]
\centerline{\parbox{13cm}{
\centerline{\includegraphics[height=2.3cm]{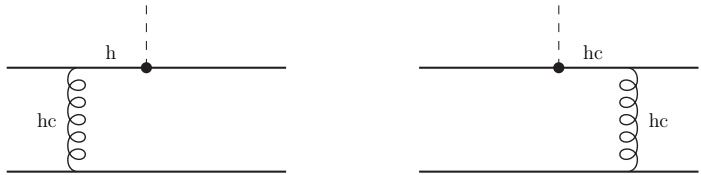}}
\caption{\label{fig:NRTree} \small \textit{Tree level diagrams. The
dot denotes the weak vertex mediating the heavy-to-light transition,
the lower line the light spectator antiquark. NR gluons from the
bound state wave functions are not drawn.}}}}
\end{figure}

We have to require at least one relativistic gluon exchange in the
large recoil case in order to rearrange the quark-antiquark pair in
the final state into a NR configuration. Consequently, the diagrams
from Figure~\ref{fig:NRTree} contribute in LO of the perturbative
expansion which imply the exchange of a hard-collinear gluon with
virtualiy
\begin{align}
(m w - m w')^2 \simeq -2\gamma m^2 = \calO(M m).
\end{align}
The result for the form factors at LO becomes
\begin{align}
\frac{F^\text{LO}_+}{1+2s} = -\frac{F^\text{LO}_-}{2s} =
\frac{F^\text{LO}_T}{2s}= \frac{f_M^\text{NR} f_m^\text{NR}}{N_c} \;
\frac{\pi\as C_F}{\gamma m^2}, \label{eq:NRFFLO}
\end{align}
where $f_M^\text{NR}$ and $f_m^\text{NR}$ are the non-relativistic
decay constants of the initial and final state mesons respectively
and we defined
\begin{align}
s\equiv\frac{M}{4\gamma m} = \frac{M^2}{M^2-q^2}=\calO(1)
\end{align}
with $s=1$ at maximum recoil $q^2=0$.

\newabs
From (\ref{eq:NRdecayconst}), we read off that the decay constants
scale as $f_M^\text{NR}\sim(m v)^{3/2}M^{-1/2}$ and
$f_m^\text{NR}\sim m v^{3/2}$ in the combined NR and HQE. We find
that the form factors scale as
\begin{align}
F^\text{LO}_i(q^2) \;\sim\; \alpha_s v^3 \left(
\frac{m}{M}\right)^{3/2} \label{eq:FFscal}
\end{align}
in agreement with the expected scaling $F_i(q^2)\sim M^{-3/2}$ from
the general discussion for heavy-to-light form factors at large
recoil \cite{MBTF2000}.

\subsection{1-loop calculation}

The tree level results derived above do not give rise to endpoint
singularities as the wave functions of the non-relativistic bound
states have vanishing support at the endpoints. For a deeper
understanding of the factorization formula we therefore have to
consider the NLO contributions. We will see below that our NLO
calculation indeed reveals the full complexity of the factorization
formula. We now switch to the technical part of the calculation and
evaluate the corresponding 1-loop diagrams following our strategy
from Section~\ref{sec:strategy}.

\subsubsection{Step 1: Set-up for loop calculation}

\begin{figure}[h!]\vspace{2mm}
\centerline{\parbox{15cm}{\includegraphics[width=15cm]{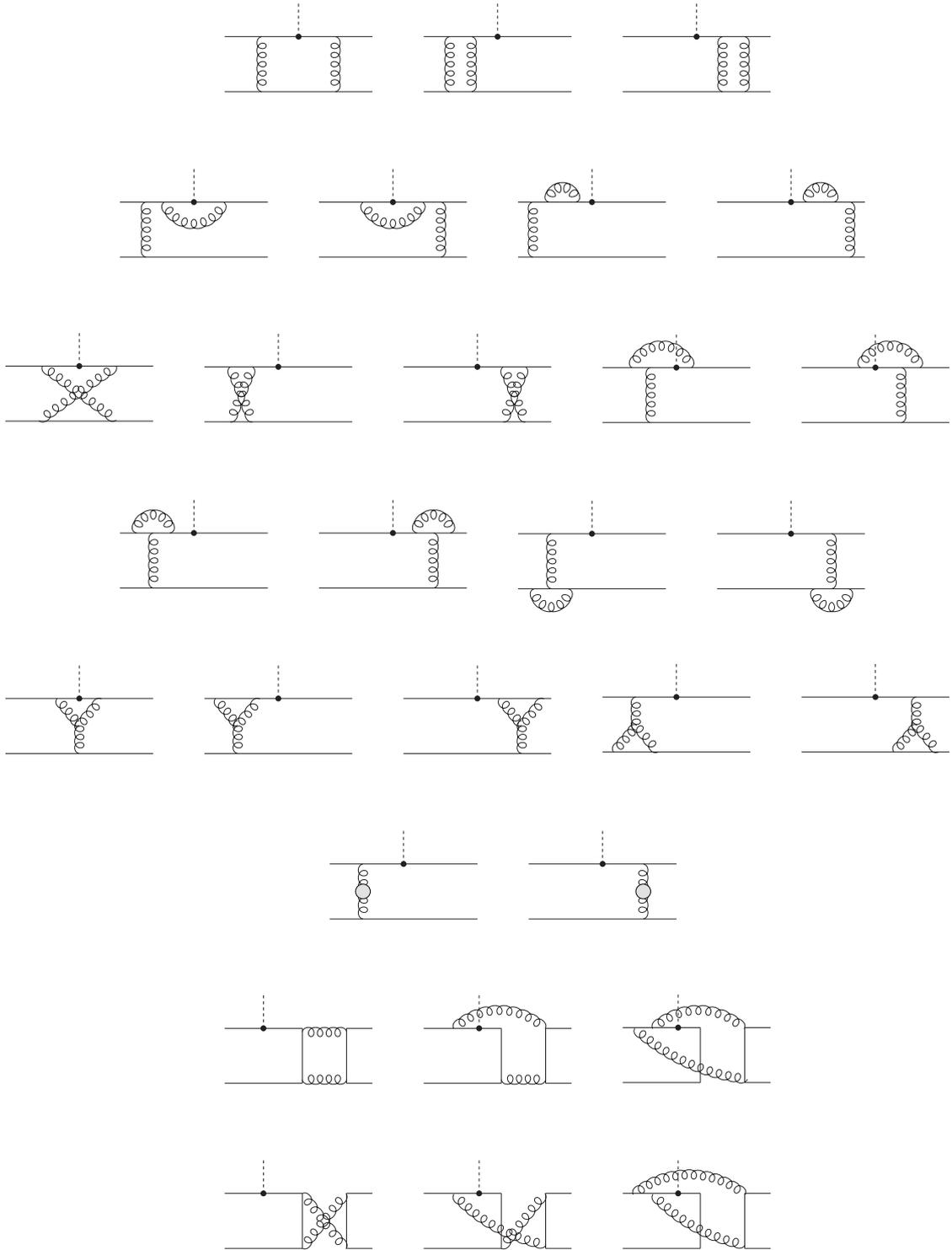}\vspace{8mm}
\caption{\label{fig:NR1loop} \small \textit{NLO diagrams. The bubble
in the diagrams from the sixth line represents the 1-loop gluon
self-energy. The diagrams from the last two lines contribute to
flavour singlet final states only.}}}}
\end{figure}

At NLO the form factors receive contributions from various 1-loop
diagrams as shown in Figure~\ref{fig:NR1loop}. The  corresponding
colour factors are summarized in Table~\ref{tab:NRColour}. All
diagrams from Figure~\ref{fig:NR1loop} can be expressed in terms of
the following denominators of particle propagators
\begin{align}
\calP_1&=\left(M w +m w-k\right)^2-M^2,  &&\calP_8=\left(M w + m w -m w'-k\right)^2-M^2,\no\\
\calP_2&=\left(M w -m w'+k\right)^2-M^2, &&\calP_9=\left(M w -k\right)^2-M^2,\no\\
\calP_3&=\left(2m w'-k\right)^2-m^2,     &&\calP_{10}=\left(2m w'-m w -k\right)^2-m^2,\no\\
\calP_4&=\left(m w' - m w +k\right)^2-m^2,&&\calP_{11}=\left(m w'-k\right)^2-m^2,\no\\
\calP_5&=k^2-m^2,                        &&\calP_{12}=k^2,\no\\
\calP_6&=\left(m w -k\right)^2,          &&\calP_{13}=\left(m w-k\right)^2-m^2,\no\\
\calP_7&=\left(m w'-k\right)^2, &&\calP_{14}=\left(2mw'-k\right)^2.
\label{eq:NRProps}
\end{align}

\subsubsection{Step 2: Reduction to Master Integrals}

\begin{figure}[b!]
\centerline{\parbox{14.8cm}{\includegraphics[width=14.8cm]{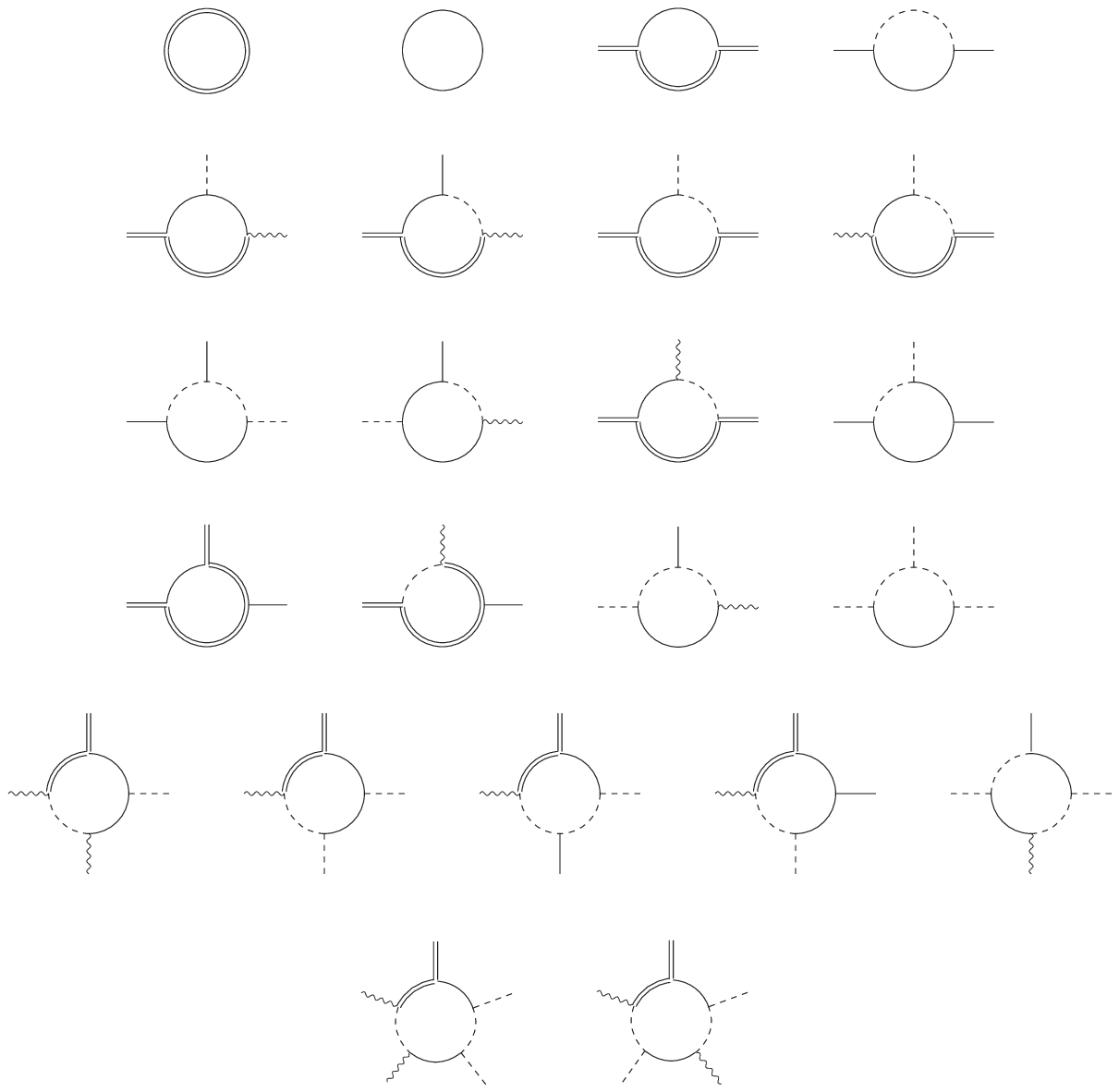}\vspace{4mm}
\caption{\label{fig:NRMIs} \small \textit{Scalar Master Integrals
that appear in the 1-loop calculation. We use dashed/solid/double
internal lines for propagators with masses $0/m/M$, respectively.
Double/solid/dashed/wavy external lines correspond to
hard/hard-collinear/collinear/soft external momenta. Notice that we
refrain from associating all external scalar products to the MIs in
this case. The figures therefore do not define the MIs unambiguously
in our representation (for the explicit definitions of the MIs we
refer to Appendix~\ref{app:MIsNR}).}}}}
\end{figure}

The reduction procedure is not as efficient in this calculation as
it was in the \mbox{2-loop} calculation from Chapter~\ref{ch:ImPart}
and~\ref{ch:RePart}. However, we were able to express about 200
scalar integrals in terms of 23 MIs which are summarized in
Figure~\ref{fig:NRMIs}. The number of MIs is comparably high in this
calculation due to the fact that we deal with three different types
of propagators (with masses $0, m ,M$) and four external momenta
(two of them linearly independent $w,w'$).

\subsubsection{Step 3: Manipulation of Dirac structures}

We project onto the NR bound states with the help of the projectors
$\calP_H$ and $\calP_L$ which we specified in
Section~\ref{sec:NRapprox}. The flavour singlet diagrams in the last
two lines of Figure~\ref{fig:NR1loop} require special care as they
involve traces like $\Tr(\gamma^\mu \gamma^\nu \gamma^\rho
\gamma^\sigma \gamma_5)$ which may invalidate the treatment of an
anticommuting $\gamma_5$ within DR. However, the flavour singlet
diagrams turn out to give finite contributions and the traces can
safely be calculated in $d=4$ dimensions.

\begin{table}[t!]
\centerline{
\parbox{13cm}{\setlength{\doublerulesep}{0.1mm}
\centerline{\begin{tabular}{|c||c|c|c|c|c|} \hline
\hspace*{1.75cm}&\hspace*{1.75cm}&\hspace*{1.75cm}&\hspace*{1.75cm}&\hspace*{1.75cm}&\hspace*{1.75cm} \\[-0.7em]
Diagram & line 1-2 & line 3-4 & line 5& line 6 & line 7-8 \\[0.3em]
\hline\hline&&&&& \\[-0.7em]
Colour & $C_F^2$ & $C_F^2-\frac{C_FN}{2}$ & $\frac{C_F N}{2}$ & $C_F$ & $\frac{C_F}{2}$\\[0.3em]
\hline
\end{tabular}} \vspace{4mm} \caption{\label{tab:NRColour}\small
\textit{Colour factors of the diagrams in Figure~\ref{fig:NR1loop}.
The normalization is chosen such that the tree diagrams from
Figure~\ref{fig:NRTree} give $C_F$.}}}}\vspace{1mm}
\end{table}

\subsubsection{Step 4: Calculation of Master Integrals}

Though most of the MIs in Figure~\ref{fig:NRMIs} correspond to
apparently simple 1-loop \mbox{3-topologies}, the calculation is
non-trivial due to the fact that the integrals involve several
distinct scales and many massive propagators. The calculation
simplifies as we only require the leading power in $m/M$ of the MIs
which we extract using the method of regions and Mellin-Barnes
techniques as described in detail in Section~\ref{sec:CalcMI}. The
analytical results of the MIs can be found in
Appendix~\ref{app:MIsNR}. \vspace*{2mm}

\subsection{Form factors in NLO}

\begin{figure}[b!]
\centerline{\parbox{13cm}{
\centerline{\includegraphics[width=2.5cm]{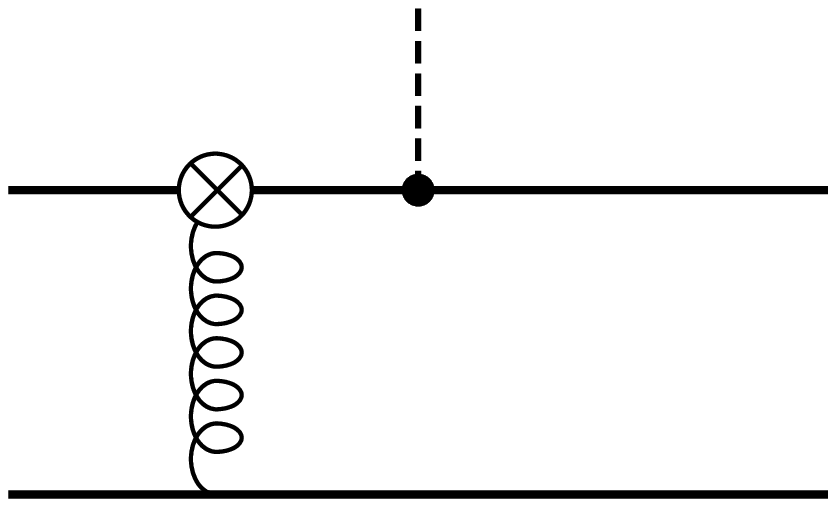}\hspace{10mm}\includegraphics[width=2.5cm]{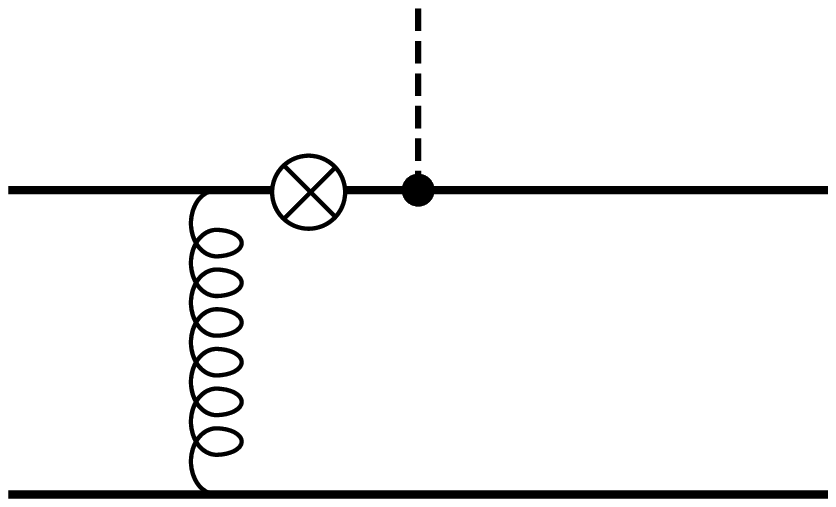}\hspace{10mm}\includegraphics[width=2.5cm]{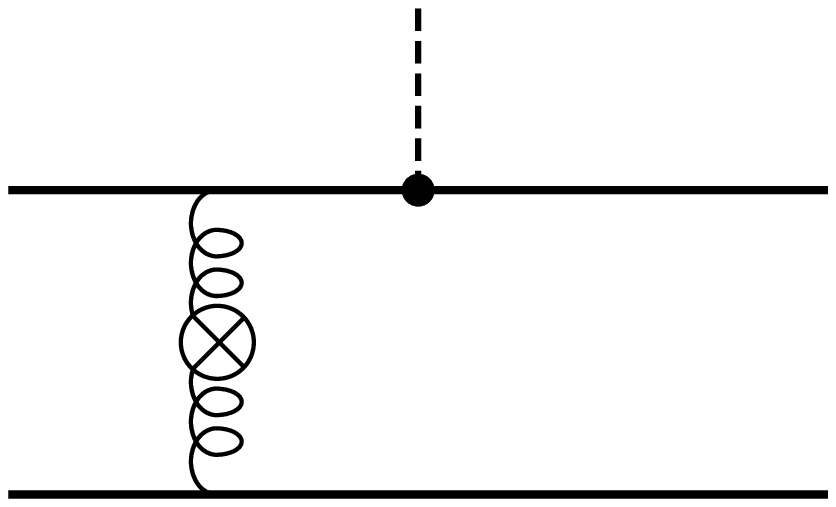}\hspace{10mm}\includegraphics[width=2.5cm]{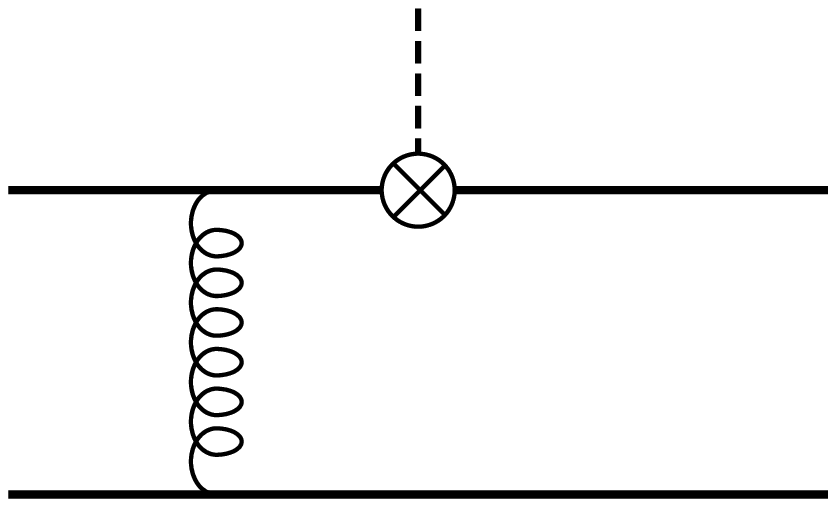}}
\caption{\label{fig:NRCTs} \small \textit{Sample of counterterm
diagrams.}}}}
\end{figure}

In NLO we have to take into account the 1-loop diagrams from
Figure~\ref{fig:NR1loop} as well as standard (tree-level)
counterterm diagrams like the ones depicted in
Figure~\ref{fig:NRCTs}. We use \MSbar-scheme renormalization
constants
\begin{align}
Z_g &= 1 - \frac{\as}{4\pi\eps} \left(\frac{11}{6} N_c - \frac13 n_f
\right) + \calO(\as^2), \no \\
Z_A &= 1 - \frac{\as}{4\pi\eps} \left(\frac23 n_f - \frac53 N_c
\right) + \calO(\as^2)
\end{align}
for the coupling constant and the gluon field, respectively. The
(non-relativistic) quark fields are conveniently treated in the
on-shell scheme with
\begin{align}
Z_q(m) &= 1 - \frac{\as C_F}{4\pi} \left(\frac{1}{\eps} +
\frac{2}{\eps_\IR} + 3 \ln \frac{\mu^2}{m^2}+4
\right) + \calO(\as^2), \no \\
Z_m(m) &=  1 - \frac{\as C_F}{4\pi} \left(\frac{3}{\eps} + 3 \ln
\frac{\mu^2}{m^2}+4 \right) + \calO(\as^2), \label{eq:NRZqZm}
\end{align}
where we indicated that the wave function renormalization constant
contains IR divergences. The renormalization of the weak vertex
involves the Z-factor of the heavy-to-light current $\bar
q\,\Gamma\, Q$ which is given by
\begin{align}
Z_\Gamma &= 1 + \frac{\as C_F}{4\pi\eps} + \calO(\as^2) \qquad
\text{for} \qquad \Gamma=\sigma^{\mu\nu},
\end{align}
whereas $Z_\Gamma =1$ for the conserved vector current with
$\Gamma=\gamma^\mu$.

\newabs
Adding up all diagrams and counterterms at NLO, all UV and IR
divergences cancel in the form factors as expected. Our explicit NLO
results are summarized in Appendix~\ref{app:NRFormFactor}. Let us
quote the result for the form factor $F_+(q^2)$ at maximum recoil
$q^2=0$ here to illustrate the structure of the NLO contributions.
We find
\begin{align}
F_+^\text{NLO}(0)
    &=\frac{f_M^\text{NR} f_m^\text{NR}}{N_c} \;
\frac{12\pi\as(\mu) C_F}{M m} \bigg\{ 1 + \frac{\as(\mu)}{4\pi}
\bigg[
      \!  \left( \frac{11}{3}N_c -\frac23 n_f \right) \! \ln \frac{2\mu^2}{m M} - \frac{10}{9} n_f + \calS_+ \no\\
    & \hspace{0.4cm} + C_F \bigg( \frac12 \, \ln^2 \frac{M}{m} + \frac{35-20\ln2}{6} \, \ln \frac{M}{m} +\frac23 \, \ln^2 2 +3 \ln 2 +\frac{7\pi^2}{9} -\frac{103}{6} \bigg) \no \\
    & \hspace{0.4cm} + N_c \bigg( \!-\frac16 \, \ln^2 \frac{M}{m} + \frac{1+\ln2}{3} \, \ln \frac{M}{m} +\frac13 \, \ln^2 2 -\frac43\, \ln 2 -\frac{5\pi^2}{36} +\frac{73}{9}
    \bigg) \bigg] \bigg\}
    \label{eq:NRFplusShort}
\end{align}
where $M$ and $m$ are to be considered as pole masses. In
(\ref{eq:NRFplusShort}) we isolated the contribution from the
flavour singlet diagrams which is given by $\calS_+=1/3 \, \ln
(M/4m)$.

\begin{figure}[t!]
\centerline{\parbox{13cm}{
\centerline{\includegraphics[width=6.5cm]{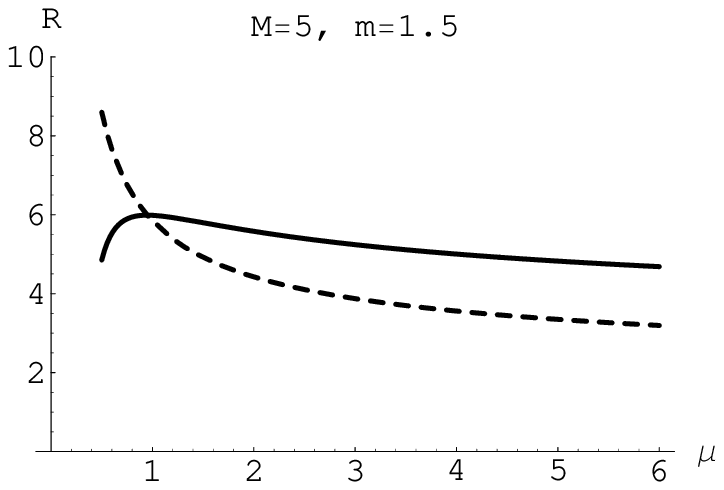}\hspace{5mm}
\includegraphics[width=6.5cm]{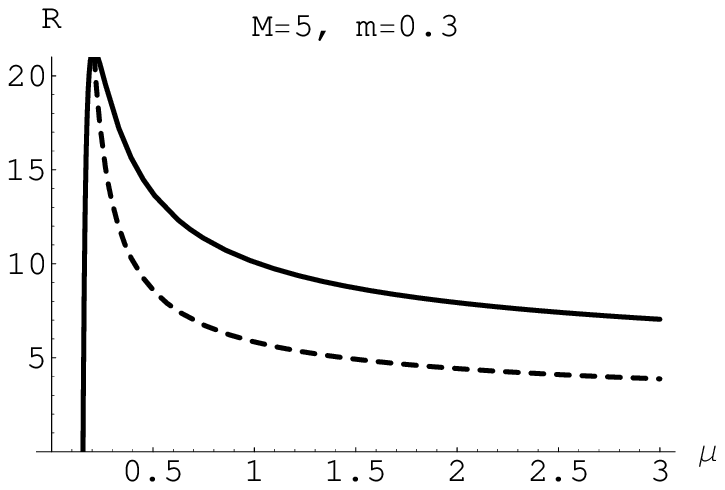}}
\caption{\label{fig:NRFPlus0} \small \textit{Renormalization-scale
dependence of the ratio $R$ for different choices of quark masses.
The solid (dashed) line denotes the NLO (LO) result. Left: Realistic
case relevant for the $B_c\to\eta_c$ transition. Right: Toy example
for the $B\to\pi$ transition.}}}}
\end{figure}

\newabs
We now investigate the residual renormalization-scale dependence of
$F_+^\text{NLO}(0)$. For illustration we will use two sets of quark
masses with
\begin{align}
&\text{set 1} :  &&M=5\,\gev, &&m=1.5\,\gev, &&\mu_{hc}=\sqrt{M m}\simeq 2.7\,\gev,\no \\
&\text{set 2} :  &&M=5\,\gev, &&m=0.3\,\gev, &&\mu_{hc}=\sqrt{M
m}\simeq 1.2 \,\gev
\end{align}
which correspond to the physical decay $B_c\to\eta_c$ and a toy
model for the $B\to\pi$ transition, respectively (in the second case
we thus drop the singlet contribution). We use 1-loop running of
$\as$ with $n_f=4$ and $\as(5\,\gev)=0.2$ and study the ratio
\begin{align}
R &=\frac{M m}{f_M^\text{NR} f_m^\text{NR}} \; F_+(0).
\end{align}
The result is illustrated in Figure~\ref{fig:NRFPlus0}. In both
cases we observe a significant enhancement from the NLO
contribution. For the $B_c\to\eta_c$ case (set 1) the enhancement of
the tree level result at the hard-collinear scale is about 35\%. We
find a substantial improvement of the scale dependence from about
40\% at LO to 20\% at NLO if we vary $\mu$ between the soft scale
$m$ and the hard scale $M$. The hierarchy of scales $M\gg m$ is not
too large in this case, such that the formally large logarithms $\ln
M/m$ do not spoil the convergence of the perturbation series too
badly.

\newabs
In the toy case for the $B\to\pi$ form factor (set 2), the soft
scale $m$ is close to the QCD scale $\LQCD$ and consequently the
convergence of the perturbative expansion breaks down at small
scales. The theoretical error due to the renormalization scale is
not under control in this case. From the conceptual viewpoint we are
particularly interested in the structure and resummation of the
formally large logarithms $\ln M/m$ in the $B\to\pi$ case where they
should be counted as $\as \ln M/m\sim 1$. We therefore examine the
origin of these logarithms in the subsequent section in detail.

\newpage
\section{Factorization Formula}
\label{sec:NRFF}

We now come to the interpretation of our explicit NLO calculation in
terms of the factorization formula for heavy-to-light form factors
at large recoil energy
\begin{align}
F_i(q^2) &\simeq H_i(q^2) \; \xi(q^2) + \int_0^\infty \! d\om
\int_0^1 \! du \;\; \phi_{B}(\om) \; T_i(u,\om; q^2) \;
\phi_{\eta}(u). \label{eq:BtoPiFFCopy}
\end{align}
We start with a closer look at the factorization formula following
\cite{BPiJet:BY}. The two terms in (\ref{eq:BtoPiFFCopy}) are
associated to the matrix elements of two distinct \SCETI~operator
structures, the so-called $A$-type and $B$-type operators (cf.
(\ref{eq:xidef}) for notation)
\begin{align}
J_A &= (\bar{\xi}_{hc} W_{hc}) \, h_v, \no \\
J_B &= (\bar{\xi}_{hc} W_{hc}) \, (W^\dagger_{hc}i \slashc{D}_{\perp
hc} W_{hc}) \, h_v.
\end{align}
The matrix element of the operator $J_A$ defines the form factor
$\xi(q^2)$ and the $H_i(q^2)$ denote the corresponding matching
coefficients from QCD to \SCETI~which include the contributions from
hard momentum fluctuations. The operator $J_B$ defines a non-local
form factor $\Xi(\tau;q^2)$ with short-distance coefficients
$C_i(q^2;\tau)$. In the first matching step the factorization
formula thus takes the form
\begin{align}
F_i(q^2) &\simeq H_i(q^2) \; \xi(q^2) + \int_0^1 \! d\tau \;\;
C_i(q^2;\tau) \; \Xi(\tau;q^2). \label{eq:BtoPiFFStep1}
\end{align}
The non-local form factor $\Xi(\tau;q^2)$ can be factorized further
in a second matching step from \SCETI~to \SCETII~ including the
effects from hard-collinear fluctuations. This results in a
convolution of a perturbative jet function $J(\tau;u,\om; q^2)$ with
leading twist distribution amplitudes given by
\begin{align}
\Xi(\tau;q^2) &= \int_0^\infty \! d\om \int_0^1 \! du \;\;
\phi_B(\om) \; J(\tau;u,\om; q^2) \; \phi_\eta(u).
\label{eq:BtoPiFFStep2}
\end{align}
Combining (\ref{eq:BtoPiFFStep1}) and (\ref{eq:BtoPiFFStep2}), we
see that the hard-scattering kernel from (\ref{eq:BtoPiFFCopy}) can
be identified as
\begin{align}
T_i(u,\om; q^2) &= \int_0^1 \! d\tau \;\; C_i(q^2;\tau) \;
J(\tau;u,\om; q^2). \label{eq:BtoPiFFTi}
\end{align}
In our analysis we calculate the leading twist distribution
amplitudes of the $B_c$ and $\eta_c$ meson at leading power in the
NR approximation and including perturbative (relativistic) effects
to NLO in QCD. We then use our explicit NLO results of the form
factors together with the known expressions for $H_i$, $C_i$ and $J$
to extract the overlap-contribution $\xi(q^2)$ in our NR set-up
which may then be analyzed further concerning its factorization
properties in \SCETII. This immediately leads us to the problem of
endpoint singularities which arise in our calculation at NLO of the
perturbative expansion. We find it instructive to demonstrate how
these endpoint singularities enter our calculation before looking at
the factorization of the heavy-to-light form factors in detail.

\subsection{Endpoint singularities}

The appearance of endpoint singularities is related to the issue of
large logarithms which we already mentioned at the end of the last
section. As these logarithms may spoil the convergence of the
perturbation series, we are particularly interested in the
resummation of such logarithms within the effective theory. The
structure of formally large logarithms is particularly complicated
in heavy-to-light decays due to the presence of various scales and
degrees of freedom.

\newabs
We encountered these logarithms in our explicit NLO calculation in
form of $\ln \gamma$ where $\gamma=\calO(M/m)$ is the large
relativistic boost between the meson rest frames (cf. our results in
Appendix~\ref{app:NRFormFactor}). We find that the origin of the
leading double logarithms is related to two different mechanisms
which we consider in detail in the remainder of this section: The
first one gives rise to (standard) Sudakov logarithms, the second
one is related to endpoint singularities.

\subsubsection{Sudakov logarithms}

\begin{figure}[b!]
\centerline{\parbox{13cm}{
\centerline{\includegraphics[width=13cm]{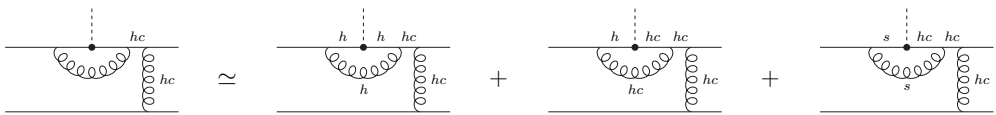}}
\caption{\label{fig:NRSudakov} \small \textit{Origin of Sudakov
logarithms (in Feynman gauge).}}}}
\end{figure}

Sudakov logarithms appear naturally in processes with
(hard-)collinear and soft degrees of freedom.. In our calculation
they arise for instance from the diagram in
Figure~\ref{fig:NRSudakov} (in Feynman gauge). Rather than looking
at the full diagram here, we profit from our reduction algorithm and
consider an underlying (scalar) MI which allows for a more
transparent presentation of the interesting aspects.

\newabs
The MI to be considered here is \vspace*{5mm}
\begin{align}
\parbox[c]{2.5cm}{\vspace{-5mm}\psfig{file=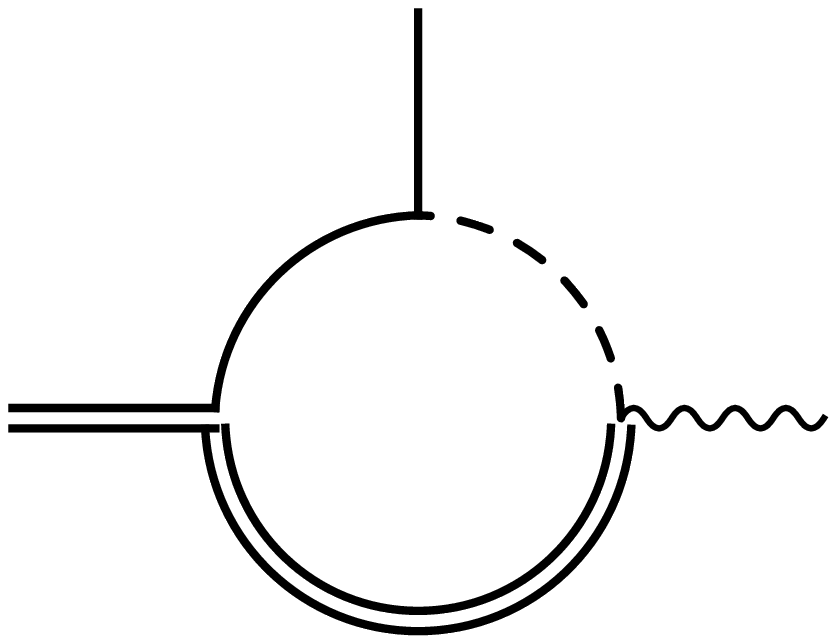,width =
2cm}} &= \;\;
    \int \, [dk] \; \frac{1}{\calP_{1}\calP_{3}\calP_{6}} \;\; \simeq\;\;
    \ln^2(4\gamma) -2\Lib(1-s) -\ln^2 s + \pi^2 + \calO(\eps) \label{eq:MInt_32Copy}
\end{align}
with the $\calP_i$ given in (\ref{eq:NRProps}). Notice that the
integral contains a double logarithm in $\gamma$ and that we
suppressed a prefactor $-1/(8\gamma m M)$ for simplicity, cf.
(\ref{eq:MInt_32}). The integral may be calculated with
Mellin-Barnes techniques as described in Section~\ref{sec:MB}.

\newabs
In order to understand the origin of the double logarithm, we
disentangle the contributions from different momentum regions of the
loop integration. Following the method of expansion by regions which
we described in Section~\ref{sec:ExpByRegions}, the MI is found to
receive leading contributions from the hard, hard-collinear and soft
region.

\newabs
The hard momentum region with $k\sim(1,1,1)$ is found to give
\vspace*{5mm}
\begin{align}
\parbox[c]{2.5cm}{\vspace{-5mm}\psfig{file=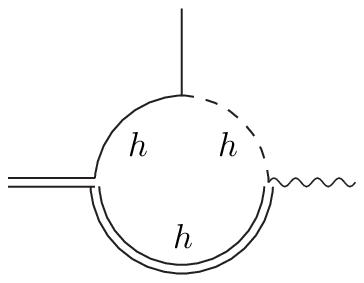,width =
2cm}} \!\!\!\!&\simeq
    \frac{1}{\eps^2} + \frac{1}{\eps} \ln
    \frac{\mu^2}{16\gamma^2m^2} + \frac{1}{2} \ln^2
    \frac{\mu^2}{16\gamma^2m^2} -2\Lib(1-s) -\ln^2 s +
    \frac{\pi^2}{6} + \calO(\eps), \label{eq:MInt_32H}
\end{align}
with double poles in $\eps$ which arise in the limits $k_\perp\to0$
and $k_+\to0$. The contribution from the hard-collinear region with
$k\sim(1,\lambda,\lambda^2)$ becomes  \vspace*{5mm}
\begin{align}
\parbox[c]{2.5cm}{\vspace{-5mm}\psfig{file=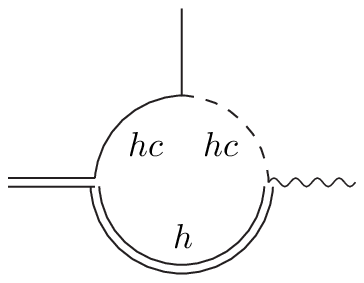,width =
2cm}} &\simeq\;\;
    -\frac{2}{\eps^2} - \frac{2}{\eps} \ln
    \frac{\mu^2}{4\gamma m^2} - \ln^2
    \frac{\mu^2}{4\gamma m^2} + \calO(\eps), \label{eq:MInt_32HC} \hspace{3.4cm}
\end{align}
where the singularities stem from $k_\perp\to\infty$ and $k_-\to0$.
Finally, the soft region with $k\sim(\lambda^2,\lambda^2,\lambda^2)$
generates poles for $k_\perp\to\infty$ and $k_+\to\infty$ and reads
\vspace*{5mm}
\begin{align}
\parbox[c]{2.5cm}{\vspace{-5mm}\psfig{file=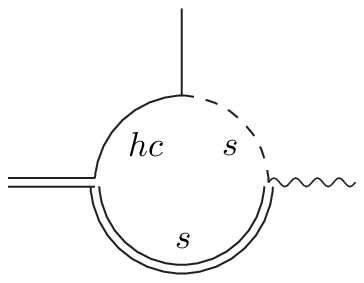,width =
2cm}} &\simeq\;\;
    \frac{1}{\eps^2} + \frac{1}{\eps} \ln
    \frac{\mu^2}{m^2} +\frac12 \ln^2
    \frac{\mu^2}{m^2} + \frac{5\pi^2}{6} + \calO(\eps). \label{eq:MInt_32S} \hspace{3.1cm}
\end{align}
Adding up these contributions, all singularities drop out and we
reproduce (\ref{eq:MInt_32Copy}).

\newabs
From the viewpoint of QCD Factorization, the hard contribution from
(\ref{eq:MInt_32H}) is to be considered as part of a
$\text{QCD}\to\SCETI$ matching calculation, the hard-collinear
effects from (\ref{eq:MInt_32HC}) contribute in the
$\SCETI\to\SCETII$ matching procedure and the soft contribution from
(\ref{eq:MInt_32S}) to the (perturbative) calculation of a
\SCETII~matrix element. Let us think of all regions to be
$\MSbar$-subtracted (the poles contain information about the
anomalous dimensions of operators in the effective theory).

We see that the first matching step in (\ref{eq:MInt_32H}) is free
of large logarithms if we choose the matching scale $\mu\sim M$. We
may then evolve the scale down to $\mu\sim \mu_{hc}$ using RGEs in
\SCETI~which implicitly resum logarithms of the type $\ln M/\mu$. At
the hard-collinear scale we perform the second matching step and we
see that the choice $\mu\sim (M m)^{1/2}$ guarantees the absence of
large logarithms in (\ref{eq:MInt_32HC}). The procedure continues
summing logarithms of the type $\ln \mu_{hc}/\mu$ with the help of
corresponding RGEs in \SCETII~and evolving the scale down to
$\mu\sim m$. In the $B\to\pi$ case the \SCETII~matrix elements have
to be calculated with the help of a non-perturbative method. As the
soft scale is still perturbative in our NR approach, the
corresponding \SCETII~matrix elements are calculable giving
(\ref{eq:MInt_32S}).

\newabs
The aforementioned factorization and resummation procedure is of
course oversimplifying as we considered a single integral instead of
operators in the effective theory. However, it allowed us to
emphasize an important feature concerning the resummation of Sudakov
logarithms: We saw that we were able to choose the factorization
scale $\mu$ such that there are no large logarithms in each matching
calculation and we pointed out that the resummation of logarithms
can be performed using standard RG-techniques varying $\mu$ between
two well-separated scales which correspond to the
\emph{virtualities} of the respective modes. We will see in the
following that part of the double logarithms in our explicit NLO
results are of conceptually different origin.

\newpage
\subsubsection{Non-factorizable logarithms}

\begin{figure}[b!]
\centerline{\parbox{13cm}{
\centerline{\includegraphics[width=13cm]{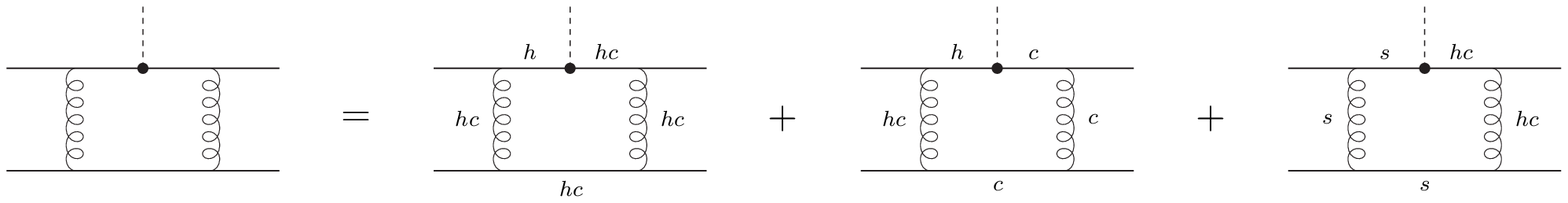}}
\caption{\label{fig:NRPentagon} \small \textit{Origin of
non-factorizable logarithms (in Feynman gauge).}}}}
\end{figure}

In a second example we consider the pentagon diagram from
Figure~\ref{fig:NRPentagon} (in Feynman gauge, see also the
discussion in \cite{GBTF05}). For simplicity, let us again focus on
a related MI here given by  \vspace*{5mm}
\begin{align}
\parbox[c]{2.5cm}{\vspace{-5mm}\psfig{file=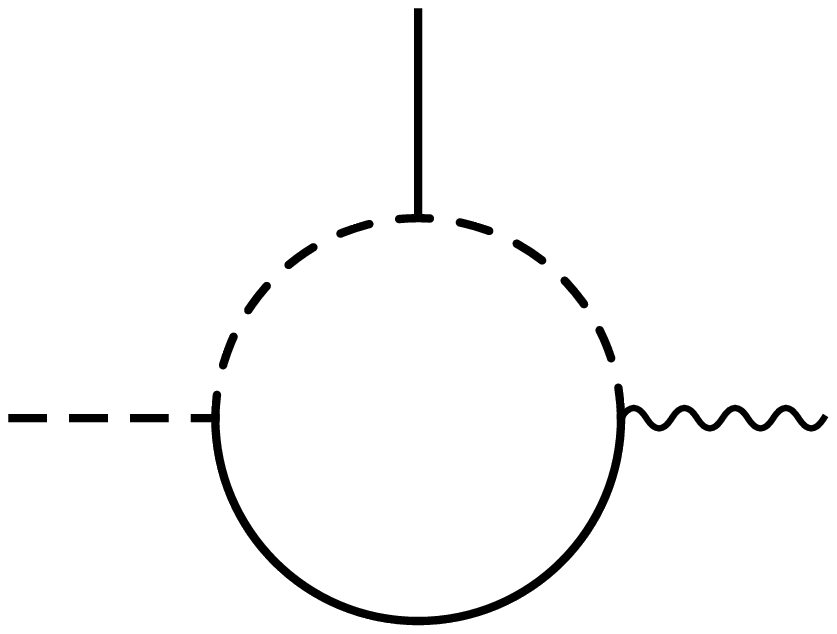,width =
2cm}} &= \;\;
    \int \, [dk] \; \frac{1}{\calP_{5}\calP_{6}\calP_{7}} \;\;
\simeq\;\;
     \frac12 \ln^2(2\gamma) + \frac{2\pi^2}{3}+  \calO(\eps)
     \hspace{2.2cm} \label{eq:MInt_313Copy}
\end{align}
which may be calculated with Mellin-Barnes techniques. Notice that
the MI gives rise to a double logarithm and that we dropped a
prefactor $-1/(2\gamma m^2)$, cf. (\ref{eq:MInt_313}).

\newabs
At leading power the MI receives contributions from the
hard-collinear, collinear and soft momentum region. The
hard-collinear contribution with $k\sim(1,\lambda,\lambda^2)$ is
found to be divergent for $k_\perp\to0$ and $k_+\to0$ and reads
\vspace*{5mm}
\begin{align}
\parbox[c]{2.5cm}{\vspace{-5mm}\psfig{file=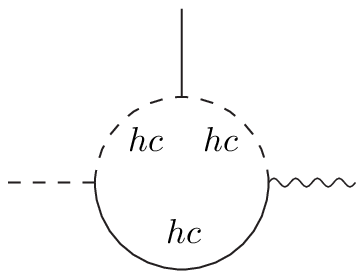,width =
2cm}} &\simeq\;\; \frac{1}{\eps^2}  +\frac{1}{\eps} \ln
    \frac{\mu^2}{2\gamma m^2} +\frac12 \ln^2
    \frac{\mu^2}{2\gamma m^2} + \calO(\eps).
    \label{eq:MInt_313HC} \hspace{3.4cm}
\end{align}
This looks very much like (\ref{eq:MInt_32HC}) but all poles are of
IR origin in this case.

\newabs
The collinear region with $k\sim(1,\lambda^2,\lambda^4)$ is found to
be divergent for $k_\perp\to\infty$ and $k_-\to0$ and similarly the
soft region with $k\sim(\lambda^2,\lambda^2,\lambda^2)$ for
$k_\perp\to\infty$ and $k_+\to0$. In contrast to what we have seen
so far, it turns out that the longitudinal integrations along $k_-$
($k_+$) in the collinear (soft) region are \emph{not} regularized in
this case (notice that DR only influences the integration over
transverse momenta in \mbox{$d_\perp=2-2\eps$} dimensions). However,
we may introduce an additional regularization procedure in order to
render these contributions finite. Following \cite{Proof
B->Pi:MBTF,GBTF05}, we apply an analytic continuation replacing
$\calP_6^{-1}\to(-\nu^2)^\delta\calP_6^{-1-\delta}$ such that
endpoint singularities for $k_-\to0$ ($k_+\to0$) show up as poles in
$1/\delta$. We thus find  \vspace*{5mm}
\begin{align}
\parbox[c]{2.5cm}{\vspace{-5mm}\psfig{file=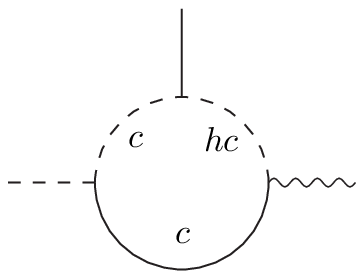,width =
2cm}} &\simeq\;\; - \left( \frac{1}{\eps} + \ln\frac{\mu^2}{m^2}
\right) \left( \frac{1}{\delta} + \ln\frac{\nu^2}{2\gamma m^2}
\right) + \frac{\pi^2}{3} + \calO(\eps),
    \label{eq:MInt_313C} \\[1.5em]
\parbox[c]{2.5cm}{\vspace{-5mm}\psfig{file=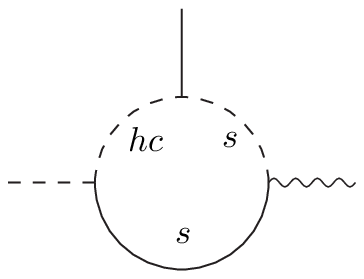,width =
2cm}} &\simeq\;\; -\frac{1}{\eps^2}  -\frac{1}{\eps} \ln
    \frac{\mu^2}{m^2} -\frac12 \ln^2
    \frac{\mu^2}{m^2} \no \\
& \hspace{0.7cm} + \left( \frac{1}{\eps} + \ln\frac{\mu^2}{m^2}
\right) \left( \frac{1}{\delta} + \ln\frac{\nu^2}{m^2} \right) +
\frac{\pi^2}{3} + \calO(\eps).
    \label{eq:MInt_313S}
\end{align}
Notice that the dependence on the ad-hoc parameters $\delta$ and
$\nu$ disappears when we consider the sum of collinear and soft
momentum regions \vspace*{5mm}
\begin{align}
& \parbox[c]{2.5cm}{\vspace{-5mm}\psfig{file=MInt_313c.ps,width =
2cm}} + \;\;\;
\parbox[c]{2.5cm}{\vspace{-5mm}\psfig{file=MInt_313s.ps,width =
2cm}} \no \\
& \hspace{0.7cm} \simeq\;\; -\frac{1}{\eps^2}  -\frac{1}{\eps} \ln
    \frac{\mu^2}{m^2} -\frac12 \ln^2
    \frac{\mu^2}{m^2}+ \left( \frac{1}{\eps} + \ln\frac{\mu^2}{m^2}
\right) \ln(2\gamma) + \frac{2\pi^2}{3} + \calO(\eps)
    \label{eq:MInt_313C+S}
\end{align}
and that we reproduce the full result from (\ref{eq:MInt_313Copy})
by adding up (\ref{eq:MInt_313HC}) and (\ref{eq:MInt_313C+S}).

\newabs
To summarize, we have seen that we \emph{cannot} disentangle
collinear and soft effects in the considered integral as the
respective contributions suffer endpoint singularities related to
the integrations over longitudinal light-cone momentum fractions
which are not regularized in DR (at fixed transverse
momenta/virtuality). However, the sum of collinear and soft momentum
region turns out to be unambiguously well-defined. We may now ask
the question if we can factorize the perturbative hard-collinear
effects in (\ref{eq:MInt_313HC}) from the remnant \SCETII~matrix
element in (\ref{eq:MInt_313C+S}) summing all formally large
logarithms into short-distance coefficient functions.

\newabs
We repeat our analysis from the last section and think of the
$\SCETI\to\SCETII$ matching calculation in (\ref{eq:MInt_313HC}) and
the \SCETII~matrix element in (\ref{eq:MInt_313C+S}) to be
\MSbar-subtracted. Similar to what we have seen in the last section,
(\ref{eq:MInt_313HC}) is free of large logarithms choosing $\mu\sim
(M m)^{1/2}$. We then evolve the scale down to $\mu\sim m$ using
RG-techniques which implicitly resum logarithms of the type $\ln
\mu_{hc}/\mu$. However, in contrast to (\ref{eq:MInt_32S}) the
\SCETII~matrix element (\ref{eq:MInt_313C+S}) is found to contain a
large logarithm $\ln(2\gamma)$ at the low scale $\mu\sim m$ (the
term vanishes in our example for the specific choice $\mu_0=m$ but
is present in general for any variation of the scale with $\mu\sim
m$). In other words, we do not resum \emph{all} large logarithms
with the help of standard RG-techniques in \SCETII.

The remnant logarithms are related to the appearance of endpoint
singularities as can be seen from the artificial decomposition into
individual collinear and soft regions in (\ref{eq:MInt_313C}) and
(\ref{eq:MInt_313S}). We see that they do not belong to the
variation of the factorization scale $\mu$ but rather to the
soft-collinear cross-talk \emph{at fixed virtuality} $\mu^2\sim
m^2$. The low-energy \SCETII~matrix element in
(\ref{eq:MInt_313C+S}) still depends on the high scale due to the
fact that the meson rest frames are related by a large boost
$\gamma=\calO(M/m)$. We conclude that these logarithms \emph{cannot}
be factorized into short-distance coefficient functions and we
therefore call them \emph{non-factorizable}.

\newabs
We emphasize in this context that we do not find contributions from
soft-collinear messenger modes in our set-up. By looking at the
scaling of soft-collinear momentum regions in the loop integrals we
find that these modes always give power-suppressed contributions and
that they are in particular not needed to describe the physics of
endpoint-singularities. The reason for this is that the light quark
mass $m$ provides a physical IR cut-off in our calculation such that
messenger modes with virtualities smaller than $m^2$ cannot
contribute to on-shell amplitudes.

\subsection{Factorization of tree level result}

For the following discussion it will be convenient to introduce a
short-hand notation for the factorization formula
(\ref{eq:BtoPiFFCopy})
\begin{align}
F_i \;&\simeq\; H_i \; \xi \;+\; \phi_B \; \otimes \; T_i \; \otimes
\;\phi_\eta, \label{eq:BtoPiFFshort}
\end{align}
where the symbol $\otimes$ represents the convolution integrals. We
write the perturbative expansion of the form factors in the form
\begin{align}
F_i \;&=\; \sum_{k=1}^{\infty} \left(\frac{\as}{4\pi}\right)^k
F_i^{(k)} \label{eq:NRFFExp}
\end{align}
and similarly for all other quantities in (\ref{eq:BtoPiFFshort}).
Notice that the perturbative expansion of the form factor starts at
$\calO(\as)$ in our set-up according to our discussion in
Section~\ref{sec:NRTree}. Since the perturbative expansion of $H_i$
starts at $\calO(1)$ and the one of $T_i$ at $\calO(\as)$, the form
factors take in LO the form
\begin{align}
F_i^{(1)}  \;&=\; H_i^{(0)}  \; \xi^{(1)}  \;+\; \phi_B^{(0)} \;
\otimes \; T_i^{(1)} \; \otimes \;\phi_\eta^{(0)}
\label{eq:BtoPiFFLO}
\end{align}
with the hard-scattering kernels given by (\ref{eq:BtoPiFFTi}) as
$T_i^{(1)} = C_i^{(0)} \otimes  J^{(1)}$. The jet function $J^{(1)}$
has first been computed by Beneke and Feldmann in \cite{MBTF2000}
using a different definition of the soft-overlap contribution
$\xi(q^2)$. Our convention corresponds to the one of Beneke and Yang
in \cite{BPiJet:BY} from which we read off
\begin{align}
H_+^{(0)} = - H_-^{(0)} = H_T^{(0)} &= \frac{C_+^{(0)}(\tau)}{s-2}
    = - \frac{C_-^{(0)}(\tau)}{s} = \frac{C_T^{(0)}(\tau)}{s} = 1, \no \\
J^{(1)}(\tau;u,\om)&= - \frac{\hat{f}_M f_m}{N_c} \; \frac{\pi^2
C_F}{\gamma m} \; \frac{\delta(\tau-\ubar)}{\om \ubar},
\label{eq:H0C0J1}
\end{align}
where $\hat{f}_M$ ($f_m$) is the decay constant of the heavy (light)
meson defined in HQET (QCD). In LO of the perturbative expansion we
simply have $\hat{f}_M=f_M^\text{NR}$, $f_m=f_m^\text{NR}$.

\newabs
In order to extract the overlap-contribution $\xi^{(1)}$ from
(\ref{eq:BtoPiFFLO}) we have to calculate the distribution
amplitudes of the $B_c$ and $\eta_c$ meson to LO in the perturbative
expansion. As we treat the quarks in the static approximation at
leading power in the NR expansion, we simply find
\begin{align}
\phi_B^{(0)}(\om) = \delta(\om-m), \qquad \phi_\eta^{(0)}(u) =
\delta(u-1/2). \label{eq:NRLOLCDA}
\end{align}
The second term in (\ref{eq:BtoPiFFLO}) can now be calculated giving
\begin{align}
\phi_B^{(0)} \; \otimes \; T_i^{(1)} \; \otimes \;\phi_\eta^{(0)} &=
- \frac{f_M^\text{NR}f_m^\text{NR}}{N_c} \; \frac{2\pi^2 C_F}{\gamma
m^2} \small \left\{ \begin{array}{ccc} s-2& &i=+\\-s
&\text{\;\;if\;\;}&i=-\\s&&i=T
\end{array}\right.
\end{align}
We may now isolate $\xi^{(1)}$ using our explicit LO results from
(\ref{eq:NRFFLO}) together with (\ref{eq:BtoPiFFLO}) and
(\ref{eq:H0C0J1}). We indeed find a universal contribution in this
way which reads
\begin{align}
\xi^{(1)} = \frac{f_M^\text{NR} f_m^\text{NR}}{N_c} \; \frac{\pi^2
C_F}{\gamma m^2} \;\;10s.
\end{align}
We have shown in some detail how to extract the soft-overlap
contribution $\xi(q^2)$ from our explicit perturbative calculation
in Section~\ref{sec:NRPT}. As the conceptually interesting aspects
related to the physics of endpoint singularities and the appearance
of non-factorizable logarithms only enter our calculation at NLO, we
proceed with a similar analysis in the following section.

\subsection{Factorization in NLO}

At NLO of the perturbative expansion the extraction of the
overlap-contribution is much more involved. We now start from
\begin{align}\label{eq:BtoPiFFNLO}
F_i^{(2)}  \;&=\; H_i^{(0)}  \; \xi^{(2)}  \;+\; H_i^{(1)}  \;
\xi^{(1)}   \\
&\hspace{6mm}+\; \phi_B^{(0)} \; \otimes \; T_i^{(2)} \; \otimes
\;\phi_\eta^{(0)}+\; \phi_B^{(1)} \; \otimes \; T_i^{(1)} \; \otimes
\;\phi_\eta^{(0)}+\; \phi_B^{(0)} \; \otimes \; T_i^{(1)} \; \otimes
\;\phi_\eta^{(1)} \no
\end{align}
with $T_i^{(2)} = C_i^{(0)} \otimes  J^{(2)}  + C_i^{(1)} \otimes
J^{(1)}$. Whereas the hard coefficient functions $H_i^{(1)}$ have
already been computed in \cite{MBTF2000}, the calculation of
$C_i^{(1)}$ and $J^{(2)}$ has been performed recently
\cite{BenekeKiyoYang,BPiJet:BH,BPiJet:K,BPiJet:BY}. In order to
extract $\xi^{(2)}$ from (\ref{eq:BtoPiFFNLO}), we still have to
consider the corrections to the distribution amplitudes
$\phi_B^{(1)}$ and $\phi_\eta^{(1)}$ and to perform the subsequent
subtractions. In the following we present our results for the
distribution amplitudes whereas we relegate the extraction of the
overlap-contribution to~\cite{GBTFinprep}.

\subsubsection{Light-cone distribution amplitude of $\eta_c$ meson}

We start with the calculation of the leading twist light-cone
distribution amplitude (LCDA) of the ''light'' $\eta_c$ meson
defined by \cite{LCDA:Pion:Def,MBTF2000}
\begin{align}
\langle \eta_c(q) | \, \bar c(y) \gamma_\mu \gamma_5 c(x) \, | 0
\rangle  \big{|}_{(x-y)^2=0}&\simeq - i f_m \, q_\mu \int_0^1 du \;
e^{i(u q \cdot y+\bar u q \cdot x)} \, \phi_\eta(u;\mu),
\label{eq:NRdefLCDAEta}
\end{align}
where we omitted an appropriate path-ordered exponential of gluon
fields which makes the definition gauge-invariant. Notice that there
is an additional leading twist two-gluon LCDA for flavour singlet
mesons which we will not consider here.

\newabs
The NR bound states are described by parton configurations with
fixed momenta which correspond to the light-cone momentum fraction
$u=1/2$ for the $\eta_c$ meson, cf. (\ref{eq:NRLOLCDA}).
Relativistic effects from collinear gluon exchange lead to
modifications: First, there is a correction from matching \SCETII~to
the NR theory. Second, there is the usual evolution under the change
of the renormalization scale \cite{ER,BL}. In particular, the
support region for the parton momenta is extended to $0<u<1$.

\begin{figure}[t!]
\centerline{\parbox{13cm}{
\centerline{\includegraphics[width=13cm]{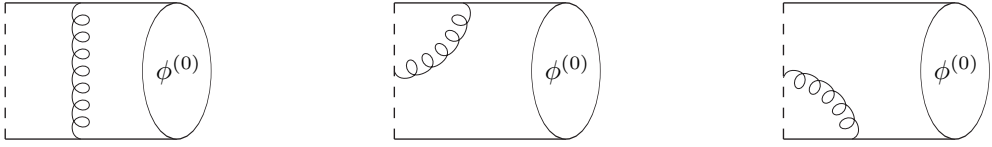}}
\caption{\label{fig:NRNLOLCDA} \small \textit{Relativistic
corrections to the light-cone distribution amplitudes. The dashes
line indicates the Wilson-line.}}}}\vspace{2mm}
\end{figure}

\newabs
The first-order relativistic corrections arise from the diagrams in
Figure~\ref{fig:NRNLOLCDA}. Apart from these 1-loop diagrams, we
have to include the wave-function renormalization of the external
quark lines. We first consider the local limit of the light-cone
matrix element (\ref{eq:NRdefLCDAEta}) which determines the
relativistic corrections to the NR decay constant. In this case, the
diagrams with the gluon attached to the Wilson-line are absent and
we find
\begin{align}
f_m &= f_m^\text{NR} \left[ 1 -6 \, \frac{\as C_F}{4\pi}  +
\calO(\as^2) \right]
\end{align}
which coincides with the result in \cite{BraatenFleming} for equal
quark masses.

\newabs
The NLO corrections to the leading twist distribution amplitude are
found to be
\begin{align}
\phi^{(1)}_\eta(u;\mu) &= 4C_F  \, \ln \frac{\mu^2}{m^2} \left[
\left( 1+\frac{1}{1/2-u} \right) u\,\theta(1/2-u) + (u
\leftrightarrow
\bar u) \right]_+ \no \\
&\;\; -4C_F  \left\{ \left[ \left( 1+\frac{1}{1/2-u} \right) \ln
(1-2u)^2 + \left( 1-\frac{1/2}{(1/2-u)^2} \right)
\right] u \, \theta(1/2-u) \right.\no \\
&\hspace{25mm} + (u \leftrightarrow \bar u) \bigg\}_+
 \label{eq:NRLCDABNLO}
\end{align}
where we have introduced plus-distributions defined by
\begin{align} \label{eq:NRLCDAEtaPlus}
\int_0^1 \! du \; [\ldots]_+ \;f(u) &\equiv \int_0^1 \! du \;
[\ldots] \; \bigg( f(u) - f(1/2) -f'(1/2) \, (u-1/2)\bigg).
\end{align}

\newabs
Notice that our result for the distribution amplitude obeys the
evolution equation
\begin{align}
\frac{d}{d \ln \mu} \; \phi_\eta(u;\mu) =  \frac{\as C_F}{\pi}
\int_0^1 dv \; V(u,v) \; \phi_\eta(v;\mu) + \calO(\as^2),
\end{align}
where $V(u,v)$ denotes the familiar Brodsky-Lepage kernel
\cite{ER,BL}
\begin{align}
V(u,v) = \left[ \left(1 + \frac{1}{v-u} \right) \frac{u}{v}\,
\theta(v-u) + \left(1 + \frac{1}{\bar v-\bar u} \right) \frac{\bar
u}{\bar v} \, \theta(u-v)
 \right]_+
\end{align}
While writing this thesis an independent calculation of the
leading-twist LCDAs of the $\eta_c$ and $J/\psi$ meson appeared
in~\cite{MaSi} where Ma and Si perform a similar analysis
factorizing perturbative effects from NRQCD matrix elements at
leading power in the NR expansion. Our result is not in agreement
with these findings which can already be seen by the fact that the
distribution amplitude in~\cite{MaSi} is not properly normalized to
unity. We will have a closer look at the origin of this discrepancy
in~\cite{GBTFinprep}.

\newabs
The factorization formula (\ref{eq:BtoPiFFNLO}) contains the
distribution amplitude $\phi_\eta^{(1)}$ in form of a convolution
with the jet function $J^{(1)}$ from (\ref{eq:H0C0J1}). We thus
require the knowledge of the $1/\ubar$ moment which is found to be
\begin{align}
\int_0^1 \! du \;\; \frac{\phi_\eta^{(1)}(u;\mu)}{\ubar} &= C_F
\left[ 2(3-2\ln2) \ln \frac{\mu^2}{m^2} +4\ln2-\frac23 \pi^2+12
\right].
\end{align}

\subsubsection{Light-cone distribution amplitude of $B_c$ meson}

The calculation of the LCDA for the $B_c$ meson goes along the same
lines as for the $\eta_c$ meson. However, important differences
arise because the heavy $b$-quark is to be treated in HQET which
modifies the divergence structure of the loop integrals. As a
consequence, the evolution equations for the LCDA of heavy mesons
\cite{LCDA:Bmeson:Evol:I,LCDA:Bmeson:Evol:II} differ from that of
light mesons \cite{ER,BL}.

\newabs
We define the two-particle LCDAs of the $B_c$ meson following
\cite{LCDA:Bmeson:Def,MBTF2000}
\begin{align}
\langle 0 | \bar{c}_\beta(z) b_\alpha(0) | B_c(p) \rangle
\big{|}_{z^2=0} &= - \frac{i \hat{f}_M(\mu) M}{4} \left[ \frac{1
+\slashb{w}}{2} \left\{ 2 \tilde{\phi}_+^B(t) +
\frac{\tilde{\phi}_-^B(t)-\tilde{\phi}_+^B(t)}{t} \slashb{z}
\right\} \gamma_5 \right]_{\alpha \beta} \label{eq:NRdefLCDABc}
\end{align}
with $p\simeq M w$, $t\equiv w\cdot z$ and $\hat{f}_M$ the decay
constant in HQET (we again omitted an appropriate Wilson-line in the
definition). The Fourier-transformed expressions, which usually
appear in factorization formulas, are given through
\begin{align}
\tilde{\phi}_\pm^B(t) = \int_0^\infty \! d\om \;\; e^{-i\om t}
\phi_\pm^B(\om).
\end{align}
In the following we focus on the distribution amplitude
$\phi_B(\om)\equiv\phi_+^B(\om)$ which enters the factorization
formula (\ref{eq:BtoPiFFCopy}). In the strict NR limit it simply
corresponds to a delta-function, cf.~(\ref{eq:NRLOLCDA}). Similar to
what we have seen in the case of the $\eta_c$ meson, we now have to
take into account relativistic effects from soft gluon exchange
according to analogous diagrams to those in
Figure~\ref{fig:NRNLOLCDA} which now extend the support region of
the distribution amplitude to $0<\om<\infty$.

\newabs
In the local limit we derive the corrections to the decay constant.
We find
\begin{align}
\hat{f}_M(\mu) &= f_M^\text{NR} \left[ 1 + \, \frac{\as C_F}{4\pi}
\left( 3\ln\frac{\mu}{m}-4 \right)+ \calO(\as^2) \right].
\end{align}
Notice that the decay constant of a heavy meson in HQET is
scale-dependent.

\newabs
The NLO corrections to the distribution amplitude read
\begin{align} \label{eq:NRLCDAB}
\phi^{(1)}_B(\om;\mu) &= 2C_F  \ln \frac{\mu^2}{(\om-m)^2} \left\{
\left[ \frac{\om
\theta(m-\om)-m\theta(\om-m)\theta(\Lambda-\om)}{m(m-\om)}
\right]_+ + \frac{\theta(\om-\Lambda)}{w-m}\right\}\no\\
&\!\!\!-C_F  \; \delta(\om-m) \left( \frac12 \ln^2
\frac{\mu^2}{(\Lambda-m)^2} +  \ln \frac{\mu^2}{m^2} +
\frac{\pi^2}{12} \right) \no \\
&\!\!\!+2C_F  \left\{ \!  \left[
\frac{(m+\om)[\om\theta(m-\om)+m\theta(\om-m)\theta(\Lambda-\om)]}{m(m-\om)^2}
\right]_+ \!\!\!+ \frac{(m+\om)\theta(\om-\Lambda)}{(w-m)^2}\right.\no\\
&\!\!\!\left.-  \delta(\om-m) \left( \ln \frac{m}{\Lambda-m}+
\frac{2m}{\Lambda-m}\right)- m \, \delta'(\om-m) \left( 2\ln
\frac{\Lambda-m}{m}+\frac{\Lambda}{m}+\frac32\right) \right\}
\end{align}
with an analogous definition of plus-distributions as
in~(\ref{eq:NRLCDAEtaPlus}). One remark is in order concerning the
dependence of our results on an ad-hoc parameter $\Lambda>m$. In the
computation of the LCDA we encounter IR singularities for $\om=m$
(similar to the $\eta_c$ case for $u=1/2$) which we regularize with
the help of adequate plus-distributions. We rewrite for instance
\begin{align} \label{eq:NRLCDABplus}
\frac{\theta(\om-m)}{(w-m)^{1+2\eps}} \; \to \;
\left[\frac{\theta(\om-m)}{w-m}\right]_+
\end{align}
This is to be understood in the sense of a distribution which is
convoluted with a kernel $f(\om)$ that is assumed to be
non-vanishing for $\om=m$ and to scale as $1/\om$ for large $\om$
(as usual in application of QCD Factorization). The point to notice
is that the (IR regular) plus-distribution gives rise to UV
singularities in the convolutions as $f(\om)\sim1/\om$ is replaced
by $f(\om)-f(m)\sim1$ for large $\om$. In order to avoid these
artificial UV divergences we introduce an ad-hoc cutoff $\Lambda>m$
in the plus-distributions such that (\ref{eq:NRLCDABplus}) is
replaced by
\begin{align}
\frac{\theta(\om-m)}{(w-m)^{1+2\eps}} \; \to \;
\left[\frac{\theta(\om-m)\theta(\Lambda-\om)}{w-m}\right]_+ \!\!\!+
\frac{\theta(\om-\Lambda)}{w-m}-
\frac{(\Lambda-m)^{-2\eps}}{2\eps_\IR} \, \delta(\om-m) +
\calO(\eps).
\end{align}
This expression is thus free of UV singularities whereas the IR
divergence has been isolated in the local term. Notice that the
dependence on the cutoff parameter drops out in the convolution with
an arbitrary kernel $f(\om)$. This can also be seen below in our
expression for the $1/\om$ moment of the LCDA in
(\ref{eq:NRLCDABmom}).

\newabs
The distribution amplitude in (\ref{eq:NRLCDAB}) obeys the evolution
equation
\begin{align} \label{eq:NRLCDABEvol}
\frac{d}{d \ln \mu} \; \phi_B(\om;\mu) =  \frac{\as C_F}{\pi}
\int_0^\infty d\om' \; \gamma_+(\om,\om';\mu) \; \phi_B(\om';\mu) +
\calO(\as^2),
\end{align}
where the anomalous dimension $\gamma_+(\om,\om';\mu)$ is given by
\cite{LCDA:Bmeson:Evol:I,LCDA:Bmeson:Evol:II}
\begin{align}
\gamma_+(\om,\om';\mu) =\left( - \ln \frac{\mu}{\om} + \frac12
\right) \delta(\om-\om') + \om \left[
\frac{\theta(\om'-\om)}{\om'(\om'-\om)} +
\frac{\theta(\om-\om')}{\om(\om-\om')} \right]_+
\end{align}
This can be verified most easily by integrating
(\ref{eq:NRLCDABEvol}) over a test function $f(\om)$ such that the
dependence of the LCDA on the cutoff parameter $\Lambda$ drops out.

\newabs
In contrast to the $\eta_c$ case, the normalization of the $B_c$
distribution amplitude is ill-defined. Imposing a hard cutoff
$\Lambda_\UV\gg m$ and expanding to first order in $m/\Lambda_\UV$,
we derive
\begin{align}
\int_0^{\Lambda_\UV} \!\!\!\!\!\!d\om \; \phi_B(\om;\mu) &\simeq 1 -
\frac{\as C_F}{4\pi} \left[ \frac12 \ln^2
\frac{\mu^2}{\Lambda_\UV^2}+\ln \frac{\mu^2}{\Lambda_\UV^2} +
\frac{\pi^2}{12} \right] + \calO(\as^2)
\end{align}
and similarly for the first moment
\begin{align}
\int_0^{\Lambda_\UV} \!\!\!\!\!\!d\om \; \om\, \phi_B(\om;\mu)
&\simeq \frac{\as C_F}{4\pi} \left[ 2\ln \frac{\mu^2}{\Lambda_\UV^2}
+6 \right] \Lambda_\UV + \calO(\as^2).
\end{align}
The last two expressions provide model-independent properties of the
distribution amplitude which have been studied within the operator
product expansion in~\cite{LCDA:Bmesom:OPE}. Our results are in
agreement with these general findings. Notice, however, that our
subleading terms in $m/\Lambda_\UV$ cannot be compared
with~\cite{LCDA:Bmesom:OPE} as we deal with massive light quarks
here.

\newabs
We finally quote our result for the $1/\om$ moment which appears in
the factorization formula (\ref{eq:BtoPiFFNLO}) in form of the
convolution with the jet function $J^{(1)}$ from (\ref{eq:H0C0J1})
\begin{align} \label{eq:NRLCDABmom}
\int_0^\infty d\om \;\; \frac{\phi_B^{(1)}(\om;\mu)}{\om} =
-\frac{C_F}{m} \left[ \frac12 \ln^2 \frac{\mu^2}{m^2} - \ln
\frac{\mu^2}{m^2} +\frac{3\pi^2}{4} -2 \right].
\end{align}
Notice the presence of a double Sudakov logarithm as in
(\ref{eq:MInt_32S}).

\part{Conclusion}

\pagestyle{fancyplain} \rhead[\uppercase{Conclusion}]{\fancyplain{}
\thepage} \lhead[\thepage]{}

\chapter*{Conclusion}
 
The dedicated study of charmless $B$ decays has become one of the
most active and promising fields in particle physics. It aims at a
precision determination of the flavour parameters in the Standard
Model and may help to reveal the nature of CP violation. On the
experimental side, the $B$ factories BaBar and Belle are
continuously accumulating larger data samples and provide us with
measurements of unprecedented precision. On the theory side, the
main challenge consists in the quantitative control of the
complicated hadronic dynamics.

QCD Factorization represents a model-independent framework to
compute hadronic matrix elements from first principles. Based on a
power expansion in $\LQCD/m_b$, it allows for a clear separation of
short- and long-distance effects in certain classes of $B$ decays.
Whereas the development of Soft-Collinear Effective Theory provided
the means to put factorization formulas onto a firmer basis, the
implementation of higher-order QCD corrections has started only
recently.

\newabs
In this thesis we have mainly addressed the most challenging class
of charmless $B$ decays: exclusive nonleptonic $B$ decays with its
most prominent example $B\to\pi\pi$. This wide class of decays is of
primary importance as most of the observables at the $B$ factories
are related to it. Our calculation considers one class of the NNLO
corrections to hadronic two-body decays within QCD Factorization.
Whereas the (1-loop) spectator scattering contributions have been
considered recently by various groups~\citer{BenekeJager,Volker}, we
have presented the first calculation of the 2-loop vertex
corrections which constituted the missing piece for a full NNLO
analysis of the topological tree amplitudes in QCD Factorization.

The knowledge of the NNLO corrections is particularly important with
respect to strong interaction phases and hence direct CP
asymmetries. As strong phases arise first at $\calO(\as)$ in QCD
Factorization, they were known so far to LO accuracy only. As in any
perturbative calculation the considered corrections are thus
necessary to eliminate large scale ambiguities of the LO result.
They may even drastically change the pattern of direct CP
asymmetries in QCD Factorization.

We indeed found that the considered corrections can exceed the
formally leading contributions whenever the latter are accompanied
by small numerical coefficients. This is the case for the imaginary
part of the colour-allowed tree amplitude which receives its
dominant contribution from the considered 2-loop vertex corrections.
The respective contribution to the colour-suppressed tree amplitude
also turned out to be sizeable. However, in absolute terms our
corrections are small. In other words, we did not encounter any
source of sizeable strong phases from the perturbative calculation.
We further remark that we have observed an accidental cancellation
between the individual NNLO contributions which resulted in a
moderate correction to the NLO result for the imaginary part of the
tree amplitudes.

The calculation of the real part of the topological tree amplitudes
is more involved. We have presented a preliminary result for the
colour-allowed tree amplitude which is based on certain technical
simplifications. More importantly, we have already tackled the most
challenging part of the calculation which consisted in the
evaluation of a large set of 2-loop integrals. We could further
verify that factorization holds at the considered order in
perturbation theory which is a non-trivial statement as the formal
all-order proof is still missing. We emphasize that this is the
result of a highly complicated subtraction procedure which can also
be considered as a very powerful cross-check of our calculation.

From the technical point of view we have presented a calculation of
2-loop hadronic matrix elements. Even in the wider sense of
perturbative corrections to weak decays, there exist only very few
calculations of this type. In the context of inclusive decays,
2-loop matrix elements have been considered for $B\to
X_s\ell^+\ell^-$~\cite{B->X_sl+l-:I,B->X_sl+l-:II} and 3-loop matrix
elements for $B\to X_s\gamma$~\cite{B->X_sgamma}. Concerning the
exclusive modes 2-loop matrix elements have been calculated for
$B\to K^* \gamma$ or $B\to \rho \ell^+\ell^-$~\cite{Seidel}.
However, our analysis represents the first calculation of 2-loop
hadronic matrix elements for the more complicated nonleptonic decays
which are mediated by a $b\to uud$ transition. For this class of
decays 2-loop hadronic matrix elements have never been calculated so
far, neither in the inclusive nor in the exclusive channel.

\newabs
In our final analysis we examined the formal factorization
properties of heavy-to-light form factors. The development of
Soft-Collinear Effective Theory has led to a deeper understanding of
the QCD dynamics in heavy-to-light transitions and the $B\to\pi$
form factor (at large recoil) has emerged as the central object in
these studies. Rather than stating a complete separation of short-
and long-distance effects, the factorization formula contains a
classification into (non-calculable) symmetry conserving
contributions, the so-called soft-overlap contribution, and
perturbative symmetry breaking effects. Remnant short-distance
effects in the soft-overlap cannot be factorized by standard means
as this would lead to endpoint-divergent convolution integrals. The
question arises if one can understand this non-factorization of soft
and collinear effects from the viewpoint of the effective theory.

In order to address this question we have considered a simplified
scenario. We investigated heavy-to-light form factors between
non-relativistic bound states which can be addressed in perturbation
theory. We have performed a NLO analysis of these form factors and
looked for an interpretation in terms of the factorization formula.
We showed that the soft-overlap contribution can be calculated in
our set-up. We have addressed the question of the origin of formally
large (Sudakov) logarithms and argued that one class of logarithms
related to endpoint-singularities cannot be resummed with standard
RG techniques. We in particular did not need to introduce
soft-collinear messenger modes to describe the physics of the
soft-overlap contribution. As a byproduct of our analysis, we have
calculated leading-twist light-cone distribution amplitudes for
non-relativistic bound states which can be applied for the
description of $B_c$ and $\eta_c$ mesons.

\newabs
To summarize, the development of QCD Factorization and
Soft-Collinear Effective Theory has substantially improved our
understanding of QCD in $B$ meson decays. To date, the QCD
Factorization predictions are in good overall agreement with the
experimental measurements. In order to perform a precision study of
$B$ decays, we have to reduce theoretical and experimental
uncertainties as much as possible. This thesis has contributed to
the former by calculating higher-order QCD corrections to exclusive
charmless $B$ decays. Apart from the perturbative corrections,
further improvements on the theory side concern the control of power
corrections and precise determinations of the hadronic input
parameters. We think that it will be exiting to confront these
improved theoretical predictions with updated experimental data in
the near future.

\part{Appendix}

\pagestyle{fancyplain}
\renewcommand{\chaptermark}[1]{\markboth{\thechapter.~\uppercase{#1}}{}}
\renewcommand{\sectionmark}[1]{\markright{\thesection.~\uppercase{#1}}}
\rhead[\leftmark]{\fancyplain{} \thepage}
\lhead[\thepage]{\rightmark}

\begin{appendix}

\chapter{Master Integrals}

In the following we present the analytical results for all MIs that
appeared in our calculations from Chapter~\ref{ch:ImPart},
\ref{ch:RePart} and~\ref{ch:NRModel}. The MIs are normalized
according to
\begin{align}
[dk] \equiv  \frac{\Gamma(1-\eps)}{i\pi^{d/2}}\; d^d k.
\end{align}
We give the results up to the order in $\eps$ that was required in
our calculations. For some simpler MIs, we obtain the results in a
closed form in $\eps$.

\section{Hadronic two-body decays I}
\label{app:MIsIm}

In the calculation from Chapter~\ref{ch:ImPart} we found 14 MIs
which are shown in Figure~\ref{fig:BtoPiPi:MIs}. The corresponding
particle propagators $\calP_i$ can be found in
(\ref{eq:BtoPiPi:Props}). Although the calculation in
Chapter~\ref{ch:ImPart} requires the knowledge of the imaginary part
only, we give the full results of the MIs here in view of our
extension in Chapter~\ref{ch:RePart}.

\subsubsection{1--loop integrals}

Some 2-loop integrals can be written as products of the following
1-loop integrals

\vspace{3mm}
\begin{align}
\hspace{8mm} \SetScale{0.13}
  \begin{picture}(23,23) (33,-5)
    \SetWidth{2.5}
    \SetColor{Black}
    \CArc(250,-22)(105,11,371)\CArc(250,-22)(97,11,371)
  \end{picture}
    \hspace{8mm}=\quad
    \int [dk] \; \frac{1}{\calP_{13}}
\hspace{10mm}=\quad
    - (m^2)´^{1-\eps} \; \Gamma(1-\eps) \Gamma(-1+\eps)
\end{align}

\vspace{3mm}
\begin{align}
\SetScale{0.13}
  \begin{picture}(40,23) (44,-10)
    \SetWidth{1.25}
    \SetColor{Black}
    \Line(50,-50)(150,-50)
    \SetWidth{2.0}
    \Line(350,-50)(450,-50)
    \Line(50,-50)(150,-50)
    \DashCArc(250,-50)(100.5,174,534){15}
  \end{picture}
    =\quad
    \int [dk] \; \frac{1}{\calP_{1}\calP_{3}}
\hspace{5.5mm}=\quad
    (u m^2)´^{-\eps} \; e^{i \pi \eps} \; \frac{\Gamma(\eps)\Gamma(1-\eps)^3}{\Gamma(2-2\eps)}
\end{align}

\newpage
\begin{align}
&\SetScale{0.13}
  \begin{picture}(40,23) (44,-10)
    \SetWidth{2.0}
    \SetColor{Black}
    \Line(50,-45)(150,-45)
    \CArc(250,-44.87)(105.13,-177.06,-2.94)\CArc(250,-44.87)(95.13,-177.06,-2.94)
    \DashCArc(250,-45)(100,-0,180){15}
    \Line(350,-45)(450,-45)
  \end{picture}
=\quad
    \int [dk] \; \frac{1}{\calP_{5}\calP_{13}}
\quad \equiv \quad
    (m^2)^{-\eps} \; \left\{ \sum_{i=-1}^{3} c_i^{(22)} \, \eps^i + \calO(\eps^4) \right\}
\end{align}

\vspace{4mm} with
\begin{align}
c_{-1}^{(22)}   &=
    1, \no \\
c_{0}^{(22)}    &=
    2+ \frac{\ubar}{u}  \ln \ubar, \no \\
c_{1}^{(22)}    &=
    4+\frac{\pi^2}{6} - \frac{\ubar}{u}
    \bigg[ \Lib (u) + \ln^2 \ubar -2 \ln \ubar \bigg], \hspace{6cm}
    \no \\
c_{2}^{(22)}    &=
    8+\frac{\pi^2}{3} - \frac{\ubar}{u}
    \bigg[ \Lic(u) -2\Sab(u) - 2 \ln \ubar \,\Lib(u)  -\frac23 \ln^3 \ubar + 2\Lib(u) + 2\ln^2 \ubar  \no \\
    &\; \quad  -\Big(4+\frac{\pi^2}{6} \Big) \ln \ubar \bigg], \no \\
c_{3}^{(22)}    &=
    16 +\frac{2\pi^2}{3} + \frac{7\pi^4}{360}  - \frac{\ubar}{u}
    \bigg[ \Lid(u)  -2 \Sbb(u) + 4 \Sac(u) -2 \Big( \Lic(u) -2 \Sab(u) \Big) \ln \ubar \no \\
    &\; \quad + 2\ln^2 \ubar \,\Lib(u) + \frac13  \ln^4 \ubar  +2 \Lic(u) -4 \Sab(u) -4 \ln \ubar \,\Lib(u) -\frac43 \ln^3 \ubar \no \\
    &\; \quad + \Big( 4+\frac{\pi^2}{6} \Big) \Big( \Lib(u) + \ln^2 \ubar -2 \ln \ubar \Big) \bigg].
\end{align}

\vspace{2mm}
\subsubsection{3--topologies}

\vspace{-2mm}
\begin{align}
  \SetScale{0.13}
  \begin{picture}(40,23) (24,-10)
    \SetWidth{2.0}
    \SetColor{Black}
    \DashCArc(250,-49)(100.72,173,533){15}
    \Line(50,-49)(150,-49)
    \Line(350,-49)(450,-49)
    \DashLine(150,-49)(350,-49){15}
  \end{picture}
=\quad \int [dk] [dl] \; \frac{1}{\calP_{1}\calP_{4}\calP_{9}}
\quad=\quad (u m^2)´^{1-2\eps} \; e^{2i \pi \eps} \;
\frac{\Gamma(-1+2\eps)\Gamma(1-\eps)^5}{\Gamma(3-3\eps)}
\end{align}

\subsubsection{4--topologies}

\vspace{1mm}
\begin{align}
  \SetScale{0.13}
  \begin{picture}(40,23) (24,3)
    \SetWidth{2.0}
    \SetColor{Black}
    \DashCArc(250,50)(100.72,173,533){15}
    \DashLine(50,50)(450,50){15}
    \Line(250,150)(250,250)
  \end{picture}
&=\quad
\int [dk] [dl] \; \frac{1}{\calP_{1}\calP_{3}\calP_{8}\calP_{9}} \hspace{8.5cm} \no \\
&=\quad
 (u m^2)´^{-2\eps} \; e^{2i \pi \eps} \;
\frac{\Gamma(2\eps)\Gamma(\eps)\Gamma(1-2\eps)^2\Gamma(1-\eps)^4}{\Gamma(2-2\eps)\Gamma(2-3\eps)}
\end{align}

\begin{align}
&  \SetScale{0.13}
  \begin{picture}(40,23) (24,3)
    \SetWidth{2.0}
    \SetColor{Black}
    \DashCArc(250,50)(100.72,173,533){15}
    \DashLine(50,50)(350,50){15}
    \Line(250,150)(250,250)
    \Line(350,55)(450,55)\Line(350,45)(450,45)
  \end{picture}
=\quad
        \int [dk] [dl] \; \frac{1}{\calP_{1}\calP_{3}\calP_{6}\calP_{10}}
\;\; \equiv \;\;
    (m^2)^{-2\eps} \; e^{2i \pi \eps} \; \left\{ \sum_{i=-2}^{2} c_i^{(42)} \, \eps^i + \calO(\eps^3) \right\}
\end{align}

\newpage
with
\begin{align}
c_{-2}^{(42)}   \, &=\;
    \frac12, \no \\
c_{-1}^{(42)}    \, &=\;
    \frac52 - \ln u, \no \\
c_{0}^{(42)}    \, &=\;
    \frac{19}{2} - \frac{\pi^2}{6} + \Lib(u) + \frac12 \Big( \ln u + \ln \ubar \Big)^2 -\frac12 \ln^2\ubar-5 \ln u, \no \\
c_{1}^{(42)}    \, &=\;
    \frac{65}{2} -\frac{5\pi^2}{6} - 3 \zeta_3 - \Lic(u) - \Sab(u) + \Big(5 - \ln \ubar\Big) \Lib(u)  - \frac16 \Big( \ln u + \ln \ubar \Big)^3  \no \\
    &\; \quad +\frac16 \ln^3\ubar+ \frac52 \Big( \ln u + \ln \ubar \Big)^2 - \frac52 \ln^2\ubar + \frac{\pi^2}{6}  \Big(\ln u+  \ln \ubar\Big) - 19 \ln u , \no \\
c_{2}^{(42)}    \, &=\;
    \frac{211}{2} -\frac{19\pi^2}{6} - \frac{13\pi^4}{120} - 15 \zeta_3 + \Lid(u) + \Sbb(u) + \Sac(u) + \frac{1}{24} \Big(\ln u + \ln \ubar\Big)^4   \no \\
    &\; \quad  -\frac{1}{24}\ln^4\ubar- \frac56 \Big(\ln u + \ln \ubar\Big)^3 + \frac56 \ln^3\ubar- \Big(5 - \ln \ubar\Big) \Big( \Lic(u) + \Sab(u) \Big)     \no \\
    &\; \quad  - \Big( 5- \frac12 \ln \ubar\Big) \ln \ubar\,\Lib(u) + 19 \Big( \Lib(u) + \frac12 \ln^2 u+ \ln u \ln \ubar \Big) - 2 \zeta_3  \ln \ubar\no \\
    &\; \quad  + \Big( 8\zeta_3  -65 \Big) \ln u + \frac{5\pi^2}{6} \Big(\ln u+  \ln \ubar\Big) - \frac{\pi^2}{12} \Big( \ln u + \ln \ubar\Big)^2.
\end{align}

\vspace{6mm}
\begin{align}
&\SetScale{0.13}
  \begin{picture}(40,23) (24,3)
    \SetWidth{2.0}
    \SetColor{Black}
    \DashLine(50,49)(350,49){15}
    \Line(245,149)(245,249)\Line(255,149)(255,249)
    \Line(350,49)(450,49)
    \DashCArc(249.58,49.42)(99.58,-179.76,89.76){15}
    \CArc(252.98,46.29)(108.01,94.44,178.5)\CArc(252.98,46.29)(98.01,94.44,178.5)
  \end{picture}\hspace{0mm}
=\quad
        \int [dk] [dl] \; \frac{1}{\calP_{4}\calP_{7}\calP_{13}\calP_{19}}
\quad \equiv \quad
    (m^2)^{-2\eps} \; \left\{ \sum_{i=-2}^{2} c_i^{(43)} \, \eps^i + \calO(\eps^3) \right\}
\end{align}

\vspace{2mm} with
\begin{align}
c_{-2}^{(43)}   \, &=\;
    \frac12, \no \\
c_{-1}^{(43)}    \, &=\;
    \frac52, \no \\
c_{0}^{(43)}    \, &=\;
    \frac{19}{2} + \frac{\pi^2}{3} + \frac{u}{2\ubar} \ln^2 u - i \pi \, \frac{u}{\ubar} \ln
    u,\no \\
c_{1}^{(43)}    \, &=\;
    \frac{65}{2} + \frac{5\pi^2}{3} - \zeta_3 - \frac{u}{\ubar} \bigg[ 2 \Lic(u)- 2 \Lib(u) \ln u  + \frac56 \ln^3 u  - \frac52 \ln^2 u - \frac{4\pi^2}{3} \ln u \no \\
    &\; \quad  - \ln^2 u \ln \ubar -2 \zeta_3 \bigg] - i \pi  \, \frac{u}{\ubar} \bigg[ 2 \Lib (u) + 2 \ln u \ln \ubar - \frac52 \ln^2 u + 5 \ln u - \frac{\pi^2}{3}
    \bigg],\no \\
c_{2}^{(43)}    \, &=\;
    \frac{211}{2} + \frac{19\pi^2}{3} + \frac{5\pi^4}{36} - 5 \zeta_3 + \frac{u}{\ubar} \bigg[ 8 \Lid(u) - 4 \Sbb(u) + 4 \Sab(u) \ln u + \frac{19}{24} \ln^4 u  \no
\end{align}
\begin{align}
    &\; \quad  - 2 \Big(5 + \ln u+ 2 \ln \ubar \Big) \Lic (u)  -\frac53 \ln^3 u\ln \ubar + \ln^2 u \ln^2\ubar + 5 \ln^2 u \ln \ubar   \no \\
    &\; \quad  - \frac{25}{6} \ln^3 u + \frac{19}{2} \ln^2 u  + 2 \Big( \zeta_3 + \ln u \,\Lib(u) \Big) \Big(5 - \ln u+ 2 \ln \ubar \Big) - \frac{47\pi^4}{90}  \no \\
    &\; \quad  + \frac{\pi^2}{3} \Big(  8 \Lib(u) -9 \ln^2 u + 8 \ln u \ln \ubar  + 20 \ln u \Big)  \bigg] + i \pi  \, \frac{u}{\ubar} \bigg[ 2 \Lic (u) -4 \Sab (u)   \no \\
    &\; \quad  +  4 \ln u \,\Lib(u) -  \Big( 5 + 2 \ln \ubar \Big) \Big( 2\Lib(u) - \frac{\pi^2}{3} \Big) - \frac{19}{6} \ln^3 u + 5 \ln^2 u \ln \ubar   \no \\
    &\; \quad  -2 \ln u \ln^2\ubar + \frac{25}{2} \ln^2 u -10 \ln u \ln \ubar - \Big( 19 + \frac{\pi^2}{3} \Big) \ln u + 2 \zeta_3 \bigg].
\end{align}
\vspace{4mm}
\begin{align}
&\SetScale{0.13}
  \begin{picture}(40,23) (24,3)
    \SetWidth{2.0}
    \SetColor{Black}
    \Line(250,155)(250,255)
    \DashLine(150,55)(450,55){15}
    \Line(50,60)(150,60)\Line(50,50)(150,50)
    \CArc(250,55.13)(105.13,-177.06,-2.94)\CArc(250,55.13)(95.13,-177.06,-2.94)
    \DashCArc(250,55)(100,-0,180){15}
  \end{picture}
=\quad
        \int [dk] [dl] \; \frac{1}{\calP_{2}\calP_{4}\calP_{13}\calP_{19}}
\quad \equiv \quad
    (m^2)^{-2\eps} \; \left\{ \sum_{i=-2}^{1} c_i^{(44)} \, \eps^i + \calO(\eps^2) \right\}
\end{align}
\vspace{2mm} with
\begin{align}
c_{-2}^{(44)}   \, &=\;
    \frac12, \no \\
c_{-1}^{(44)}    \, &=\;
    \frac52 - \ln u + i \pi, \no \\
c_{0}^{(44)}    \, &=\;
    \frac{19}{2} - \frac{5\pi^2}{6} + \frac{1}{\ubar} \bigg[ \Lic(u) - \ln u \,\Lib(u) - \frac12 \ln^2 u \ln \ubar +\frac{1-2u}{2} \ln^2 u  - \zeta_3 \no \\
    &\; \quad -\Big(5\ubar-\frac{\pi^2}{6} \Big) \ln u \bigg] + i \pi \bigg[ 5 + \frac{1}{\ubar} \Big( \Lib(u) + \ln u \ln \ubar -(1-2u) \ln u -\frac{\pi^2}{6} \Big) \bigg], \no \\
c_{1}^{(44)}    \, &=\;
    \frac{65}{2} -\frac{25\pi^2}{6} - 4 \zeta_3 + \frac{1}{\ubar} \bigg[ \frac{59\pi^4}{360}+ 2 \Lid(u) +3 \Sbb(u) -3 \ln u \Big( \Lic(u)+ \Sab(u) \Big) \no \\
    &\; \quad + \Big( 3\ln \ubar +2(1+u) \Big) \Big( \Lic(u) - \ln u \,\Lib(u) \Big) +2 \ln^2 u \,\Lib(u) + \frac56 \ln^3 u \ln \ubar \no \\
    &\; \quad -\frac34 \ln^2 u \ln^2\ubar - (1+u) \ln^2 u \ln \ubar -\frac{1-6u}{6} \ln^3 u - \frac{7\pi^2}{6} \Lib(u)  -19 \ubar \ln u \no \\
    &\; \quad + \frac{5(1-2u)}{2} \ln^2 u - \Big( \frac12 \ln^2 u + 7 \ln u \ln \ubar -2(2-5u) \ln u\Big) \frac{\pi^2}{6} + \Big( 6 \ln u \no \\
    &\; \quad - 3\ln \ubar -4 \Big) \zeta_3 \bigg] + i \pi \bigg[ 19 + \frac{\pi^2}{3} + \frac{1}{\ubar} \bigg( 3 \Lic(u) + 3 \Sab(u) - 6 \zeta_3 -\frac52 \ln^2 u \ln \ubar \no \\
    &\; \quad - \Big( 4 \ln u - 3 \ln \ubar \Big) \Lib(u) + \frac32 \ln u \ln^2\ubar + \frac{1-6u}{2} \ln^2 u -5(1-2u)\ln u \no \\
    &\; \quad + 2(1+u) \Big( \Lib(u) + \ln u \ln \ubar \Big) + \Big( \ln u -3 \ln \ubar -4 \Big) \frac{\pi^2}{6} \bigg) \bigg].
\end{align}

\newpage
\begin{align}
&\SetScale{0.13}
  \begin{picture}(40,23) (24,3)
    \SetWidth{2.0}
    \SetColor{Black}
    \Line(250,155)(250,255)
    \DashLine(150,55)(450,55){15}
    \Line(50,60)(150,60)\Line(50,50)(150,50)
    \CArc(250,55.13)(105.13,-177.06,-2.94)\CArc(250,55.13)(95.13,-177.06,-2.94)
    \DashCArc(250,55)(100,-0,180){15}
    \SetWidth{0.5}
    \Vertex(180,125){9.9}
  \end{picture}
=\quad
        \int [dk] [dl] \; \frac{1}{\calP_{2}^2\calP_{4}\calP_{13}\calP_{19}}
\quad \equiv \quad
    (m^2)^{-1-2\eps} \; \left\{ \sum_{i=-2}^{2} c_i^{(45)} \, \eps^i + \calO(\eps^3) \right\}
\end{align}
\vspace{3mm} with
\begin{align}
c_{-2}^{(45)}   \, &=\;
    -\frac{1}{u}, \no \\
c_{-1}^{(45)}    \, &=\;
    -\frac{1}{u} \Big( 2 - \ln u + i \pi \Big), \no \\
c_{0}^{(45)}    \, &=\;
    \frac{1}{u} \bigg[ \frac{\pi^2}{6} -4 - \frac{1-2u}{2\ubar} \ln^2 u + 2 \ln u
    - i \pi \Big( 2 - \frac{1-2u}{\ubar} \ln u \Big) \bigg], \no \\
c_{1}^{(45)}    \, &=\;
    \frac{1}{u} \bigg\{ \frac{\pi^2}{3} -8 - 5\zeta_3 + \frac{1}{\ubar} \bigg[ (1-2u) \Big( \Lic (u) - \ln u \,\Lib (u) -\frac12 \ln^2 u \ln \ubar - \ln^2 u \Big) \no \\
    &\; \quad + \frac{1-6u}{6} \ln^3 u + \Big( 4 \ubar- \frac{(1-10u)\pi^2}{6} \Big) \ln u + \zeta_3 \bigg] - i \pi \bigg[ 4 + \frac{1-6u}{2\ubar} \ln^2 u \no \\
    &\; \quad -\frac{1-2u}{\ubar} \Big( \Lib (u) + \ln u \ln \ubar + 2 \ln u - \frac{\pi^2}{6} \Big) \bigg] \bigg\}     , \no \\
c_{2}^{(45)}    \, &=\;
    \frac{1}{u} \bigg\{ \frac{2\pi^2}{3} + \frac{17\pi^4}{60} -16 -10 \zeta_3 + \frac{1}{\ubar} \bigg[2(1+7u) \Lid (u) - \frac{1-22u}{24} \ln^4 u   \no \\
    &\; \quad -3(1+2u) \ln u \,\Lic(u) + (2-u) \ln^2 u \,\Lib(u)  + \frac{(1-36u)\pi^2}{12} \ln^2 u  + (3-4u)   \no \\
    &\; \quad \, \Big( \Sbb(u) - \ln u \,\Sab (u) + \ln \ubar ( \Lic(u) - \ln u  \,\Lib (u) )  -\frac14 \ln^2 u \ln^2 \ubar - \zeta_3 \ln \ubar \Big) \no \\
    &\; \quad +2(1-2u) \Big( \Lic(u) - \ln u \,\Lib(u) + \frac{5}{12} \ln^3 u \ln \ubar -\frac12 \ln^2 u \ln \ubar - \ln^2 u \Big) \no \\
    &\; \quad - \frac{(7-16u)\pi^2}{6} \Big( \Lib(u) + \ln u \ln \ubar \Big) + \Big( 8 \ubar - \frac{(1-10u)\pi^2}{3} + 4 (1+u) \zeta_3\Big) \ln u \no \\
    &\; \quad + \frac{1-6u}{3} \ln^3 u + 2 \zeta_3 - \frac{17\pi^4}{40} \bigg] + i \pi \bigg[  \frac{1}{\ubar} \bigg( 3(1+2u) \Lic (u) -2(2-u) \ln u \,\Lib(u) \no \\
    &\; \quad + (3-4u) \Big( \Sab(u)+\ln \ubar \,\Lib(u) + \frac12 \ln u \ln^2\ubar - \frac{\pi^2}{6} \ln \ubar\Big) + \frac{1-22u}{6}\ln^3 u \no \\
    &\; \quad + (1-2u) \Big( 2 \Lib(u) -\frac52 \ln^2 u \ln \ubar +2 \ln u \ln \ubar + 4 \ln u \Big) - (1-6u) \ln^2 u \no \\
    &\; \quad + \frac{(1-8u)\pi^2}{6} \ln u + \frac{\pi^2}{3} - 8\zeta_3 \bigg) + 4 \zeta_3 -8 - \frac{2\pi^2}{3} \bigg] \bigg\}.
\end{align}
\vspace{8mm}

\newpage
The following MIs can be written in terms of the variable $\xi\equiv
z^2/u$ and
\begin{align}
\eta &\equiv \frac12 \left( 1- \sqrt{1+4\xi} \right), \no \\
\bar \eta &\equiv 1-\eta = \frac12 \left( 1+ \sqrt{1+4\xi} \right).
\end{align}

\vspace{2mm}
\begin{align}
& \SetScale{0.13}
  \begin{picture}(40,23) (24,3)
    \SetWidth{2.0}
    \SetColor{Black}
    \Line(250,150)(250,250)
    \DashLine(350,50)(450,50){15}
    \DashCArc(250,50)(100,-0,180){15}
    \SetWidth{2.0}
    \PhotonArc(250,50)(100,-180,0){5}{8}
    \Photon(150,50)(350,50){5}{5}
    \SetWidth{2.0}
    \DashLine(50,50)(150,50){15}
  \end{picture}
=\quad
        \int [dk] [dl] \; \frac{1}{\calP_{1}\calP_{3}\calP_{22}\calP_{23}}
\quad \equiv \quad
    (z^2 m^2)^{-2\eps} \; \left\{ \sum_{i=-2}^{0} c_i^{(46)} \, \eps^i + \calO(\eps) \right\}
\end{align}
with
\begin{align}
c_{-2}^{(46)}   \, &=\;
    \frac12, \no \\
c_{-1}^{(46)}    \, &=\;
    \frac52 + \ln \xi  + i \pi, \no \\
c_{0}^{(46)}    \, &=\;
    \frac{19}{2} - \frac{5\pi^2}{6}
    + 4 \xi \Big( \zeta_3 - \Sab (\eta) + \Lic(\eta) -\ln \bar \eta  \,\Lib(\eta) +\frac12 \ln^2 \xi \ln \bar \eta   - \ln \xi  \ln^2 \bar \eta  \no \\
    &\; \quad -\frac{1}{12} \ln^3 \xi +\frac12 \ln^3\bar \eta \Big) - (1 - 2 \eta ) \Big( 2\Lib (\eta) + 2 \ln \xi  \ln \bar \eta- \ln^2\bar \eta\Big) + \bar \eta \ln^2\xi \no \\
    &\; \quad + 5 \ln \xi  + \Big( 4 \xi \ln \xi  - 8 \xi \ln \bar \eta +4\eta \Big) \frac{\pi^2}{6}
    + i \pi \bigg[ 5 - \xi \Big( \ln \xi  - 2 \ln \bar \eta \Big)^2 + 2 \bar \eta \ln \xi  \no \\
    &\; \quad -2 (1-2\eta) \ln \bar \eta \bigg].
\end{align}

\vspace{5mm}
\begin{align}
& \SetScale{0.13}
  \begin{picture}(40,23) (24,3)
    \SetWidth{2.0}
    \SetColor{Black}
    \Line(250,150)(250,250)
    \DashLine(350,50)(450,50){15}
    \DashCArc(250,50)(100,-0,180){15}
    \SetWidth{0.5}
    \Vertex(180,120){9.9}
    \SetWidth{2.0}
    \PhotonArc(250,50)(100,-180,0){5}{8}
    \Photon(150,50)(350,50){5}{5}
    \SetWidth{2.0}
    \DashLine(50,50)(150,50){15}
  \end{picture}
=\quad
        \int [dk] [dl] \; \frac{1}{\calP_{1}\calP_{3}^2\calP_{22}\calP_{23}}
\;\; \equiv \;\;
    (z^2 m^2)^{-1-2\eps} \; \left\{ \sum_{i=-2}^{0} c_i^{(47)} \, \eps^i + \calO(\eps) \right\}
\end{align}
with
\begin{align}
c_{-2}^{(47)}   \, &=\;
    -\xi, \no \\
c_{-1}^{(47)}    \, &=\;
    -\xi \Big(\ln \xi  + i \pi \Big), \no \\
c_{0}^{(47)}    \, &=\;
    -\xi \bigg[ 4 - (1-2\eta) \Big( 2 \Lib(\eta) + 2 \ln \xi  \ln \bar \eta  - \ln^2 \bar\eta\Big) + \bar \eta \ln^2 \xi + 2 \ln \xi  -\frac{2\pi^2}{3} \bar \eta  \no \\
    &\; \quad \qquad+ 2 i \pi \Big( 1 + \bar \eta \ln \xi  -(1-2\eta) \ln \bar \eta \Big) \bigg].
\end{align}

\newpage
\subsubsection{5--topologies}

\vspace{-3mm}
\begin{align}
&  \SetScale{0.13}
  \begin{picture}(45,23) (24,20)
  \SetScale{0.15}
    \SetWidth{3.0}
    \SetColor{Black}
    \DashLine(50,150)(450,150){15}
    \Line(245,250)(245,350)\Line(255,250)(255,350)
    \DashCArc(249.58,150.42)(99.58,-179.76,89.76){15}
    \CArc(252.98,147.29)(108.01,94.44,178.5)\CArc(252.98,147.29)(98.01,94.44,178.5)
    \DashLine(250,50)(250,-50){15}
    \SetWidth{2.0}
    \CArc(196,96)(70.94,-158.5,-111.5)
    \CArc(304,96)(70.94,-68.5,-21.5)
    \Text(21,2)[r]{\tiny $\frac{\bar{u}}{2}$}
    \Text(53,2)[l]{\tiny $\frac{u}{2}$}
  \end{picture}
=\quad
        \int [dk] [dl] \; \frac{1}{\calP_{4}\calP_{7}\calP_{9}\calP_{13}\calP_{19}}
\;\; \equiv \;\;
    (m^2)^{-1-2\eps} \; \left\{ \sum_{i=-1}^{0} c_i^{(51)} \, \eps^i + \calO(\eps) \right\}
\end{align}
with
\begin{align}
c_{-1}^{(51)}   \, &=\;
    \zeta_3 + \Lic (u) - \Sab(u) - \ln \ubar \,\Lib(u) - \frac16 \Big( \ln u- \ln \ubar \Big)^3 - \frac16 \ln^3\ubar  \no \\
    &\; \quad - \Big( \ln u- \ln \ubar \Big) \frac{\pi^2}{6} + i \pi \bigg[ \frac{\pi^2}{3} - \Lib(u) + \frac12 \ln^2 u - \ln u \ln \ubar \bigg]\no \\
c_{0}^{(51)}    \, &=\;
    - \frac{41\pi^4}{72} + 13 \Lid(u) -6 \Sbb(u) -\Sac(u) - \Big(11 \ln u + 6 \ln \ubar \Big) \Lic(u) + \frac38 \ln^4 u \no \\
    &\; \quad  + \Big( 11 \ln u - \ln \ubar \Big) \Sab(u) - \frac56 \ln^3 u \ln \ubar + \frac32 \ln^2 u \ln^2 \ubar - \frac16 \ln u \ln^3 \ubar \no \\
    &\; \quad  + \frac12 \Big( 5\ln^2 u +12\ln u \ln \ubar - \ln^2 \ubar - 5 \Lib(u) \Big) \Lib(u) - \zeta_3 \Big( 11 \ln u -7 \ln \ubar \Big) \no \\
    &\; \quad  + \Big( 26 \Lib(u) - 4\ln^2 u +16 \ln u \ln \ubar + \ln^2 \ubar \Big) \frac{\pi^2}{12} + i \pi \bigg[ 2\zeta_3 + 7 \Lic(u) - 2 \Sab(u) \no \\
    &\; \quad  - \Big( \ln u + 2 \ln \ubar \Big) \Lib(u) -\frac32 \ln^3 u  + \frac52 \ln^2 u  \ln \ubar - \ln u \ln^2 \ubar \no \\
    &\; \quad - \Big( 5 \ln u - \ln \ubar\Big) \frac{\pi^2}{3} \bigg].
\end{align}

\vspace{5mm}
\begin{align}
&  \SetScale{0.13}
  \begin{picture}(45,23) (24,20)
    \SetScale{0.15}
    \SetWidth{3.0}
    \SetColor{Black}
    \DashLine(50,150)(450,150){15}
    \Line(245,250)(245,350)\Line(255,250)(255,350)
    \DashCArc(249.58,150.42)(99.58,-179.76,89.76){15}
    \CArc(252.98,147.29)(108.01,94.44,178.5)\CArc(252.98,147.29)(98.01,94.44,178.5)
    \DashLine(250,50)(250,-50){15}
    \SetWidth{2.0}
    \CArc(196,96)(70.94,-158.5,-111.5)
    \CArc(304,96)(70.94,-68.5,-21.5)
    \Text(53,2)[l]{\tiny $\frac{\bar{u}}{2}$}
    \Text(21,3)[r]{\tiny $0$}
  \end{picture}
=\quad
        \int [dk] [dl] \; \frac{1}{\calP_{6}\calP_{7}\calP_{8}\calP_{13}\calP_{19}}
\;\; \equiv \;\;
    (m^2)^{-1-2\eps} \; \left\{ \sum_{i=-1}^{0} c_i^{(52)} \, \eps^i + \calO(\eps) \right\}
\end{align}
with
\begin{align}
c_{-1}^{(52)}   \, &=\;
    \frac{1}{\ubar} \bigg[ \zeta_3 - \Sab(u) + i \pi \bigg( \frac{\pi^2}{6} - \Lib(u) \bigg) \bigg]\hspace{7cm}\no \\
c_{0}^{(52)}    \, &=\;
    \frac{1}{\ubar} \bigg[ - \frac{29\pi^4}{180} - 2 \Sbb(u) - 5 \Sac(u) + \frac{4\pi^2}{3} \Lib(u) \no \\
    &\; \quad~~~~~+ i \pi \bigg( 7 \zeta_3 -2 \Lic(u) - 5 \Sab(u) \bigg) \bigg]
\end{align}

\newpage
\subsubsection{6--topologies}

The non-planar massless 6-denominator integral has been considered
in~\citer{Gonsalves,Kramer}. Recently it has been formulated in
terms of hypergeometric functions which can be expanded to arbitrary
order in $\eps$ with the help of the {\sc Mathematica} package {\sc
HypExp}~\cite{MImassless}. Our result
\begin{align}
&\SetScale{0.13}
  \begin{picture}(45,23) (24,-10)
    \SetWidth{1.25}
    \SetColor{Black}
    \Line(50,-50)(150,-50)
    \SetWidth{2.0}
    \Line(50,-50)(150,-50)
    \DashCArc(250,-50)(100,90,270){15}
    \DashLine(250,50)(450,50){15}
    \DashLine(250,-150)(450,-150){15}
    \DashLine(250,50)(350,-150){15}
    \DashLine(350,50)(250,-150){15}
  \end{picture}
=\quad
        \int [dk] [dl] \; \frac{1}{\calP_{1}\calP_{2}\calP_{3}\calP_{8}\calP_{9}\calP_{10}}\no\\
&\hspace{2.3cm} =\quad
        (u m^2)^{-2-2e} \; e^{2i\pi e} \; \bigg\{ \frac{1}{\eps^4} -\frac{5\pi^2}{6\eps^2}-\frac{27\zeta_3}{\eps}-\frac{23\pi^4}{36}
+ \calO(\eps) \bigg\}
\end{align}
is in agreement with the previous findings.

\newpage
\section{Hadronic two-body decays II}
\label{app:MIsRe}

Our analysis in Chapter~\ref{ch:RePart} involves the calculation of
the MIs from Figure~\ref{fig:BtoPiPi:MIsRe}. The corresponding
particle propagators $\calP_i$ can be found in
(\ref{eq:BtoPiPi:Props}). So far we have not yet calculated the MIs
with an internal charm quark (wavy lines) which will be considered
in~\cite{GBRePart} (these amount to four MIs out of 22).

\subsubsection{3--topologies}

\vspace{-5mm}
\begin{align}
& \SetScale{0.13}
  \begin{picture}(45,23) (24,-10)
    \SetWidth{2.0}
    \SetColor{Black}
    \Line(50,-45)(150,-45)
    \CArc(250,-44.87)(105.13,-177.06,-2.94)\CArc(250,-44.87)(95.13,-177.06,-2.94)
    \DashCArc(250,-45)(100,-0,180){15}
    \Line(350,-45)(450,-45)
    \DashLine(150,-45)(350,-45){15}
  \end{picture}
=\quad
    \int [dk] [dl] \;  \frac{1}{\calP_{8}\calP_{13}\calP_{17}}
\quad \equiv \quad
    (m^2)^{1-2\eps} \; \left\{ \sum_{i=-2}^{2} d_i^{(31)} \, \eps^i + \calO(\eps^3) \right\}
\end{align}

\vspace{-2mm} with
\begin{align}
d_{-2}^{(31)}    &=
    \frac12, \no \\
d_{-1}^{(31)}    &=
    \frac32 - \frac{u}{4}, \no \\
d_{0}^{(31)}     &=
    3 + \frac{\pi^2}{3} - \frac{13u}{8} + \Lib (u) -
    \frac{1-u^2}{2u} \ln \ubar , \no \\
d_{1}^{(31)}    &=
    \frac{15}{4} + \pi^2 - \zeta_3 - \Big( \frac{115}{16} + \frac{\pi^2}{6} \Big) u
    + \Lic(u) +4 \Sab (u) + \frac{1+6u-2u^2}{2u} \Lib (u) \no \\
    &\; \quad + \frac{1-u^2}{u} \ln^2 \ubar - \frac{13(1-u^2)}{4u} \ln \ubar, \no \\
d_{2}^{(31)}    &=
    -\frac{21}{8} + 2\pi^2+\frac{5\pi^4}{36} - 3 \zeta_3 -
    \Big(\frac{865}{32}+\frac{13\pi^2}{12}- \frac{\zeta_3}{2} \Big) u
    + \Lid (u) -8 \Sbb (u) \no \\
    &\; \quad +16 \Sac (u) + \frac{1+6u-2u^2}{2u} \Lic(u)
    - \frac{4-12u-2u^2}{u} \Sab (u) + 3 \Lib^2(u) \no \\
    &\; \quad - \frac{1-u^2}{u} \bigg[ 3 \ln \ubar \, \Lib(u) +\frac43 \ln^3 \ubar - \frac{13}{2}
    \ln^2 \ubar + \Big( \frac{115}{8} + \frac{\pi^2}{3} \Big) \ln \ubar
    \bigg] \no \\
    &\; \quad + \frac{1}{4u} \bigg[ 13  + \Big( 24 +
    \frac{8\pi^2}{3} \Big) u -26 u^2 \bigg] \Lib(u).
\end{align}

\vspace{2mm}
\begin{align}
&  \SetScale{0.13}
  \begin{picture}(45,23) (24,-10)
    \SetWidth{2.0}
    \SetColor{Black}
    \Line(50,-35)(150,-35)
    \CArc(250,-34.87)(105.13,-177.06,-2.94)\CArc(250,-34.87)(95.13,-177.06,-2.94)
    \DashCArc(250,-35)(100,-0,180){15}
    \SetWidth{0.5}
    \Vertex(250,65){9.9}
    \SetWidth{2.0}
    \Line(350,-35)(450,-35)
    \DashLine(150,-35)(350,-35){15}
  \end{picture}
=\quad
    \int [dk] [dl] \;  \frac{1}{\calP_{8}^2\calP_{13}\calP_{17}}
\quad \equiv \quad
    (m^2)^{-2\eps} \; \left\{ \sum_{i=-2}^{2} d_i^{(32)} \, \eps^i + \calO(\eps^3) \right\}
\end{align}

\vspace{-2mm} with
\begin{align}
d_{-2}^{(32)}    &=
    -\frac12, \hspace{12cm} \no
\end{align}
\begin{align}
d_{-1}^{(32)}    &=
    -\frac32 - \frac{\ubar}{u} \ln \ubar ,\no \\
d_{0}^{(32)}     &=
    - \frac92 - \frac{\pi^2}{3} + \frac{1-2u}{u} \, \Lib(u) + \frac{\ubar}{u} \Big(2\ln^2 \ubar-3 \ln \ubar \Big) , \no \\
d_{1}^{(32)}    &=
    \zeta_3 - \frac{27}{2} - \pi^2 -\frac{4(2-u)}{u} \, \Sab(u) + \frac{1-2u}{u} \Big( \Lic(u) +3 \Lib(u) \Big) \no \\
    &\; \quad - \frac{\ubar}{u} \bigg[ 6 \ln \ubar \, \Lib(u) + \frac83 \ln^3 \ubar - 6 \ln^2 \ubar + \Big( 9 + \frac{2\pi^2}{3} \Big) \ln \ubar \bigg], \no \\
d_{2}^{(32)}    &=
    3\zeta_3 - \frac{81}{2} - 3\pi^2 - \frac{5\pi^4}{36} -3 \Lib^2 (u) - \frac{12(2-u)}{u} \Sab(u) \no \\
    &\; \quad + \frac{1-2u}{u} \bigg[ \Lid(u) - 8 \Sbb(u) +16 \Sac(u) + 3
    \Lic(u) + \Big( 9 + \frac{2\pi^2}{3} \Big) \Lib(u) \bigg] \no \\
    &\; \quad + \frac{\ubar}{u} \bigg[ 24 \ln \ubar \, \Sab(u) -6
    \ln \ubar \, \Lic(u) + 12 \ln^2 \ubar \, \Lib(u) + \frac83 \ln^4
    \ubar - 18 \ln \ubar \, \Lib(u) \no \\
    &\; \quad - 8 \ln^3 \ubar + \Big( 18 +
    \frac{4\pi^2}{3} \Big) \ln^2 \ubar - \Big( 27+2\pi^2 -2 \zeta_3
    \Big) \ln \ubar \bigg] .
\end{align}

\vspace{6mm} The massive sunrise integral can be found
in~\cite{sunrise}. Our result for the equal mass case is in
agreement with these findings and reads \vspace{6mm}
\begin{align}
& \SetScale{0.13}
  \begin{picture}(45,23) (24,-10)
    \SetWidth{2.0}
    \SetColor{Black}
    \Line(50,-35)(450,-35)\Line(50,-45)(450,-45)
    \CArc(250,-39.87)(105.13,-177.06,-2.94)\CArc(250,-39.87)(95.13,-177.06,-2.94)
    \CArc(250,-40.13)(105.13,2.94,177.06)\CArc(250,-40.13)(95.13,2.94,177.06)
  \end{picture}
=\quad
        \int [dk] [dl] \; \frac{1}{\calP_{13}\calP_{14}\calP_{15}}\no\\
&\hspace{2.3cm} =\quad
        (m^2)^{1-2\eps} \;  \bigg\{ \frac{3}{2\eps^2} +
        \frac{17}{4\eps} + \Big( \frac{59}{8} + \frac{\pi^2}{2}
        \Big) + \Big( \frac{65}{16} + \frac{11\pi^2}{4} \Big) \eps \\
&\hspace{4.8cm}
        + \Big( - \frac{1117}{32} + \frac{89\pi^2}{8} - 8 \ln 2 \,
        \pi^2 + \frac{\pi^4}{10} + 28 \zeta_3 \Big) \eps^2
        + \calO(\eps^3) \bigg\} \no
\end{align}
\vspace{2mm}

\subsubsection{4--topologies}

\vspace{3mm}
\begin{align}
&  \SetScale{0.13}
  \begin{picture}(45,23) (24,3)
    \SetWidth{2.0}
    \SetColor{Black}
    \DashLine(50,49)(450,49){15}
    \Line(250,149)(250,249)
    \DashCArc(249.58,49.42)(99.58,-179.76,89.76){15}
    \CArc(252.98,46.29)(108.01,94.44,178.5)\CArc(252.98,46.29)(98.01,94.44,178.5)
  \end{picture}
=\quad
    \int [dk] [dl] \;  \frac{1}{\calP_{2}\calP_{5}\calP_{13}\calP_{17}}
\quad \equiv \quad
    (m^2)^{-2\eps} \; \left\{ \sum_{i=-2}^{2} d_i^{(41)} \, \eps^i + \calO(\eps^3) \right\}
\end{align}
with
\begin{align}
d_{-2}^{(41)}    &=
    \frac12,\hspace{13cm} \no
\end{align}
\begin{align}
d_{-1}^{(41)}    &=
    \frac52 + \frac{\ubar}{u} \ln \ubar, \no \\
d_{0}^{(41)}    &=
    \frac{19}{2} + \frac{\pi^2}{3} - \frac{\ubar}{u} \bigg[ \Lib(u) + \frac32 \ln^2 \ubar - 5 \ln \ubar
    \bigg], \no \\
d_{1}^{(41)}    &=
    \frac{65}{2} + \frac{5\pi^2}{3} - \zeta_3 - \frac{\ubar}{u} \bigg[ \Lic(u) -3 \Sab(u) - 3\ln \ubar \, \Lib(u) - \frac32 \ln^3 \ubar + 5 \Lib(u) \no \\
    &\; \quad+ \frac{15}{2} \ln^2 \ubar - \Big( 19 + \frac{2\pi^2}{3} \Big) \ln \ubar
    \bigg],  \no \\
d_{2}^{(41)}    &=
    \frac{211}{2} + \frac{19\pi^2}{3} + \frac{5\pi^4}{36} - 5
    \zeta_3 - \frac{\ubar}{u} \bigg[ \Lid(u) -3 \Sbb(u) + 9 \Sac(u)
    - 3 \ln \ubar \, \Lic(u) \no \\
    &\, \quad + 9 \ln \ubar \, \Sab(u) + \frac92 \ln^2
    \ubar \, \Lib(u) + \frac98 \ln^4 \ubar + 5 \Lic(u) - 15 \Sab(u)
    - \frac{15}{2} \ln^3 \ubar \no \\
    &\, \quad  -15 \ln \ubar \, \Lib(u) + \Big( 19 +
    \frac{2\pi^2}{3} \Big) \Big( \Lib(u) + \frac32 \ln^2 \ubar \Big)
    - \Big( 65 + \frac{10\pi^2}{3}-2\zeta_3 \Big) \ln \ubar \bigg].
\end{align}

\vspace{5mm}
\begin{align}
& \SetScale{0.13}
  \begin{picture}(45,23) (24,3)
    \SetWidth{2.0}
    \SetColor{Black}
    \DashLine(150,49)(450,49){15}
    \Line(250,149)(250,249)
    \DashCArc(249.58,49.42)(99.58,-179.76,89.76){15}
    \CArc(252.71,46.02)(108.01,91.5,175.56)\CArc(252.71,46.02)(98.01,91.5,175.56)
    \Line(50,54)(145,54)\Line(50,44)(145,44)
  \end{picture}
=\quad
    \int [dk] [dl] \;  \frac{1}{\calP_{5}\calP_{8}\calP_{9}\calP_{13}}
\quad \equiv \quad
    (m^2)^{-2\eps} \; \left\{ \sum_{i=-2}^{2} d_i^{(42)} \, \eps^i + \calO(\eps^3) \right\}
\end{align}
with
\begin{align}
d_{-2}^{(42)}    &=
    \frac12, \no \\
d_{-1}^{(42)}    &=
    \frac52 + \frac{\ubar}{u} \ln \ubar, \no \\
d_{0}^{(42)}    &=
    \frac{19}{2} + \frac{\pi^2}{2} - \frac{\ubar}{u} \bigg[ 2
    \Lib(u) + 2 \ln^2 \ubar -5 \ln \ubar \bigg], \no \\
d_{1}^{(42)}    &=
    \frac{65}{2} + \frac{5\pi^2}{2} + 4\zeta_3 - \frac{\ubar}{u}
    \bigg[ 4 \Lic(u) -8 \Sab(u) + 10 \Lib(u) - \frac83 \ln^3 \ubar \no \\
    &\, \quad
    -8 \ln \ubar \, \Lib(u) + 10 \ln^2 \ubar - (19 + \pi^2) \ln
    \ubar \bigg], \no \\
d_{2}^{(42)}    &=
    \frac{211}{2} + \frac{19\pi^2}{2} + \frac{\pi^4}{2} + 20 \zeta_3
    - \frac{\ubar}{u} \bigg[ 8 \Lid(u) -16 \Sbb(u) +32 \Sac(u) + \frac83 \ln^4
    \ubar \no \\
    &\, \quad -16 \ln \ubar \, \Lic(u) +32 \ln \ubar \, \Sab(u) + 16 \ln^2 \ubar \,
    \Lib(u) +20 \Lic(u) - 40 \ln \ubar \, \Lib(u) \no \\
    &\, \quad  -40 \Sab(u)- \frac{40}{3} \ln^3 \ubar +2(19+\pi^2) (\Lib(u) + \ln^2 \ubar ) -
    ( 65 + 5 \pi^2 + 8 \zeta_3 ) \ln \ubar \bigg].
\end{align}

\vspace{5mm}
\begin{align}
&   \SetScale{0.13}
  \begin{picture}(45,23) (24,3)
    \SetWidth{2.0}
    \SetColor{Black}
    \DashLine(50,49)(150,49){15}
    \Line(245,149)(245,249)\Line(255,149)(255,249)
    \DashCArc(249.58,49.42)(99.58,-179.76,89.76){15}
    \CArc(252.71,46.02)(108.01,91.5,175.56)\CArc(252.71,46.02)(98.01,91.5,175.56)
    \Line(155,54)(350,54)\Line(155,44)(350,44)
    \Line(350,49)(450,49)
  \end{picture}
=\quad
    \int [dk] [dl] \;  \frac{1}{\calP_{5}\calP_{8}\calP_{14}\calP_{15}}
\quad \equiv \quad
    (m^2)^{-2\eps} \; \left\{ \sum_{i=-2}^{1} d_i^{(43)} \, \eps^i + \calO(\eps^2) \right\}
\end{align}
with
\begin{align}
d_{-2}^{(43)}    &=
    \frac12, \no \\
d_{-1}^{(43)}    &=
    \frac52, \no \\
d_{0}^{(43)}    &=
    \frac{1}{\ubar} \bigg[ \frac{19\ubar}{2} -2 \zeta_3 +2 \Lic(u) + \ubar
    \, \Lib(u)\bigg], \no \\
d_{1}^{(43)}    &=
    \frac{1}{\ubar} \bigg[ \frac{65\ubar}{2} - \frac{7\pi^4}{180} +4 \Sbb(u)  +3 \Lib^2(u) + 4 \ln \ubar \, \Lic(u) +(5-
    3 u) (\Lic(u)-\zeta_3) \no \\
    &\;\; \quad -2\ubar \ln \ubar \, \Lib(u) + \Big( 5 \ubar -
    \frac{\pi^2}{3} \Big) \Lib(u) + \Big( \frac{\pi^2}{3} \ubar -4
    \zeta_3 \Big) \ln \ubar \bigg].
\end{align}

\vspace{5mm}
\begin{align}
&  \SetScale{0.13}
  \begin{picture}(45,23) (24,3)
    \SetWidth{2.0}
    \SetColor{Black}
    \DashLine(50,49)(150,49){15}
    \Line(245,149)(245,249)\Line(255,149)(255,249)
    \DashCArc(249.58,49.42)(99.58,-179.76,89.76){15}
    \CArc(252.71,46.02)(108.01,91.5,175.56)\CArc(252.71,46.02)(98.01,91.5,175.56)
    \Line(155,54)(350,54)\Line(155,44)(350,44)
    \Line(350,49)(450,49)
    \SetWidth{0.5}
    \Vertex(320,119){10}
  \end{picture}
=\quad
    \int [dk] [dl] \;  \frac{1}{\calP_{5}\calP_{8}^2\calP_{14}\calP_{15}}
\quad \equiv \quad
    (m^2)^{-1-2\eps} \; \left\{ \sum_{i=-2}^{2} d_i^{(44)} \, \eps^i + \calO(\eps^3) \right\}
\end{align}
with
\begin{align}
d_{-2}^{(44)}    &=
    -\frac12, \no \\
d_{-1}^{(44)}    &=
    \frac{1+u}{2u} + \frac{1+u^2}{2u^2} \ln \ubar, \no \\
d_{0}^{(44)}    &=
    -2 - \frac{\pi^2}{6} + \frac{1-3u^2}{2u^2} \Lib(u)
    -\frac{1+u^2}{u^2} \ln^2 \ubar + \frac{(1-4u-u^2)}{2u^2} \ln
    \ubar, \no \\
d_{1}^{(44)}    &=
    2\zeta_3 + \frac{2(1+u)}{u} + \frac{(3+u)\pi^2}{6u} +
    \frac{1}{u^2} \bigg[ 2 (3+u^2) \Sab(u)
    +2 (1+2u^2) \ln \ubar \, \Lib(u) \no \\
    &\; \quad - \frac{1+5u^2}{2} \Lic(u) + \frac{4(1+u^2)}{3} \ln^3
    \ubar + \frac{(1+u)(1+3u)}{2} \, \Lib(u) \no \\
    &\; \quad -(1-4u-u^2) \ln^2 \ubar +
    \Big( 2(1+u^2)+ (3+u^2) \frac{\pi^2}{6} \Big) \ln \ubar \bigg],
    \no \\
d_{2}^{(44)}    &=
    -8 -\frac{2\pi^2}{3} + \frac{\pi^4}{180} + \frac{2(2-u)}{u}
    \zeta_3 +
    \frac{1}{u^2} \bigg[ \frac{1-3u^2}{2} \Lid(u) - 2(3-5u^2)
    \Sbb(u) \no \\
    &\; \quad +2(1-9u^2) \Sac(u) -2(1-2u^2)\ln \ubar \, \Lic(u)
    -12(1+u^2)\ln \ubar \, \Sab(u) \no
\end{align}
\begin{align}
    &\; \quad -3 u^2 \Lib(u)^2 -(5+7u^2) \ln^2
    \ubar \, \Lib(u) - \frac{4(1+u^2)}{3} \ln^4 \ubar
    -\frac{(1+5u)\ubar}{2} \Lic(u) \no \\
    &\; \quad +2(3-6u-u^2) \Sab(u)
    +2(1-5u-2u^2) \ln \ubar \, \Lib(u) + \frac{4(1-4u-u^2)}{3} \ln^3\ubar\no \\
    &\; \quad + \Big(
    2(1-3u^2)+(3-5u^2)\frac{\pi^2}{6} \Big) \Lib(u) - \Big( 4(1+u^2)
    +(5+3u^2)\frac{\pi^2}{6} \Big) \ln^2 \ubar \no \\
    &\; \quad + \Big(
    2(1-4u-u^2)+(3-10u-u^2)\frac{\pi^2}{6} +2(2-u^2)\zeta_3 \Big)
    \ln \ubar \bigg].
\end{align}

\vspace{3mm}
\begin{align}
&  \SetScale{0.13}
  \begin{picture}(45,23) (24,3)
    \SetWidth{2.0}
    \SetColor{Black}
    \DashLine(50,49)(150,49){15}
    \Line(250,149)(250,249)
    \DashCArc(249.58,49.42)(99.58,-179.76,89.76){15}
    \CArc(252.71,46.02)(108.01,91.5,175.56)\CArc(252.71,46.02)(98.01,91.5,175.56)
    \Line(155,54)(350,54)\Line(155,44)(350,44)
    \DashLine(350,49)(450,49){15}
  \end{picture}
=\quad
    \int [dk] [dl] \;  \frac{1}{\calP_{3}\calP_{6}\calP_{14}\calP_{15}}
\quad \equiv \quad
    (m^2)^{-2\eps} \; \left\{ \sum_{i=-2}^{1} d_i^{(45)} \, \eps^i + \calO(\eps^2) \right\}
\end{align}

\vspace{-4mm} with
\begin{align}
d_{-2}^{(45)}    &=
    \frac12, \no \\
d_{-1}^{(45)}    &=
    \frac52 + \frac{\ubar}{u} \ln \ubar,\no \\
d_{0}^{(45)}    &=
    \frac{19}{2} + \frac{\pi^2}{6} + \frac{5\ubar}{u} \ln \ubar - \frac{1}{u} \bigg[ 2 \Sab(u) +
    \ln \ubar \, \Lib(u) + \Lib(u) + \ubar \ln^2 \ubar -
    \frac{\pi^2}{6} \ln \ubar \bigg],\no \\
d_{1}^{(45)}    &=
    \frac{65}{2} + \frac{5\pi^2}{6} - \frac{1}{u} \bigg[ 2 \Sbb(u) +
    3 \Sac(u) - \ln \ubar \, \Lic(u) - \frac12 \ln^2 \ubar \,
    \Lib(u) - \Lib(u)^2 \no \\
    &\; \quad +(1-2u) \Lic(u) +2(1+2u) \Sab(u) +2u \ln
    \ubar \, \Lib(u) - \frac{2\ubar}{3} \ln^3 \ubar \no \\
    &\; \quad+ \Big(
    5+\frac{\pi^2}{6} \Big) \Lib(u) + \Big( 5\ubar +
    \frac{\pi^2}{12} \Big) \ln^2 \ubar - \Big( 19\ubar +(2-u)
    \frac{\pi^2}{3} - \zeta_3 \Big) \ln \ubar \bigg].
\end{align}

\vspace{1mm}
\begin{align}
&   \SetScale{0.13}
  \begin{picture}(45,23) (24,3)
    \SetWidth{2.0}
    \SetColor{Black}
    \DashLine(50,60)(150,60){15}
    \Line(250,160)(250,260)
    \DashCArc(249.58,60.42)(99.58,-179.76,89.76){15}
    \CArc(252.71,57.02)(108.01,91.5,175.56)\CArc(252.71,57.02)(98.01,91.5,175.56)
    \Line(155,65)(350,65)\Line(155,55)(350,55)
    \SetWidth{0.5}
    \Vertex(250,-40){10}
    \SetWidth{2.0}
    \DashLine(350,60)(450,60){15}
  \end{picture}
=\quad
    \int [dk] [dl] \;  \frac{1}{\calP_{3}^2\calP_{6}\calP_{14}\calP_{15}}
\quad \equiv \quad
    (m^2)^{-1-2\eps} \; \left\{ \sum_{i=-1}^{1} d_i^{(46)} \, \eps^i + \calO(\eps^2) \right\}
\end{align}

\vspace{-4mm} with
\begin{align}
d_{-1}^{(46)}    &=
    \frac{1}{u} \, \Lib(u), \no \\
d_{0}^{(46)}    &=
    \frac{1}{u} \bigg[ 4 \Sab(u) - \Lic(u) + \ln \ubar \, \Lib(u) -
    \frac{\pi^2}{6} \ln \ubar \bigg], \no \\
d_{1}^{(46)}    &=
     \frac{1}{u} \bigg[ 3 \Lid(u) +7 \Sac(u) -4 \Sbb(u) - \ln \ubar
     \, \Lic(u) - \frac12 \ln^2 \ubar \, \Lib(u) + \frac{\pi^2}{2}
     \Lib(u) \no \\
    &\;\; \quad+ \frac{\pi^2}{12} \ln^2 \ubar + \zeta_3 \ln \ubar
     \bigg].
\end{align}

The following MIs give rise to HPLs with parameter $-1$

\vspace{4mm}
\begin{align}
&  \SetScale{0.13}
  \begin{picture}(45,23) (24,3)
    \SetWidth{2.0}
    \SetColor{Black}
    \DashLine(50,54)(150,54){15}
    \Line(250,154)(250,254)
    \Line(155,59)(450,59)\Line(155,49)(450,49)
    \CArc(249.58,54.42)(104.58,-179.76,89.76)\CArc(249.58,54.42)(94.58,-179.76,89.76)
    \DashCArc(252.5,51.5)(102.53,91.4,178.6){15}
  \end{picture}
=\quad
    \int [dk] [dl] \;  \frac{1}{\calP_{6}\calP_{13}\calP_{14}\calP_{15}}
\quad \equiv \quad
    (m^2)^{-2\eps} \; \left\{ \sum_{i=-2}^{1} d_i^{(47)} \, \eps^i + \calO(\eps^2) \right\}
\label{eq:MI:HPLext1}
\end{align}

\vspace{-4mm} with
\begin{align}
d_{-2}^{(47)}    &=
    \frac12, \no \\
d_{-1}^{(47)}    &=
    \frac52 + \frac{\ubar}{u} \ln \ubar, \no \\
d_{0}^{(47)}     &=
    \frac{19}{2} + \frac{\pi^2}{6} + \frac{1}{\ubar} \bigg[ 2
    \Lic(u) -2 \zeta_3 \bigg] -  \frac{1}{u} \bigg[ 2\Lib(u) + \ubar
    \ln^2 \ubar -5 \ubar \ln \ubar \bigg], \no \\
d_{1}^{(47)}     &=
    \frac{65}{2} + \frac{5\pi^2}{6} + \frac{1}{\ubar} \bigg[ 6
    \Lid(u) -4 \Sbb(u) + 4 \Lib(u)^2 + 4\ln \ubar \, \Lic(u)  - 2(2+u) \zeta_3 \no \\
    &\, \quad -8 H(0,-1,0,1;u)  -
    \frac{2\pi^2}{3} H(0,-1;u) + \frac{\pi^4}{60} -
    \frac{2(2+u)(1-2u)}{u} \Lic(u)  \no \\
    &\, \quad - \Big( \frac{10\ubar}{u} +
    \frac{2\pi^2}{3} \Big) \Lib(u)+ \Big(  \frac{19\ubar^2}{u} -
    \frac{2\ubar \pi^2}{3} -4 \, \zeta_3 \Big) \ln \ubar
    \bigg] + \frac{1}{u} \bigg[ 4 \ln \ubar \, \Lib(u) \no \\
    &\, \quad + 4 \Sab(u)
    +\ubar \Big( \frac23 \ln^3 \ubar-5 \ln^2 \ubar \Big) +(1+u)
    \Big( 4 H(-1,0,1;u) + \frac{\pi^2}{3} H(-1;u) \Big) \bigg].
\end{align}

\vspace{3mm}
\begin{align}
&  \SetScale{0.13}
  \begin{picture}(45,23) (24,3)
    \SetWidth{0.5}
    \SetColor{Black}
    \Vertex(180,124){10}
    \SetWidth{2.0}
    \DashLine(50,54)(150,54){15}
    \Line(250,154)(250,254)
    \Line(155,59)(450,59)\Line(155,49)(450,49)
    \CArc(249.58,54.42)(104.58,-179.76,89.76)\CArc(249.58,54.42)(94.58,-179.76,89.76)
    \DashCArc(252.5,51.5)(102.53,91.4,178.6){15}
  \end{picture}
=\quad
    \int [dk] [dl] \;  \frac{1}{\calP_{6}^2\calP_{13}\calP_{14}\calP_{15}}
\;\; \equiv \;\;
    (m^2)^{-1-2\eps} \; \left\{ \sum_{i=-2}^{1} d_i^{(48)} \, \eps^i + \calO(\eps^2) \right\}
\end{align}

\vspace{-4mm} with
\begin{align}
d_{-2}^{(48)}    &=
    \frac{1}{\ubar}, \no \\
d_{-1}^{(48)}    &=
    - \frac{1+u}{u \ubar} \ln \ubar, \no \\
d_{0}^{(48)}    &=
    \frac{1}{\ubar} \bigg[ 4  + \frac{1+u}{u
    \ubar} \Big(2 \Lib(u) + \ubar \ln^2\ubar -2 \ubar \ln \ubar - \frac{u \pi^2}{3}\Big)
    \bigg],  \no \\
d_{1}^{(48)}    &=
    \frac{1}{\ubar^2} \bigg\{ (3\ln 2-1) \frac{4\pi^2}{3} -2(u+8)
    \zeta_3 + \frac{1+u}{u} \bigg[ 4(1+u) \Lic(u)  -4 \ln
    \ubar \, \Lib(u) \no \\
    &\, \quad -4 \Sab(u) - \frac23 \ubar \ln^3 \ubar +4 \Lib(u) +2 \ubar
    \ln^2 \ubar - \Big( 4\ubar -\frac{2u\pi^2}{3}  \Big) \ln \ubar \no \\
    &\, \quad
    -4(1+u) H(-1,0,1;u)-\frac{(1+u)\pi^2}{3} H(-1;u) \bigg]
    \bigg\}.
\end{align}

\vspace{5mm}
\begin{align}
& \SetScale{0.13}
  \begin{picture}(45,23) (24,3)
    \SetWidth{2.0}
    \SetColor{Black}
    \DashLine(50,49)(150,49){15}
    \Line(250,149)(250,249)
    \DashCArc(249.58,49.42)(99.58,-179.76,89.76){15}
    \CArc(252.71,46.02)(108.01,91.5,175.56)\CArc(252.71,46.02)(98.01,91.5,175.56)
    \Line(155,54)(450,54)\Line(155,44)(450,44)
  \end{picture}
=\quad
    \int [dk] [dl] \;  \frac{1}{\calP_{5}\calP_{11}\calP_{13}\calP_{14}}
\quad \equiv \quad
    (m^2)^{-2\eps} \; \left\{ \sum_{i=-2}^{1} d_i^{(49)} \, \eps^i + \calO(\eps^2) \right\}
\end{align}

\vspace{-4mm} with
\begin{align}
d_{-2}^{(49)}    &=
    \frac12, \no \\
d_{-1}^{(49)}    &=
    \frac52 + \frac{\ubar}{u} \ln \ubar, \no \\
d_{0}^{(49)}    &=
    \frac{19}{2} + \frac{\pi^2}{3} + \frac{1}{\ubar} \bigg[ 2
    \zeta_3 - \Lic(u) - \Sab(u) + \Lic (u-1) -\ln \ubar \, \Lib(u-1) \no \\
    &\, \quad -\frac12 \Big(\ln^2 \ubar+\pi^2\Big) \ln(2-u) \bigg] - \frac{1}{u} \bigg[ (2-u)
    \Lib(u) + \frac{3\ubar}{2} \ln^2 \ubar -5 \ubar \ln \ubar \bigg], \no \\
d_{1}^{(49)}    &=
    \frac{65}{2} + \frac{5\pi^2}{3} - \frac{1}{\ubar} \bigg[ 3
    \Lid(u) -4 \Sbb(u) +12 \Sac(u) +2 \ln \ubar \, \Lic(u) +2 \ln \ubar \,
    \Sab(u) \no \\
    &\, \quad + 3 \Lib(u)^2 +\frac{\pi^2}{2} \Lib(u) -4 \zeta_3 \ln
    \ubar -(3-u) \zeta_3 - \frac{161\pi^4}{720} + \frac{\pi^2}{3}
    \ln^2 2 - \frac13 \ln^4 2  \no \\
    &\, \quad-7 \zeta_3 \ln2 - 8 \Lid({\textstyle\frac12}) - 12 \Lid(u-1) - \Big( 4\ln^2 \ubar -u \ln \ubar +
    \frac{11\pi^2}{6} \Big) \Lib(u-1) \no \\
    &\, \quad +10 \ln \ubar \, \Lic(u-1) -\Big( \ln^3 \ubar -
    \frac{u}{2} \ln^2 \ubar + \frac{5\pi^2}{6} \ln \ubar -7 \zeta_3
    -\frac{u\pi^2}{2} \Big) \ln (2-u) \no \\
    &\, \quad -u \, \Lic(u-1) -2 {\cal{A}}(u) + \frac{4-5u+2u^2}{u} \Lic(u)
    - \frac{6-12u+5u^2}{u} \Sab(u) \no \\
    &\, \quad - \frac{6-10u+4u^2}{u} \ln \ubar \, \Lib(u) \bigg]
    + \frac{1}{u} \bigg[ \frac{5\ubar}{3} \ln^3 \ubar -5 (2-u)
    \Lib(u) \no \\
    &\, \quad- \frac{15-16u}{2} \ln^2 \ubar + \Big( 19 \ubar
    +(5-7u)\frac{\pi^2}{6} \Big) \ln \ubar \bigg] -\frac12 \ln^2
    \ubar,
\end{align}
where we defined an auxiliary function ${\cal{A}}(u)$ which is given
by the integral
\begin{align}
{\cal{A}}(u) &\equiv \int_0^u du' \frac{1}{2-u'} \; H(1,0,1;u').
\label{eq:auxA}
\end{align}

\newpage
\begin{align}
&  \SetScale{0.13}
  \begin{picture}(45,23) (24,3)
    \SetWidth{2.0}
    \SetColor{Black}
    \DashLine(50,49)(150,49){15}
    \Line(250,149)(250,249)
    \DashCArc(249.58,49.42)(99.58,-179.76,89.76){15}
    \CArc(252.71,46.02)(108.01,91.5,175.56)\CArc(252.71,46.02)(98.01,91.5,175.56)
    \Line(155,54)(450,54)\Line(155,44)(450,44)
    \SetWidth{0.5}
    \Vertex(320,119){9.9}
  \end{picture}
=\quad
    \int [dk] [dl] \;  \frac{1}{\calP_{5}^2\calP_{11}\calP_{13}\calP_{14}}
\;\; \equiv \;\;
    (m^2)^{-1-2\eps} \; \left\{ \sum_{i=-2}^{2} d_i^{(410)} \, \eps^i + \calO(\eps^3) \right\}
\end{align}

\vspace{-4mm} with
\begin{align}
d_{-2}^{(410)}    &=
    \frac{1}{\ubar}, \no \\
d_{-1}^{(410)}    &=
    \frac{1}{\ubar} \bigg[ 2 - \frac{1+u}{u} \ln \ubar \bigg], \no \\
d_{0}^{(410)}    &=
    \frac{1}{\ubar} \bigg[ 4 + \frac{2}{u} \Lib(u) +
    \frac{3-u^2}{u(2-u)} \ln^2 \ubar - \frac{2(1+u)}{u} \ln \ubar +
    \frac{(7-5u)\pi^2}{(2-u)6} \bigg], \no \\
d_{1}^{(410)}    &=
    \frac{1}{\ubar} \bigg\{ 8 + \frac{7\ubar\,\zeta_3}{2-u}  +\frac{(7-5u)\pi^2}{(2-u)3}
    + \frac{2}{u} \bigg[ (2-u) \Lic(u) +2 \Lib(u) -2(1+u) \ln
    \ubar \bigg] \no \\
    &\, \quad -\frac{2}{u(2-u)} \bigg[ (6-3u-u^2) \Sab(u) +2(3-2u)
    \ln \ubar \, \Lib(u) + \frac{5-2u-u^2}{3} \ln^3 \ubar \no \\
    &\, \quad -(3-u^2)
    \ln^2 \ubar + (5-u-2u^2)\frac{\pi^2}{6} \ln \ubar \bigg] \bigg\}, \no \\
d_{2}^{(410)}    &=
    \frac{1}{\ubar} \bigg\{ 16 + \frac{14 \ubar\, \zeta_3}{2-u} + \frac{2}{u}
    \bigg[ (4-3u) \Lid(u) -2(3-2u) \Sbb(u) +2(2-u) \Lic(u) \no \\
    &\, \quad + \Big( 4 + (5-3u) \frac{\pi^2}{6} \Big) \Lib(u) -4(1+u) \ln \ubar \bigg]
    + \frac{2}{u(2-u)} \bigg[ \frac{9-6u-u^2}{6} \ln^4 \ubar \no \\
    &\, \quad +(20-19u+3u^2)\Sac(u) -
    \frac{2+u-3u^2}{2} \Lib(u)^2 +2(5-4u) \ln^2 \ubar \, \Lib(u)\no \\
    &\, \quad
    -4(3-2u) \ln \ubar \, \Lib(u) -(14-13u+u^2) \ln \ubar \,\Lic(u)
    -2(6-3u-u^2) \Sab(u) \no \\
    &\, \quad+(20-15u-u^2) \ln \ubar \, \Sab(u) -
    \frac{2(5-2u-u^2)}{3} \ln^3 \ubar  -(5-u-2u^2) \frac{\pi^2}{3}
    \ln \ubar \no \\
    &\, \quad + \Big( 2(3-u^2) +(5-3u-u^2)
    \frac{\pi^2}{3} \Big) \ln^2 \ubar
    +\frac{(17+47u-60u^2)\pi^4}{120}
    +\frac{u(7-5u)\pi^2}{3}\no \\
    &\, \quad +\ubar (2+u) \Big( 3 \Lid(u-1) -2\ln \ubar\, \Lic(u-1)
    + \frac{\ln^2 \ubar +\pi^2}{2} \Lib(u-1) \Big) \no \\
    &\, \quad- \ubar (3+u)
    \zeta_3 \ln \ubar \bigg] \bigg\}.
\end{align}

\newpage
\begin{align}
&  \SetScale{0.13}
  \begin{picture}(45,23) (24,3)
    \SetWidth{2.0}
    \SetColor{Black}
    \DashLine(50,60)(150,60){15}
    \Line(250,160)(250,260)
    \DashCArc(249.58,60.42)(99.58,-179.76,89.76){15}
    \CArc(252.71,57.02)(108.01,91.5,175.56)\CArc(252.71,57.02)(98.01,91.5,175.56)
    \Line(155,65)(450,65)\Line(155,55)(450,55)
    \SetWidth{0.5}
    \Vertex(250,-40){9.9}
  \end{picture}
=\quad
    \int [dk] [dl] \;  \frac{1}{\calP_{5}\calP_{11}^2\calP_{13}\calP_{14}}
\;\; \equiv \;\;
    (m^2)^{-1-2\eps} \; \left\{ \sum_{i=-1}^{1} d_i^{(411)} \, \eps^i + \calO(\eps^2) \right\}
\end{align}

\vspace{-4mm} with
\begin{align}
d_{-1}^{(411)}    &=
    \frac{1}{\ubar} \bigg[ \frac{\pi^2}{6} - \Lib(u) \bigg], \no \\
d_{0}^{(411)}    &=
    \frac{1}{\ubar} \bigg[ \zeta_3 -2 \Lic(u) + \Sab(u) + 2 \ln
    \ubar \, \Lib(u) - \frac{\pi^2}{3} \ln \ubar - \Lic(u-1) \no \\
    &\, \quad + \ln \ubar \, \Lib(u-1) +\frac12 \ln^2 \ubar \, \ln
    (2-u) + \frac{\pi^2}{2} \ln (2-u) \bigg], \no \\
d_{1}^{(411)}    &=
    \frac{1}{\ubar} \bigg[ \frac{35\pi^4}{144} + \frac{\pi^2}{3}
    \ln^2 2 - \frac13 \ln^4 2 -7 \zeta_3 \,\ln2 - 8
    \Lid({\textstyle\frac12}) -4 \Lid(u) +6 \Sbb(u) \no \\
    &\, \quad +2 \Sac(u) -2
    \ln^2 \ubar \, \Lib(u) +6 \ln \ubar \, \Lic(u) - 2 {\cal{A}}(u)
    - \frac{5\pi^2}{6} \Lib(u) + \frac{\pi^2}{3} \ln^2 \ubar \no \\
    &\, \quad -6
    \zeta_3 \ln \ubar -6 \Lid(u-1) +6 \ln \ubar \, \Lic(u-1) -3 \ln^2 \ubar \, \Lib(u-1)  \no \\
    &\, \quad -\frac{5\pi^2}{6} \Lib(u-1) - \Big(\ln^3 \ubar + \frac{5\pi^2}{6} \ln \ubar - 7 \zeta_3\Big)
    \ln(2-u) \bigg].
\label{eq:MI:HPLext2}
\end{align}
The auxiliary function ${\cal{A}}(u)$ can be found in
(\ref{eq:auxA}).

\subsubsection{5--topologies}

\begin{align}
&  \SetScale{0.13}
  \begin{picture}(45,23) (24,3)
    \SetWidth{2.0}
    \SetColor{Black}
    \CArc(250,55.13)(105.13,-177.06,-2.94)\CArc(250,55.13)(95.13,-177.06,-2.94)
    \DashCArc(250,55)(100,-0,180){15}
    \DashLine(250,255)(250,-45){15}
    \Line(350,60)(450,60)\Line(350,50)(450,50)
    \Line(50,55)(150,55)
  \end{picture}
=\;\;
    \int [dk] [dl] \;  \frac{1}{\calP_{5}\calP_{8}\calP_{9}\calP_{13}\calP_{14}}
\;\; \equiv \;\;
    (m^2)^{-1-2\eps} \; \left\{ \sum_{i=0}^{0} d_i^{(51)} \, \eps^i + \calO(\eps) \right\}
\end{align}

\vspace{-4mm} with
\begin{align}
d_{0}^{(51)}    &=
    - \frac{1}{\ubar} \bigg[ \Lid(u) +4 \Sbb(u) - \frac12 \Lib^2(u)
    - \frac{\pi^2}{2} \Lib(u) + \frac{3\pi^4}{40} \bigg].
\end{align}

\begin{align}
&  \SetScale{0.13}
  \begin{picture}(45,23) (24,3)
    \SetWidth{2.0}
    \SetColor{Black}
    \CArc(250,55.13)(105.13,-177.06,-2.94)\CArc(250,55.13)(95.13,-177.06,-2.94)
    \DashCArc(250,55)(100,-0,180){15}
    \DashLine(250,255)(250,-45){15}
    \Line(350,60)(450,60)\Line(350,50)(450,50)
    \Line(50,55)(150,55)
    \SetWidth{0.5}
    \Vertex(320,125){9.9}
  \end{picture}
=\;\;
    \int [dk] [dl] \;  \frac{1}{\calP_{5}\calP_{8}^2\calP_{9}\calP_{13}\calP_{14}}
\;\; \equiv \;\;
    (m^2)^{-2-2\eps} \; \left\{ \sum_{i=-2}^{2} d_i^{(52)} \, \eps^i + \calO(\eps^3) \right\}
\end{align}

\vspace{-4mm} with
\begin{align}
d_{-2}^{(52)}    &=
    -\frac{1}{12\ubar^2},  \no \\
d_{-1}^{(52)}    &=
    - \frac{1}{6\ubar^2} \bigg[ 1-2\ln \ubar \bigg], \no \\
d_{0}^{(52)}    &=
    - \frac{1}{6\ubar^2} \bigg[ \frac{(8+3u) \pi^2}{6} - 7 - 6 u  +6
    \Lib(u) + 4 \ln^2 \ubar - 2(5-3u) \ln \ubar \bigg], \no \\
d_{1}^{(52)}    &=
    - \frac{1}{6\ubar^2} \bigg[ 25 + 54 u - \frac{16\pi^2}{3} +
    (11-3u) \zeta_3 + 6 (1-2u) \Lic(u) -12(3+u) \Sab(u)  \no \\
    &\, \quad - 24 \ln \ubar \, \Lib(u) -
    \frac{16}{3} \ln^3 \ubar-6(2-u) \Lib(u) +8 \ln^2 \ubar  \no \\
    &\, \quad +2 \Big(
    41-15 u - \frac{(8+3u)\pi^2}{3} \Big) \ln \ubar \bigg], \no \\
d_{2}^{(52)}    &=
    - \frac{1}{6\ubar^2} \bigg[  \frac{2(35-12u)\pi^2}{3} -67 -330 u
    +  \frac{(179-93u)\pi^4}{90} -38 \zeta_3 +\frac{16}{3} \ln^4
    \ubar \no \\
    & \;\; +6(1-10u) \Lid(u)-12(1+5u) \Sbb(u)+24(7+3u) \Sac(u) -
    \frac{8(1+3u)}{3} \ln^3 \ubar \no \\
    & \;\;
    -24(1-2u) \ln \ubar \, \Lic(u) +48 (3+u) \ln  \ubar \, \Sab(u)
    -6(1-2u) \Lib^2(u) \no \\
    & \;\; +48 \ln^2 \ubar \, \Lib(u) -6(2-9u) \Lic(u) +24(2+3u)
    \Sab(u) +12 (3-u) \ln \ubar \, \Lib(u) \no \\
    & \;\; +6 \Big( 4-9u + (1+3u)\frac{\pi^2}{3} \Big) \Lib(u) - \frac43
    \Big( 
    69+9u-(8+3u)\pi^2 \Big) \ln^2 \ubar \no \\
    & \;\; -\frac23 \Big( 645-171u -4(5+3u)\pi^2 +6(11-3u)\zeta_3\Big)
    \ln\ubar \bigg].
\end{align}

\begin{align}
&  \SetScale{0.13}
  \begin{picture}(45,23) (24,13)
    \SetScale{0.1}
    \SetWidth{2.0}
    \SetColor{Black}
    \DashCArc(250,150)(100.72,173,533){15}
    \Line(50,155)(350,155)\Line(50,145)(350,145)
    \DashLine(350,150)(450,150){15}
    \DashLine(250,50)(250,-50){15}
    \DashLine(250,350)(250,250){15}
    \CArc(304,96)(70.94,-68.5,-21.5)
    \CArc(304,204)(70.94,21.5,68.5)
    \Text(37,2)[l]{\tiny $\frac{u}{2}$}
    \Text(37,28)[l]{\tiny $0$}
  \end{picture}
=\;\;
    \int [dk] [dl] \;  \frac{1}{\calP_{2}\calP_{5}\calP_{7}\calP_{8}\calP_{18}}
\;\; \equiv \;\;
    (m^2)^{-1-2\eps} \; \left\{ \sum_{i=-2}^{0} d_i^{(53)} \, \eps^i + \calO(\eps) \right\}
\end{align}

\vspace{-4mm} with
\begin{align}
d_{-2}^{(53)}    &=
    \frac{1}{\ubar} \bigg[ \Lib(u) - \frac{\pi^2}{6} \bigg], \no \\
d_{-1}^{(53)}    &=
    \frac{1}{\ubar} \bigg[ 2 \Lic(u) + 4 \Sab(u) - 6\zeta_3 \bigg], \no \\
d_{0}^{(53)}    &=
    \frac{1}{\ubar} \bigg[ 2 \Lid(u) -4 \Sbb(u) + 16 \Sac(u) + 3
    \Lib^2(u) + \frac{2\pi^2}{3} \Lib(u) - \frac{23\pi^4}{60}
    \bigg].
\end{align}

\newpage
\subsubsection{6--topologies}

\begin{align}
&  \SetScale{0.13}
  \begin{picture}(45,23) (24,-8)
    \SetWidth{2.0}
    \SetColor{Black}
    \DashLine(250,55)(350,-140){15}
    \SetWidth{1.25}
    \Line(50,-45)(150,-45)
    \SetWidth{2.0}
    \Line(50,-45)(150,-45)
    \DashLine(250,55)(450,55){15}
    \DashCArc(252.5,-47.5)(102.53,91.4,178.6){15}
    \CArc(252.5,-42.5)(107.53,-178.6,-91.4)\CArc(252.5,-42.5)(97.53,-178.6,-91.4)
    \DashLine(350,55)(250,-140){10}
    \Line(250,-140)(450,-140)\Line(250,-150)(450,-150)
  \end{picture}
=\;\;
    \int [dk] [dl] \;  \frac{1}{\calP_{5}\calP_{8}\calP_{9}\calP_{11}\calP_{13}\calP_{14}}
\;\; \equiv \;\;
    (m^2)^{-2-2\eps} \; \left\{ \sum_{i=-4}^{0} d_i^{(61)} \, \eps^i + \calO(\eps) \right\}
\end{align}

\vspace{-4mm} with
\begin{align}
d_{-4}^{(61)}    &=
    \frac{1}{12\ubar^2}, \no \\
d_{-3}^{(61)}    &=
    - \frac{1}{3\ubar^2} \ln \ubar, \no \\
d_{-2}^{(61)}    &=
    \frac{1}{6\ubar^2} \bigg[ 4\ln^2 \ubar - \frac{5\pi^2}{12} \bigg], \no \\
d_{-1}^{(61)}    &=
    \frac{1}{6\ubar^2} \bigg[ -\frac{89\zeta_3}{2} -
    \frac{16}{3}\ln^3\ubar + \frac{5\pi^2}{3} \ln \ubar \bigg], \no \\
d_{0}^{(61)}    &=
    \frac{1}{6\ubar^2} \bigg[  6\mathcal{C}_0 +48\ln \ubar \,
    \Lic(u) + 24 \Lib^2(u) +\frac{16}{3} \ln^4 \ubar -
    \frac{10\pi^2}{3} \ln^2 \ubar + 130 \zeta_3 \ln \ubar \bigg],
\end{align}
where we have introduced a constant $\mathcal{C}_0$, which arises in the
calculation of the boundary condition for the 6-topology MI. We could
not find an analytical expression for $\mathcal{C}_0$\footnote{We thank
  R.~Bonciani and A.~Ferroglia for   pointing out an error in an earlier
  version of this thesis.}. With the help of our implementation of the
method of sector decomposition, we obtain $\mathcal{C}_0=-60.2\pm0.1$.

\newpage
\section{Heavy-to-light form factors}
\label{app:MIsNR}

The MIs in our calculation from Chapter~\ref{ch:NRModel} are shown
in Figure~\ref{fig:NRMIs}. The corresponding propagators $\calP_i$
can be found in (\ref{eq:NRProps}) which involve two external
momenta $w$ and $w'$ with $w^2={w'}^2=1$ and $\gamma\equiv w\cdot
w'=\calO(M/m)$. We extract the leading power of the MIs in $m/M$ and
write $s\equiv M/(4\gamma m)=\calO(1)$.

\subsubsection{1--topologies}

\vspace{-2mm}
\begin{align}
\parbox[c]{1.5cm}{\psfig{file=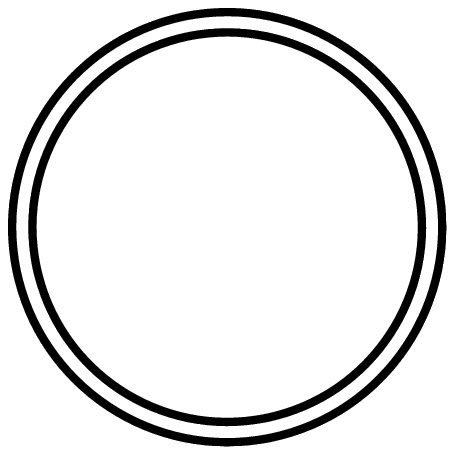,width =
1cm}} &=\;\;
    \int \, [dk] \; \frac{1}{\calP_{1}}  \hspace*{8.5cm}\no \\
&=\;\;
    - (M^2)´^{1-\eps} \; \Gamma(1-\eps) \Gamma(-1+\eps)
\end{align}
\begin{align}
\parbox[c]{1.5cm}{\psfig{file=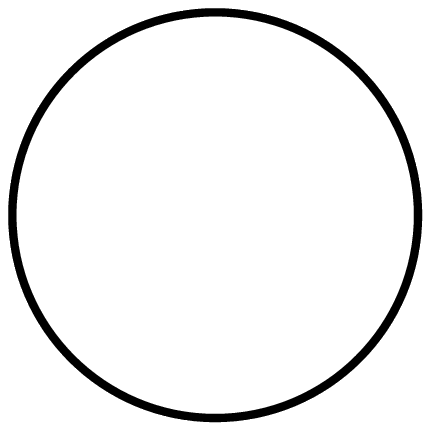,width =
1cm}} &=\;\;
    \int \, [dk] \; \frac{1}{\calP_{3}} \hspace*{8.5cm}\no \\
&=\;\;
    - (m^2)´^{1-\eps} \; \Gamma(1-\eps) \Gamma(-1+\eps)
\end{align}

\vspace{2mm}
\subsubsection{2--topologies}

\vspace{-2mm}
\begin{align}
\parbox[c]{2.5cm}{\psfig{file=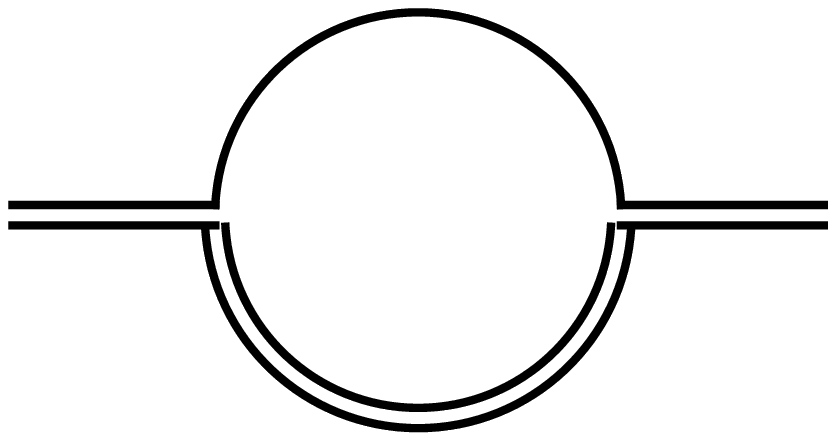,width =
2cm}} &= \;\;
    \int \, [dk] \; \frac{1}{\calP_{1}\calP_{3}} \hspace*{9cm}\no \\
&\simeq\;\;
    (M^2)´^{-\eps} \left\{ \frac{1}{\eps} - \frac{1}{s-1} \ln s +2 + \left[4 + \frac{\pi^2}{6}
    \right.    \right.
    \\
& \hspace{25mm}  \left. \left.+ \frac{1}{s-1} \left( \Lib(1-s)
-\frac12\ln^2 s-2 \ln s \right)\right] \eps + \calO(\eps^2) \no
    \right\}
\end{align}
\begin{align}
\parbox[c]{2.5cm}{\psfig{file=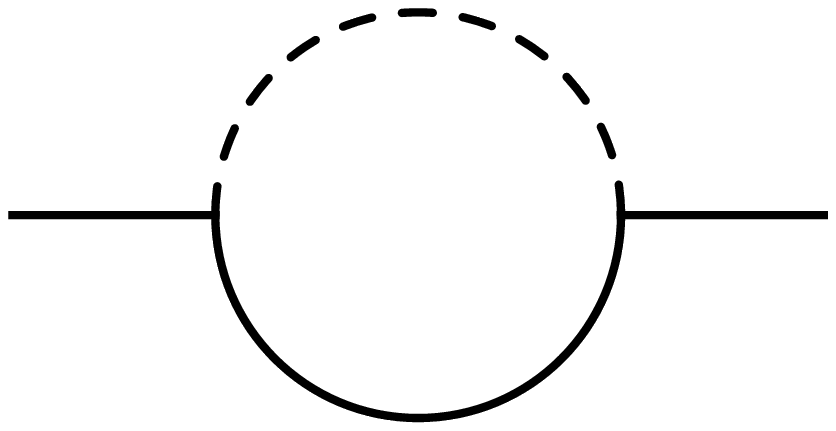,width =
2cm}} &= \;\;
    \int \, [dk] \; \frac{1}{\calP_{3}\calP_{6}}  \hspace*{9cm}\no \\
&\simeq\;\;
    (4\gamma m^2)´^{-\eps} \left\{ \frac{1}{\eps} +2 + 4 \eps+ \calO(\eps^2)
    \right\}
\end{align}

\vspace{2mm}
\subsubsection{3--topologies}

\vspace{3mm}
\begin{align}
\parbox[c]{2.5cm}{\vspace{-5mm}\psfig{file=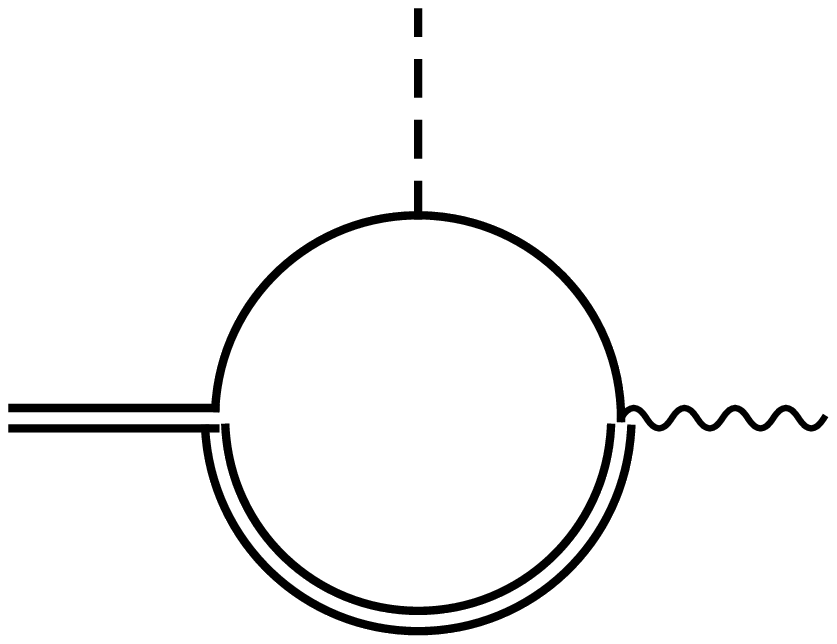,width =
2cm}} &= \;\;
    \int \, [dk] \; \frac{1}{\calP_{1}\calP_{3}\calP_{5}}  \\
&\simeq\;\;
    -\,\frac{1}{8\gamma m M} \; \bigg\{ 2 \ln^2(4\gamma) -2\Lib(1-s) -\ln^2 s + \frac{8\pi^2}{3} + \calO(\eps)
    \bigg\} \no \hspace{0.8cm}
\end{align}
\vspace{5mm}
\begin{align}
\parbox[c]{2.5cm}{\vspace{-5mm}\psfig{file=MInt_32.eps,width =
2cm}} &= \;\;
    \int \, [dk] \; \frac{1}{\calP_{1}\calP_{3}\calP_{6}}  \hspace*{8.6cm}\no \\
&\simeq\;\;
    -\,\frac{1}{8\gamma m M} \; \bigg\{ \ln^2(4\gamma) -2\Lib(1-s) -\ln^2 s + \pi^2 + \calO(\eps)
    \bigg\}\label{eq:MInt_32}
\end{align}
\vspace{5mm}
\begin{align}
\parbox[c]{2.5cm}{\vspace{-5mm}\psfig{file=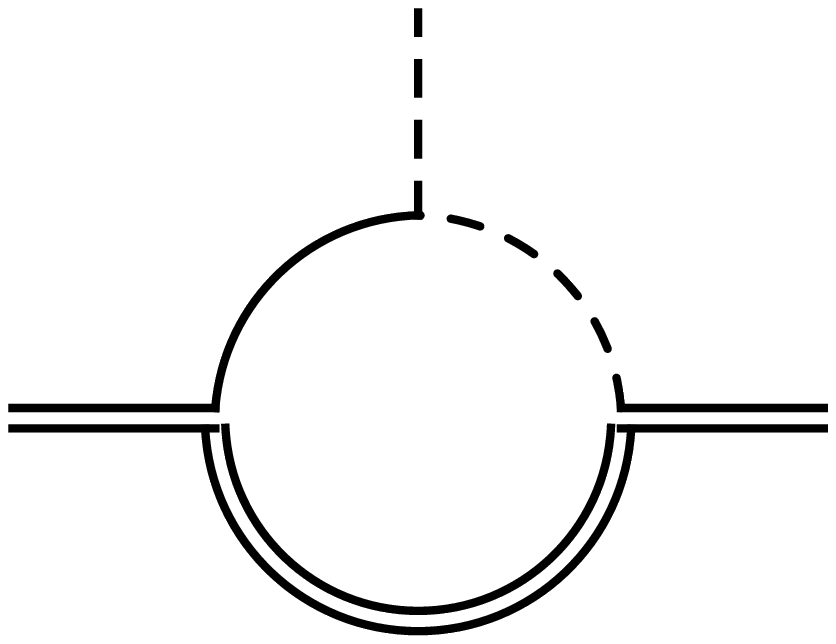,width =
2cm}} &= \;\;
    \int \, [dk] \; \frac{1}{\calP_{1}\calP_{3}\calP_{7}}  \hspace*{8.6cm}\no \\
&\simeq\;\;
    -\,\frac{1}{4\gamma m M} \; \bigg\{ 4\ln 2\ln \gamma + 2 \ln2 \ln
    s +2\Lib(1-2s) -2\Lib(1-s) \no \\
& \hspace{30mm} +7 \ln^2 2 + \frac{\pi^2}{3}+ \calO(\eps)
    \bigg\}
\end{align}
\vspace{0mm}
\begin{align}
\parbox[c]{2.5cm}{\vspace{-5mm}\psfig{file=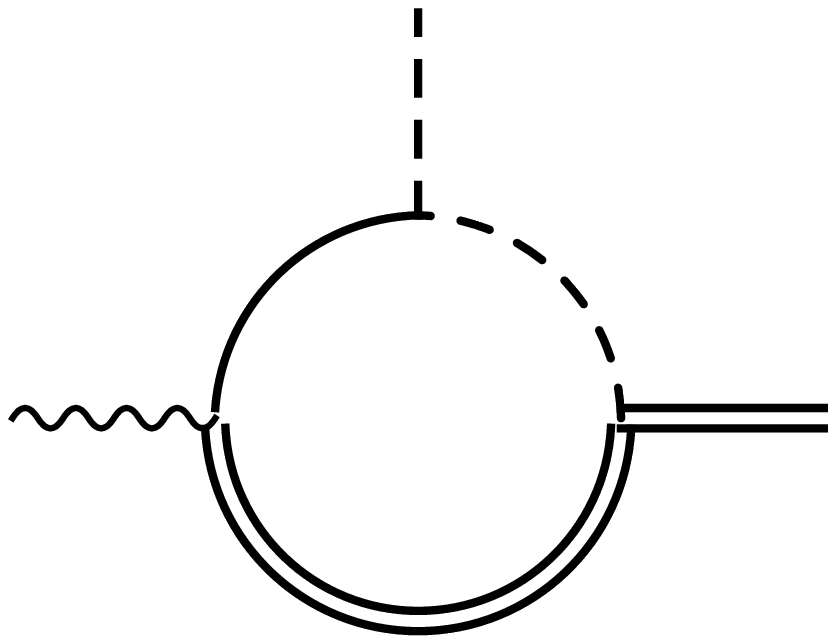,width =
2cm}} &= \;\;
    \int \, [dk] \; \frac{1}{\calP_{1}\calP_{5}\calP_{7}}  \\
&\simeq\;\;
    -\,\frac{1}{4\gamma m M} \; \bigg\{ 2\ln^2(2\gamma) -2\Lib(1-2s) -\ln^2(2s) + \frac{7\pi^2}{3}  + \calO(\eps)
    \bigg\} \no \hspace{0.3cm}
\end{align}
\vspace{5mm}
\begin{align}
\parbox[c]{2.5cm}{\vspace{-5mm}\psfig{file=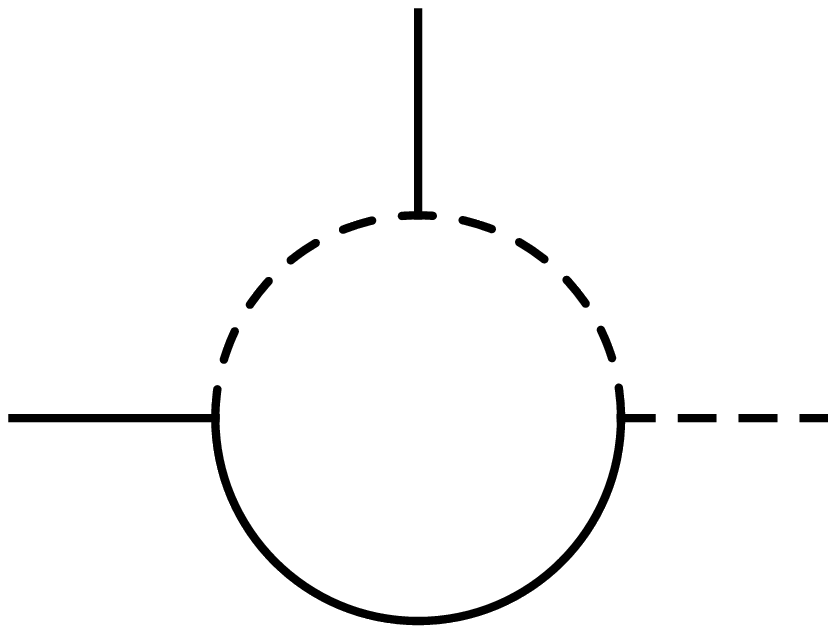,width =
2cm}} &= \;\;
    \int \, [dk] \; \frac{1}{\calP_{3}\calP_{6}\calP_{7}}  \hspace*{8.6cm}\no \\
&\simeq\;\;
    -\,\frac{1}{2\gamma m^2} \; \bigg\{ \ln2 \ln\gamma + \frac32
    \ln^2 2 + \frac{\pi^2}{6}  + \calO(\eps)
    \bigg\}
\end{align}
\vspace{5mm}
\begin{align}
\parbox[c]{2.5cm}{\vspace{-5mm}\psfig{file=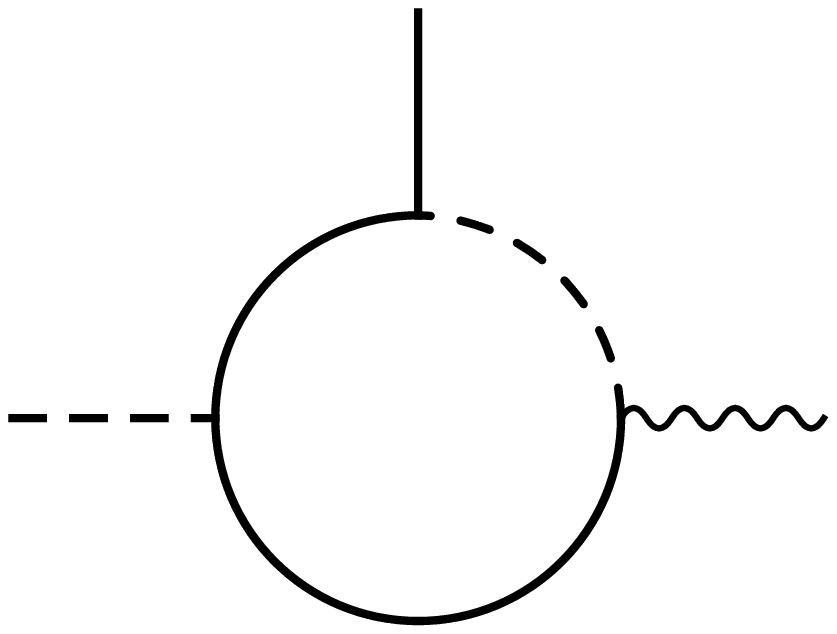,width =
2cm}} &= \;\;
    \int \, [dk] \; \frac{1}{\calP_{3}\calP_{5}\calP_{6}}  \hspace*{8.6cm}\no \\
&\simeq\;\;
    -\,\frac{1}{8\gamma m^2} \; \bigg\{\ln^2(4\gamma) + \frac{5\pi^2}{3}  + \calO(\eps)
    \bigg\}
\end{align}
\vspace{5mm}
\begin{align}
\parbox[c]{2.5cm}{\vspace{-5mm}\psfig{file=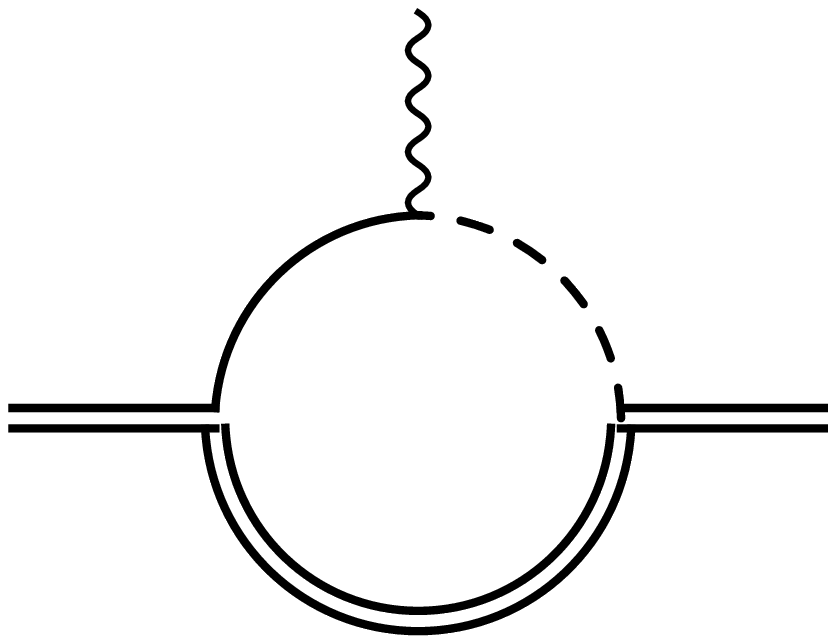,width =
2cm}} &= \;\;
    \int \, [dk] \; \frac{1}{\calP_{2}\calP_{5}\calP_{6}}  \hspace*{8.6cm}\no \\
&\simeq\;\;
    \frac{1}{2\gamma m M} \; \bigg\{-2\ln(2\gamma) + \frac{1}{2s-1} \ln(2s)  -2 +  \calO(\eps)
    \bigg\}
\end{align}
\vspace{5mm}
\begin{align}
\parbox[c]{2.5cm}{\vspace{-5mm}\psfig{file=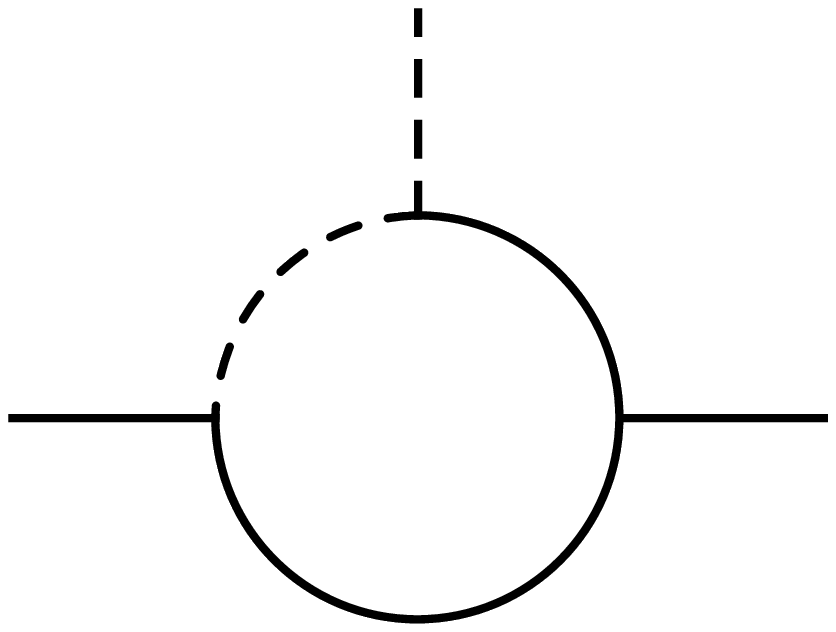,width =
2cm}} &= \;\;
    \int \, [dk] \; \frac{1}{\calP_{4}\calP_{5}\calP_{7}}  \hspace*{8.6cm}\no \\
&\simeq\;\;
    -\, \frac{1}{2\gamma m^2} \; \bigg\{\ln2 \ln\gamma + \frac12 \ln^2 2 + \frac{\pi^2}{6} +  \calO(\eps)
    \bigg\}
\end{align}
\vspace{5mm}
\begin{align}
\parbox[c]{2.5cm}{\vspace{-5mm}\psfig{file=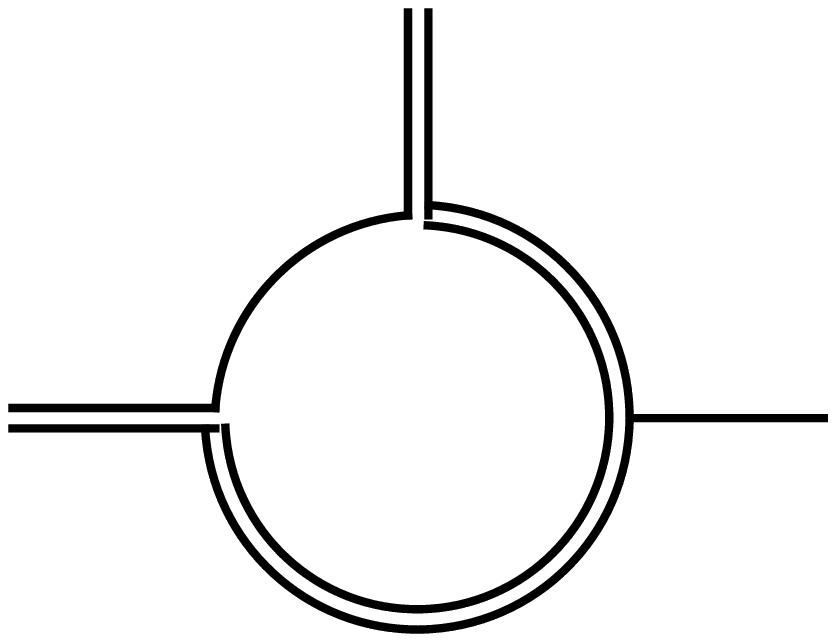,width =
2cm}} &= \;\;
    \int \, [dk] \; \frac{1}{\calP_{8}\calP_{9}\calP_{11}}   \\
&\simeq\;\;
    \frac{1}{4\gamma m M} \; \bigg\{2\Lib(1-2s) - 2\Lib(1-s) +2 \ln2 \ln s + \ln^2 2  +  \calO(\eps)
    \bigg\} \no \hspace{0cm}
\end{align}
\vspace{5mm}
\begin{align}
\parbox[c]{2.5cm}{\vspace{-5mm}\psfig{file=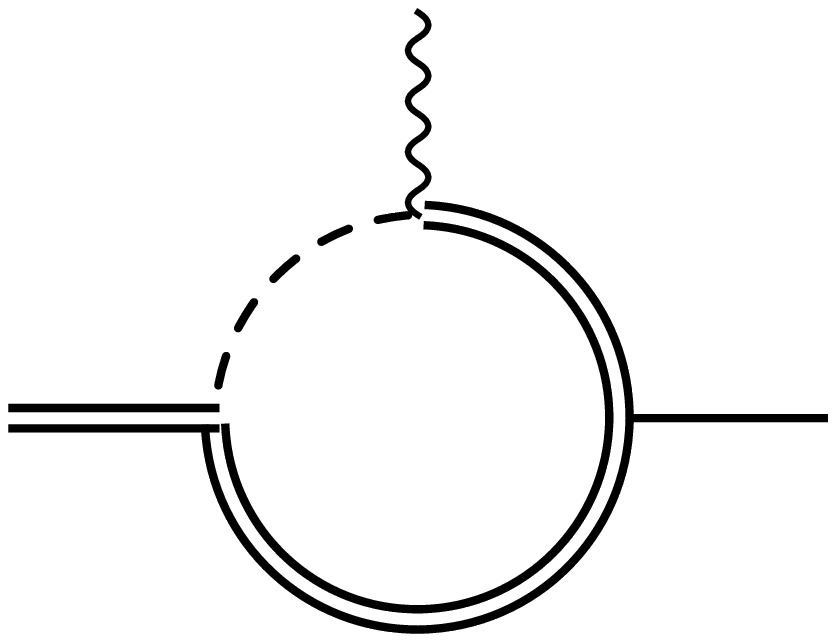,width =
2cm}} &= \;\;
    \int \, [dk] \; \frac{1}{\calP_{8}\calP_{9}\calP_{12}}  \hspace*{8.4cm}\no \\
&\simeq\;\;
    -\, \frac{1}{4\gamma m M} \; \bigg\{2\Lib(1-2s) + \ln^2(2s) + \frac{\pi^2}{3}+  \calO(\eps)
    \bigg\}
\end{align}
\vspace{5mm}
\begin{align}
\parbox[c]{2.5cm}{\vspace{-5mm}\psfig{file=MInt_313.eps,width =
2cm}} &= \;\;
    \int \, [dk] \; \frac{1}{\calP_{5}\calP_{6}\calP_{7}}  \hspace*{8.6cm}\no \\
&\simeq\;\;
    -\,\frac{1}{2\gamma m^2} \; \bigg\{\frac12 \ln^2(2\gamma) + \frac{2\pi^2}{3}+  \calO(\eps)
    \bigg\}\label{eq:MInt_313}
\end{align}
\vspace{5mm}
\begin{align}
\parbox[c]{2.5cm}{\vspace{-5mm}\psfig{file=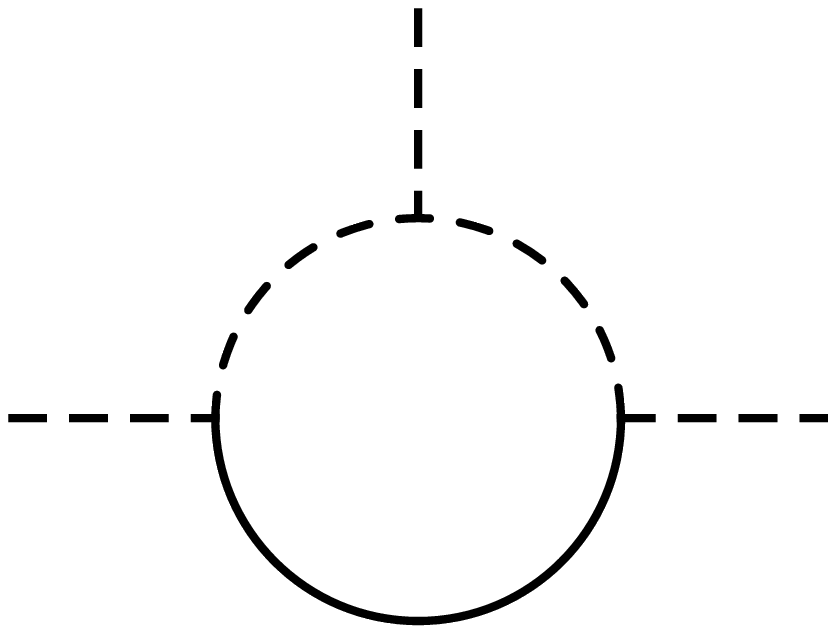,width =
2cm}} &= \;\;
    \int \, [dk] \; \frac{1}{\calP_{11}\calP_{12}\calP_{15}}  \hspace*{8.1cm}\no \\
&=\;\;
    \frac{1}{2m^2} \; \bigg\{2\ln 2 - i \pi +  \calO(\eps)
    \bigg\}
\end{align}

\vspace{2mm}
\subsubsection{4--topologies}

\vspace{-2mm}
\begin{align}
\parbox[c]{2.5cm}{\psfig{file=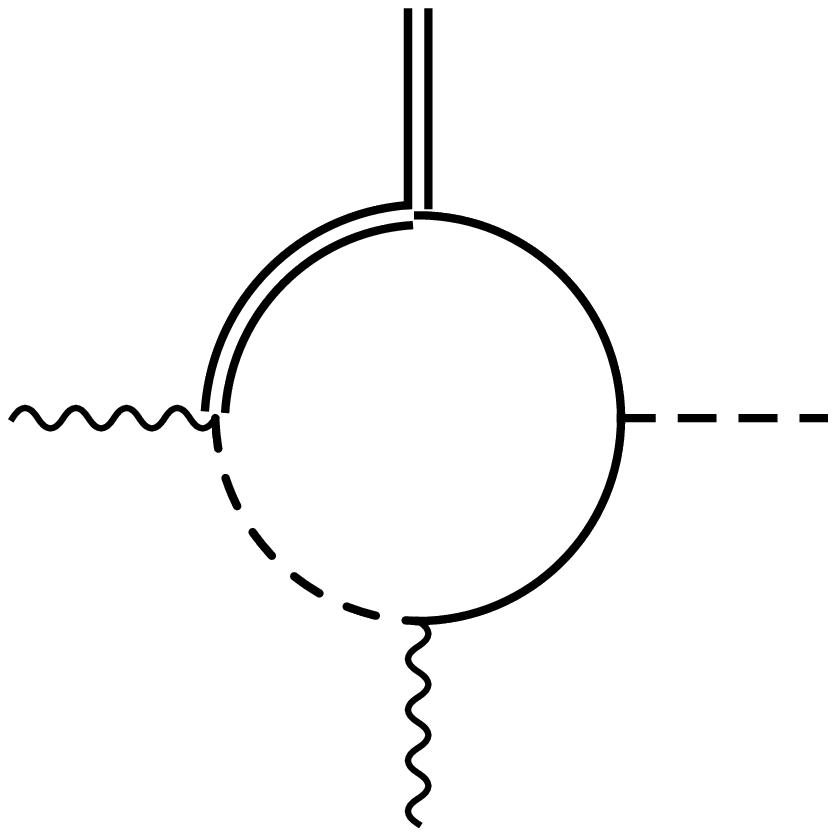,width =
2cm}} &= \;\;
    \int \, [dk] \; \frac{1}{\calP_{1}\calP_{3}\calP_{5}\calP_{6}}  \hspace*{8.1cm}\no \\
&\simeq\;\;
    \frac{1}{4\gamma m^3 M} \; (m^2)^{-\eps} \bigg\{\frac{1}{2\eps} -1 +  \calO(\eps)
    \bigg\}
\end{align}
\begin{align}
\parbox[c]{2.5cm}{\psfig{file=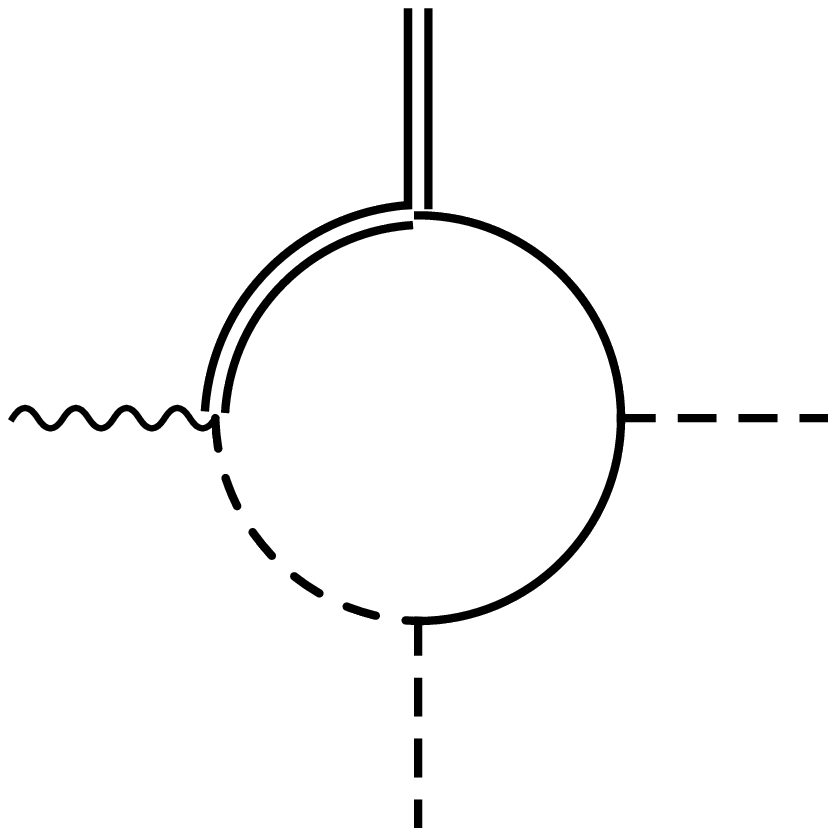,width =
2cm}} &= \;\;
    \int \, [dk] \; \frac{1}{\calP_{1}\calP_{3}\calP_{5}\calP_{7}}  \hspace*{8.1cm}\no \\
&\simeq\;\;
    \frac{1}{4\gamma m^3 M} \; (m^2)^{-\eps} \bigg\{\frac{1}{\eps} +2\ln 2-2 +  \calO(\eps)
    \bigg\}
\end{align}
\begin{align}
\parbox[c]{2.5cm}{\psfig{file=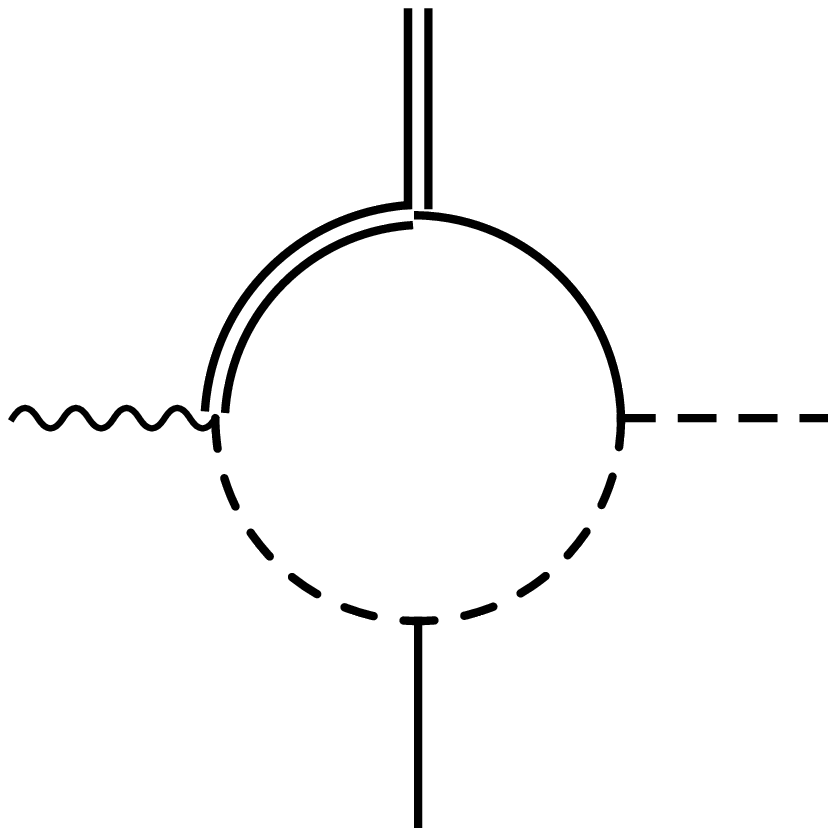,width =
2cm}} &= \;\;
    \int \, [dk] \; \frac{1}{\calP_{1}\calP_{3}\calP_{6}\calP_{7}}  \hspace*{8.1cm}\no \\
&\simeq\;\;
    \frac{1}{4\gamma^2 m^3 M} \;  \bigg\{\ln (4\gamma)+1 +  \calO(\eps)
    \bigg\}
\end{align}
\begin{align}
\parbox[c]{2.5cm}{\psfig{file=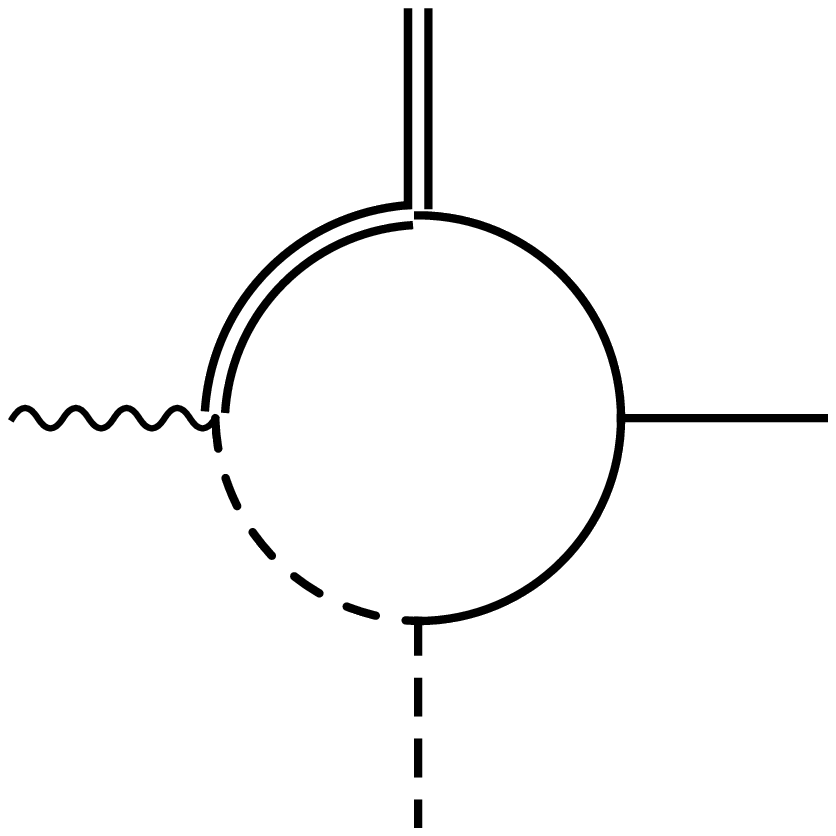,width =
2cm}} &= \;\;
    \int \, [dk] \; \frac{1}{\calP_{2}\calP_{4}\calP_{5}\calP_{7}}  \\
&\simeq\;\;
    -\,\frac{1}{8\gamma^2 m^3 M} \;  (m^2)^{-\eps} \bigg\{\frac{\ln (2\gamma)}{\eps} -\frac12 \ln^2(4\gamma) + \ln^2 2 + \frac{\pi^2}{6}+ \calO(\eps)
    \bigg\}\no \hspace{2mm}
\end{align}
\begin{align}
\parbox[c]{2.5cm}{\psfig{file=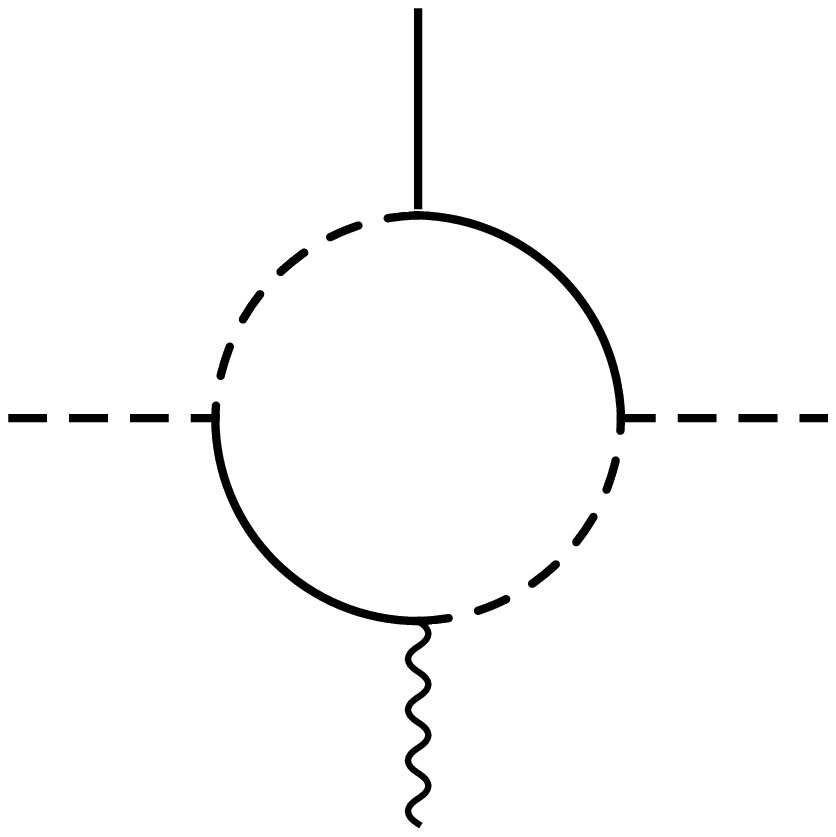,width =
2cm}} &= \;\;
    \int \, [dk] \; \frac{1}{\calP_{4}\calP_{5}\calP_{6}\calP_{7}}  \hspace*{8.1cm}\no \\
&\simeq\;\;
    -\,\frac{1}{4\gamma^2 m^4} \; (m^2)^{-\eps} \bigg\{\frac{\ln (2\gamma)}{\eps} - \ln^2\gamma + \ln^2 2 - \frac{\pi^2}{6}+ \calO(\eps)
    \bigg\}
\end{align}

\vspace{0mm}
\subsubsection{5--topologies}

\vspace{0mm}
\begin{align}
\parbox[c]{2.5cm}{\vspace{-3mm}\psfig{file=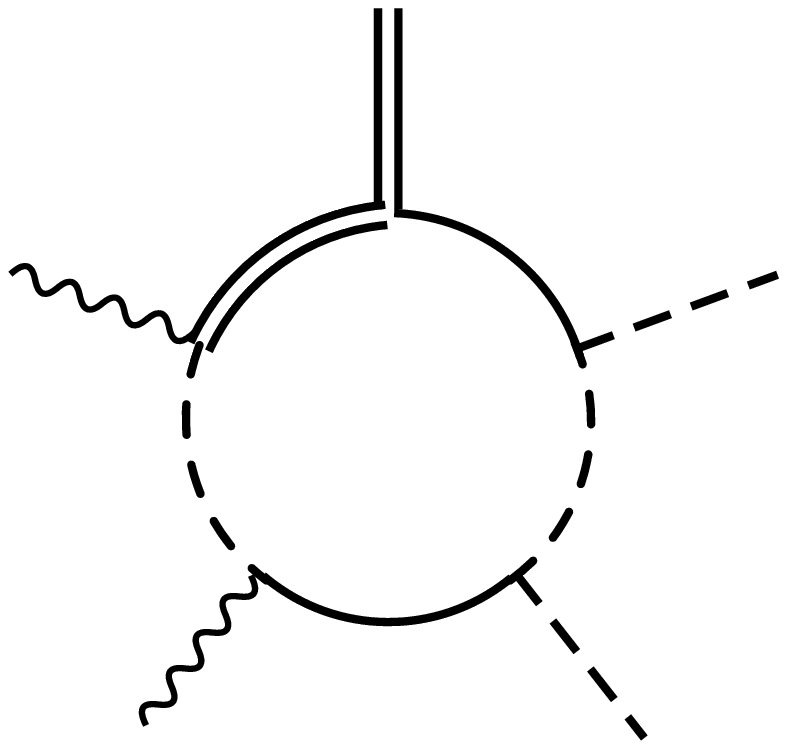,width =
2cm}} &= \;\;
    \int \, [dk] \; \frac{1}{\calP_{1}\calP_{3}\calP_{5}\calP_{6}\calP_{7}}  \hspace*{7.6cm}\no \\
&\simeq\;\;
    -\,\frac{1}{8\gamma^2 m^5 M} \; (m^2)^{-\eps} \bigg\{\frac{3}{2\eps} + \ln (16\gamma) -2+ \calO(\eps)
    \bigg\}
\end{align}
\vspace{3mm}
\begin{align}
\parbox[c]{2.5cm}{\vspace{-3mm}\psfig{file=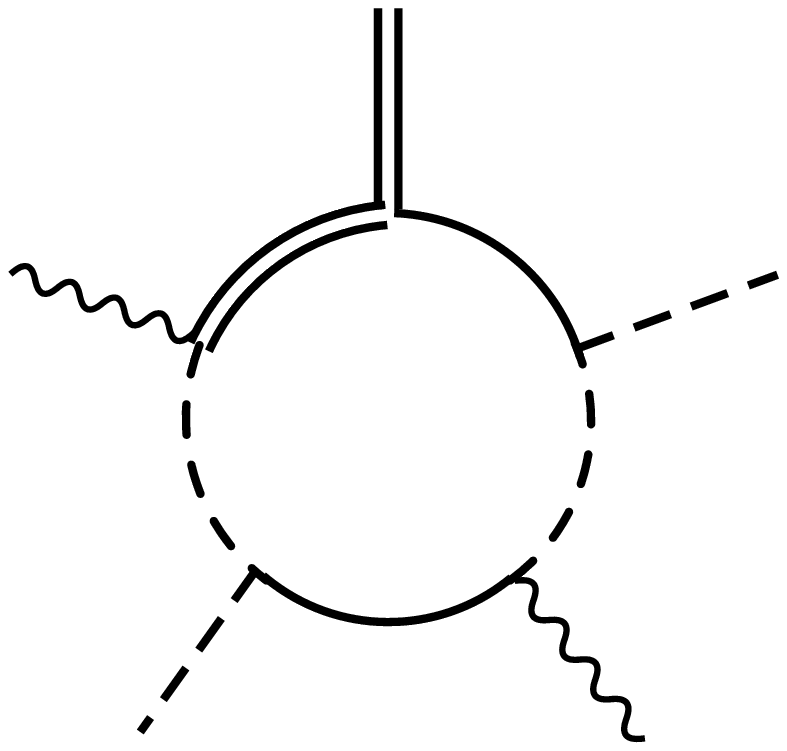,width =
2cm}} &= \;\;
    \int \, [dk] \; \frac{1}{\calP_{2}\calP_{4}\calP_{5}\calP_{6}\calP_{7}}\\
&\simeq\;\;
    \frac{1}{16\gamma^3 m^5 M} \; (m^2)^{-\eps}
    \bigg\{\frac{3\ln(2\gamma)}{\eps} -\frac52 \ln^2 \gamma
    +2(1-\ln 2)\ln \gamma \no \\
& \hspace{45mm}
    +\ln^2 2+4\ln 2+2-\frac{\pi^2}{6} + \calO(\eps)
    \bigg\} \hspace{1.4cm} \no
\end{align}

\chapter{Form factors in NLO}
\label{app:NRFormFactor}
 
We summarize our NLO results for the transition form factors
$F_+(q^2)$, $F_-(q^2)$ and $F_T(q^2)$ which can be defined in
analogy to (\ref{eq:defFF}) for the $B_c\to\eta_c$ transition. All
results are expressed in terms of pole masses $M\equiv m_b$ and
$m\equiv m_c$ and correspond to the leading power in the HQE in
$m/M$. The relation between the momentum tramsfer $q^2$ and the
relativistic boost $\gamma=\calO(M/m)$ can be found in
(\ref{eq:boost}). We further introduce $s\equiv M/(4\gamma
m)=\calO(1)$ and refer to (\ref{eq:NRFFLO}) for the LO expressions.

\begin{align}
\frac{F_+^\text{NLO}}{F_+^\text{LO}} &=1 + \frac{\as}{4\pi} \bigg\{
        \left(\frac{11}{3}N_c-\frac23 n_f \right) \ln \frac{\mu^2}{2\gamma m^2} - \frac{10}{9} n_f + \calS_+ \no\\
& + C_F \bigg[
        \frac{3s-2}{2(1+2s)} \,\ln^2\gamma +\frac{3(9s+4-4\ln2)}{2(1+2s)}
        \,\ln \gamma +2 \Lib(1-s) + \ln^2 s \no \\
    & \hspace{1.2cm}
        + \frac{s(24s^2-20s+5)}{(1+2s)(1-2s)^2} \,\ln(2s) -\frac{3s+7}{1+2s} \,\ln^2
        2 + \frac{21s+10}{1+2s}\,\ln2 \no \\
    & \hspace{1.2cm}
        +\frac{(17s-8)\pi^2}{6(1+2s)} + \frac{170s^2-171s+52}{18(1-2s)(1+2s)} \bigg] \no \\
& + \left(C_F - \frac{N_c}{2} \right) \bigg[
        \frac{s}{1+2s} \, \ln^2 \gamma - \frac{2(s+(s-2)\ln2)}{1+2s} \, \ln
        \gamma - \frac{2s^2}{1+2s}\, \ln^2 \left(\frac{s}{2}\right) \no \\
    & \hspace{2.7cm}- \frac{2(4s^2+s-2)}{1+2s} \, \Lib(1-2s) + \frac{2(2s^2+s-2)}{1+2s} \, \Lib(1-s)\no \\
    & \hspace{2.7cm} + \frac{2(s+(2+3s)(1-2s)^2\ln 2)}{(1-2s)(1+2s)} \ln(2s) + \frac{2(5s^2-2s+2)}{1+2s} \ln^2 2\no \\
    & \hspace{2.7cm} + \frac{2(5s-2)}{1+2s} \, \ln 2 - \frac{(2s^2-s-2)\pi^2}{3(1+2s)}
    - \frac{2(85+134s)}{9(1+2s)} \bigg] \bigg\},
\end{align}
where $\calS_+$ refers to the contribution from the flavour singlet
diagrams which reads
\begin{align}
\calS_+ &= \frac{s}{1+2s} \, \ln \gamma - \frac{2(s-1)}{1+2s} \, \ln
2 + \frac{(s-1)\pi^2}{3(1+2s)}.
\end{align}
\begin{align}
\frac{F_-^\text{NLO}}{F_-^\text{LO}} &=1 + \frac{\as}{4\pi} \bigg\{
        \left(\frac{11}{3}N_c-\frac23 n_f \right)  \ln \frac{\mu^2}{2\gamma m^2} - \frac{10}{9} n_f + \calS_- \no\\
& + C_F \bigg[
        \frac34 \,\ln^2\gamma +\frac{27}{4} \,\ln \gamma +2
        \Lib(1-s) + \ln^2 s - \frac32 \,\ln^2 2 \no \\
    & \hspace{1.2cm}
        + \frac{24s^4-80s^3+125s^2-77s+15}{2(1-s)^2(1-2s)^2} \, \ln
        s + \frac{3(18s^2-20s+5)}{(1-2s)^2}\, \ln 2 \no \\
    & \hspace{1.2cm}+
        \frac{17\pi^2}{12} - \frac{170s^2-57s+13}{36(1-s)(1-2s)}\bigg] \no \\
& + \left(C_F - \frac{N_c}{2} \right) \bigg[
        \frac12 \, \ln^2 \gamma - (1+\ln2) \ln \gamma -(1+4s)
        \Lib(1-2s) \no \\
    & \hspace{2.7cm}+(1+2s) \Lib(1-s) -s\ln^2(4s) -(3-2s)\ln^2 2 \no \\
    & \hspace{2.7cm}- \frac{s+1+(1-s)(1-2s)\ln2}{(1-s)(1-2s)} \, \ln
    s- \frac{2(5s-1)}{1-2s} \, \ln2\no \\
    & \hspace{2.7cm}+(1-2s)\frac{\pi^2}{6}-\frac{134}{9} \bigg] \bigg\}
\end{align}
with the flavour singlet contribution given by
\begin{align}
\calS_- &= \frac12 \ln \gamma - \ln 2+ \frac{\pi^2}{6}.
\label{eq:NRSingMinus}
\end{align}
\begin{align}
\frac{F_T^\text{NLO}}{F_T^\text{LO}} &=1 + \frac{\as}{4\pi} \bigg\{
        \left(\frac{11}{3}N_c-\frac23 n_f \right)  \ln \frac{\mu^2}{2\gamma m^2} - \frac{10}{9} n_f + \calS_T \no\\
& + C_F \bigg[ - \ln \frac{\mu^2}{M^2} + \frac34 \,\ln^2\gamma
+\frac{27}{4} \,\ln \gamma +2
        \Lib(1-s) + \ln^2 s - \frac32 \,\ln^2 2 +
        \frac{17\pi^2}{12}\no \\
    & \hspace{1.2cm} - \frac{24s^3-28s^2+13s-2}{2(1-s)(1-2s)^2} \, \ln
        s + \frac{54s^2-58s+15}{(1-2s)^2}\, \ln 2  + \frac{170s-103}{36(1-2s)}\bigg] \no \\
& + \left(C_F - \frac{N_c}{2} \right) \bigg[ \frac12 \, \ln^2 \gamma
-
    (1+\ln2) \ln \gamma -(1-4s) \Lib(1-2s) \no \\
    & \hspace{2.7cm}+(1-2s) \Lib(1-s) +s\ln^2(4s) -(3+2s)\ln^2 2 - \frac{2s}{1-2s} \, \ln2\no \\
    & \hspace{2.7cm}+ \frac{s-1-(1-s)(1-2s)\ln2}{(1-s)(1-2s)} \, \ln
    s  +(1+2s)\frac{\pi^2}{6}-\frac{134}{9}\bigg] \bigg\}
\end{align}
and the flavour singlet contribution is found to be the same as in
(\ref{eq:NRSingMinus}), i.e $\calS_T=\calS_-$. Notice that the
tensor form factor $F_T(q^2)$ has a residual scale dependence of
$\calO(\as^2)$ in NLO due to the fact that the tensor current is not
conserved.

\end{appendix}

\pagestyle{fancyplain}
\rhead[\uppercase{Bibliography}]{\fancyplain{} \thepage}
\lhead[\thepage]{}

\newpage
\pagestyle{empty}

\addcontentsline{toc}{chapter}{Acknowledgements}
\vspace*{2.2cm}
 
{\Huge\bf Acknowledgements}

\vspace{1.4cm}

In the end I would like to thank everybody who helped this thesis
coming into being.

\newabs
First of all I would like to thank my advisor, Prof.~Gerhard
Buchalla, for being an outstanding teacher. I am grateful to his
continuous help and guidance throughout my time in Munich. It was
always a great pleasure to work with him. I certainly remember the
many fruitful discussions we had and I highly value his permanent
encouragement during my work.

\newabs
Next I would like to thank Dr.~Thorsten Feldmann for his exclusive
support. The collaboration with him has been very pleasant and
certainly very fruitful. I could in particular learn a lot about
Soft-Collinear Effective Theory from him. I also would like to thank
Thorsten for his assistance in arranging my stay at CERN.

\newabs
I am grateful to Volker Pilipp and Matth\"aus Bartsch for numerous
stimulating discussions. I also thank Gerhard, Thorsten, Volker and
Matth\"aus for proofreading parts of the manuscript. I am indebted
to Tobias Huber for discussions related to the calculation of
loop-integrals.

\newabs
Finally, I would like to thank my parents for their unconditional
support. Last, but definitely not least, I thank Conchita for her
patience and permanent encouragement.

\newabs
This work was supported by the German-Israeli Foundation for
Scientific Research and Development under Grant G-698-22.7/2001.

\end{document}